\documentclass[letterpaper,aps,rmp,twocolumn,notitlepage,nofootinbib,10pt]{revtex4-1}

\usepackage{amssymb,amsmath,latexsym}
\usepackage{amsthm}
\usepackage[usenames,dvipsnames]{xcolor}
\usepackage{nicefrac}
\usepackage{enumitem}
\usepackage{graphicx}
\usepackage{mathrsfs}
\usepackage[sans]{dsfont} 
\usepackage{overpic}
\usepackage{subfigure}

\usepackage[utf8]{inputenc}
\usepackage[T1]{fontenc}

\usepackage[ colorlinks = true, 
             linkcolor = blue,
             urlcolor  = blue,
             citecolor = red,
             anchorcolor = green,
]{hyperref} 

\long\def\ca#1\cb{} 

\newcommand{\1}{\mathds{1}}  
\newcommand{\id}{\mathcal{I}}  
\newcommand{\ad}{^{\dagger}}  
\newcommand{\ot}{\otimes} 

\newcommand{\abs}[2][\Biggerthanbig]
{\ifthenelse{\isundefined{#1}}{\left| #2 \right|}{#1| #2 #1|}}
\newcommand{\norm}[2][\Biggerthanbig]
{\ifthenelse{\isundefined{#1}}{\left\| #2 \right\|}{#1\| #2 #1\|}}

\newcommand{\cD}{\mathcal{D}}  
\newcommand{\cV}{\mathcal{V}}  
\newcommand{\cH}{\mathcal{H}}  
\newcommand{\cP}{\mathcal{P}}  
\newcommand{\cM}{\mathcal{M}}  
\newcommand{\cR}{\mathcal{R}}  
\newcommand{\cS}{\mathcal{S}}  
\newcommand{\cE}{\mathcal{E}}  
\newcommand{\cF}{\mathcal{F}}  
\newcommand{\cX}{\mathcal{X}}  
\newcommand{\cZ}{\mathcal{Z}}  

\newcommand{\bW}{\mathbb{W}}   
\newcommand{\bX}{\mathbb{X}}   
\newcommand{\bY}{\mathbb{Y}}   
\newcommand{\bZ}{\mathbb{Z}}   

\newcommand{\al}{\alpha }
\newcommand{\bt}{\beta }

\newcommand{\sg}{\sigma }

\newcommand{\om}{\omega }

\newcommand{\ddd}{d }

\newcommand{\bbR}{\mathbb{R}}  

\newcommand{\ket}[1]{|#1\rangle}  
\newcommand{\bra}[1]{\langle #1|}
\newcommand{\proj}[1]{|#1\rangle\!\langle #1|}
\newcommand{\avg}[1]{\langle #1\rangle }
\newcommand{\colo}{\,\hbox{:}\,}              
\newcommand{\dya}[1]{\ket{#1}\!\bra{#1}}
\newcommand{\dyad}[2]{\ket{#1}\!\bra{#2}}        
\newcommand{\ip}[2]{\langle #1|#2\rangle}      
\newcommand{\expct}[2]{\langle #1|#2|#1\rangle} 

\DeclareMathOperator{\tr}{tr}  

\newcommand{\bin}{\text{bin}}  
\newcommand{\eps}{\varepsilon}  

\newcommand{\guess}{\text{guess}}
\DeclareMathOperator\arccosh{arccosh}

\newtheorem{example}{Example}
\newtheorem*{example*}{Example}

\newcommand*{\sD}{\textsf{D}}
\newcommand*{\sE}{\textsf{E}}
\newcommand*{\sN}{\textsf{N}}

\begin{document}

\title{Entropic Uncertainty Relations and their Applications}

\author{Patrick J.~Coles}
\email{pcoles@uwaterloo.ca}
\affiliation{Institute for Quantum Computing and Department of Physics and Astronomy,
University of Waterloo, 
N2L3G1 Waterloo, Ontario,
Canada}

\author{Mario Berta}
\email{berta@caltech.edu}
\affiliation{Institute for Quantum Information and Matter,
California Institute of Technology, Pasadena, CA 91125, USA}

\author{Marco Tomamichel}
\email{marco.tomamichel@sydney.edu.au}
\affiliation{School of Physics,
The University of Sydney,
Sydney, NSW 2006,
Australia}

\author{Stephanie Wehner}
\email{s.d.c.wehner@tudelft.nl}
\affiliation{QuTech,
Delft University of Technology,
2628 CJ Delft,
Netherlands}


\begin{abstract}
Heisenberg's uncertainty principle forms a fundamental element of quantum mechanics. Uncertainty relations in terms of entropies were initially proposed to deal with conceptual shortcomings in the original formulation of the uncertainty principle and, hence, play an important role in quantum foundations. More recently, entropic uncertainty relations have emerged as the central ingredient in the security analysis of almost all quantum cryptographic protocols, such as quantum key distribution and two-party quantum cryptography. This review surveys entropic uncertainty relations that capture Heisenberg's idea that the results of incompatible measurements are impossible to predict, covering both finite- and infinite-dimensional measurements. These ideas are then extended to incorporate quantum correlations between the observed object and its environment, allowing for a variety of recent, more general formulations of the uncertainty principle. Finally, various applications are discussed, ranging from entanglement witnessing to wave-particle duality to quantum cryptography.
\end{abstract}

\maketitle  

\tableofcontents

\bigskip


\section{Introduction}\label{sec:introduction}

Quantum mechanics has revolutionized our understanding of the world. Relative to classical mechanics, the most dramatic change in our understanding is that the quantum world\,---\,our world\,---\,is inherently unpredictable. 

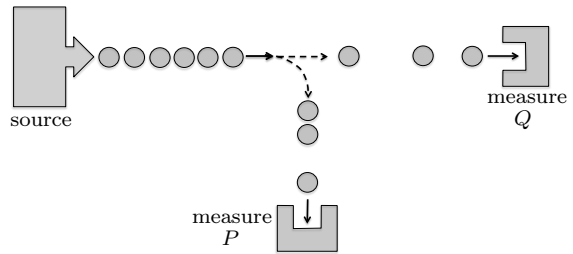
\begin{figure}[tbp]
\begin{center}
\vspace{0.4cm}
\begin{overpic}[width=7.2cm]{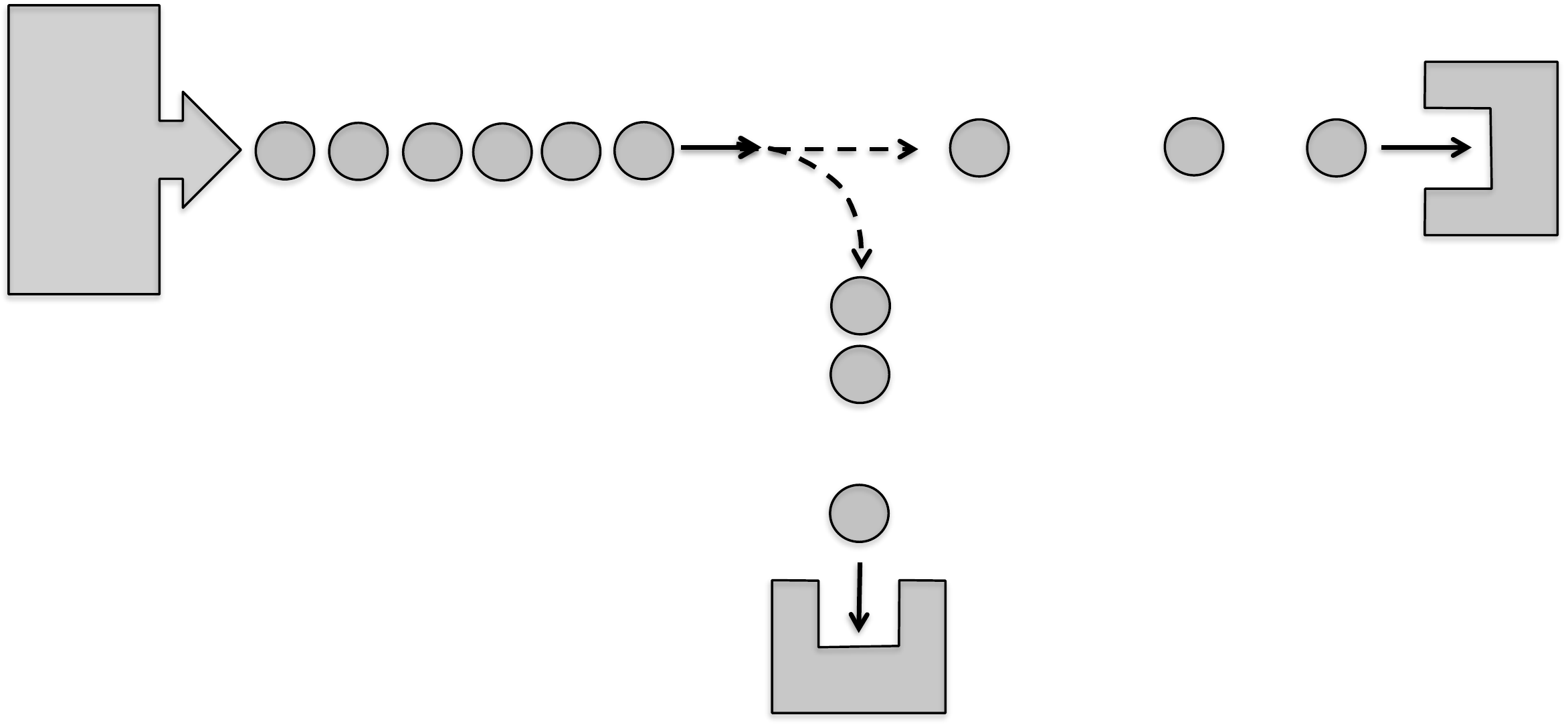}
\put(0,24.1){\footnotesize source}
\put(88,28.3){\footnotesize measure}
\put(93,24.2){\footnotesize $Q$}
\put(33.5,6.2){\footnotesize measure}
\put(39,2){\footnotesize $P$}
\end{overpic}
\caption{Physical scenario relevant to preparation uncertainty relations. Each incoming particle is measured using either measurement $P$ or measurement $Q$, where the choice of the measurement is random. An uncertainty relation says we cannot predict the outcomes of both $P$ and $Q$. If we can predict the outcome of $P$ well, then we are necessarily uncertain about the about the outcome of measurement $Q$, and vice versa.}
\label{figpreparationuncertainty}
\end{center}
\end{figure}

By far the most famous statement of unpredictability is Heisenberg's uncertainty principle~\cite{heisenberg27}, which we treat here as a statement about preparation uncertainty. Roughly speaking, it states that it is impossible to prepare a quantum particle for which both position and momentum are sharply defined. Operationally, consider a source that consistently prepares copies of a quantum particle in the same way, as shown in Fig.~\ref{figpreparationuncertainty}. For each copy, suppose we randomly measure either its position or its momentum (but we never attempt to measure both quantities for the same particle\footnote{Section~\ref{sec:scope} briefly notes other uncertainty principles that involve consecutive or joint measurements.}).  We record the outcomes and sort them into two sequences associated with the two different measurements. The uncertainty principle states that it is impossible to predict both the outcome of the position and the momentum measurements: at least one of the two sequences of outcomes will be unpredictable. More precisely, the better such a preparation procedure allows one to predict the outcome of the position measurement, the more uncertain the outcome of the momentum measurement will be, and vice versa.

An elegant aspect of quantum mechanics is that it allows for simple quantitative statements of this idea, i.e., constraints on the predictability of observable pairs like position and momentum. These quantitative statements are known as uncertainty relations. It is worth noting that Heisenberg's original argument, while conceptually enlightening, was heuristic. The first, rigorously-proven uncertainty relation for position $Q$ and momentum $P$ is due to \textcite{kennard27}. It establishes that (see also the work of~\textcite{weyl28})
\begin{align}\label{eqnKennardUR}
\sigma (Q) \sigma (P) \geq \frac{\hbar}{2}\,,
\end{align}
where $\sigma (Q)$ and $\sigma (P)$ denote the standard deviation of the position and momentum, respectively, and $\hbar$ is the reduced Planck constant. 

We now know that Heisenberg's principle applies much more generally, not only to position and momentum. Other examples of pairs of observables obeying an uncertainty relation include the phase and excitation number of a harmonic oscillator, the angle and the orbital angular momentum of a particle, and orthogonal components of spin angular momentum. In fact, for arbitrary observables\footnote{More precisely, Robertson's relation refers to observables with bounded spectrum.}, $X$ and $Z$, \textcite{robertson29} showed that~
\begin{align}\label{eqnRobertsonUR}
\sigma (X) \sigma (Z) \geq \frac{1}{2}|\expct{\psi}{[X,Z]}|\,,
\end{align}
where $[\cdot, \cdot]$ denotes the commutator. Note a distinct difference between~\eqref{eqnKennardUR} and~\eqref{eqnRobertsonUR}: the right-hand side of the former is a constant whereas that of the latter can be state-dependent, an issue that we will discuss more in Sec.~\ref{sec:stddev}.

These relations have a beauty to them and also give conceptual insight. Equation~\eqref{eqnKennardUR} identifies $\hbar$ as a fundamental limit to our knowledge. More generally~\eqref{eqnRobertsonUR} identifies the commutator as the relevant quantity for determining how large the knowledge trade-off is for two observables. One could argue that a reasonable goal in our studies of uncertainty in quantum mechanics should be to find simple, conceptually insightful statements like these. 

If this problem was only of fundamental importance, it would be a well-motivated one. Yet in recent years there is new motivation to study the uncertainty principle. The rise of quantum information theory has led to new applications of quantum uncertainty, for example in quantum cryptography. In particular quantum key distribution is already commercially marketed and its security crucially relies on Heisenberg's uncertainty principle. (We will discuss various applications in Sec.~\ref{sec:app}.)
There is a clear need for uncertainty relations that are directly applicable to these technologies.

In the above uncertainty relations, \eqref{eqnKennardUR} and \eqref{eqnRobertsonUR}, uncertainty has been quantified using the standard deviation of the measurement results. This is, however, not the only way to express the uncertainty principle. It is instructive to consider what preparation uncertainty means in the most general setting. Suppose we have prepared a state $\rho$ on which we can perform two (or more) possible measurements labeled by $\theta$. Let us use $x$ to label the outcomes of such measurement. We can then identify a list of (conditional) probabilities
\begin{align}
	S_{\rho} = \big\{ p(x|\theta)_{\rho} \big\}_{x,\theta} \,,
\end{align}
where $p(x|\theta)_{\rho}$ denotes the probability of obtaining measurement outcome $x$ when performing the measurement $\theta$ on the state $\rho$. Quantum mechanics predicts restrictions on the set $S_\rho$ of allowed conditional probability distributions that are valid for all or a large class of states~$\rho$. Needless to say, there are many ways to formulate such restrictions on the set of allowed distributions.

In particular, information theory offers a very versatile, abstract framework that allows us to formalize notions like uncertainty and unpredictability. This theory is the basis of modern communication technologies and cryptography and has been successfully generalized to include quantum effects. The preferred mathematical quantity to express uncertainty in information theory is entropy. Entropies are functionals on random variables and quantum states that aim to quantify their inherent uncertainty. Amongst a myriad of such measures, we mainly restrict our attention to the Boltzmann--Gibbs--Shannon entropy \cite{boltzmann1872,gibbs1876,shannon48} and its quantum generalization, the von Neumann entropy~\cite{vonneumann32}. Due to their importance in quantum cryptography, we will also consider R{\'e}nyi entropic measures \cite{renyi61} such as the min-entropy. Entropy is a natural measure of uncertainty, perhaps even more natural than the standard deviation, as we argue in Sec.~\ref{sec:stddev}.

Can the uncertainty principle be formulated in terms of entropy?  This question was first brought up by~\textcite{everett57} and answered in the affirmative by \textcite{hirschman57} who considered the position and momentum observables, formulating the first \emph{entropic uncertainty relation}. This was later improved by \textcite{beckner75, biaynicki75}, who obtained the relation\footnote{More precisely, the right-hand side of \eqref{eqnPosMomEUR1} should be $\log (e \pi \hbar/ (l_Q l_P))$, where $l_Q$ and $l_P$ are length and momentum scales, respectively, chosen to make the argument of the logarithm dimensionless. Throughout this review, all logarithms are base 2.}
\begin{align}\label{eqnPosMomEUR1}
h(Q)+h(P) \geq \log (e \pi \hbar)\,,
\end{align}
where $h$ is the differential entropy (defined in~\eqref{eqnDifferentialEntropy1} below). \textcite{biaynicki75} also showed that \eqref{eqnPosMomEUR1} is stronger than, and hence implies, Kennard's relation~\eqref{eqnKennardUR}.

The extension of the entropic uncertainty relation to observables with finite spectrum\footnote{More precisely, the relation applies to non-degenerate observables on a finite-dimensional Hilbert space (see Sec.~\ref{sctpreliminaries}).} was given by \textcite{deutsch83}, and later improved by \textcite{maassen88} following a conjecture by \textcite{kraus87}. The result of \textcite{maassen88} is arguably the most well-known entropic uncertainty relation. It states that
\begin{align}\label{eqnMUEUR1}
H(X) + H(Z) \geq \log\frac{1}{c},
\end{align}
where $H$ is Shannon's entropy (see Sec.~\ref{sec:pre-entropy} for definition), and $c$ denotes the maximum overlap between any two eigenvectors of the $X$ and $Z$ observables. Just as~\eqref{eqnRobertsonUR} established the commutator as an important parameter in determining the uncertainty tradeoff for standard deviation, \eqref{eqnMUEUR1} established the maximum overlap $c$ as a central parameter in entropic uncertainty.

While these articles represent the early history of entropic uncertainty relations, there has recently been an explosion of work on this topic. One of the most important recent advances concerns a generalization of the uncertainty paradigm that allows the measured system to be correlated to its environment in a non-classical way. Entanglement between the measured system and the environment can be exploited to reduce the uncertainty of an observer (with access to the environment) below the usual bounds. 

\begin{figure}[tbp]
\begin{center}
\begin{overpic}[width=8.1cm]{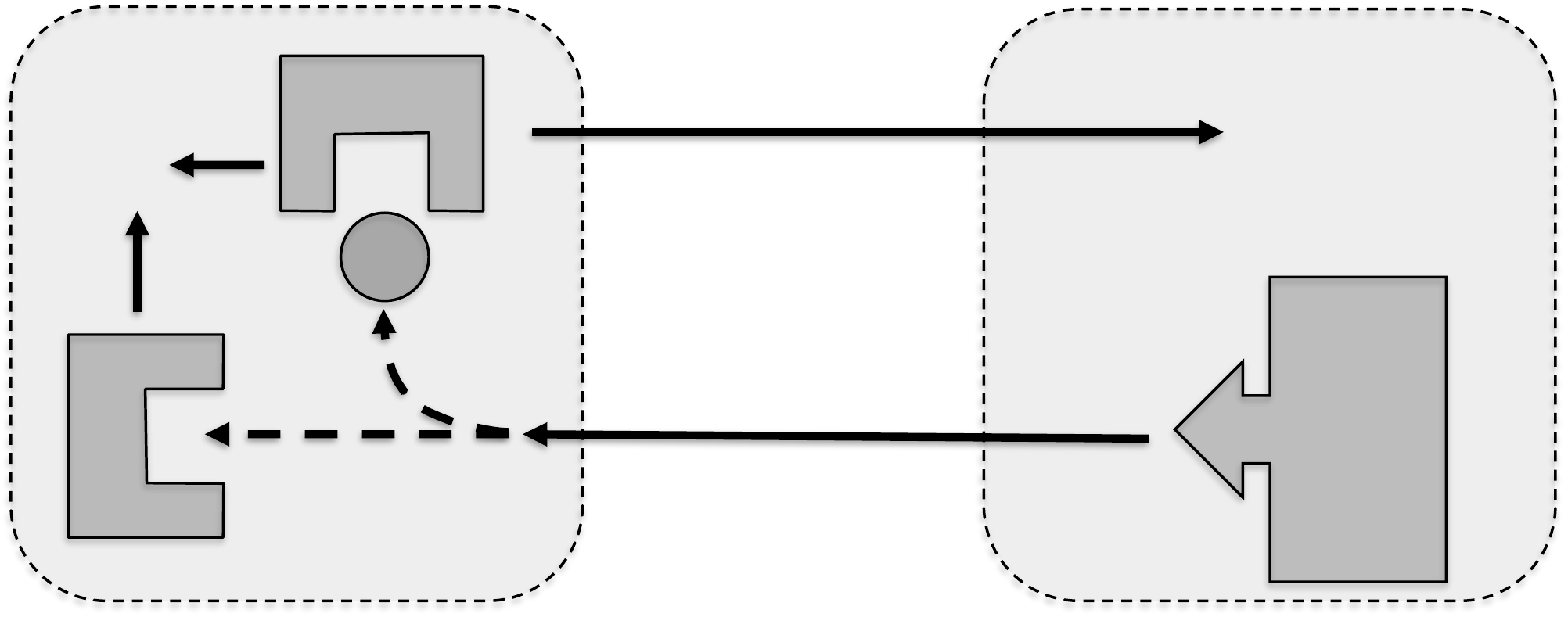}
\put(19.2,32){\footnotesize $\Theta=\bZ$}
\put(5.7,6.4){\rotatebox{90}{\footnotesize $\Theta=\bX$}}
\put(85.3,31){\footnotesize Bob}
\put(4,34){\footnotesize Alice}
\put(7,27.4){\footnotesize $K$}
\put(49,13){\footnotesize $A$}
\put(49,32){\footnotesize $\Theta$}
\put(84,11.5){\footnotesize $\rho_{A}$}
\put(85.2,26){\footnotesize $K$?}
\vspace{1cm}
\end{overpic}
\caption{Diagram showing a guessing game with players Alice and Bob. First, Bob prepares $A$ in state $\rho_{A}$ and sends it to Alice. Second, Alice measures either $\bX$ or $\bZ$ with equal probability and stores the measurement choice in the bit $\Theta$. Third, Alice stores the measurement outcome in bit $K$ and reveals the measurement choice $\Theta$ to Bob. Bob's task is to guess $K$ (given $\Theta$). Entropic uncertainty relations like the Maassen-Uffink relation~\eqref{eqnMUEUR1} can be understood as fundamental constraints on the optimal guessing probability.}
\label{figguessingnomemory}
\end{center}
\end{figure}
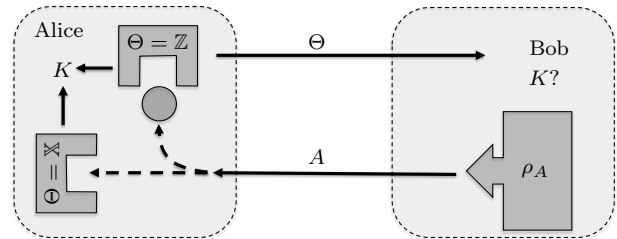

To explain this extension, let us introduce a modern formulation of the uncertainty principle as a so-called guessing game, which makes such extensions of the uncertainty principle natural and highlights their relevance for quantum cryptography. As outlined in Fig.~\ref{figguessingnomemory}, we imagine that an observer, Bob, can prepare an arbitrary state $\rho_A$ which he will send to a referee, Alice. Alice then randomly chooses to perform one of two (or more) possible measurements, where we will use $\Theta$ to denote her choice of measurement. She records the outcome, $K$. Finally, she tells Bob the choice of her measurement, i.e., she sends him $\Theta$. Bob's task is to guess Alice's measurement outcome $K$ (given $\Theta$).

The uncertainty principle tells us that if Alice makes two incompatible measurements, then Bob cannot guess Alice's outcome with certainty for both measurements. This corresponds precisely to the notion of preparation uncertainty. It is indeed intuitive why such uncertainty relations form an important ingredient in proving the security of quantum cryptographic protocols, as we will explore in detail in Sec.~\ref{sec:app}. In the cryptographic setting $\rho_A$ will be sent by an adversary trying to break a quantum cryptographic protocol. If Alice's measurements are incompatible, there is no way for the adversary to know the outcomes of both possible measurements with certainty - no matter what state he prepares.

The formulation of uncertainty relations as guessing games also makes it clear that there is an important twist to such games: What if Bob prepares a bipartite state $\rho_{AB}$ and sends only the $A$ part to Alice? That is, what if Bob's system is correlated with Alice's? Or, adopting the modern perspective of information, what if Bob has a non-trivial amount of \emph{side information} about Alice's system? Traditional uncertainty relations implicitly assume that Bob has only classical side information. For example, he may possess a classical description of the state $\rho_A$ or other details about the preparation. However, modern uncertainty relations---for example those derived by~\textcite{berta10} improving on work by \textcite{christandl05} and \textcite{renes09}---allow Bob to have \emph{quantum} rather than classical information about the state. As was already observed by~\textcite{epr35}, Bob's uncertainty can vanish in this case (in the sense that he can correctly guess Alice's measurement outcome $K$ in the game described above). 

We will devote Sec.~\ref{sctMemory} to such modern uncertainty relations. It is these relations that will be of central importance in quantum cryptography, where the adversary may have gathered \emph{quantum} and not just classical information during the course of the protocol that may reduce his uncertainty.


\subsection{Scope of this review}\label{sec:scope}

Two survey articles partially discuss the topic of entropic uncertainty relations. \textcite{birula10} take a physics perspective and cover continuous variable entropic uncertainty relations and some discretized measurements. In contrast, \textcite{wehner09} take an information-theoretic perspective and discuss entropic uncertainty relations for discrete (finite) variables with an emphasis on relations that involve more than two measurements.

These reviews predate many recent advances in the field. For example, both reviews do not cover entropic uncertainty relations that take into account quantum correlations with the environment of the measured system. Moreover, applications of entropic uncertainty relations are only marginally discussed in both of these reviews. Here, we discuss both physical and information-based applications. We therefore aim to give a comprehensive treatment of all of these topics in one reference, with the hope of benefiting some of the quickly emerging technologies that exploit quantum information.

There is an additional aspect of the uncertainty principle known as \emph{measurement uncertainty}, see, e.g., \textcite{ozawa03,hall04,busch07,busch14b}. This includes (1) joint measurability, the concept that there exists pairs of observables that cannot be measured simultaneously, and (2) measurement disturbance, the concept that there exist pairs of observables for which measuring one causes a disturbance of the other. Measurement uncertainty is a debated topic of current research. We focus our review article on the concept of preparation uncertainty, although we briefly mention entropic approaches to measurement uncertainty in Sec.~\ref{sctmeasurementuncertainty}.


\section{Relation to Standard Deviation Approach}\label{sec:stddev}


Traditional formulations of the uncertainty principle, for example the ones due to Kennard and Robertson, measure uncertainty in terms of the standard deviation. In this section we argue why we think entropic formulations are preferable. For further discussion we refer to~\textcite{uffinkThesis}.


\subsection{Position and momentum uncertainty relations}

For the case of position and momentum observables, the strength of the entropic formulation can be seen from the fact that the entropic uncertainty relation in \eqref{eqnPosMomEUR1} is stronger, and in fact implies, the standard deviation relation~\eqref{eqnKennardUR}.
Following~\textcite{biaynicki75}, we formally show that
\begin{align}
h(Q)+h(P) \geq \log (e \pi )\quad \implies \quad \sigma (Q) \sigma (P) \geq \frac{1}{2}
\end{align}
for all states, where here and henceforth in this article we work in units such that $\hbar = 1$. Let us consider a random variable $Q$ governed by a probability density $\Gamma(q)$, and
the differential entropy 
\begin{align}\label{eqnDifferentialEntropy1}
h(Q) = - \int_{-\infty}^{\infty} \Gamma(q) \log \Gamma(q) dq\,.
\end{align}
In the following we assume that this quantity is finite. Gaussian probability distributions,
\begin{align}\label{eqnGaussianDist1}
\Gamma(q) = \frac{1}{\sqrt{2 \pi \sigma (Q)^2 }}\exp \bigg(\frac{-(q-\overline{q})^2}{2\sigma (Q)^2} \bigg) \,,
\end{align}
where $\overline{q}$ denotes the mean, are special in the following sense: for a fixed standard deviation $\sigma (Q)$, distributions of the form of \eqref{eqnGaussianDist1} maximize the entropy in \eqref{eqnDifferentialEntropy1}. It is a simple exercise to show this, e.g., using variational calculus with Lagrange multipliers.

It is furthermore straightforward to insert \eqref{eqnGaussianDist1} into \eqref{eqnDifferentialEntropy1} to calculate the entropy of a Gaussian distribution
\begin{align}\label{eqnDifferentialEntropyGaussian}
h(Q) = \log \sqrt{2\pi e \sigma (Q)^2}\quad\text{(Gaussian)} \,.
\end{align}
Since Gaussians maximize the entropy, the following inequality holds in general
\begin{align}\label{eqnEntropyToStdevRelation}
h(Q) \leq \log \sqrt{2\pi e \sigma (Q)^2}\quad\text{(in general)} \,.
\end{align}

Now consider an arbitrary quantum state for a particle's translational degree of freedom, which gives rise to random variables $P$ and $Q$ for the position and momentum, respectively. Let us insert the resulting relations into \eqref{eqnPosMomEUR1} to find
\begin{align}\label{eqnProofOfKennard}
\log (2\pi e \sigma (Q) \sigma (P)) & = \log \sqrt{2\pi e \sigma (Q)^2} +\log \sqrt{2\pi e \sigma (P)^2}\\
&\geq h(Q)+h(P)\\
&\geq \log (e \pi ) \,.
\end{align}
By comparing the left- and right-hand sides of \eqref{eqnProofOfKennard} and noting that the logarithm is a monotonic function, we see that \eqref{eqnProofOfKennard} implies \eqref{eqnKennardUR}, and hence so does \eqref{eqnPosMomEUR1}.

It is worth noting that \eqref{eqnEntropyToStdevRelation} is a strict inequality if the distribution is non-Gaussian, and hence \eqref{eqnPosMomEUR1} is strictly stronger than \eqref{eqnKennardUR} if the quantum state is non-Gaussian. While quantum mechanics textbooks often present \eqref{eqnKennardUR} as \emph{the} fundamental statement of the uncertainty principle, it is clear that \eqref{eqnPosMomEUR1} is stronger and yet not much more complicated. Furthermore, as discussed in Sec.~\ref{sctMemory} the entropic formulation is more robust, allowing the relation to be easily generalized to situations involving correlations with the environment.


\subsection{Finite spectrum uncertainty relations}

As noted above, both the standard deviation and the entropy have been applied to formulate uncertainty relations for observables with a finite spectrum. However, it is largely unclear how the most popular formulations, Robertson's~\eqref{eqnRobertsonUR} and Maassen-Uffink's~\eqref{eqnMUEUR1}, are related. It remains an interesting open question whether there exists a formulation that unifies these two formulations. However, there is an important difference between~\eqref{eqnRobertsonUR} and~\eqref{eqnMUEUR1} in that the former has a bound that depends on the state, while the latter only depends on the two observables.

\begin{example}\label{exRobertson}
Consider~\eqref{eqnRobertsonUR} for the case of a spin-1/2 particle, where $X = \dyad{0}{1}+\dyad{1}{0}$ and $Z=\dya{0}-\dya{1}$, corresponding to the $x$- and $z$-axes of the Bloch sphere. Then the commutator is proportional to the $Y$ Pauli operator and the right-hand side of \eqref{eqnRobertsonUR} reduces to $(1/2)\avg{Y}$. Hence, \eqref{eqnRobertsonUR} gives a trivial bound for all states that lie in the $xz$ plane of the Bloch sphere. For the eigenstates of $X$ and $Z$, this bound is tight since one of the two uncertainty terms is zero, and hence the trivial bound is a (perhaps undesirable) consequence of the fact that the left-hand side involves a \emph{product} (rather than a sum) of uncertainties. However, for any other states in the $xz$ plane, neither uncertainty is zero. This implies that \eqref{eqnRobertsonUR} is \emph{not tight} for these states.
\end{example}

This example illustrates a weakness of Robertson's relation for finite-dimensional systems\,---\,it gives trivial bounds for certain states, even when the left-hand side is non-zero. \textcite{schroedinger30} slightly strengthened Robertson's bound by adding an additional state-dependent term that helps to get rid of the artificial trivial bound discussed in Ex.~\ref{exRobertson}. Likewise, \textcite{Maccone14} recently proved a state-dependent bound on the sum (not the product) of the two variances, and this bound also removes the trivial behavior of Robertson's bound. Furthermore, one still may be able to obtain a non-vanishing state-independent bound using standard deviation uncertainty measures in the finite-dimensional case. For example, \textcite{busch14} considered the qubit case and obtained a state-independent bound on the sum of the variances.

The state-dependent nature of Robertson's bound was noted, e.g., by \textcite{deutsch83} and used as motivation for entropic uncertainty relations, which do not suffer from this weakness. However, the above discussion suggests that this issue might be avoided while still using standard deviation as the uncertainty measure. On the other hand, there are more important issues that we now discuss.


\subsection{Advantages of entropic formulation}

From a practical perspective, a crucial advantage of entropic uncertainty relations are their applications throughout quantum cryptography. However, let us now mention several other
reasons why we think that the entropic formulation of the uncertainty principle is advantageous over the standard deviation formulation.


\subsubsection{Counterintuitive behavior of standard deviation}

While the standard deviation is, of course, a good measure of deviation from the mean, its interpretation as a measure of uncertainty has been questioned. It has been pointed out by several authors, for example by~\textcite{birula10}, that the standard deviation behaves somewhat strangely for some simple examples.

\begin{example}\label{exStddev1}
Consider a spin-1 particle with equal probability $\Pr (s_z)=1/3$ to have each of the three possible values of $Z$-angular momentum, $s_z \in \{-1,0,1\}$. The standard deviation of the $Z$-angular momentum is $\sigma (Z) = \sqrt{2/3}$. Now suppose we gain information about the spin such that we now know that it definitely does not have the value $s_z=0$. The new probability distribution is $\Pr (1) = \Pr (-1) =1/2$, $\Pr(0)=0$. We might expect the uncertainty to decrease, since we have gained information about the spin, but in fact the standard deviation increases, the new value being $\sigma (Z) = 1$.
\end{example}

We remark that the different behavior of standard deviation and entropy for spin angular momentum was recently highlighted by \textcite{dammeier15}, in the context of states that saturate the relevant uncertainty relation.

\textcite{birula10} noted an example for a particle's spatial position that is analogous to the above example.

\begin{figure}
\begin{center}
\includegraphics[width=3in]{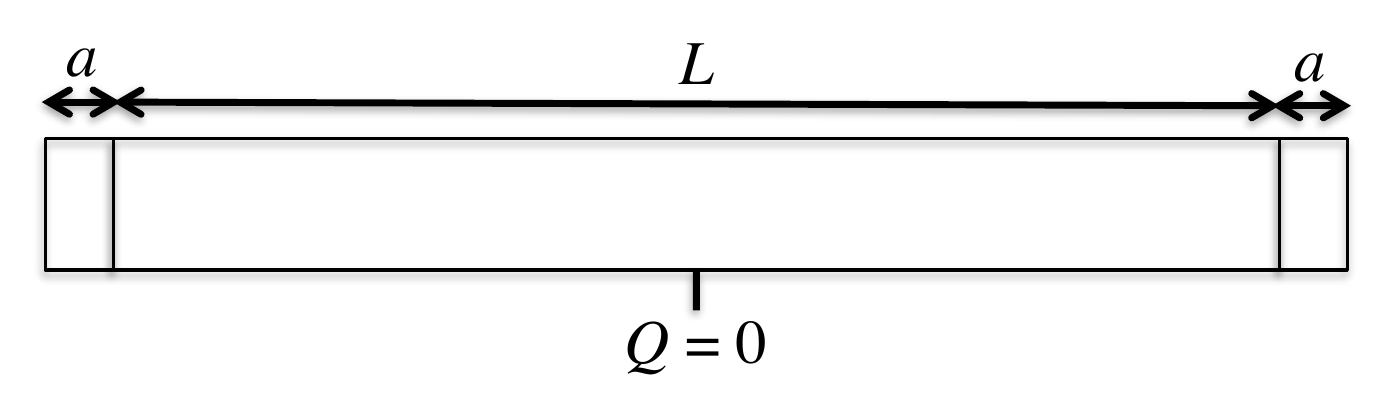}
\caption{Illustration for Ex.~\ref{exStddev2}, where a particle is initially confined to the two small boxes at the end and excluded from the long middle box. Then the particle is allowed to go free into the middle box.}
\label{FigStdDev}
\end{center}
\end{figure}

\begin{example}\label{exStddev2}
Consider a long box of length $L$, centered at $Q=0$, with two small boxes of length $a$ attached to the two ends of the long box, as depicted in Fig.~\ref{FigStdDev}. Suppose we know that a classical particle is confined to the two small end boxes, i.e., with equal probability it is one of the two small boxes. The standard deviation of the position is $\sigma (Q) \approx L/2$, assuming that $L \gg a$. Now suppose the barriers that separate the end boxes from the middle box are removed, and the particle is allowed to move freely between all three boxes. Intuitively one might expect that the uncertainty of the particle's position is now larger, since we now know nothing about where the particle is inside the three boxes. However, the new standard deviation is actually smaller: $\sigma (Q) \approx L/\sqrt{12}$.
\end{example}

Entropies on the other hand do not have this counterintuitive behavior, due to properties discussed below. Finally, let us note a somewhat obvious issue that, in some cases, a quantitative label (and hence the standard deviation) does not make sense, as illustrated in the following example.

\begin{example}\label{exStddev3}
Consider a neutrino's flavor, which is often modeled as a three-outcome observable with outcomes ``electron'', ``muon'', or ``tau''. As this is a non-quantitative observable, the standard deviation does not make sense in this context. Nevertheless, it is of interest to quantify the uncertainty about the neutrino flavor, i.e., how difficult it is to guess the flavor, which is naturally captured by the notion of entropy.
\end{example}


\subsubsection{Intuitive entropic properties}

\textcite{deutsch83} emphasized that the standard deviation can change under a simple relabeling of the outcomes. For example, if one were to assign quantitative labels to the outcomes in Ex.~\ref{exStddev3} and then relabel them, the standard deviation would change. In contrast, the entropy is invariant under relabeling of outcomes, because it naturally captures the amount of \emph{information} about a measurement outcome.

Furthermore, there is a nice monotonic property of entropy in the following sense. Suppose one does a random relabeling of the outcomes. One can think of this as a relabeling plus added noise, which naturally tends to spread the probability distribution out over the outcomes. Intuitively, a relabeling with the injection of randomness should never decrease the uncertainty. This property\,---\,non-decreasing under random relabeling\,---\,was highlighted by \textcite{friedland13} as a desirable property of an uncertainty measure. Indeed, entropy satisfies this property. On the other hand, the physical process in Ex.~\ref{exStddev2} can be modeled mathematically as a random relabeling. Hence, we see the contrast in behavior between entropy and standard deviation. 

Monotonicity under random relabeling is actually a special case of an even more powerful property. Think of the random relabeling as due to the fact that the observer is denied access to an auxiliary register that stores the information about which relabeling occurred. If the observer had access to the register, then their uncertainty would remain the same, but without access their uncertainty could potentially increase, but never decrease! More generally, this idea (that losing access to an auxiliary system cannot reduce one's uncertainty) is a desirable and powerful property of uncertainty measures known as the \textit{data-processing inequality}. It is arguably a defining property of entropy measures, or more precisely, conditional entropy measures as discussed in Sec.~\ref{sec:conditionalentropy}. Furthermore this property is central in proving entropic uncertainty relations \cite{colbeck11}.


\subsubsection{Framework for correlated quantum systems}

Entropy provides a robust mathematical framework that can be generalized to deal with correlated quantum systems. For example, the entropy framework allows us to discuss the uncertainty of an observable from the perspective of an observer who has access to part of the environment of the system, or to quantify quantum correlations like entanglement between two quantum systems. This requires measures of conditional uncertainty, namely conditional entropies. We highlight the utility of this framework in Sec.~\ref{sctMemory}. A similar framework for standard deviation has not been developed.


\subsubsection{Operational meaning and information applications}

Perhaps the most compelling reason to consider entropy as the uncertainty measure of choice is that it has operational significance for various information-processing tasks. The standard deviation, in contrast, does not play a significant role in information theory. This is because entropy abstracts from the physical representation of information, as one can see from the following example.

\begin{figure}
\subfigure[\,low entropy distribution]{\includegraphics[width=0.45\columnwidth]{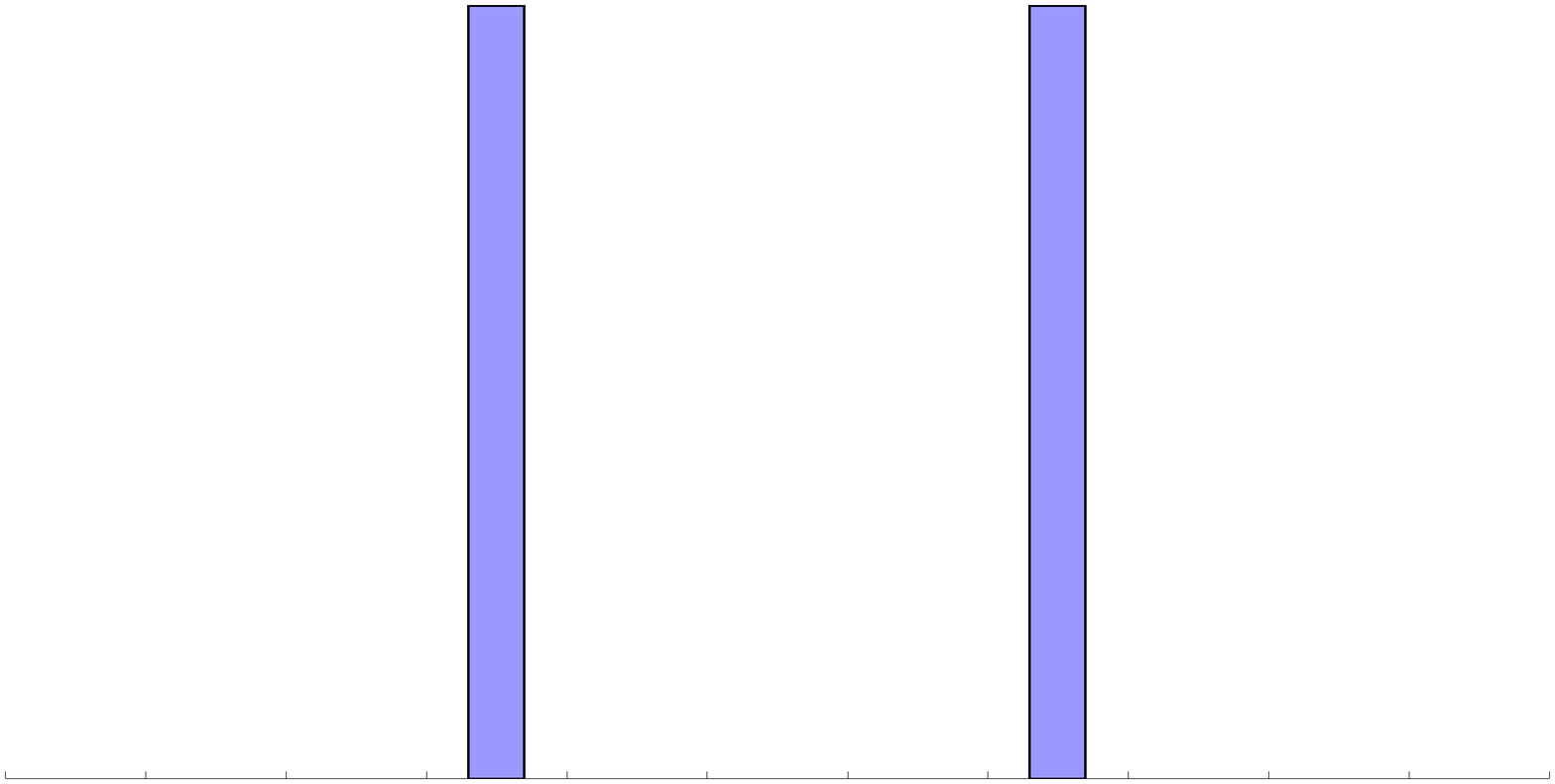}\label{fig:dista}}
\subfigure[\,high entropy distribution]{\includegraphics[width=0.45\columnwidth]{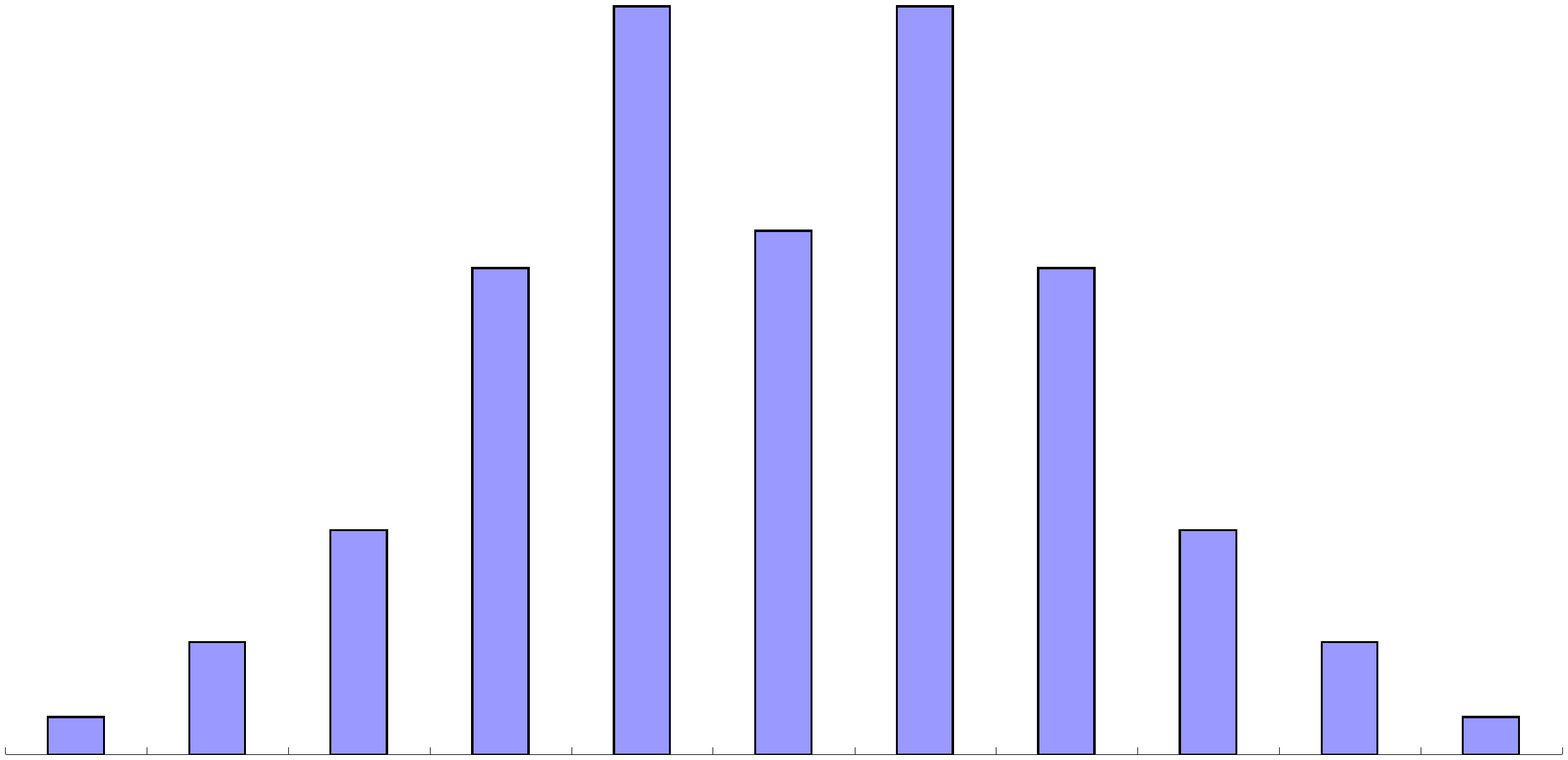}\label{fig:distb}}
\caption{Two probability distributions with the same standard deviation but different entropy, as explained in Ex.~\ref{ex:std/ent}.} 
\label{fig:dist}
\end{figure}

\begin{example}\label{ex:std/ent}
Consider the two probability distributions in Fig.~\ref{fig:dist}. They have the same standard deviation but different entropy. The distribution in Fig.~\ref{fig:dista} has one bit of entropy since only two events are possible and occur with equal probability. If we want to record data from this random experiment this will require exactly one bit of storage per run. On the other hand, the distribution in Fig.~\ref{fig:distb} has approximately 3 bits of entropy and the recorded data cannot be compressed to less than 3 bits per run. Clearly, entropy has operational meaning in this context while standard deviation fails to distinguish these random experiments. 
\end{example}

Entropies have operational meaning for tasks such as randomness extraction (extracting perfect randomness from a partially random source) and data compression (sending minimal information to someone to help them guess the output of a partially random source). It is precisely these operational meanings that make entropic uncertainty relations useful for proving the security of quantum key distribution and other cryptographic tasks. We discuss such applications in Sec.~\ref{sec:app}. 

The operational significance of entropy allows one to frame entropic uncertainty relations in terms of guessing games (see Sec.~\ref{sec:games_cl} and~\ref{sec:bipartiteguessing}). These are simple yet insightful tasks where, e.g., one party is trying to guess the outcome of another party's measurements (see the description in Fig.~\ref{figguessingnomemory}). Such games make it clear that the uncertainty principle is not just abstract mathematics; rather it is relevant to physical tasks that can be performed in a laboratory. 



\section{Uncertainty without a Memory System}\label{sec:finite_dimensional}

Historically, entropic uncertainty relations were first studied for position and momentum observables. However, to keep the discussion mathematically simple we begin here by introducing entropic uncertainty relations for finite-dimensional quantum systems, and we defer the discussion of infinite dimensions to Sec.~\ref{sec:infinite-dim}. It is worth noting that many physical systems of interest are finite-dimensional, such as photon polarization, neutrino flavor, and spin angular momentum.

In this section, we consider uncertainty relations for a single system $A$. That is, there is no memory system. We emphasize that all uncertainty relations \emph{with} a memory system can also be applied to the situation \emph{without}.


\subsection{Entropy measures}\label{sec:pre-entropy}

Let us consider a discrete \emph{random variable} $X$ distributed according to the probability distribution $P_X$. We assume that $X$ takes values in a finite set $\mathcal{X}$. For example, this set could be binary values $\{ 0, 1 \}$ or spin states $\{ \uparrow, \downarrow \}$. In general, we will associate the random variable $X$ with the outcome of a particular measurement. This random variable can take values $X = x$, where $x$ is a specific instance of a measurement outcome that can be obtained with probability $P_X(X=x)$. However, entropies only depend on the probability law $P_X$ and not on the specific labels of the elements in the set $\mathcal{X}$. Thus, we will in the following just assume this set to be of the form $[d] := \{ 1, 2, 3, \ldots, d\}$, where $d = |X|$ stands for the cardinality of the set $\mathcal{X}$.


\subsubsection{Surprisal and Shannon entropy}

Following~\textcite{shannon48}, we first define the \emph{surprisal} of the event $X=x$ distributed according to $P_X$ as $- \log {P_X(x)}$, often also referred to as \emph{information content}. As its name suggests, the information content of $X=x$ gets larger when the event $X=x$ is less likely, i.e., when $P_X(x)$ is smaller. In particular, deterministic events have no information content at all, 
which is indeed intuitive since we learn nothing by observing an event that we are assured will happen with certainty. 
In contrast, the information content of very unlikely events can get arbitrarily large. 
Based on this intuition, the \emph{Shannon entropy} is defined as
\begin{align}\label{eq:shannon}
H(X) := \sum_x P_X(x) \log \frac{1}{P_X(x)}\,,
\end{align}
and quantifies the \textit{average information content} of $X$. It is therefore a measure of the uncertainty of the outcome of the random experiment described by $X$.
The Shannon entropy is by far the best-known measure of uncertainty, and it is the one most commonly used to express uncertainty relations.


\subsubsection{R\'enyi entropies}

However, for some applications it is important to consider other measures of uncertainty that give more weight to events with high or low information content, respectively. For this purpose we employ a generalization of the Shannon entropy to a family of entropies introduced by~\textcite{renyi61}. The family includes several important special cases which we will discuss individually.
These entropies have found many applications in cryptography and information theory (see Sec.~\ref{sec:app}) and have convenient mathematical properties.\footnote{Another family of entropies that are often encountered are the Tsallis entropies~\cite{tsallis88}. They have not found an operational interpretation in cryptography or information theory. Thus, we defer the discussion of Tsallis entropies until Sec.~\ref{sec:other-entropies}.}

The \emph{R\'enyi entropy} of order $\alpha$ is defined as
\begin{align}\label{eq:renyi-class}
  H_{\alpha}(X) :=& \frac{1}{1-\alpha} \log \sum_x P_X(x)^{\alpha} \notag\\
   & \textrm{for} \quad \alpha \in (0, 1) \cup (1,\infty) \,,
\end{align}
and as the corresponding limit for $\alpha \in \{0, 1, \infty\}$. For $\alpha = 1$ the limit yields the Shannon entropy\footnote{It is a simple exercise to apply L'H\^opital's rule to \eqref{eq:renyi-class} in the limit $\al \to 1$.}, and the R\'enyi entropies are thus a proper generalization of the Shannon entropy.

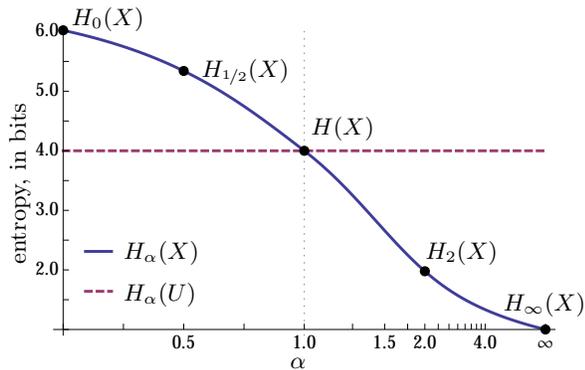
\begin{figure}
  \begin{overpic}[width=0.8\columnwidth]{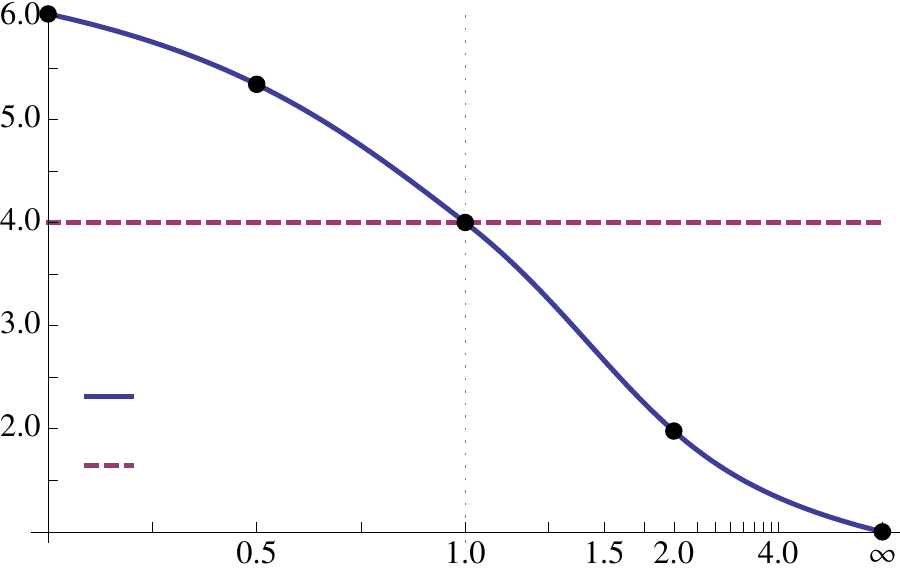}
    \put(-5,17){\rotatebox{90}{entropy, in bits}}
    \put(49,-3){$\alpha$}
    \put(17,18){$H_{\alpha}(X)$}
    \put(17,10){$H_{\alpha}(U)$}
    \put(7,63){$H_{0}(X)$}
    \put(32,53){$H_{\nicefrac12}(X)$}
    \put(75,18){$H_{2}(X)$}
    \put(90,8){$H_{\infty}(X)$}
    \put(53,42){$H(X)$}
  \end{overpic}
  \caption{R\'enyi entropies of $X$ with probability distribution as in Ex.~\ref{ex:renyi} with $|X|=65$ compared to a uniform random variable $U$ on $4$ bits.}
  \label{fig:renyi}
\end{figure}

The R\'enyi entropies are monotonically decreasing as a function of $\alpha$. Entropies with $\alpha > 1$ give more weight to events with high surprisal. The \emph{collision entropy}, $H_{\rm coll}:=H_2$, is given by
\begin{align}\label{eqncollision1}
H_{\rm coll}(X) =& - \log p_{\rm coll}(X) \,, \quad \textrm{where} \notag\\
p_{\rm coll}(X) :=& \sum_x P_X(x)^2 
\end{align}
is the collision probability, i.e., the probability that two independent instances of $X$ are equal. The \emph{min-entropy} $H_{\min}:=H_{\infty}$, is of special significance in many applications. It characterizes the optimal probability of correctly guessing the value of $X$ in the following sense
\begin{align}\label{eq:guess-class}
H_{\min}(X) =& - \log p_{\rm guess}(X) \,,\quad \textrm{where} \notag\\
p_{\rm guess}(X) :=& \max_x P_X(x) \,.
\end{align}
Clearly, the optimal guessing strategy is to bet on the most likely value of $X$, and the winning probability is then given by the maximum in~\eqref{eq:guess-class}. The min-entropy can also be seen as the minimum surprisal of $X$.

The R\'enyi entropies with $\alpha < 1$ give more weight to events with small surprisal. Noteworthy examples are the \emph{max-entropy}, $H_{\max}:=H_{\nicefrac12}$, and
\begin{align}
H_0(X) = \log \big|\{x : P_X(x) > 0\}\big|\,,
\end{align}
where the latter is simply the logarithm of the support of $P_X$. 


\subsubsection{Examples and properties}\label{sec:cl-ent-prop}

For all the R\'enyi entropies, $H_{\alpha}(X) = 0$ if and only if the distribution is perfectly peaked, i.e., $P_X(x) = 1$ for some particular value $x$. On the other hand, the distribution $P_X(x) = |X|^{-1}$ is uniform if and only if the entropy takes its maximal value $H_{\alpha}(X) = \log |X|$.

The R\'enyi entropies can take on very different values depending on the parameter~$\alpha$ as the following example, visualized in Fig.~\ref{fig:renyi}, shows.

\begin{example}\label{ex:renyi}
Consider a distribution of the form
\begin{align}\label{eq:distex}
P_X(x) &= \begin{cases} \frac12 & \textrm{for } x = 1 \\ \frac{1}{2(|X|-1)} & \textrm{else} \end{cases} 
\end{align}
so that we have
\begin{align}
H_{\min}(X) &= \log 2 \,, \quad \textrm{whereas} \notag\\
H(X) &= \log 2 + \frac12 \log (|X|-1)
\end{align}
is arbitrarily large as $|X| \geq 2$ increases. This is of particular relevance in cryptographic applications where $H_{\min}(X)$\,---\,and not $H(X)$\,---\,characterizes how difficult it is to guess a secret $X$.
As we will see later, $H_{\min}(X)$ determines precisely the number of random bits that can be obtained from $X$.
\end{example}

Consider two probability distributions, $P_X$ and $Q_Y$, and define $d = \max \{ |X|, |Y| \}$. Now let us reorder the probabilities in $P_X$ into a vector $P_X^{\downarrow}$ such that $P_X^{\downarrow}(1) \geq P_X^{\downarrow}(2) \geq \ldots \geq P_X^{\downarrow}(d)$, padding with zeros if necessary. Analogously arrange the probabilities in $Q_Y$ into a vector $Q_Y^{\downarrow}$.
We say $P_X$ majorizes $Q_Y$ and write $P_X \succ Q_Y$ if 
\begin{align}\label{eq:MajorDef}
\sum_{x=1}^{y} P_X^{\downarrow}(x) \geq \sum_{x=1}^{y} Q_Y^{\downarrow}(x), \quad \textrm{for all } y \in [d]\,.
\end{align}
Intuitively, the fact that $P_X$ majorizes $Q_Y$ means that $P_X$ is less spread out than $Q_Y$. For example, the distribution $\{1, 0, \ldots, 0\}$ majorizes every other distribution, while the uniform distribution $\{ |X|^{-1}, \ldots, |X|^{-1} \}$ is majorized by every other distribution.

One of the most fundamental properties of the R\'enyi entropies is that they are Schur-concave~\cite{marshall11}, meaning that they satisfy
\begin{align}\label{eq:SchurCon}
H_{\alpha}(X) \leq H_{\alpha}(Y) \qquad \textrm{if } P_X \succ Q_Y . 
\end{align}
This has an important consequence. Let $Y = f(X)$ for some (deterministic) function $f$. In other words, $Y$ is obtained by \emph{processing} $X$ using the function $f$. The random variable $Y$ is then governed by the push forward $Q_Y$ of $P_X$, that is
\begin{align}
Q_Y(y) = \sum_{x : f(x) = y} P_X(x) \,.
\end{align}
Clearly $P_X \prec Q_Y$ and thus we have $H_{\alpha}(X) \geq H_{\alpha}(Y)$. This corroborates our intuition that the input of a function is at least as uncertain as its output. If $Z$ is just a reordering of $X$, or more generally if $f$ is injective, then the two entropies are equal.

Finally we note that if two random variables $X$ and $Y$ are independent, we have
\begin{align}
H_{\alpha}(XY) = H_{\alpha}(X) + H_{\alpha}(Y) \,.
\end{align}
This property is called \emph{additivity}. 


\subsection{Preliminaries}\label{sctpreliminaries}

\subsubsection{Physical setup}\label{sctsetup}

The physical setup used throughout the remainder of this section is as follows. We consider a quantum system, $A$, that is measured in either one of two (or more) bases. The initial state of the system $A$ is represented by a \emph{density operator}, $\rho_A$, or more formally a positive semidefinite operator with unit trace acting on a finite-dimensional Hilbert space $A$. The measurements, for now, are given by two \emph{orthonormal bases} of $A$. An orthonormal basis is a set of unit vectors in $A$ that are mutually orthogonal and span the space $A$. The two bases are denoted by sets of rank-$1$ projectors,
\begin{align}
\bX=\Big\{\proj{\bX^{x}}\Big\}_x\quad\mathrm{and}\quad \bZ=\Big\{\proj{\bZ^{z}}\Big\}_z\,.
\end{align}
We use projectors to keep the notation consistent as we will later consider more general measurements. This induces two random variables, $X$ and $Z$, corresponding to the measurement outcomes that result from measuring in the bases $\bX$ and $\bZ$, respectively. These are governed by the following probability laws, given by the Born rule. We have
\begin{align}\label{eq:mu-pm-def}
P_{X}(x)= \bra{\bX^{x}}\rho_A \ket{\bX^{x}} \quad \mathrm{and}\quad P_{Z}(z)= \bra{\bZ^{z}}\rho_A \ket{\bZ^{z}}\,,
\end{align}
respectively. We also note that $|X| = |Z| = d$, which is the dimension of the Hilbert space $A$.


\subsubsection{Mutually unbiased bases (MUBs)}\label{sctMUBdefinition}

Before delving into uncertainty relations, let us consider pairs of observables such that perfect knowledge about observable $\bX$ implies complete ignorance about observable $\bZ$. We say that such observables are unbiased, or mutually unbiased. For any finite-dimensional space there exist pairs of orthonormal bases that satisfy this property. More precisely, two orthonormal bases $\bX$ and $\bZ$ are mutually unbiased bases (MUBs) if
\begin{align}\label{eqnMUBdef}
|\ip{\bX^{x}}{\bZ^{z}}|^2 = \frac{1}{d},\quad \forall x,z\,.
\end{align}
In addition, a set of $n$ orthonormal bases $\{\bX_j\}$ is said to be a set of $n$ MUBs if each basis $\bX_j$ is mutually unbiased to every other basis $\bX_k$, with $k\neq j$, in the set.

\begin{example}\label{exPauli}
For a qubit the eigenvectors of the Pauli operators,
\begin{align}\label{eqnPaulidef}
\sigma_{\bX}&:=\dyad{0}{1}+\dyad{1}{0}\\
\sigma_{\bY}&:=-i\dyad{0}{1}+i\dyad{1}{0}\\
\sigma_{\bZ}&:=\dya{0}-\dya{1}
\end{align}
form a set of 3 MUBs.
\end{example}

In App.~\ref{app:mub} we discuss constructions for sets of MUBs in higher dimensional spaces. We also point to~\cite{durt10} for a review on this topic.


\subsection{Measuring in two orthonormal bases}\label{sctMU}

\subsubsection{Shannon entropy}

Based on the pioneering work by \textcite{deutsch83} and following a conjecture of \textcite{kraus87}, \textcite{maassen88} formulated entropic uncertainty relations for measurements of two complementary observables. Their best known relation uses the Shannon entropy to quantify uncertainty. It states that, for any state $\rho_A$,
\begin{align}\label{eq:shannonUR}
H(X)+H(Z)\geq\log\frac{1}{c}=:q_{\rm MU}\,,
\end{align}
where the measure of incompatibility is a function of the \emph{maximum overlap} of the two measurements, namely
\begin{align}\label{eq:overlap}
c=\max_{x,z} c_{xz}, \quad\text{where }c_{xz}= |\ip{\bX^{x}}{\bZ^{z}}|^2\,.
\end{align}
Note that $q_{\rm MU}$ is state-independent, i.e., independent of the initial state $\rho_A$. This is in contrast to Robertson's bound in~\eqref{eqnRobertsonUR}.

The bound $q_{\rm MU}$ is non-trivial as long as $\bX$ and $\bZ$ do not have any vectors in common. In this case,~\eqref{eq:shannonUR} shows that for any input density matrix there is some uncertainty in at least one of the two random variables $X$ and $Z$, quantified by the Shannon entropies $H(X)$ and $H(Z)$, respectively. In general we have
\begin{align}
\frac{1}{d}\leq c\leq1\quad\text{and hence }\quad0\leq q_{\rm MU}\leq\log d\,.
\end{align}
For the extreme case that $\bX$ and $\bZ$ are MUBs, as defined in \eqref{eqnMUBdef}, the overlap matrix $[c_{xz}]$ is flat: $c_{xz}=1/d$ for all $x$ and $z$, and the lower bound on the uncertainty then becomes maximal
\begin{align}
H(X)+H(Z) \geq\log d\,.
\end{align}
Note that this is a necessary and sufficient condition: $c=1/d$ if and only if the two bases are MUBs. Hence, MUBs uniquely give the strongest uncertainty bound here.

For general observables $\bX$ and $\bZ$ the overlap matrix is not necessarily flat and the asymmetry of the matrix elements $c_{xz}$ is quantified in~\eqref{eq:overlap} by taking the maximum over all $x,z$. In order to see why the maximum entry provides some (fairly coarse) measure of the flatness of the whole matrix, note that if the maximum entry of the overlap matrix is $1/d$, then all entries in the matrix must be $1/d$. Alternative measures of incompatibility will be discussed in Secs.~\ref{sec:tighter_bounds} and \ref{sec:tighter_bounds_d}.


\subsubsection{R\'enyi entropies}\label{sec:renyi_muffink}

\textcite{maassen88} also showed that the above relation~\eqref{eq:shannonUR} holds more generally in terms of R\'enyi entropies. For any $\alpha,\beta \geq \frac12$ with $1/\alpha+1/\beta=2$, we have
\begin{align}\label{eq:mu_relation}
H_{\alpha}(X) + H_{\beta}(Z) \geq q_{\rm MU}\,.
\end{align}
It is easily checked that the relation~\eqref{eq:shannonUR} in terms of the Shannon entropy is recovered for $\alpha=\beta=1$. For $\alpha\to\infty$ with $\beta\to1/2$ we get another interesting special case of~\eqref{eq:mu_relation} in terms of the min- and max-entropy
\begin{align}\label{eq:minmaxUR}
H_{\min}(X) + H_{\max}(Z) \geq q_{\rm MU}\,.
\end{align}
Since the min-entropy characterizes the probability of correctly guessing the outcome $X$, it is this type of relation that becomes most useful for applications in quantum cryptography and quantum information theory (see Sec.~\ref{sec:app}).


\subsubsection{Proof of Maassen-Uffink}\label{sctMUproof}

The original proof of~\eqref{eq:mu_relation} by Maassen and Uffink makes use of the Riesz-Thorin interpolation theorem (see, e.g., \cite{bergh76}). Recently an alternative proof was formulated by~\textcite{coles10, colbeck11} using the monotonicity of the relative entropy under quantum channels. The latter approach is illustrated in App.~\ref{app:shannon_proof}, where we prove the special case of the Shannon entropy relation~\eqref{eq:shannonUR}. The proof is simple and straightforward. Hence, we highly recommend the interested reader to study App.~\ref{app:shannon_proof}. The R\'enyi entropy relation \eqref{eq:mu_relation} follows from a more general line of argument given in App.~\ref{sec:memory_proof}.


\subsubsection{Tightness and extensions}

Given the simple and appealing form of the Maassen-Uffink relations~\eqref{eq:mu_relation} a natural question to ask is how tight these relations are. It is easily seen that if $\bX$ and $\bZ$ are MUBs, then they are tight for any of the states $\rho_A =\proj{\bX^{x}}$ or $\rho_A =\proj{\bZ^{z}}$. Thus, there cannot exist a better state-independent bound if $\bX$ and $\bZ$ are MUBs. However, for general orthonormal bases $\bX$ and $\bZ$ the relations~\eqref{eq:mu_relation} are not necessarily tight. This issue is addressed in the following subsections, where we also note that \eqref{eq:shannonUR} can be tightened for mixed states $\rho_A$ with a state-dependent bound.

Going beyond orthonormal bases, the above relations can be extended to more general measurements, as discussed in Sec.~\ref{sec:arbitrary_povm}. Finally, another interesting extension considers more than two observables (which in some cases leads to tighter bounds for two observables), as discussed in Sec.~\ref{sec:mult_obs}.


\subsubsection{Tighter bounds for qubits}\label{sec:tighter_bounds}

Various attempts have been made to strengthen the Maassen--Uffink bound, particularly in the Shannon-entropy form~\eqref{eq:shannonUR}. Let us begin by first discussing improvements upon~\eqref{eq:shannonUR} in the qubit case and then move on to arbitrary dimensions.

For qubits the situation is fairly simple since the overlap matrix $[c_{xz}]$ only depends on a single parameter, which we can take as the maximum overlap $c = \max_{x,z} c_{xz}$. Hence, the goal is to find the largest function of $c$ that still lower-bounds the entropic sum. Significant progress along these lines was made by~\textcite{sanchez98}, who noted that the Maassen-Uffink bound, $q_{\rm MU}$, could be replaced by the stronger bound
\begin{align}\label{eqnqSR}
q_{\rm SR}:= h_{\bin} \bigg( \frac{1+\sqrt{2c-1}}{2}  \bigg)\,.
\end{align}
Here, $h_{\bin}(p) := -p \log p - (1-p)\log (1-p)$ denotes the binary entropy.

\begin{figure}
\begin{center}
\begin{overpic}[width=2.6in]{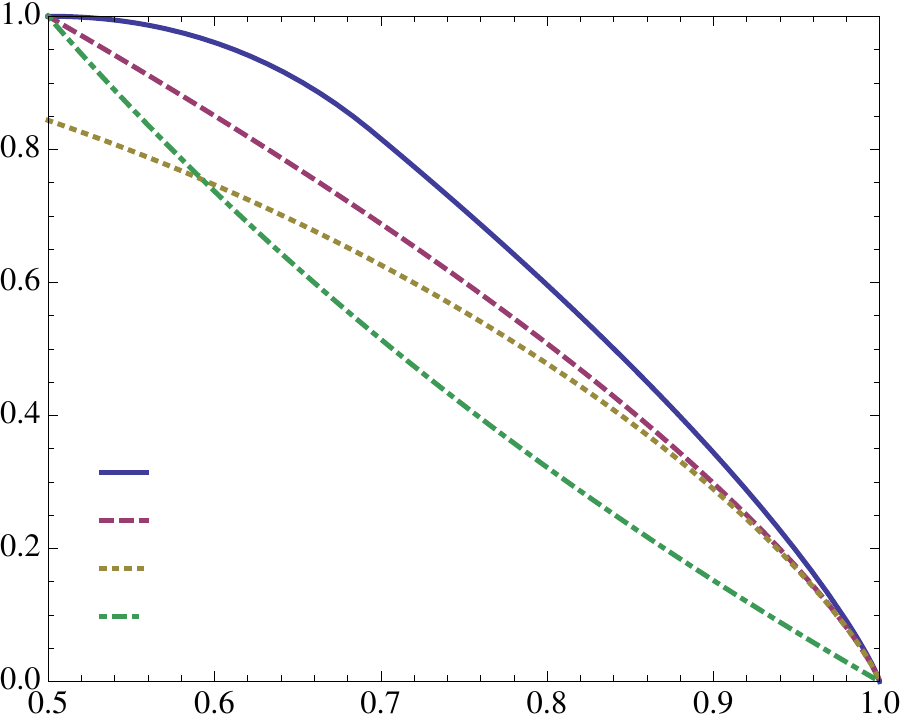}
\put(19,27){$q_{\textrm{opt}}${\footnotesize, Eq.~\eqref{eqnGhir1}}}
\put(19,21.7){$q_{\textrm{SR}}${\footnotesize, Eq.~\eqref{eqnqSR}}}
\put(19,16.3){$q_{\textrm{maj}}${\footnotesize, Eq.~\eqref{eq:vn-majo}}}
\put(19,11){$q_{\textrm{MU}}${\footnotesize, Eq.~\eqref{eq:shannonUR}}}
\put(-7,10){\rotatebox{90}{\footnotesize uncertainty, $q = H(X) + H(Z)$}}
\put(40,-4){\footnotesize overlap, $c$}
\put(57,59){\footnotesize allowed region}
\end{overpic}
\vspace{0.17cm}
\caption{Plot of various literature bounds on entropic uncertainty for qubit orthonormal bases, as a function of the maximum overlap $c$. The region above $q_{\rm opt}$ contains  pairs $(c,q)$ that can be achieved by quantum mechanics.}
\label{fgrqubit}
\end{center}
\end{figure}
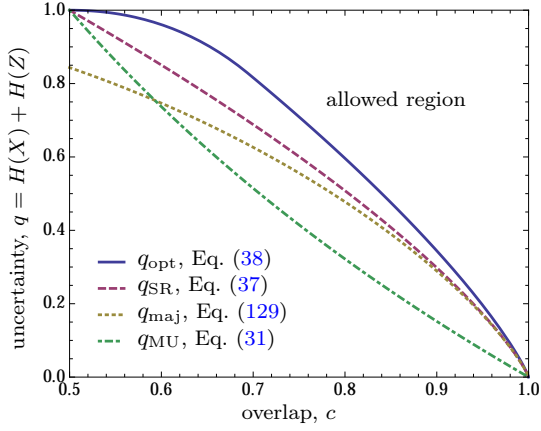

Later work by \textcite{ghirardi03} attempted to find the optimal bound. They simplified the problem to a single-parameter optimization as
\begin{align}\label{eqnGhir1}
q_{\rm opt}:= \min_{\theta} \left( h_{\bin} \left( \frac{1+\cos \theta}{2}  \right) + h_{\bin} \left( \frac{1+\cos (\alpha - \theta)}{2}  \right) \right)
\end{align}
where $\alpha := 2 \arccos \sqrt{c}$. While it is straightforward to perform this optimization, \textcite{ghirardi03} noted that an analytical solution could only be found for $c \gtrsim 0.7$. They showed that this analytical bound is given by
\begin{align}
q_{\rm G}:= 2 h_{\bin} (b) , \quad c \gtrsim 0.7\,,&\\
\label{eqnbdef}\text{where }b:=\bigg( \frac{1+\sqrt{c}}{2}\bigg)&\,.
\end{align}
Fig.~\ref{fgrqubit} shows a plot of $q_{\rm opt}$, $q_{\rm SR}$, and $q_{\rm MU}$. In addition, this plot also shows the bound $q_{\rm maj}$ obtained from a majorization technique discussed in Sec.~\ref{sec:major}.

For pairs of R\'enyi entropies $H_{\alpha}$ and $H_{\beta}$ in~\eqref{eq:mu_relation}, \textcite{zozor13} and \textcite{abdelkhalek15} completely characterized the amount of uncertainty in the qubit case.

\subsubsection{Tighter bounds in arbitrary dimension}\label{sec:tighter_bounds_d}

Extending the qubit result from~\eqref{eqnGhir1}, \textcite{devicente08} found an analytical bound in the large overlap (i.e., large $c$) regime
\begin{align}\label{eqndVSR}
q_{\rm dVSR}:= 2 h_{\bin} (b)\quad\mathrm{for}\quad c\gtrsim 0.7\,,
\end{align}
which is stronger than the MU bound over this range, and they also obtained a numerical improvement over MU for the range $1/2 \leq c \lesssim 0.7$. 

However, the situation for $d>2$ is more complicated than the qubit case. For $d>2$ the overlap matrix $[c_{xz}]$ depends on more parameters than simply the maximum overlap $c$. Recent work has focused on exploiting these other overlaps to improve upon the MU bound. For example, \textcite{coles14} derived a simple improvement on $q_{\rm MU}$ that captures the role of the second-largest entry of $[c_{xz}]$, denoted $c_2$, with the bound
\begin{align}
q_{\rm CP}:= \log\frac{1}{c}+ \frac{1}{2}(1-\sqrt{c})\log\frac{c}{c_2}\,.
\end{align}
Consider the following qutrit example where $q_{\rm CP} > q_{\rm MU}$.

\begin{example}\label{exQutrit}
Let $d=3$ and consider the two orthonormal bases $\bX$ and $\bZ$ related by the unitary transformation,
\begin{align}\label{eqnU3930}
U=\begin{pmatrix}
1/\sqrt{3} &  1/\sqrt{3}& 1/\sqrt{3}  \\
1/\sqrt{2}  & 0 & -1/\sqrt{2}\\
1/\sqrt{6} & -\sqrt{2/3} &1/\sqrt{6} 
\end{pmatrix}\,.
\end{align}
We have $q_{\rm MU} = \log (3/2)\approx 0.58$ while $q_{\rm CP} \approx 0.64$.
\end{example}

More recently, a bound similar in spirit to $q_{\rm CP}$ was obtained by~\textcite{rudnicki14}, of the form
\begin{align}
q_{\rm RPZ}:= \log\frac{1}{c} - \log \left( b^2 +\frac{c_2}{c}(1-b^2)\right) \,.
\end{align}
Note that $q_{\rm RPZ} \geq q_{\rm MU}$. However, there is no clear relation between $q_{\rm CP}$ and $q_{\rm RPZ}$.

For arbitrary pairs of entropies $H_{\alpha}$ and $H_{\beta}$, \textcite{abdelkhalek15} give conditions on the minimizing state of~\eqref{eq:mu_relation}. In particular, the minimizing state is pure and real. For measurements in the standard and Fourier basis, further conditions are obtained.


\subsubsection{Tighter bounds for mixed states}

Notice that~\eqref{eq:shannonUR} can be quite loose for mixed states. For example, if $\rho_A = \1 /d$, then the left-hand side of~\eqref{eq:shannonUR} is $2 \log d$, whereas the right-hand side is at most $\log d$. This looseness can be addressed by introducing a state-dependent bound that gets larger as $\rho_A$ becomes more mixed. 
The mixedness of $\rho_A$ can be quantified by the von Neumann entropy $H(\rho_A)$, which we also denote by $H(A)_{\rho}$, defined by
\begin{align}\label{eqnvonneumann1}
H(\rho_A):= - \tr\big[\rho_A \log \rho_A\big] =  \sum_{j} \lambda_j \log \frac{1}{\lambda_j} \,,
\end{align}
where an eigenvalue decomposition of the state is given by $\rho_A = \sum_j \lambda_j \proj{\phi_j}_A$. Note that $0 \leq H(\rho_A) \leq \log d$, where $H(\rho_A) = 0$ for pure states and $H(\rho_A) = \log d$ for maximally mixed states. In the literature, the von Neumann entropy is sometimes also denoted using $S(A) = H(A)$. However, here we will follow the more common convention in quantum information theory. We note that the entropy never decreases when applying a projective measurement $\bX=\left\{\proj{\bX^{x}}\right\}_x$ to $\rho_A$, that is,
\begin{align}
H(\rho_A)\leq H(X)_P\quad\mathrm{with}\quad P_{X}(x)= \bra{\bX^{x}}\rho_A \ket{\bX^{x}}\,.
\end{align}
Equation~\eqref{eq:shannonUR} was strengthened for mixed states by~\textcite{berta10}, with the bound
\begin{align}\label{eq:shannonURmixed}
H(X)+H(Z) \geq q_{\rm MU}  + H(\rho_A) \,.
\end{align}
A proof of \eqref{eq:shannonURmixed} is given in App.~\ref{app:shannon_proof}; see also \textcite{frank12-2} for a direct matrix analysis proof. When $\bX$ and $\bZ$ are MUBs, this bound is tight for any state $\rho_A$ that is diagonal in either the $\bX$ or $\bZ$ basis.


\subsection{Arbitrary measurements}\label{sec:arbitrary_povm}

Many interesting measurements are not of the orthonormal basis form. For example, coarse-grained (degenerate) projective measurements are relevant to probing macroscopic systems. Also, there are other measurements that are informationally complete in the sense that their statistics allow one to reconstruct the density operator.

The most general description of measurements in quantum mechanics is that of positive operator-valued measures (POVMs). A POVM on a system $A$ is a set of positive semidefinite operators $\{\bX^{x}\}$ that sum to the identity, $\sum_x \bX^x = \1_A$. The number of POVM elements in the set can be much larger or much smaller than the Hilbert space dimension of the system. Physically, a POVM can be implemented as a projective measurement on an enlarged Hilbert space, e.g., as a joint measurement on the system of interest with an ancilla system.

For two POVMs $\bX = \{ \bX^{x}\}_x$ and $\bZ = \{\bZ^{z} \}_z$, the general Born rule now induces the distributions
\begin{align}
P_X(x) = \tr\big[\rho_A \bX^x\big] \quad \textrm{and} \quad
P_Z(z) = \tr\big[\rho_A \bZ^z\big] \,. \label{eq:povm-pm-def}
\end{align}
\textcite{krishna01} proposed an incompatibility measure for POVMs using the operator norm. Namely, they considered
\begin{align}\label{eqncjkPOVM}
c= \max_{x,z} c_{xz}\quad\mathrm{with}\quad c_{xz} = \Big\|\sqrt{\bX^{x}}\sqrt{\bZ^{z}} \Big\|^2\,,
\end{align}
where $\| \cdot \|$ denotes the operator norm (i.e., the maximal singular value). Using this measure they generalized~\eqref{eq:shannonUR} to the case of POVMs. That is, we still have
\begin{align}\label{eqnKPpovm}
H(X)+H(Z) \geq\log\frac{1}{c} \,,
\end{align}
but now using the generalized version of $c$ in~\eqref{eqncjkPOVM}. More recently, \textcite{mythesis} noted that an alternative generalization to POVMs is obtained by replacing $c$ with
\begin{align}\label{eqncPOVMprime}
c' := \min \left\{ \max_x  \bigg\| \sum_z \bZ^{z} \bX^{x} \bZ^{z} \bigg\|,\ \max_z  \bigg\| \sum_x \bX^{x} \bZ^{z} \bX^{x}\bigg\| \right\}\,,
\end{align}
and the author conjectured that $c'$ always provides a stronger bound than $c$.

Indeed this conjecture was proved by~\textcite{coles14}:
\begin{align}\label{eqncPOVMlemma}
\bigg\| \sum_z \bZ^{z} \bX^{x} \bZ^{z} \bigg\| \leq \max_{z} c_{xz}\,.
\end{align}
Hence, $c' \leq c$, implying that $\log(1/c')$ provides a stronger bound on entropic uncertainty than $\log(1/c)$.

\begin{example}\label{exPOVM}
Consider two POVMs given by
\begin{align}
\bX = \bZ = \frac12 \Big\{\dya{0}, \dya{1}, \dya{+}, \dya{-}\Big\}\,.
\end{align}
For these POVMs we find $c = 1/4$, but $c' = 3/16$ is strictly smaller.
\end{example}

Interestingly, a general POVM can have a non-trivial uncertainty relation on its own. That is, for some POVM $\bX$, there may not exist any state $\rho_A$ that has $H(X)=0$. \textcite{krishna01} noted this and derived the single POVM uncertainty relation
\begin{align}\label{eqnKPsinglepovm}
H(X)  \geq - \log \max_x \Big\| \bX^{x} \Big\|\,.
\end{align}
In fact the proof is straightforward: simply apply \eqref{eqnKPpovm} to the case where $\bZ = \{\1 \}$ is the trivial POVM. The relation \eqref{eqnKPsinglepovm} can be further strengthened by applying this approach to $c'$ in \eqref{eqncPOVMprime}, instead of $c$.


\subsection{State-dependent measures of incompatibility}\label{sec:state-dep}

In most uncertainty relations we have encountered so far, the measure of incompatibility, for example the overlap~$c$, is a function of the measurements employed but is independent of the quantum state prior to measurement. The sole exception is the strengthened Maassen-Uffink relation in~\eqref{eq:shannonURmixed} where the lower bound is the sum of an ordinary, state-independent measure of incompatibility and the entropy of $\rho_A$. In the following, we review some uncertainty relations that use measures of incompatibility that are \emph{state dependent}.

It was shown by~\textcite{haenggi11} that the Maassen-Uffink relation~\eqref{eq:shannonUR} also holds when the overlap $c$ is replaced by an effective overlap, denoted $c^*$. Informally, $c^*$ is given by the average overlap of the two measurements on different subspaces of the Hilbert space, averaged over the probability of finding the state in the subspace. We refer the reader to the paper mentioned above for a formal definition of $c^*$. Here, we discuss a simple example showing that state-dependent uncertainty relations can be significantly tighter.

\begin{example}\label{exStateDepend}
Let us apply one out of two projective measurements, either in the orthonormal basis\footnote{The diagonal states are $\ket{\pm} = (\ket{0} \pm \ket{1})/\sqrt{2}$.}
\begin{align}
\Big\{ \ket{0}, \ket{1}, \ket{\!\perp} \Big\}\quad\mathrm{or}\quad\Big\{
\ket{+}, \ket{-}, \ket{\!\perp} \Big\}\,,
\end{align}
on a state $\rho$ which has the property that `$\perp$' is measured with probability at most $\eps$. The Maassen-Uffink relation~\eqref{eq:shannonUR} gives a trivial bound as the overlap of the two bases is $c = 1$ due to the vector $\ket{\!\perp}$ that appears in both bases. Still, our intuitive understanding is that the uncertainty about the measurement outcome is high as long as $\eps$ is small. The \emph{effective overlap}~\cite{haenggi11} captures this intuition: 
\begin{align}
c^* = (1 - \eps) \frac{1}{2} + \eps\,.
\end{align} 
This formula can be interpreted as follows: with probability $1-\eps$ we are in the subspace spanned by $\ket{0}$ and $\ket{1}$, where the overlap is $1/2$, and with probability $\eps$ we measure $\perp$ and have full overlap.
\end{example}

An alternative approach to state-dependent uncertainty relations was introduced by~\textcite{coles14}. They showed that the factor $q_{\rm MU} = \log (1/c)$ in the Maassen-Uffink relation~\eqref{eq:shannonUR} can be replaced by the state-dependent factor
\begin{align}\label{eqncolespianistatedepend}
q(\rho_A) &:= \max \{q_{X}(\rho_A), q_{Z}(\rho_A)\}\,, \quad \text{where}\\
 q_{X}(\rho_A) &:= \sum_x P_X(x) \log \frac{1}{\max_z c_{xz} }, 
\end{align} 
and $q_{Z}(\rho_A)$ is defined analogously to $q_{X}(\rho_A)$, but with $x$ and $z$ interchanged. Here, $P_X(x)$ and $c_{xz}$ are given by \eqref{eq:mu-pm-def} and \eqref{eq:overlap}, respectively. Note that this strengthens the Maassen-Uffink bound, $q(\rho_A)\geq q_{\rm MU}$, since averaging $\log (1/\max_z c_{xz})$ over all $x$ is larger than minimizing it over all $x$. In many cases $q(\rho_A)$ is significantly stronger than $q_{\rm MU}$.

Recently, \textcite{kaniewski14} derived entropic uncertainty relations in terms of the \emph{effective anti-commutator} of arbitrary binary POVMs $\bX = \{\bX^0, \bX^1\}$ and $\bZ = \{ \bZ^0, \bZ^1\}$. Namely, the quantity
\begin{align}\label{eqneffectiveanticommutator}
\eps^* &= \frac12 \tr \big[ \rho [ O_{\bX}, O_{\bZ} ]_{+} \big] = \frac12 \tr \big[ \rho (O_{\bX} O_{\bZ} + O_{\bZ} O_{\bX}) \big]\,, \notag\\
& \textrm{with} \quad O_{\bX} = \bX^0 - \bX^1 \quad \textrm{and} \quad O_{\bZ} = \bZ^0 - \bZ^1
\end{align}
binary observables corresponding to the POVMs $\bX$ and $\bZ$, respectively. In \eqref{eqneffectiveanticommutator}, we use the notation $[ \cdot, \cdot ]_{+}$ to denote the anti-commutator. We note that $\eps^* \in [-1, 1]$. This results, for example, in the following uncertainty relation for the Shannon entropy:
\begin{align}\label{eq:kaniewski}
H(X) + H(Z) \geq h_{\rm bin} \left( \frac{1+\sqrt{|\eps^*|}}{2} \right)\,.
\end{align}
We refer the reader to~\textcite{kaniewski14} for similar uncertainty relations in terms of R\'enyi entropies as well as extensions to more than two measurements. Finally, for measurements acting on qubits, we find that $|\eps^*| = 2c - 1$, and~\eqref{eq:kaniewski} hence reduces to the Sanchez-Ruiz bound~\eqref{eqnqSR}.


\subsection{Relation to guessing games}\label{sec:games_cl}

Let us now explain in detail how some of the relations above can be interpreted in terms of a guessing game. We elaborate on the brief discussion of guessing games in Sec.~\ref{sec:introduction}, and we refer the reader back to Fig.~\ref{figguessingnomemory} for an illustration of the game. 

The game is as follows. Suppose that Bob prepares system $A$ in state $\rho_{A}$. He then sends $A$ to Alice, who randomly performs either the $\bX$ or $\bZ$ measurement. The measurement outcome is a bit denoted as $K$, and Bob's task is to guess $K$, given that he received the basis choice denoted by $\Theta\in\{\theta_{\bX},\theta_{\bZ}\}$ from Alice.

We can rewrite the Maassen-Uffink relation~\eqref{eq:shannonUR} in the following way such that the connection to the above guessing game becomes transparent. Denote the standard basis on $A$ as $\{\ket{k}\}_{k=1}^d$, and let $U_{\bX}$ and $U_{\bZ}$ respectively be unitaries that map this basis to the $\bX$ and $\bZ$ bases, i.e.,
\begin{align}
\ket{\bX^k} = U_{\bX} \ket{k}\quad\mathrm{and}\quad\ket{\bZ^k} = U_{\bZ} \ket{k}\,.
\end{align}
Then, we have
\begin{align}
\frac{1}{2}\Big(H(K|\Theta=\theta_{\bX})+H(K|\Theta=\theta_{\bZ})\Big)\geq\frac{1}{2}q_{\rm MU}\,,
\end{align}
with the conditional probability distribution
\begin{align}\label{eq:cond_prob}
P_{K|\Theta=\theta_{\bX}}(k):=\bra{k}U_{\bX}^\dagger\rho U_{\bX}\ket{k}\;\;\mathrm{for}\;\;k \in\{1,\ldots,d\}
\end{align}
and similarly for $\theta_{\bZ}$. Alternatively we can also write this as
\begin{align}\label{eq:multiple_m-u}
H(K|\Theta)\geq\frac{1}{2}q_{\rm MU}\;\;\text{with}\;\;\Theta\in\left\{\theta_{\bX},\theta_{\bZ}\right\}\,,
\end{align}
in terms of the conditional Shannon entropy
\begin{align}\label{eq:cshannon_chain111}
H(K|\Theta)&:=H(K\Theta)-H(\Theta)\\
&=\frac{1}{2}\Big(H(K|\Theta=\theta_{\bX})+H(K|\Theta=\theta_{\bX})\Big)\label{eq:cshannon_chain2}
\end{align}
of the bipartite distribution
\begin{align}\label{eqnkthetadistribution}
P_{K\Theta}(k,\theta_j):=\frac{1}{2}\bra{k}U_j^\dagger\rho U_j\ket{k}\;\;\mathrm{with}\;\;&k\in \{1,\ldots,d\} \notag\\
&j \in \{\bX, \bZ \}\,.
\end{align}
That is, each measurement labeled $\theta_j$ is chosen with equal probability $1/2$ and we condition on this choice. Notice that the form in~\eqref{eq:multiple_m-u} is connected to the guessing game in Fig.~\ref{figguessingnomemory}. Regardless of the state $\rho_A$ that Bob prepares, the uncertainty relation \eqref{eq:multiple_m-u} implies that he will not be able to perfectly guess $K$ if $q_{\rm MU}>0$. In this sense, the Maassen-Uffink relation is a \textit{fundamental constraint on one's ability to win a guessing game}.

Actually, in the context of guessing games, the min-entropy is more operationally relevant than the Shannon entropy. For example, a diligent reading of~\textcite{deutsch83} reveals the relation
\begin{align}\label{eqnguessingprobproduct}
p_{\rm guess}(X) \cdot p_{\rm guess}(Z) \leq b^2 \,,
\end{align}
for orthonormal bases $\bX$ and $\bZ$, where $b$ is defined in~\eqref{eqnbdef}. This relation gives an upper bound on the product of the guessing probabilities (or equivalently, a lower bound on the sum of the min-entropies) associated with $X$ and $Z$. However, to make a more explicit connection to the guessing game described above, one would like to upper bound the sum (or average) of the guessing probabilities, namely the quantity
\begin{align}\label{eqnguessingprobaverage}
p_{\rm guess}(K| \Theta) =  \frac{1}{2}\Big(p_{\rm guess}(K|\Theta=\theta_{\bX})+p_{\rm guess}(K|\Theta=\theta_{\bZ})\Big) \,.
\end{align}
Indeed, the quantity~\eqref{eqnguessingprobaverage} can easily be upper-bounded as~\cite{schaffner07}
\begin{align}
p_{\rm guess}(K| \Theta) &\leq b 
\quad\text{or equivalently}\label{eqnguessingprobaverage2}\\
H_{\min}(K| \Theta) &\geq \log \frac{1}{b}\ .\label{eqnguessingprobaverage3}
\end{align}

\begin{example}
For the Pauli qubit measurements $\{\sigma_{\bX},\sigma_{\bZ}\}$ the min-entropy uncertainty relation~\eqref{eqnguessingprobaverage3} becomes
\begin{align}\label{eqnguessingprobaveragequbit3}
H_{\min}(K| \Theta) \geq \log \frac{2\sqrt{2}}{1+\sqrt{2}}\,.
\end{align}
\end{example}

We emphasize that $p_{\rm guess}(K| \Theta)$ is precisely the probability for winning the game described in Fig.~\ref{figguessingnomemory}. Hence, the entropic uncertainty relation~\eqref{eqnguessingprobaverage3} gives \emph{the} fundamental limit on winning the game. Finally, we remark that~\eqref{eqnguessingprobaverage3} is stronger than Deutsch's relation~\eqref{eqnguessingprobproduct}, due to the following argument. For the min-entropy, conditioning on the measurement choice is defined as
\begin{align}
H_{\min}(K|\Theta)&:=-\log\left(\frac{1}{2}\sum_{j=1,2}2^{-H_{\min}(K|\Theta=\theta_j)}\right)\\
&\neq H_{\min}(K\Theta)-H_{\min}(\Theta)\quad \text{(in general)}\,,\notag
\end{align}
in contrast to the Shannon entropy in~\eqref{eq:cshannon_chain111}. However, in analogy to~\eqref{eq:cshannon_chain2}, we have
\begin{align}H_{\min}(K|\Theta)\leq\frac{1}{2}\sum_{j=1,2}H_{\min}(K|\Theta=\theta_j)\,.
\end{align}
due to the concavity of the logarithm. For a general discussion of conditional entropies we point to Sec.~\ref{sec:conditionalentropy}.


\subsection{Multiple measurements}\label{sec:mult_obs}

So far we have only considered entropic uncertainty relations quantifying the complementarity of two measurements. However, there is no fundamental reason for restricting to this setup, and in the following we discuss the more general case of $L$ measurements. We mostly focus on special sets of measurements that generate strong uncertainty relations. This is of particular interest for various applications in quantum cryptography (see Sec.~\ref{sec:bounded_noisy}).

The notation introduced for guessing games in Sec.~\ref{sec:games_cl} is particularly useful in the multiple measurements setting. In this notation, for larger sets of measurements we are interested in finding lower bounds of the form
\begin{align}\label{eq:multiple_generic}
H(K|\Theta)\geq f(\Theta,\rho_A)>0\;\;\text{with}\;\;\Theta\in\left\{\theta_1,\ldots,\theta_L\right\}\,,
\end{align}
where, similarly to~\eqref{eqnkthetadistribution},
\begin{align}
P_{K\Theta}(k,\theta_j):=\frac{1}{L}\bra{k}U_j^\dagger\rho U_j\ket{k}\;\;\mathrm{with}\;\;&k \in \{1,\ldots,d\} \notag\\
&j \in\{1,\ldots,L\}\,.
\end{align}
Again the left-hand side of~\eqref{eq:multiple_generic} might alternatively be written as
\begin{align}\label{eq:cshannon_chain}
H(K|\Theta)=\frac{1}{L}\sum_{j=1}^{L}H(K|\Theta=\theta_j)\,,
\end{align}
where the conditional probability distribution $P_{K|\Theta=\theta_j}$ is defined analogously to~\eqref{eq:cond_prob}.


\subsubsection{Bounds implied by two measurements}\label{sec:two_implied}

It is important to realize that, e.g., the Maassen-Uffink relation~\eqref{eq:shannonUR} already implies bounds for larger sets of measurements. This is easily seen by just applying \eqref{eq:shannonUR} to all possible pairs of measurements and adding the corresponding lower bounds.

\begin{example}
For the qubit Pauli measurements we find by an iterative application of the tightened Maassen-Uffink bound~\eqref{eq:shannonURmixed} for the measurement pairs $\{\sigma_{\bX},\sigma_{\bY}\}$, $\{\sigma_{\bX},\sigma_{\bZ}\}$, and $\{\sigma_{\bY},\sigma_{\bZ}\}$ that
\begin{align}\label{eq:pauli_mu}
H(K|\Theta)\geq\frac{1}{2}+\frac{1}{2}H(\rho_A)\;\;\mathrm{with}\;\;\Theta\in\left\{\sigma_{\bX},\sigma_{\bY},\sigma_{\bZ}\right\}\,.
\end{align}
\end{example}

The goal of this section is to find uncertainty relations that are stronger than any bounds that can be derived directly from relations for two measurements.


\subsubsection{Complete sets of MUBs}\label{sec:mub_relations}

A promising candidate for deriving strong uncertainty relations are complete sets of MUBs, i.e., sets of $d+1$ MUBs (which we only know to exist in certain dimensions, see Appendix~\ref{app:mub} for elaboration). Consider the qubit case in the following example.

\begin{example}
For the qubit Pauli measurements, we have from~\textcite{sanchez95,sanchez98} that
\begin{align}\label{eq:mub_qubitpauli}
H(K|\Theta)\geq\frac{2}{3}\;\;\mathrm{with}\;\;\Theta\in\left\{\sigma_{\bX},\sigma_{\bY},\sigma_{\bZ}\right\}\,.
\end{align}
Moreover, from~\textcite{coles10} we can add an entropy dependent term on the right-hand side,
\begin{align}\label{eq:mub_qubitpauli2}
H(K|\Theta)\geq\frac{2}{3}+\frac{1}{3}H(\rho_A)\;\;\mathrm{with}\;\;\Theta\in\left\{\sigma_{\bX},\sigma_{\bY},\sigma_{\bZ}\right\}\,.
\end{align}
Note that \eqref{eq:mub_qubitpauli2} is never a worse bound than~\eqref{eq:pauli_mu} which just followed from the tightened Maassen-Uffink relation for two measurements~\eqref{eq:shannonURmixed}. Moreover, the relation~\eqref{eq:mub_qubitpauli} becomes an equality for any eigenstate of the Pauli measurements, while~\eqref{eq:mub_qubitpauli2} becomes an equality for any state $\rho_A$ that is diagonal in the eigenbasis of one of the Pauli measurements.
\end{example}


More generally, for a full set of $d+1$ MUBs in dimension $d$, \textcite{larsen90,ivanovic92,sanchez93} showed that,
\begin{align}\label{eq:mub_full}
H(K|\Theta)\geq\log(d+1)-1\;\;\text{with}\;\;\Theta\in\{\theta_1,\ldots,\theta_{d+1}\}\,.
\end{align}
This is a strong bound since the entropic term on the left-hand side can become at most $\log d$ for any number and choice of measurements. The relation~\eqref{eq:mub_full} can be derived from an uncertainty equality for the collision entropy $H_{\rm coll}$. Namely, for any quantum state $\rho_A$ on a $d$-dimensional system and a full set of $d+1$ MUBs, we have~\cite{ivanovic92,brukner99,ballester07}
\begin{align}\label{eq:h_2meta_noqsi}
H_{\rm coll}(K|\Theta)=\log(d+1)-\log\left(2^{-H_{\rm coll}(\rho_A)}+1\right)&\notag\\
\text{with}\;\;\Theta\in\{\theta_1,\ldots,\theta_{d+1}\}&\,,
\end{align}
where for the collision entropy the conditioning on the measurement choice is defined as
\begin{align}
H_{\rm coll}(K|\Theta)&:=-\log\left(\frac{1}{L}\sum_{j=1}^{L}2^{-H_{\rm coll}(K|\Theta=\theta_j)}\right)\label{eq:collision_cond}\\
&\neq H_{\rm coll}(K\Theta)-H_{\rm coll}(\Theta)\quad \text{(in general)}\notag\,.
\end{align}
We refer to Sec.~\ref{sec:conditionalentropy} for a general discussion about conditional entropies. Moreover, the quantum collision entropy is a measure for how mixed the state $\rho_A$ is and defined as
\begin{align}
H_{\rm coll}(\rho_A):=-\log\tr\left[\rho^2_A\right]\,.
\end{align}
We emphasize that since~\eqref{eq:h_2meta_noqsi} is an equality it is tight for every state. By the concavity of the logarithm we also have in analogy to the Shannon entropy~\eqref{eq:cshannon_chain},
\begin{align}\label{eq:h2_equalitync}
H_{\rm coll}(K|\Theta)\leq\frac{1}{d+1}\sum_{j=1}^{d+1}H_{\rm coll}(K|\Theta=\theta_j)\,.
\end{align}

\begin{example}
For the qubit Pauli measurements, \eqref{eq:h_2meta_noqsi} yields
$H_{\rm coll}(K|\Theta)=\log3-\log\left(2^{-H_{\rm coll}(\rho_A)}+1\right)$
with $\Theta\in\{\sigma_{\bX},\sigma_{\bY},\sigma_{\bZ}\}$.
\end{example}

The uncertainty relation~\eqref{eq:mub_full} for the Shannon entropy follows from~\eqref{eq:h_2meta_noqsi} by at first only considering pure states, i.e., states with $H_{\rm coll}(\rho_A)=0$, and using that the R\'enyi entropies are monotonically decreasing as a function of the parameter $\alpha$ (note that the collision entropy corresponds to $\alpha=2$ and the Shannon entropy to $\alpha=1$). For mixed states $\rho_A$ we can extend this in a second step by taking the eigendecomposition and making use of the concavity of the Shannon entropy. For later purposes we note that it is technically often accessible to work with the collision entropy $H_{\rm coll}$ (even when ultimately interested in uncertainty relations in terms of other entropies).

The uncertainty relation~\eqref{eq:mub_full} was improved for $d$ even to~\cite{sanchez95},
\begin{align}\label{eq:mub_d+1}
H(K|\Theta)\geq\frac{1}{d+1}\left(\frac{d}{2}\log\left(\frac{d}{2}\right)+\left(\frac{d}{2}+1\right)\log\left(\frac{d}{2}+1\right)\right)&\notag\\
\text{with}\;\;\Theta\in\{\theta_1,\ldots,\theta_{d+1}\}&\,,
\end{align}
Note that this relation generalizes the qubit result in~\eqref{eq:mub_qubitpauli} to arbitrary dimensions.

Furthermore, uncertainty relations for a full set of $L = d+1$ MUBs can also be expressed in terms of the extrema of Wigner functions~\cite{wooters07,mandayam10}.


\subsubsection{General sets of MUBs}\label{sec:mub_relations2}

At first glance, one might think that measuring in mutually unbiased bases always results in a large amount of uncertainty. Somewhat surprisingly, 
this is not the case.  In fact, \textcite{ballester07} show that for $d=p^{2l}$ with $p$ prime and $l\in\mathbb{N}$, there exist up to $L=p^l+1$ many MUBs together with a state $\rho_A$ for which
\begin{align}\label{eq:ballester}
H(K|\Theta)=\frac{\log d}{2}\;\;\text{with}\;\;\Theta\in\{\theta_1,\ldots,\theta_{L}\}\,. 
\end{align}
That is, we observe no more uncertainty than if we had just considered two incompatible measurements. While a certain amount of mutual unbiasedness is therefore a necessary condition for strong uncertainty relations, it is in general not sufficient.

For smaller sets of $L<d+1$ MUBs we immediately get a weak bound from an iterative application of the Maassen-Uffink relation~\eqref{eq:shannonUR} for MUBs,
\begin{align}\label{eq:L_mu}
H(K|\Theta)\geq\frac{\log d}{2}\;\;\text{with}\;\;\Theta\in\{\theta_1,\ldots,\theta_{L}\}\,.
\end{align}
It turns out that the bound~\eqref{eq:L_mu} cannot be improved much in general, as the following example shows.

\begin{example}
In $d=3$, \textcite{wehner09} showed that there exists a set of $L=3$ MUBs together with a state $\rho_A$ such that
$H(K|\Theta)=1$ for $\Theta\in\{\theta_1,\theta_2,\theta_3\}$.
This only allows a relatively weak uncertainty relation. \textcite{wu09} showed that
\begin{align}\label{eq:wu_qubit}
H(K|\Theta)\geq\frac{8}{9}\approx0.89\;\;\mathrm{with}\;\;\Theta\in\{\theta_1,\theta_2,\theta_3\}\,.
\end{align}
This is slightly stronger than the lower bound from~\eqref{eq:L_mu}:
\begin{align}
H(K|\Theta)\geq\frac{\log3}{2}\approx0.79\;\;\mathrm{with}\;\;\Theta\in\{\theta_1,\theta_2,\theta_3\}\,.
\end{align}
\end{example}

Generally this only allows relatively weak uncertainty relations if $L<d+1$. \textcite{wu09} showed that
\begin{align}\label{eq:wu_master}
H_{\rm coll}(K|\Theta)\geq-\log\frac{d\cdot2^{-H_{\rm coll}(\rho_A)}+L-1}{L\cdot d}&\notag\\
\text{with}\;\;\Theta\in\{\theta_1,\ldots,\theta_{L}\}&\,.
\end{align}
This implies in particular the Shannon entropy relation~\cite{azarchs04},
\begin{align}\label{eq:azarchs}
H(K|\Theta)\geq-\log\frac{d+L-1}{L\cdot d}\;\;\text{with}\;\;\Theta\in\{\theta_1,\ldots,\theta_{L}\}\,,
\end{align}
see also~\textcite{wehner09} for an elementary proof. For comparison, with $L=d=3$, \eqref{eq:azarchs} yields
\begin{align}
H(K|\Theta)\geq\log\frac{9}{5}\approx0.85\;\;\mathrm{with}\;\;\Theta\in\{\theta_1,\theta_2,\theta_3\}\,,
\end{align}
which is between~\eqref{eq:L_mu} and~\eqref{eq:wu_qubit}. Additional evidence that general sets of less than $d+1$ MUBs in dimension $d$ only generate weak uncertainty relations is presented by~\textcite{ambainis10,divincenzo04,ballester07}. Many of the findings also extend to the setting of approximate mutually unbiased bases~\cite{hayden04b}. 


In terms of the min-entropy, \textcite{mandayam10} show that for measurements in $L$ possible MUBs the following two bounds hold
\begin{align}
\frac{1}{L}\sum_{\theta=1}^L H_{\min}(K|\Theta= \theta) \geq - \log \left[\frac{1}{d}\left(1 + \frac{d-1}{\sqrt{L}}\right)\right]\ ,\\
\frac{1}{L}\sum_{\theta=1}^L H_{\min}(K|\Theta= \theta) \geq - \log \left[\frac{1}{L}\left(1 + \frac{L-1}{\sqrt{d}}\right)\right]\ .
\end{align}
Each of these is better in certain regimes, and the latter can indeed be tight. 
They also study uncertainty relations for certain classes of MUBs that exhibit special symmetry properties. It remains an interesting topic to study uncertainty relations for MUBs and in Sec.~\ref{sec:mum} we present some related results of~\textcite{kalev07}.


\subsubsection{Measurements in random bases}\label{sec:random_bases}

Another candidate for strong uncertainty relations are sets of measurements that are chosen at random.\footnote{By ``at random'' we mean according to the Haar measure on the unitary group, see, e.g., \cite{hayden04b} for more details.} Extending on the previous results of~\textcite{hayden04b}, \textcite{fawzi11} show that in dimension $d$ there exist any number of $L>2$ measurements and a universal constant $C$ (independent of $d$ and $L$) such that,
\begin{align}\label{eq:fawzi_probabilistic}
H(K|\Theta)\geq\log d\cdot\left(1-\sqrt{\frac{1}{L}\cdot C\log(L)}\right)-g(L)&\notag\\
\mathrm{with}\;\;\Theta\in\{\theta_1,\dots,\theta_L\}\,,
\end{align}
with the fudge term $g(L)=O\left(\log\left(L/\log(L)\right)\right)$. Note that for any set of $L$ measurements there exists a state such that
\begin{align}\label{eq:many_converse}
H(K|\Theta)\leq\log d\cdot\left(1-\frac{1}{L}\right)\;\;\mathrm{with}\;\;\Theta\in\{\theta_1,\dots,\theta_L\}\,.
\end{align}
Hence, the relation~\eqref{eq:fawzi_probabilistic} is already reasonably strong. However, very recently~\eqref{eq:fawzi_probabilistic} was improved by proving a conjecture stated by~\textcite{wehner09}. Namely, \textcite{adamczak14} showed that in dimension $d$ there exist any number of $L>2$ measurements and a universal constant $D$ (independent of $d$ and $L$) such that,
\begin{align}\label{eq:adamczak14_multiple}
H(K|\Theta)\geq\log d\cdot\left(1-\frac{1}{L}\right)-D&\notag\\
\mathrm{with}\;\;\Theta\in\{\theta_1,\dots,\theta_L\}&\,.
\end{align}
We emphasize that this matches the upper bound~\eqref{eq:many_converse} up to the constant $D$.

The downside with the relations~\eqref{eq:fawzi_probabilistic} and~\eqref{eq:adamczak14_multiple}, however, is that the measurements are not explicit. This is an issue for applications. In particular, it is computationally inefficient to sample from the Haar measure. \textcite{fawzi11} showed that the measurements in their relation~\eqref{eq:fawzi_probabilistic} can be made explicit and efficient if the number $L$ of measurements is small enough. More precisely, for $n$ qubits (with $n$ sufficiently large) and $\eps>0$, there exists a constant $C$ and a set of
\begin{align}
L\leq\left(n/\eps\right)^{C\log(1/\eps)}
\end{align}
measurements generated by unitaries computable by quantum circuits of size $O(\mathrm{polylog}n)$ such that
\begin{align}\label{eq:locking_ex}
H(K|\Theta)\geq n\cdot(1-2\eps)-h_{\bin}(\eps)\;\;\mathrm{with}\;\;\Theta\in\{\theta_1,\dots,\theta_L\}\,,
\end{align}
where $h_{\bin}$ denotes the binary entropy. The relation~\eqref{eq:locking_ex} will be the basis for the information locking schemes presented in Sec.~\ref{sec:locking}.


\subsubsection{Product measurements on multiple qubits}\label{sec:product_measurements_no_memory}

For applications in cryptography we usually need uncertainty relations for measurements that can be implemented locally, so-called product measurements. For example, for an $n$-qubit state we are interested in uncertainty relations for the set of $2^n$ different measurements given by measuring each qubit independently in one of the two Pauli bases $\sigma_{\bX}$ or $\sigma_{\bZ}$. These are often called BB84 measurements due to the work of \textcite{bb84}. Using the Maassen-Uffink bound~\eqref{eq:shannonUR} for two measurements iteratively we immediately find
\begin{align}\label{eq:bb84_h1}
H(K^n|\Theta^n)\geq n\cdot\frac{1}{2}\;\;\mathrm{with}\;\;\Theta^n\in\{\theta_1,\ldots,\theta_{2^n}\}\,.
\end{align}
This relation is already tight since there exist states that achieve equality.

For cryptographic applications, the relevant measure is often not the Shannon entropy but the min-entropy. The one qubit relation~\eqref{eqnguessingprobaveragequbit3} is easily extended to $n$ qubits as
\begin{align}\label{eq:schaffner_min}
H_{\min}(K^n|\Theta^n)\geq-n\cdot\log\left(\frac{1}{2}+\frac{1}{2\sqrt{2}}\right)\approx n\cdot0.22&\notag\\
\;\;\mathrm{with}\;\;\Theta^n\in\{\theta_1,\ldots,\theta_{2^n}\}&\,.
\end{align}
Again there exist states that achieve equality. More generally~\textcite{nelly12} find for $n$ qubit BB84 measurements and the R\'enyi entropy of order $\alpha\in(1,2]$,
\begin{align}\label{eq:bb84_halpha}
H_\alpha(K^n|\Theta^n)\geq n\cdot\frac{\alpha-\log\left(1+2^{\alpha-1}\right)}{\alpha-1}&\notag\\
\;\;\mathrm{with}\;\;\Theta^n\in\{\theta_1,\ldots,\theta_{2^n}\}&\,,
\end{align}
where the conditioning is given as (see App.~\ref{app:renyi}),\footnote{We emphasize that unlike in the unconditional case $H_2(K|\Theta)\neq H_{\rm coll}(K|\Theta)$ and hence~\eqref{eq:collision_cond} is different from~\eqref{eq:alpha_basiscond} for $\alpha=2$.}
\begin{align}\label{eq:alpha_basiscond}
H_\alpha(K|\Theta)=\frac{\alpha}{1-\alpha}\log\left(\frac{1}{L}\sum_{j=1}^L2^{\frac{1-\alpha}{\alpha}H_\alpha(K|\Theta=\theta_j)}\right)\,.
\end{align}

Similarly, we find for the set of $3^n$ different measurements given by measuring each qubit independently in one of the three Pauli bases $\sigma_{\bX}$, $\sigma_{\bY}$, or $\sigma_{\bZ}$ that
\begin{align}\label{eq:sixstate_uncond}
H(K^n|\Theta^n)\geq n\cdot\frac{2}{3}\;\;\mathrm{with}\;\;\Theta^n\in\{\theta_1,\ldots,\theta_{3^n}\}\,,
\end{align}
Following~\textcite{bruss98} these measurements are often called six-state measurements. The uncertainty relation~\eqref{eq:sixstate_uncond} is the extension of~\eqref{eq:mub_qubitpauli} from one to $n$ qubits. More general relations in terms of R\'enyi entropies were again derived by~\textcite{nelly12}.

Approximate extensions of all these relations when the measurements are not exactly given by the Pauli measurements $\{\sigma_{\bX},\sigma_{\bY},\sigma_{\bZ}\}$ are discussed by~\textcite{kaniewski14}. We note that some extensions of the $n$ qubit relations discussed above will be crucial for applications in two-party cryptography (Sec.~\ref{sec:bounded_noisy}).


\subsubsection{General sets of measurements}\label{sec:general_multi_measurements}

\textcite{liu15} give entropic uncertainty relations for general sets of measurements. Their bounds are qualitatively different than just combining~\eqref{eq:shannonUR} iteratively and sometimes become strictly stronger in dimension $d>2$. For simplicity we only state the case of $L=3$ measurements (in any dimension $d\geq2$),
\begin{align}\label{eq:liu_multiple}
H(K|\Theta)\geq\frac{1}{3}\log\frac{1}{m}+\frac{2}{3}H(\rho_A)&\notag\\
\mathrm{with}\;\;\Theta\in\{V^{(1)},V^{(2)},V^{(3)}\}&\,,
\end{align}
and the multiple overlap constant
\begin{align}\label{eq:b_constant}
m:=\max_k\left(\sum_j\max_i\left(c\left(v_i^1,v_j^2\right)\right)\cdot c\left(v_j^2,v_k^3\right)\right)\,,
\end{align}
and $\{\ket{v_i^1}\}$, $\{\ket{v_j^2}\}$, $\{\ket{v_k^3}\}$ are the eigenvectors of $V^{(1)}$, $V^{(2)}$, $V^{(3)}$, respectively.

\begin{example}
For a qubit and the full set of $3$ MUBs given by the Pauli measurements this gives
\begin{align}
H(K|\Theta)\geq\frac{1}{3}+\frac{2}{3}H(\rho_A) \;\;\mathrm{with}\;\;\Theta\in\{\sigma_{\bX},\sigma_{\bY},\sigma_{\bZ}\}\,.
\end{align}
This bound is, however, weaker than~\eqref{eq:pauli_mu} and~\eqref{eq:mub_qubitpauli2}. On the other hand, of course the whole point of the bound~\eqref{eq:liu_multiple} is that in contrast to~\eqref{eq:pauli_mu} and~\eqref{eq:mub_qubitpauli2} it can be applied to any set of $L=3$ measurements (in arbitrary dimension).
\end{example}

We refer to~\cite{liu15} for a fully worked out example where their bound can become stronger than any bounds implied by two measurement relations.


\subsubsection{Anti-commuting measurements}\label{sec:anti-commuting}

As already noted in Sec.~\ref{sec:arbitrary_povm}, many interesting measurements are not of the orthonormal basis form, but are more generally described by POVMs. One class of such measurements that generate maximally strong uncertainty relations are sets of anti-commuting POVMs with only two possible measurement outcomes. In more detail, we consider a set $\{\bX_1,\ldots,\bX_L\}$ of binary POVMs $\bX_j=\{\bX_j^0,\bX_j^1\}$ that generate binary observables
\begin{align}\label{eq:higher_pauli}
O_j:=\bX_j^0-\bX_j^1\;\;\mathrm{with}\;\;  [O_j,O_k ]_{+}=2\delta_{jk}\,,
\end{align}
where, as in \eqref{eqneffectiveanticommutator}, $[\cdot , \cdot ]_{+}$ denotes the anti-commutator.\footnote{An example of such anti-commuting sets in the case of $L=3$ is provided by the qubit Pauli operators $\{\sigma_{\bX},\sigma_{\bY},\sigma_{\bZ}\}$.} The goal is then to find lower bounds on entropies of the form $H(K|\Theta)$ with
\begin{align}\label{eq:povm_notation}
P_{K\Theta}(k,\bX_j):=\frac{1}{L}\tr\left[\bX_j^k\rho_A\right]\;\;\mathrm{with}\;\;&k\in \{0,1\}\notag\\
&j \in \{1,\ldots,L\} \,.
\end{align}
For simplicity we only discuss the case of $n$ qubit states for which we have sets of up to $2n+1$ many binary anti-commuting POVMs.\footnote{This is obtained by the unique Hermitian representation of the Clifford algebra via the Jordan-Wigner transformation~\cite{dietz06}.} \textcite{wehnerwinter08} then show that
\begin{align}
&H(K|\Theta)\geq1-\frac{1}{L}\;\;\mathrm{with}\;\;k \in \{0,1\}\notag\\
&\text{for any subset $\Theta\subseteq\{\bX_1,\ldots,\bX_{2n+1}\}$ of size $L$}\,.
\end{align}
These relations are tight and reduce for the $L=3$ qubit Pauli measurements $\{\sigma_{\bX},\sigma_{\bY},\sigma_{\bZ}\}$ to the bound~\eqref{eq:mub_qubitpauli}. Similarly~\textcite{wehnerwinter08} also find for the collision entropy,
\begin{align}\label{eq:anti_h2}
\frac{1}{L}\sum_{\bX_j\in\Theta}H_{\rm coll}(K|\Theta=\bX_j)\geq1-\log\left(1+\frac{1}{L}\right)\,,
\end{align}
and the min-entropy,
\begin{align}
\frac{1}{L}\sum_{\bX_j\in\Theta}H_{\min}(K|\Theta=\bX_j)\geq1-\log\left(1+\frac{1}{\sqrt{L}}\right)\,.
\end{align}
These relations are again tight. Note, however, that the average over the basis choice is outside of the logarithm, whereas for the collision and the min-entropy the average is more naturally inside of the logarithm as, e.g., in~\eqref{eq:h_2meta_noqsi} and in~\eqref{eq:schaffner_min}.

\begin{example}
For the $L=3$ qubit case~\eqref{eq:anti_h2} reduces to
\begin{align}
\frac{1}{3}\sum_{j=X,Y,Z}^{}H_{\rm coll}(K|\Theta=\sigma_j)\geq\log3-1\,,
\end{align}
which, as seen by~\eqref{eq:h2_equalitync}, is generally weaker than the corresponding bound implied by~\eqref{eq:h_2meta_noqsi},
\begin{align}
H_{\rm coll}(K|\Theta)\geq\log3-1\;\;\text{with}\;\;\Theta\in\{\sigma_{\bX},\sigma_{\bY},\sigma_{\bZ}\}\,.
\end{align}
\end{example}

Finally, we point to~\textcite{versteeg09} for the connection of the uncertainty relations described in this section to Bell inequalities.


\subsubsection{Mutually unbiased measurements}\label{sec:mum}

In Sec.~\ref{sec:mub_relations} we discussed how full sets of $d+1$ MUBs give rise to strong uncertainty relations, see, e.g., \eqref{eq:mub_full}. However, for general dimension $d$ we do not know if a full set of $d+1$ MUBs always exists (see App.~\ref{app:mub} for a discussion). \textcite{kalev07} offer the following generalization of MUBs to measurements that are not necessarily given by a basis. Two POVMs $\bX=\{\bX^x\}_{x=1}^d$ and $\bZ=\{\bZ^z\}_{z=1}^d$ on a $d$-dimensional quantum system are mutually unbiased measurements (MUMs) if for some $\kappa\in(1/d,1]$,
\begin{align}\label{eq:MUMdef}
&\tr\left[\bX^x\right]=1,\;\;\tr\left[\bZ^z\right]=1,\;\;\tr\left[\bX^x\bZ^z\right]=\frac{1}{d}\quad \forall x,z\\
&\tr\left[\bX^x\bX^{x'}\right]=\delta_{xx'}\cdot\kappa+(1-\delta_{xx'})\frac{1-\kappa}{d-1}\quad\forall x,x'\notag\\
&\text{and similarly for $z,z'$}\,.
\end{align}
In addition, a set of POVMs $\{\bX_1,\ldots,\bX_n\}$ of said form is called a set of MUMs if each POVM $\bX_j$ is mutually unbiased to each other POVM $\bX_k$, with $k\neq j$, in the set.

A straightforward example are again MUBs for which $\kappa=1$.\footnote{The trivial example for which each POVM element is the maximally mixed state $\1/d$ is excluded because this would correspond to $\kappa=1/d$.} The crucial observation of~\textcite{kalev07} is that in any dimension $d$ a full set of $d+1$ MUMs exists (see their paper for the explicit construction). Moreover, every full set of $d+1$ MUMs gives rise to a strong uncertainty relation,
\begin{align}\label{eq:mum_d+1}
H(K|\Theta)\geq\log(d+1)-\log\left(1+\kappa\right)&\notag\\
\text{with}\;\;\Theta\in\{\bX_1,\ldots,\bX_{d+1}\}&\,,
\end{align}
where the notation is as introduced in~\eqref{eq:povm_notation}. This is in full analogy with~\eqref{eq:mub_full} for a full set of $d+1$ MUBs. Tighter and state dependent versions of~\eqref{eq:mum_d+1} as well as extensions to R\'enyi entropies can be found in~\cite{rastegin15,chen15}


\subsection{Fine-grained uncertainty relations}\label{sec:fineGrained}

So far we have expressed uncertainty in terms of the von Neumann entropy and the R\'enyi entropies of the probability distribution induced by the measurement. Recall, however, that any restriction on the set of allowed probability distributions over measurement outcomes can be understood as an uncertainty relation, and hence there are many ways of formulating such restrictions. Thus, while generally the R\'enyi entropies determine the underlying probability distribution of the measurement outcomes uniquely,\footnote{To see this, note that the cumulant generating function of the random variable $Z = -\log P_X(X)$ can be expressed in terms of the R\'enyi entropy of $X$, namely $g_{Z}(s) = H_{1+s}(X)$. The cumulants of $Z$ and hence the distributions of $Z$ and $X$ are thus fully determined by the R\'enyi entropy in a neighborhood around $\alpha = 1$.} it is interesting to ask whether we can formulate more refined versions of uncertainty relations.

Suppose we perform $L$ measurements labeled $\Theta$ on a preparation $\rho_A$, where each measurement has $N$ outcomes. Fine-grained uncertainty relations~\cite{oppenheim10} consist of a set of $N^L$ equations which state that for all states we have
\begin{align}
\sum_{\theta = 1}^L P_{\Theta}(\Theta = \theta) P_X(X = x_{\theta}|\Theta = \theta) \leq \zeta_{x_{1},\ldots,x_{L}}\,,
\end{align}
for all combinations of measurement outcomes $x_1,\ldots,x_L$ that are possible for the $L$ different measurements. Here, $P_{\Theta}(\Theta = \theta)$ is the probability of choosing measurement labeled $\Theta = \theta$, and $0 \leq \zeta_{x_1,\ldots,x_L} \leq 1$. 

Note that whenever $\zeta_{x_1,\ldots,x_L} < 1$, then we observe some amount of uncertainty, since it implies that we cannot simultaneously have $P_X(X = x_{\theta}|\Theta = \theta) = 1$ for all $\theta$.
We remark that fine-grained uncertainty relations naturally give a lower bound on the min-entropy since
\begin{align}
2^{-H_{\rm min}(X|\Theta)} &= \sum_{\theta=1}^L P_{\Theta}(\Theta = \theta) \max_{x_{\theta}} P_X(X=x_{\theta}|\Theta = \theta) \\
&\leq - \log \max_{x_1,\ldots,x_L} \zeta_{x_1,\ldots,x_L}\,.
\end{align}
However, fine-grained uncertainty relations are strictly more informative and are also closely connected to Bell non-locality~\cite{oppenheim10}. While not the topic of this survey, a number of extensions of these fine-grained uncertainty relations are known~\cite{dey13,rastegin14b,ren14}.
 

\subsection{Majorization approach to entropic uncertainty}\label{sec:major}

Another way to capture uncertainty relations that relates directly to entropic ones is given by the majorization approach. Instead of sums of probabilities, we here look at products. The idea to derive entropic uncertainty relations via a majorization relation was pioneered by~\textcite{partovi11} and later extended and clarified independently by~\textcite{puchala13} and \textcite{friedland13}. Let us recall the distributions $P_X$ and $P_Z$ resulting from the measurements $\bX$ and $\bZ$, respectively, of the state $\rho_A$ as in~\eqref{eq:povm-pm-def}. We denote by ${P}_X^{\downarrow}$ and ${P}_Z^{\downarrow}$ the corresponding reordered vectors such that the probabilities are ordered form largest to smallest.


\subsubsection{Majorization approach}

The main objective of this section is to find a vector that majorizes the tensor product of the two probability vectors ${P}_X^{\downarrow}$ and ${P}_Z^{\downarrow}$. Namely, we are looking for a probability distribution $\nu = \{\nu(1), \nu(2), \ldots, \nu(|X| |Z|)\}$ such that\footnote{Recall the definition of majorization in Sec.~\ref{sec:cl-ent-prop}.}
\begin{align}\label{eq:majo-ucr}
{P}_X^{\downarrow} \times {P}_Z^{\downarrow} \prec \nu \qquad \textrm{holds for all } \rho \in \cS(\cH)\,.
\end{align}
Such a relation gives a bound on how spread out the product distribution ${P}_X^{\downarrow} \times {P}_Z^{\downarrow}$ must be. A simple and instructive example of a probability distribution $\nu$ satisfying the above relation can be constructed as follows. Consider the largest probability in the product distribution in~\eqref{eq:majo-ucr}, given by
\begin{align}
p_1 := {P}_X^{\downarrow}(1) \cdot {P}_Z^{\downarrow}(1) = p_{\rm guess}(X) \cdot p_{\rm guess}(Z)\,.
\end{align}
We know that $p_1$ is always bounded away from $1$ if the two measurements are incompatible, since it cannot be that both measurements have a deterministic outcome. For example, recall that we have \eqref{eqnguessingprobproduct} from \textcite{deutsch83}, which gives
\begin{align}\label{eqnguessingprobproduct1111111}
p_1 = p_{\rm guess}(X) \cdot p_{\rm guess}(Z) \leq b^2 =: \nu_1 \,,
\end{align}
where $b$ was defined in \eqref{eqnbdef}. As such, it is immediately clear that the vector $\nu^1 = \{ \nu_{1}, 1- \nu_{1}, 0, \ldots, 0 \}$ satisfies~\eqref{eq:majo-ucr} and in fact constitutes a simple but non-trivial uncertainty relation.

Going beyond this observation, the works of~\textcite{friedland13} and~\textcite{puchala13} both present an explicit method to construct a sequence of vectors $\{ \nu^k \}_{k = 1}^{|X|-1}$ of the form
\begin{align}
\nu^k = \{ \nu_1, \nu_2 - \nu_1, \ldots,1 - \nu_{k-1}, 0, \ldots, 0 \} \,,
\end{align}
with $\nu^k \prec \nu^{k-1}$, that satisfy~\eqref{eq:majo-ucr} and lead to tighter and tighter uncertainty relations. The expressions for $\nu_k$ are given in terms of an optimization problem and become gradually more difficult as $k$ increases. We refer the reader to these papers for details on the construction.


\subsubsection{From majorization to entropy}

Entropic uncertainty relations for R\'enyi entropy follow directly from the majorization relation above due to the fact that the R\'enyi entropy is Schur concave and additive. This implies that
\begin{align}\label{eq:majo-to-ent}
{P}_X^{\downarrow} \times {P}_Z^{\downarrow} \prec \nu \implies H_{\alpha}(X) + H_{\alpha}(Z) \geq H_{\alpha}(V)\,, 
\end{align}
where $V$ is a random variable distributed according to the law $\nu$. These uncertainty relations have a different flavor than the Maassen-Uffink relations in~\eqref{eq:mu_relation} since they provide a bound on the sum of the R\'enyi entropy of the same parameter. As a particular special case for $\alpha = \infty$, we get back Deutsch's uncertainty relation~\cite{deutsch83},
\begin{align}
H(X) + H(Z) &\geq H_{\min}(X) + H_{\min}(Z)\\
&\geq - 2 \log b =: q_{\rm D} \,,
\end{align}
where the first inequality follows by the monotonicity of the R\'enyi entropy in the parameter $\alpha$. However, an immediate improvement on this relation can be obtained by applying~\eqref{eq:majo-to-ent} directly for $\alpha = 1$, which yields
\begin{align}\label{eq:vn-majo}
H(X) + H(Z) \geq h_{\rm bin}  (b^2) =: q_{\rm maj} \,.
\end{align}
See Fig.~\ref{fgrqubit} for a comparison of this to other bounds.


\subsubsection{Measurements in random bases}

An interesting special case for which a majorization based approach gives tighter bounds is for measurements in two bases $\bX$ and $\bZ$ related by a random unitary. Intuitively, we would expect such bases to be complementary. More precisely, for any measurement in a fixed basis $\bX$ and $\bZ$ related by a unitary drawn from the Haar measure on the unitary group, \textcite{adamczak14} showed that for the Masseen-Uffink bound~\eqref{eq:shannonUR} we have with probability going to one for $d\to\infty$,
\begin{align}
H(X)+H(Z)\geq\log d-\log\log d\,.
\end{align}
However, they also show that a majorization based approach yields the tighter estimate
\begin{align}
H(X)+H(Z)\geq\log d-C_1\,,
\end{align}
where $C_1>0$ is some constant. This is close to optimal since we have that with probability going to one for $d\to\infty$~\cite{adamczak14},
\begin{align}
\log d-C_0\geq H(X)+H(Z)\,,
\end{align}
for some constant $C_0>0$. It is an open question to determine the exact asymptotic behavior, i.e., the constant $C\in(C_0,C_1)$ that gives a lower and an upper bound.


\subsubsection{Extensions}

The majorization approach has also been extended to cover general POVMs and more than two measurements~\cite{rastegin16, friedland13}. Moreover, a recent paper~\cite{rudnicki14} discusses a related method, based on finding a vector that majorizes the ordered distribution $({P}_X \cup {P}_Z)^{\downarrow}$, where $P_X \cup P_Z$ is simply the concatenation of the two probability vectors. This yields a further improvement on~\eqref{eq:vn-majo}. Finally, an extension to uncertainty measures that are not necessarily Schur concave but only monotonic under doubly stochastic matrices was presented in~\cite{narasimhachar15}.



\section{Uncertainty given a Memory System}\label{sctMemory}

The uncertainty relations presented thus far are limited in the following sense: they do not allow the observer to have access to \textit{side information}. Side information, also known as \textit{memory}, might help the observer to better predict the outcomes of the $\bX$ and $\bZ$ measurements. It is therefore a fundamental question to ask: does the uncertainty principle still hold when the observer has access to a memory system? If so, what form does it take?

The uncertainty principle in the presence of memory is important for cryptographic applications and witnessing entanglement (Sec.~\ref{sec:app}). For example, in quantum key distribution, an eavesdropper may gather some information, store it in her memory, and then later use that memory to try to guess the secret key. It is crucial to understand whether the eavesdropper's memory allows her to break a protocol's security, or whether security is maintained. This is where general uncertainty relations that allow for memory are needed.

Furthermore, such uncertainty relations are also important for basic physics. For example, the quantum-to-classical transition is an area of physics where one tries to understand why and how quantum interference effects disappear on the macroscopic scale. This is often attributed to decoherence, where information about the system of interest $S$ flows out to an environment $E$ \cite{zurek03}. In decoherence, it is important to quantify the tradeoff between the flow of one kind of information, say $\bZ$, to the environment versus the preservation of another kind of information, say $\bX$, within the system~$S$. Here, one associates $\bX$ with the ``phase'' information that is responsible for quantum interference. Hence, one can see how this ties back into the quantum-to-classical transition, since loss of $\bX$ information would destroy the quantum interference pattern. In this discussion, system $E$ plays the role of the memory, and hence uncertainty relations that allow for memory are essentially uncertainty relations that allow the system to interact with an environment. We will discuss this more in Sec.~\ref{sec:waveparticle}, in the context of interferometry experiments.


\subsection{Classical versus quantum memory}\label{sctMemoryIntro}

With this motivation in mind, we now consider two different types of memories. First, we discuss the notion of a \textit{classical memory}, i.e., a system $B$ that has no more than classical correlations with the system $A$ that is to be measured.

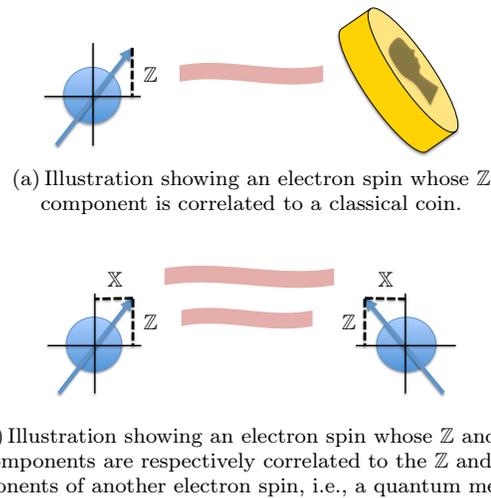
\begin{figure}[tbp]
\begin{center}
\subfigure[\,Illustration showing an electron spin whose $\bZ$ component is correlated to a classical coin.]{
\hspace{1cm}\begin{overpic}[width=5.6cm]{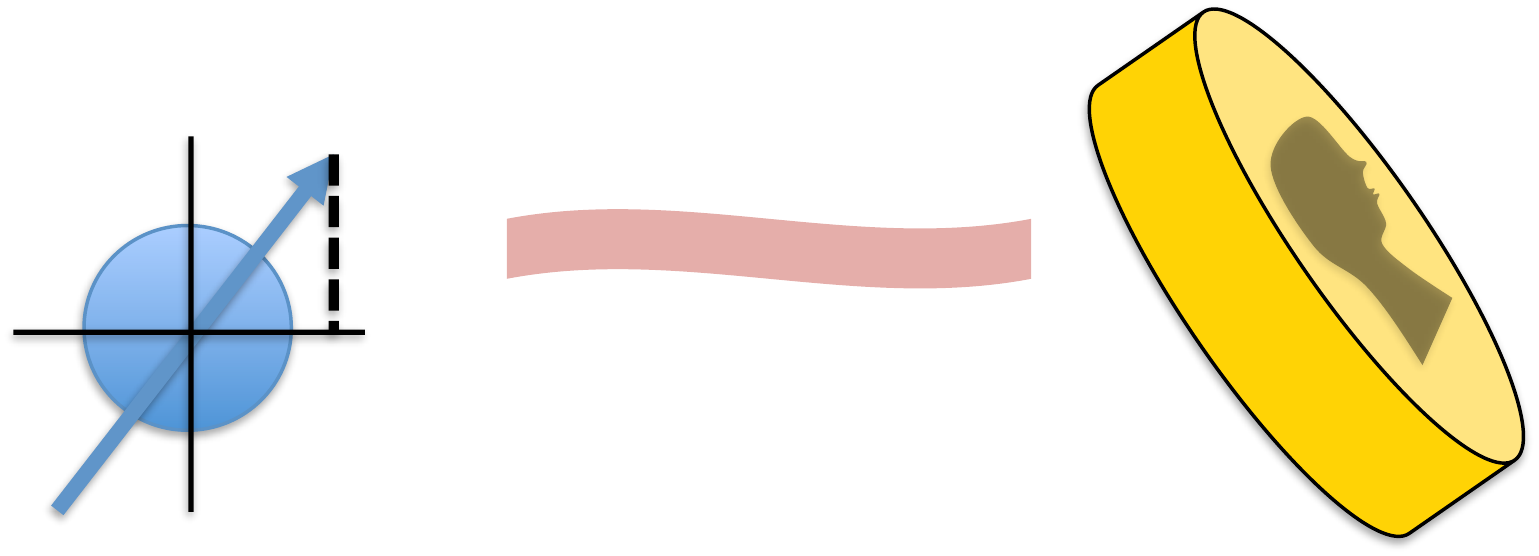}
\put(24.4,18){\footnotesize $\bZ$}
\end{overpic}\hspace{1cm}
}
\subfigure[\,Illustration showing an electron spin whose $\bZ$ and $\bX$ components are respectively correlated to the $\bZ$ and $\bX$ components of another electron spin, i.e., a quantum memory.]{
\hspace{1cm}\begin{overpic}[width=6cm]{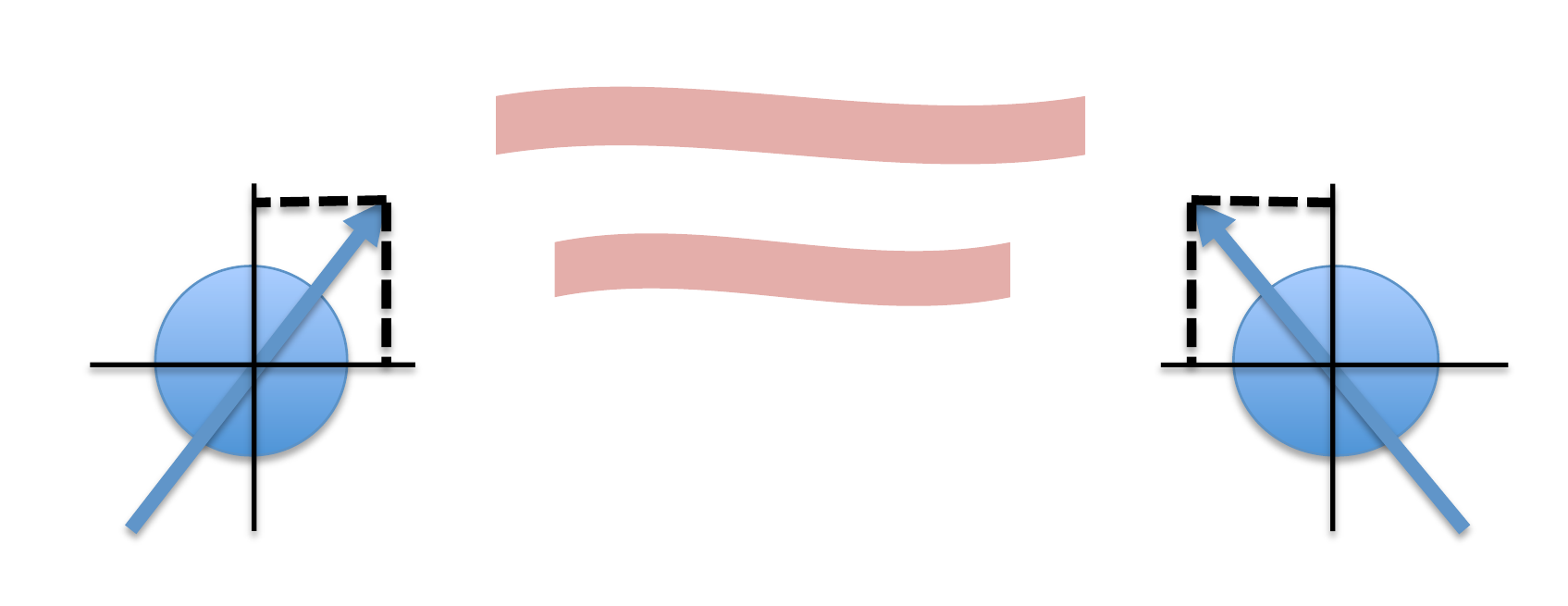}
\put(19,28.5){\footnotesize $\bX$}
\put(27,19){\footnotesize $\bZ$}
\put(79,28.5){\footnotesize $\bX$}
\put(71,19){\footnotesize $\bZ$}
\end{overpic}\hspace{1cm}
}
\vspace{-0.3cm}
\caption{Comparison of classical and quantum memory.}
\label{figclassicalquantummemory}
\end{center}
\end{figure}

\begin{example}\label{ex:classical_coin1}
Consider a spin-$1/2$ particle $A$ and a (macroscopic) coin $B$ as depicted in Fig.~\ref{figclassicalquantummemory}(a). Suppose that we flip the coin to determine whether or not we prepare $A$ in the spin-up state $\ket{0}$ or the spin-down state $\ket{1}$. Denoting the basis $\bZ=\{ \ket{0}, \ket{1}\}$ we see that $B$ is perfectly correlated to this basis. That is, 
before the measurement of $A$ the joint state is
\begin{align} 
\rho_{AB} &= \frac{1}{2} \left( \proj{0}_A \otimes \rho_B^0+ \proj{1}_A \otimes \rho_B^1 \right)\,, \notag \\ 
&\text{where }\tr \left[ \rho_B^0 \rho_B^1 \right] =0\,.\end{align}
Hence, if the observer has access to $B$ then he can perfectly predict the outcome of the $\bZ$ measurement on $A$. On the other hand, if we keep $B$ hidden from the observer, then he can only guess the outcome of the $\bZ$ measurement on $A$ with probability $1/2$. 
\end{example}

We conclude from Ex.~\ref{ex:classical_coin1} that indeed, having access to $B$ reduces the uncertainty about $\bZ$. However, notice that a classical memory $B$ provides no help to the observer if he tries to guess the outcome of a measurement on $A$ that is complementary to $\bZ$. Consider now a more general memory, one that can have any kind of correlations with system $A$ allowed by quantum mechanics. This is called a \textit{quantum memory} or \textit{quantum side information} (and includes classical memory as a special case). We remark that quantum memories are becoming an experimental reality (see, e.g., \textcite{julsgaard04}).

\begin{example}\label{ex:quantum_coin}
Consider two spin-1/2 particles $A$ and $B$ that are maximally entangled, say in the state
\begin{align}\label{eq:psi_intro}
\ket{\psi}_{AB} =\frac{1}{\sqrt{2}}\big(\ket{00}_{AB}+\ket{11}_{AB}\big)\,.
\end{align}
This is depicted in Fig.~\ref{figclassicalquantummemory}(b). Like for the classical memory in Ex.~\ref{ex:classical_coin1}, giving the observer access to $B$ allows him to perfectly predict the outcome of a $\bZ$ measurement on $A$ (by just measuring the $\bZ$ observable on $B$). But in contrast to the case with classical memory, $B$ can also be used to predict the outcome of a complementary measurement $\bX = \{\ket{+},\ket{-} \}$, with $\ket{\pm} = (\ket{0}\pm\ket{1})/\sqrt{2}$, on $A$. This follows by rewriting the maximally entangled state~\eqref{eq:psi_intro} as
\begin{align}
\ket{\psi}_{AB} = \frac{1}{\sqrt{2}}\big(\ket{\!+\!+}_{AB}+\ket{\!-\!-}_{AB}\big)\,,
\end{align}
which implies that the observer can simply measure the $\bX$ basis on $B$ to guess $\bX$ on $A$.
\end{example}

The idea described in Ex.~\ref{ex:quantum_coin} dates back to the famous EPR paper~\cite{epr35} and raises the question of whether we can still find nontrivial bounds on the uncertainty of complementary measurements when conditioning on quantum memory. In the rest of Sec.~\ref{sctMemory} we analyze this interplay between uncertainty and quantum correlations quantitatively and present entropic uncertainty relations that allow the observer to have access to (quantum) memory. For that we first introduce measures of \textit{conditional entropy}.


\subsection{Background: Conditional entropies}\label{sec:conditionalentropy}

\subsubsection{Classical-quantum states}

Our main goal here is to describe the entropy of a measured (and thus classical) random variable from the perspective of an observer who possesses a quantum memory. For this purpose, consider a classical register correlated with a quantum memory, modeled by
a joint \emph{classical quantum (cq) state} 
\begin{align}\label{eq:cq}
\rho_{XB} = \sum_x P_X(x)\, \proj{x}_X \otimes \rho_B^x\,.
\end{align}
Here, $\rho_B^x$ is the quantum state of the memory system $B$ conditioned on the event $X\!=\!x$. Formally, \emph{quantum states} or \emph{density operators} are positive semidefinite operators with unit trace acting on the Hilbert space $B$. In order to represent the joint system $XB$ in the density operator formalism we also introduced an auxiliary Hilbert space $X$ with fixed orthonormal basis $\{ \ket{x}_X \}_x$.


\subsubsection{Classical quantum entropies}

The interpretation of the min-entropy from~\eqref{eq:guess-class} in terms of the optimal guessing probability gives a natural means to generalize the min-entropy to the setting with quantum memory. Clearly, an observer with access to the quantum memory $B$ can measure out $B$ to improve his guess. The optimal guessing probability for such an observer is then given by the optimization problem
\begin{align}\label{eq:guess}
p_{\rm guess}(X|B) :=& \max_{\bX_B}\sum_x P_X(x) \tr\big[ \bX_B^x \rho_B^x \big]\,,\notag\\
&\textrm{where $\bX_B$ is a POVM on $B$.}
\end{align}
Consequently, the conditional min-entropy is defined as~\cite{renner05,koenig08},
\begin{align}\label{eqncqminentropydef}
H_{\min}(X|B) := - \log p_{\rm guess}(X|B)\,.
\end{align}
This is our first measure of conditional entropy. It quantifies the uncertainty of the classical register $X$ from the perspective of an observer with access to the quantum memory (or side information) $B$. The more difficult it is to guess the value of $X$, the smaller is the guessing probability and the higher is the conditional min-entropy.

The collision entropy from~\eqref{eqncollision1} can likewise be interpreted in terms of a guessing probability. Consider the following generalization of the collision entropy to the case where the observer has a quantum memory $B$~\cite{buhrman08},
\begin{align}\label{eq:h2_cq}
H_{\rm coll}(X|B) := - \log p_{\rm guess}^{\rm pg}(X|B) \,.
\end{align}
Here, the pretty good guessing probability is given by
\begin{align}\label{eq:guesspg}
p_{\rm guess}^{\rm pg}(X|B) :=&   \sum_x P_X(x) \tr\Big[ \Pi_B^x \rho_B^x \Big] \, ,\notag\\
&\textrm{where }\Pi_B^x =  P_X(x) \rho_B^{-1/2}\rho_B^x\rho_B^{-1/2}\,.
\end{align}
The $\Pi_B^x$ are POVM elements corresponding to the so-called pretty good measurement. The name is due to the fact that this measurement is close to optimal, in the sense that~\cite{hausladen94},
\begin{align}\label{eqnpgclosetooptimal}
p^2_{\rm guess}(X|B)\leq p_{\rm guess}^{\rm pg}(X|B)\leq p_{\rm guess}(X|B)\,.
\end{align}
That is, if the optimal guessing probability is close to one, then so is the pretty good guessing probability. Hence, $H_{\rm coll}(X|B)$ quantifies how well Bob can guess $X$ given that he performs the pretty good measurement on $B$. In particular this also implies that
\begin{align}\label{eq:hmin_h2}
H_{\min}(X|B)\leq H_{\rm coll}(X|B)\leq 2H_{\min}(X|B)\,.
\end{align}

Finally, consider the Shannon entropy $H(X)$, whose quantum counterpart $H(\rho)$ is the von Neumann entropy as defined in~\eqref{eqnvonneumann1}. The von Neumann entropy of $X$ conditioned on a quantum memory $B$ is defined as
\begin{align}\label{eqnvncqentropy}
H(X|B) := H(\rho_{XB}) - H(\rho_B) \,.
\end{align}
where $\rho_{XB}$ is given by~\eqref{eq:cq}, and
\begin{align}
\rho_B = \tr_X\big[\rho_{XB}\big] = \sum_x P_X(x) \rho_B^x\,.
\end{align}
Although $H(X|B)$ does not have a direct interpretation as a guessing probability, it does have an operational meaning in information theory. For example, if Alice samples from the distribution $P_X$ and Bob possesses system $B$, then $H(X|B)$ is the minimal information that Alice must send to Bob in order for Bob to determine the value of $X$. (More precisely, $H(X|B)$ is the minimal rate in bits per copy that Alice must send to Bob, in the asymptotic limit of many copies of the state $\rho_{XB}$~\cite{devetak03}.)


\subsubsection{Quantum entropies}

The classical-quantum conditional entropy is merely a special case of the quantum conditional entropy. It is useful to introduce the latter here, since the quantum conditional entropy will play an important role in the following.

In the simplest case, the von Neumann conditional entropy of an arbitrary bipartite state $\rho_{AB}$ with $\rho_B = \tr_A(\rho_{AB})$, takes the form
\begin{align}\label{eqnconditionalvn1}
H(A|B) := H(\rho_{AB}) - H(\rho_B) \,.
\end{align}
We remark that, in general, fully quantum conditional entropy can be negative.\footnote{This should not concern us further here; a consistent interpretation of negative entropies is possible in the context of quantum information processing~\cite{horodecki06} and also in thermodynamics~\cite{delrio11}.} This is a signature of entanglement. In fact, the quantity $-H(A|B)$, commonly known as coherent information, provides a lower bound on the distillable entanglement~\cite{devetak05}. We will discuss this connection further around~\eqref{eqnDevWin} below.

The fully quantum min-entropy also has a connection to entanglement. Namely, it can be written as,
\begin{align}\label{eqnfullyquantumminentropy}
H_{\min}(A|B) := - \log \big(d_A\cdot F(A|B)\big)\,,
\end{align}
where
\begin{align}\label{eq:entanglement_fidelity}
F(A|B) := \max_{\cE:B\to A'} F\Big((\id\ot \cE)(\rho_{AB}) , \dya{\phi_{AA'}}\Big)\notag&\\
\text{with the fidelity $F(\rho,\sigma):=\left(\tr\left[\sqrt{\sqrt{\rho}\sigma\sqrt{\rho}}\right]\right)^2$}&
\end{align}
from~\cite{uhlmann85}, $\ket{\phi_{AA'}}$ a maximally entangled state of dimension $|A|$, and the maximization over all quantum channels $\cE$ that map $B$ to $A'$.  One can think of $F(A|B)$ as the recoverable entanglement fidelity. In that sense, $-H_{\min}(A|B)$ quantifies how close the state is to a maximally entangled state.

The fully quantum collision entropy can also be related to a recoverable entanglement fidelity, in close analogy to the discussion above for the classical-quantum case. Namely, we have~\cite{berta13},
\begin{align}\label{eq:h2_qq}
H_{\rm coll}(A|B) := - \log \big(d_A\cdot F^{\rm pg}(A|B)\big)\,,
\end{align}
where 
\begin{align}
F^{\rm pg}(A|B) := F\big((\id\ot \cE^{\rm pg})(\rho_{AB}) , \dya{\phi_{AA'}}\big)\,.
\end{align}
Here, $\cE^{\rm pg}$ is the pretty good recovery map, whose action on an operator $O$ is given by
\begin{align}
\cE^{\rm pg}(O) = \left(\tr_B \left[(\1 \ot \rho_B^{-1/2} O \rho_B^{-1/2}) \rho_{A'B} \right]\right)^T\,,
\end{align}
where $(\cdot)^T$ denotes the transpose map, and $\rho_{A'B} = \rho_{AB}$, with system $A'$ being isomorphic to system $A$.\footnote{One can verify that $\cE^{\rm pg}$ is a valid quantum operation because it is completely positive and trace preserving map (assuming $\rho_B$ is full rank).} In analogy to~\eqref{eqnpgclosetooptimal}, the pretty good recovery map is close to optimal~\cite{barnum02},
\begin{align}
F^2(A|B)\leq F^{\rm pg}(A|B) \leq F(A|B)\,.
\end{align}
As in the classical case, the above conditional entropies emerge as special cases of R\'enyi entropies~\cite{lennert13}. We discuss this connection in Appendix~\ref{app:renyi}.


\subsubsection{Properties of conditional entropy}

Section~\ref{sec:cl-ent-prop} discussed properties of entropies, which are special cases of conditional entropies with trivial conditioning system. Here, we mostly discuss properties of the conditional von Neumann entropy $H(A|B)$, and only note that similar properties also hold for other conditional entropies such as $H_{\min}(A|B)$ and $H_{\rm coll}(A|B)$ (or more generally R\'enyi entropies).

Firstly, the conditional entropy reduces to the unconditional entropy for product states. That is, for bipartite states of the form $\rho_{AB}= \rho_A \ot \rho_B$, we have $H(A|B) = H(A)$. Secondly, note that the entropy of a classical-quantum state is non-negative,
\begin{align}
H(X|B)\geq 0\quad\text{for $X$ a classical register.}
\end{align}
In contrast, as noted above, the fully quantum entropy $H(A|B)$ can be negative.

A fundamental property is the so-called data-processing inequality. It says that the uncertainty of $A$ conditioned on some system $B$ never goes down if one processes system $B$, i.e., acts on $B$ with a quantum channel $\cE:B\to B'$. That is~\cite{lieb73},
\begin{align}\label{eq:data_processing}
H(A|B) \leq H(A|B')\,.
\end{align}
This includes the case where system $B= B_1B_2$ is bipartite and the processing corresponds to discarding a subsystem, say $B_2$. In this case the data-processing inequality takes the form $H(A|B) \leq H(A|B_1)$. This inequality is intuitive in the sense that having access to more information can never increase the uncertainty.

Another useful property of conditional entropies is related to the monogamy of entanglement. This corresponds to the idea that the more $A$ is entangled with $B$ the less $A$ is entangled with a purifying system $C$. Suppose that $C$ is a system that purifies $\rho_{AB}$, i.e., $\rho_{ABC} = \dya{\psi}$. Then, we have 
\begin{align}\label{eqndualentropy1}
H(A|B) = - H(A|C)\,.
\end{align}
Typically one associates entanglement with a negative conditional entropy, and indeed as discussed above, the coherent information (the negative of the conditional entropy) lower bounds the distillable entanglement. In this sense, the relation in~\eqref{eqndualentropy1} captures the intuition of monogamy of entanglement. It implies that if $\rho_{AB}$ has a negative conditional entropy, then $\rho_{AC}$ must have a positive conditional entropy. So there is a trade-off between the entanglement present in $\rho_{AB}$ and in $\rho_{AC}$.

The relation in~\eqref{eqndualentropy1} is called the duality relation, as it relates an entropy to its \textit{dual} entropy. As we have seen the von Neumann entropy is dual to itself but in general the duality relation involves two different entropies. For example, the min-entropy is dual to the max-entropy,
\begin{align}\label{eqndualentropy3}
H_{\max}(A|B):=-H_{\min}(A|C)\,.
\end{align}
We take~\eqref{eqndualentropy3} as the definition of the max-entropy, although an explicit expression in terms of the marginal $\rho_{AB}$ can be derived~\cite{koenig08}. More generally, the duality relation for the R\'enyi entropy family is given in App.~\ref{pre:prop}.


\subsection{Classical memory uncertainty relations}\label{sctCMemory}

We now have all the measures at hand to discuss uncertainty relations that allow for a memory system. Naturally, we begin with the simplest case of a classical memory. 
It turns out that uncertainty relations that allow for classical memory are often easy to derive from the uncertainty relations without memory, particularly for the Shannon entropy~\cite{hall95}. Consider the conditional Shannon entropy, which can be written as
\begin{align}\label{eqnClassicalMem1}
H(X|Y) = H(XY) - H(Y)= \sum_y P_{Y}(y) H(X | Y=y)\,.
\end{align}
Now consider some generic Shannon entropy uncertainty relation for measurements $\bX^n$ and quantum states $\rho_A$:
\begin{align}\label{eq:shannon_generic}
\sum_n H(X_n) \geq q\quad\mathrm{where}\quad P_{X_n}(x)=\bra{\bX^x_n}\rho_A\ket{\bX^x_n}&\notag\\
\quad\text{and $q>0$ state-independent}&\,.
\end{align}
The goal is to extend this to quantum-classical states $\rho_{AY}$ where the classical memory $Y$ holds some information about the preparation of the quantum marginal
\begin{align}
\rho_{AY} = \sum_y P_Y(y)\, \rho_A^y\otimes\proj{y}_Y\quad\text{with distributions}&\notag\\
P_{X_nY}(x,y)=P_Y(y)\bra{\bX^x_n}\rho_A^y\ket{\bX^x_n}&\,.
\end{align}
However, assuming that the uncertainty relation~\eqref{eq:shannon_generic} holds for all quantum states, it holds in particular for each conditional state $\rho_A^y$ associated with $Y=y$ in the classical memory $Y$. Averaging over $y$ gives 
\begin{align}\label{eqnClassicalMem2}
\sum_y P_{Y}(y) \sum_n H(X_n | Y=y)  \geq \sum_y P_{Y}(y) q = q\,.
\end{align}
Hence, we find by~\eqref{eqnClassicalMem1} that
\begin{align}\label{eqnClassicalMem3}
\sum_n H(X_n) \geq q \quad \implies \quad \sum_n H(X_n | Y) \geq q\,.
\end{align}
That is, any Shannon entropy uncertainty relation of the form~\eqref{eq:shannon_generic} implies a corresponding uncertainty relation in terms of the conditional Shannon entropy of the form~\eqref{eqnClassicalMem3}. Note that the conditional version~\eqref{eqnClassicalMem3} even provides a stronger bound, since by the data-processing inequality~\eqref{eq:data_processing} conditioning on side information can only reduce uncertainty.

\begin{example}
Consider a bipartite state $\rho_{AB}$, where Alice will measure system $A$ in one of two bases $\bX$ or $\bZ$ and Bob will measure system $B$ in the basis $\bY$. Then, the Maassen-Uffink relation~\eqref{eq:shannonUR} implies
\begin{align}\label{eq:mu_clmemory}
H(X|Y)+H(Z|Y)\geq q_{\rm MU}\,,
\end{align}
for the distribution
\begin{align}
P_{XY}(x,y)=\bra{\bX^x\otimes\bY^y}\rho_{AB}\ket{\bX^x\otimes\bY^y}\,,
\end{align}
and analogously $P_{ZY}(z,y)$.
\end{example}

It is worth noting that the classical memory $Y$ can be considered multipartite, say, of the form $Y = Y_1 Y_2 ... Y_n$~\cite{cerf02,renes09}. Since by the data-processing inequality~\eqref{eq:data_processing} discarding subsystems of $Y$ can never reduce the uncertainty, \eqref{eqnClassicalMem3} implies that
\begin{align}\label{eqnClassicalMem4}
\sum_n H(X_n) \geq q \quad \implies \quad \sum_n H(X_n | Y_n) \geq q\,.
\end{align}

\begin{example}
Consider a tripartite state $\rho_{ABC}$, where Alice will measure system $A$ in one of two bases $\bX$ or $\bZ$, Bob will measure system $B$ in the basis $\bY_B$, and the third party Charlie will measure system $C$ in the basis $\bY_C$. Then, the Maassen-Uffink relation~\eqref{eq:shannonUR} implies
\begin{align}\label{eqnClassicalMem55}
H(X|Y_B)+H(Z|Y_C) \geq q_{\rm MU}\,.
\end{align}
This relation is reminiscent of the scenario in quantum key distribution. Namely, if Alice and Bob verify that $H(X|Y_B)$ is close to zero, then \eqref{eqnClassicalMem55} implies that Charlie is fairly ignorant about $Z$. That is, $H(Z|Y_C)$ is roughly $q_{\rm MU}$ or larger. We emphasize, however, that~\eqref{eqnClassicalMem55} cannot be used to prove security against general quantum memory eavesdropping attacks (see Sec.~\ref{sec:qkd}). 
\end{example}


\subsection{Bipartite quantum memory uncertainty relations}\label{sec:bipartite}

\subsubsection{Guessing game with quantum memory}\label{sec:bipartiteguessing}

\begin{figure}[tbp]
\begin{center}
\begin{overpic}[width=8.1cm]{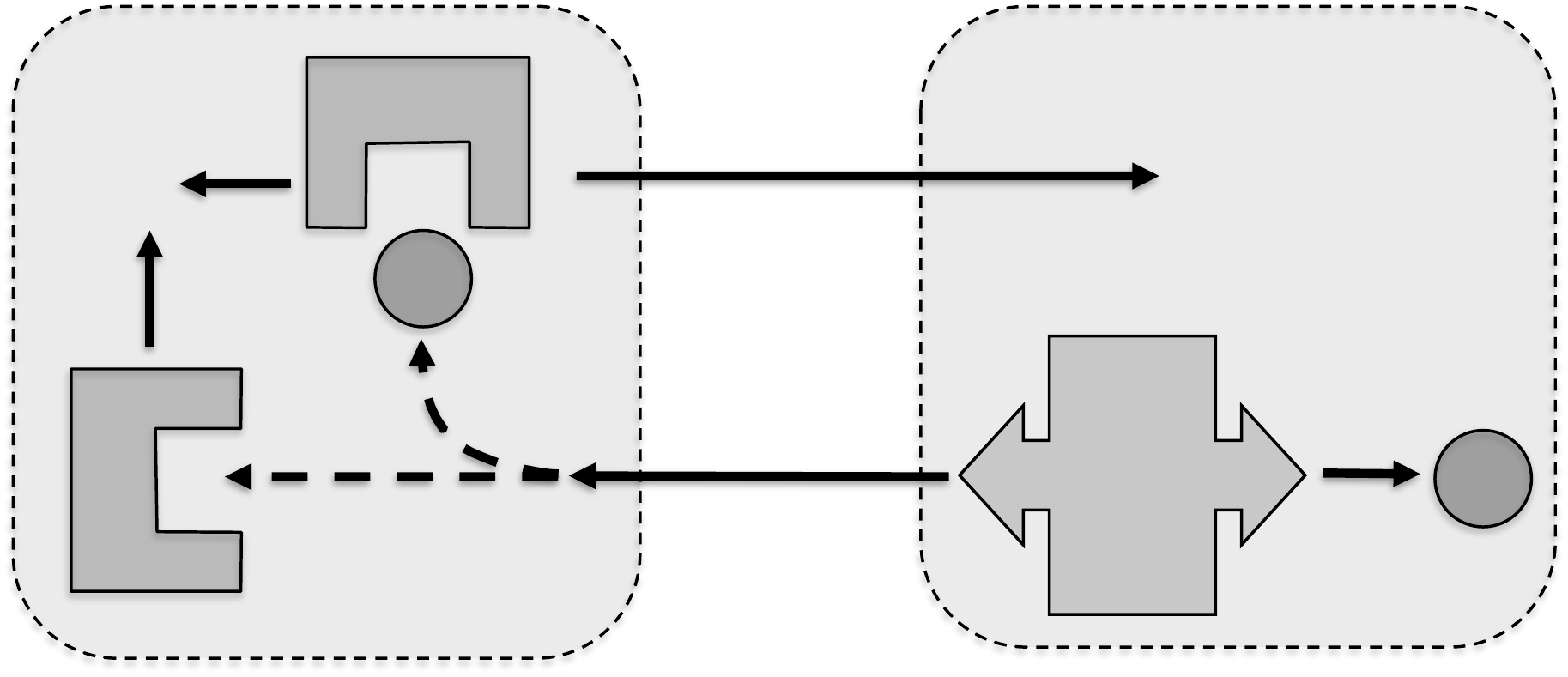}
\put(21.4,35.8){\footnotesize $\Theta=\bZ$}
\put(5.7,7.2){\rotatebox{90}{\footnotesize $\Theta=\bX$}}
\put(4,36){\footnotesize Alice}
\put(8,29.8){\footnotesize $K$}
\put(85.3,32){\footnotesize Bob}
\put(49,13.8){\footnotesize $A$}
\put(49,32.8){\footnotesize $\Theta$}
\put(69,12){\footnotesize $\rho_{AB}$}
\put(86.6,14.4){\footnotesize $B$}
\put(85.2,27){\footnotesize $K$?}
\vspace{1cm}
\end{overpic}
\caption{Diagram showing the guessing game in the presence of a quantum memory system. First, Bob prepares $AB$ in state $\rho_{AB}$, and then sends system $A$ to Alice. Second, Alice performs either the $\bX$ or $\bZ$ measurement on $A$, and then announces the measurement choice $\Theta$ to Bob. Bob's task is to correctly guess $K$. The question is thus: how uncertain is Bob about Alice's measurement outcome $K$, given that he has access to $B$ and $\Theta$?}
\label{figbipartitememory}
\end{center}
\end{figure}
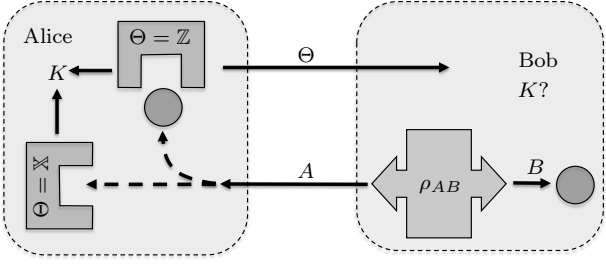

Let us now make explicit what the guessing game (see Section~\ref{sec:games_cl}) looks like when we allow quantum memory.
Specifically, the rules of the game now allow Bob to keep a \emph{quantum} memory system in order to help him guess Alice's measurement outcome. 
This is illustrated in Fig.~\ref{figbipartitememory}:
\begin{enumerate}
\item Bob prepares a bipartite quantum system $AB$ in a state $\rho_{AB}$.
He sends system $A$ to Alice while he keeps system $B$.
\item Alice performs one of two possible measurements, $\bX$ or $\bZ$, on $A$ and stores the outcome in the classical register $K$.
She communicates her choice to Bob.
\item Bob's task is to guess $K$.
\end{enumerate}
Note that in this game, Bob can make an \textit{educated} guess based on his quantum memory $B$.

\begin{example}\label{ex:bipartite:guessing}
Let the $A$ system be one qubit and Alice's two measurements given by $\sigma_{\bX}$ and $\sigma_{\bZ}$. Then, Bob can win the game with probability one by preparing the maximally entangled state and using the strategy from Ex.~\ref{ex:quantum_coin}.
\end{example}

This example illustrates the power of a quantum memory, and in particular of one that is \textit{entangled} with the system being measured. At first sight, this might seem to violate the usual notion of the uncertainty principle. However, it does not. What it illustrates is that the usual formulations of the uncertainty principle, such as the Robertson relation~\eqref{eqnRobertsonUR} or the Maassen-Uffink relation~\eqref{eq:shannonUR}, are not about conditional uncertainty. The relations~\eqref{eqnRobertsonUR} and~\eqref{eq:shannonUR} are perfectly valid but limited in this sense.

\subsubsection{Measuring in two orthonormal bases}

Let us first discuss how the Maassen-Uffink relation~\eqref{eq:shannonUR} can be extended to the setup when the observer has a quantum memory. Note that Ex.~\ref{ex:quantum_coin} and~\ref{ex:bipartite:guessing} illustrate that the bound in the uncertainty relation must become trivial in the case where Bob's memory is \textit{maximally entangled} to Alice's system. On the other hand, we know that the bound must be non-trivial when Bob has no memory, since this corresponds to the situation covered by~\eqref{eq:shannonUR}. Likewise if Bob has a memory that is only \textit{classically} correlated to Alice's system, then we already saw in~\eqref{eq:mu_clmemory} that the Maassen-Uffink relation can be extended. Therefore, it becomes clear that we need a \textit{state-dependent} extension: a bound that becomes weaker as Bob's memory is more entangled with Alice's system. Indeed, \textcite{berta10} proved the following uncertainty relation. For any bipartite state $\rho_{AB}$ and any orthonormal bases $\bX$ and $\bZ$,
\begin{align}\label{eqnBipartiteMemory13}
H(X|B)+H(Z|B)\geq q_{\rm MU}+H(A|B)\,,
\end{align}
with $q_{\rm MU}$ as in~\eqref{eq:shannonUR}. Here, the conditional entropy $H(X|B)$ is evaluated on the classical quantum state
\begin{align}\label{eq:cq_postmeas}
\rho_{XB}=\sum_x\proj{x}_{X}\otimes\big(\bra{\bX^x}\otimes\1_B\big)\rho_{AB}\big(\ket{\bX^x}\otimes\1_B\big)\,,
\end{align}
and similarly for $H(Z|B)$. The classical quantum conditional entropies $H(X|B)$ and $H(Z|B)$ quantify Bob's uncertainty about $X$ and $Z$ respectively, given that Bob has access to the quantum memory $B$.

The quantity $H(A|B)$ on the right-hand side of~\eqref{eqnBipartiteMemory13} makes the bound state-dependent. We already mentioned around~\eqref{eqnconditionalvn1} that $-H(A|B)$ is a quantifier of the entanglement present in $\rho_{AB}$. For maximally entangled states we have $-H(A|B) = \log d_A$, whereas for all separable (i.e., non-entangled) states we have $H(A|B)\geq0$.

\begin{example}
Let us explore in more detail how the bound~\eqref{eqnBipartiteMemory13} behaves for some illustrative cases:
\begin{enumerate}
\item For maximally entangled states we get
\begin{align}
q_{\rm MU}+H(A|B)=q_{\rm MU}-\log d_A\leq 0\,,
\end{align}
and hence the bound becomes trivial. This is as expected from the guessing game example discussed in Section~\ref{sec:bipartiteguessing}.
\item For the case when Bob has no memory (i.e., $B$ is trivial), \eqref{eqnBipartiteMemory13} reduces to~\eqref{eq:shannonURmixed},
\begin{align}
H(X)+H(Z)\geq q_{\rm MU}+H(\rho_A)\,.
\end{align}
This is the strengthened Maassen-Uffink relation for mixed states,
\item If $B$ is not entangled with $A$ (i.e., the state is separable), then $H(A|B)\geq 0$. Hence, we obtain
\begin{align}
\label{eqnBipartiteMemory142321}
H(X|B)+H(Z|B)\geq q_{\rm MU}\,.
\end{align}
\end{enumerate}
\end{example}

This last example illustrates that~\eqref{eqnBipartiteMemory13} has applications for entanglement witnessing. More precisely, note that by the data-processing inequality~\eqref{eq:data_processing}, \eqref{eqnBipartiteMemory13} also implies 
\begin{align}\label{eqnBipartiteMemory14}
H(X|Y_B)+H(Z|W_B)\geq q_{\rm MU}+H(A|B)&\notag\\
\text{with $\bY_B$ and $\bW_B$ measurements on $B$.}&
\end{align}
Now violating the $q_{\rm MU}$ lower bound in~\eqref{eqnBipartiteMemory142321} implies that the state $\rho_{AB}$ must have been entangled. We discuss this in detail in Sec.~\ref{sec:entwit}.

Using the following extension of the notation from Sec.~\ref{sec:games_cl} to quantum memory,
\begin{align}\label{eq:ext_notation}
\rho_{K\Theta B}:=\frac{1}{2}&\sum_{k}\sum_{j=X,Z}\proj{k}_{K}\otimes\proj{j}_{\Theta}\notag\\
&\otimes\Big(\bra{k}U_j^\dagger\otimes1_B\Big)\rho_{AB}\Big(U_j\ket{k}\otimes1_B\Big)\,,
\end{align}
we can rewrite~\eqref{eqnBipartiteMemory13} as
\begin{align}\label{eqnBipartiteMemory19}
H(K|B\Theta)\geq\frac{1}{2}\left(q_{\rm MU}+H(A|B)\right)\,.
\end{align}
This is the extension of~\eqref{eq:multiple_m-u} to quantum memory. Writing the relation in this way also makes a connection to the guessing game discussed in Sec.~\ref{sec:bipartiteguessing}, see Fig.~\ref{figbipartitememory}. We point to Sec.~\ref{sec:bb84six_qm} for a partial extension of~\eqref{eqnBipartiteMemory19} in terms of the more operational min-entropy.

Let us take a step back and look at the history that led up to the uncertainty relation~\eqref{eqnBipartiteMemory13}. Arguably the first work on uncertainty relations with quantum memory was by~\textcite{christandl05}. Their formulation was restricted to bases that are related by the Fourier matrix but their work captures similar intuition as~\eqref{eqnBipartiteMemory13}. The main difference, however, is that their relations are formulated for quantum channels rather than for quantum states. We discuss quantum channel uncertainty relations in Sec.~\ref{sctQchannelmemory}.

\textcite{renes09} gave the first quantum memory uncertainty relation in terms of the quantum state perspective. However, instead of bipartite states $\rho_{AB}$, they considered tripartite states $\rho_{ABC}$.\footnote{More precisely, \textcite{renes09} did establish a bipartite uncertainty relation\,---\,a special case of \eqref{eqnBipartiteMemory13} where the $\bX$ and $\bZ$ bases are related by the Fourier matrix. But they focused their discussion primarily on the tripartite formulation.}  We discuss entropic uncertainty relations for tripartite states in the next subsection (Sec.~\ref{scttripartiteQMemory}). Moreover, there is a close connection between tripartite and bipartite uncertainty relations. In fact, as we will discuss in Sec.~\ref{scttripartiteQMemory}, \textcite{renes09} conjectured a tripartite uncertainty relation that is equivalent to~\eqref{eqnBipartiteMemory13}. Sec.~\ref{scttripartiteQMemory} also discusses the proof of quantum memory uncertainty relations such as \eqref{eqnBipartiteMemory13}, and notes that the tripartite formulation of \eqref{eqnBipartiteMemory13} naturally generalizes to the R\'enyi entropy family.


\subsubsection{Arbitrary measurements}\label{sec:bipartitePOVMs}

Here, we discuss some generalizations of \eqref{eqnBipartiteMemory13} for arbitrary measurements. Recall from Sec.~\ref{sec:arbitrary_povm} that the Maassen-Uffink relation generalizes to POVMs with the overlap $c$ given by~\eqref{eqncjkPOVM}. In contrast, \eqref{eqnBipartiteMemory13} holds with $c$ as in~\eqref{eqncjkPOVM} if one of the POVMs has rank-one elements \cite{coles10}, but it does not hold for general POVMs. This can be remedied in two ways. The approach by~\textcite{frank12} leads to a relation of the form~\eqref{eqnBipartiteMemory13} using a weaker complementarity factor. We have
\begin{align}
H(X | B) + H( Z | B) \geq\log\frac{1}{c''} + H(A|B)& \label{eqncPOVMdbprime}
\\
\textrm{where} \quad c'' = \max_{x,z } \tr\big[\bX^x \bZ^z\big]&\,.\label{eqncPOVMdbprime2}
\end{align}
Note that $c'' \geq c$ in general and that $c''$ reduces to $c$ for measurements in bases. However, one may argue that the form~\eqref{eqncPOVMdbprime}--\eqref{eqncPOVMdbprime2} is not the most natural one if we consider general projective measurements or POVMs. This is best explained by means of an example~\cite{furrer13}.

\begin{example}\label{ex:povm}
Consider a quantum system $A$ comprised of two qubits, $A_1$ and $A_2$, where $A_1$ is maximally entangled with a second qubit, $B$, and $A_2$ is in a fully mixed state, in product with $A_1$ and $B$. We employ rank-two projective measurements $\bX_{A_1}$ and $\bZ_{A_1}$ which measure $A_1$ in two MUBs and leave $A_2$ intact. Analogously, we employ $\bX_{A_2}$ and $\bZ_{A_2}$ which measure $A_2$ in two MUBs and leave $A_1$ intact. Evaluating the terms of interest for the measurement pairs $\{\bX_{A_1},\bZ_{A_1}\}$ and $\{\bX_{A_2},\bZ_{A_2}\}$ yields $c = \frac{1}{2}$ and $c'' = 1$ in both cases. Moreover, we find that
\begin{align}
H(A|B) = H(A_1|B) + H(A_2) = -1 + 1 = 0\,.
\end{align}
Hence, the right hand side of the Frank and Lieb relation~\eqref{eqncPOVMdbprime} vanishes for both measurement pairs. Indeed, if the maximally entangled system $A_1$ is measured, we find that
\begin{align}
H(X|B) + H(Y|B) = 0\,,
\end{align}
and the bound in~\eqref{eqncPOVMdbprime} becomes an equality for the measurement pair $\{\bX_{A_1},\bZ_{A_1}\}$. On the other hand, if $A_2$ is measured instead, we find that
\begin{align}
H(X|B) + H(Y|B) = 2\,,
\end{align}
and the bound is far from tight for the measurement pair $\{\bX_{A_2},\bZ_{A_2}\}$.
\end{example}  

Examining this example, it is clear that the expected uncertainty depends strongly on which of the two systems is measured. More generally, it depends on how much entanglement is consumed in the measurement process. However, this information is not taken into account by the overlaps $c$ or $c''$, nor by the entanglement of the initial state as measured by $H(A|B)$. Example~\ref{ex:povm} suggests that~\eqref{eqnBipartiteMemory13} can be generalized by considering the difference in entanglement of the state before and after measurement. In fact, \textcite{mythesis} shows the bipartite uncertainty relation
\begin{align}\label{eq:mtbipartite}
& H(X | B) + H( Z | B) \geq \log
\frac{1}{c'} + H(A|B) \notag\\
& \qquad \qquad \quad - \min \Big\{ H(A'|XB),\, H(A'|ZB) \Big\}\,,
\end{align}
with $c'$ given by~\eqref{eqncPOVMprime}. The entropy $H(A'|XB)$ is evaluated for the post-measurement state
\begin{align}
\rho_{XA'B} &= \sum_x \proj{x}_X \otimes (\bX_A^x \otimes \1_B) \rho_{AB} (\bX_A^x \otimes \1_B)\,,
\end{align}
and similarly for $H(A'|ZB)$. (We use $A' = A$ to denote the system $A$ after measurement to avoid confusion.) Notably, the term $H(A'|XB)$ vanishes for a measurement given by a basis since in this case the state of $A'$ is pure conditioned on $X$.

\newtheorem*{expovmcontinued}{Example~\ref{ex:povm} (continued)}
\begin{expovmcontinued}
It is straightforward to see that if $A_1$ ($A_2$) is measured, the average entanglement left in the post-measurement state measured by the von Neumann entropy is given by $H(A_2|B)$ ($H(A_1|B)$). Hence,~\eqref{eq:mtbipartite} turns into
\begin{align}
H(X|B)+H(Y|B)\geq\log \frac{1}{c}+\big(H(A|B)-H(A'|B)\big)\,,
\end{align}
where $A'$ corresponds to $A_2$ ($A_1$). This inequality is tight for both measurements.
\end{expovmcontinued}


\subsubsection{Multiple measurements}\label{sec:qmult_obs}

The basic goal here is to lift some of the relations in Sec.~\ref{sec:mult_obs} to quantum memory. 
A general approach for deriving such relations is provided in~\cite{dupuis15}. As in the unconditional case (cf.~Sec.~\ref{sec:two_implied}), relations for two measurements already imply bounds for larger sets of measurements. For example, supposing $A$ is a qubit and considering the Pauli measurements on $A$, we find by the simple iterative application of the bound~\eqref{eqnBipartiteMemory13} for the measurement pairs $\{\sigma_{\bX},\sigma_{\bY}\}$, $\{\sigma_{\bX},\sigma_{\bZ}\}$, and $\{\sigma_{\bY},\sigma_{\bZ}\}$ that
\begin{align}\label{eq:multiple_qubitqsi}
H(K|B\Theta)\geq\frac{1}{2}+\frac{1}{2}H(A|B)\;\;\mathrm{with}\;\;\Theta\in\left\{\sigma_{\bX},\sigma_{\bY},\sigma_{\bZ}\right\}\,.
\end{align}
Here, we use the following extension of the notation from Sec.~\ref{sec:mult_obs} to quantum memory,
\begin{align}
\rho_{K\Theta B}:=\frac{1}{3}&\sum_{k=1,2}\sum_{j=X,Y,Z}\proj{k}_{K}\otimes\proj{j}_{\Theta}\notag\\
&\otimes\Big(\bra{k}U_j^\dagger\otimes1_B\Big)\rho_{AB}\Big(U_j\ket{k}\otimes1_B\Big)\,.
\end{align}
Note that alternatively the left-hand side of~\eqref{eq:multiple_qubitqsi} might also be written as
\begin{align}
H(K|B\Theta)=\frac{1}{3}\Big(&H(K|B\Theta=\sigma_{\bX}) +H(K|B\Theta=\sigma_{\bY})\notag\\
&+H(K|B\Theta=\sigma_{\bZ})\Big)\,.
\end{align}
where
\begin{align}
&\rho_{KB|\Theta=\sigma_{\bX}}\notag\\
&:=\sum_{k=1,2}\proj{k}_{K}\otimes\Big(\bra{k}U_X^\dagger\otimes1_B\Big)\rho_{AB}\Big(U_X\ket{k}\otimes1_B\Big)\,,
\end{align}
and similarly for $\sigma_{\bY},\sigma_{\bZ}$. The goal in this subsection will be to find uncertainty relations that are stronger than any bounds that can be derived directly from relations for two measurements.


\subsubsection{Complex projective two-designs}\label{sec:mub_qsi}

\textcite{berta13} showed that the uncertainty equality~\eqref{eq:h_2meta_noqsi} in terms of the collision entropy for a full set of MUBs also holds with quantum memory. That is, for any bipartite state $\rho_{AB}$ with a full set of $d+1$ MUBs on the $d$-dimensional $A$-system,
\begin{align}\label{eq:h2_bcw}
H_{\rm coll}(K|B\Theta)=\log(d+1)-\log\left(2^{-H_{\rm coll}(A|B)}+1\right)&\notag\\
\text{with}\;\;\Theta\in\{\theta_1,\ldots,\theta_{d+1}\}&\,.
\end{align}
Here, as in~\eqref{eq:ext_notation}, we use the notation,
\begin{align}
\rho_{K\Theta B}:=\frac{1}{d+1}&\sum_{k=1}^{d}\sum_{j=1}^{d+1}\proj{k}_{K}\otimes\proj{j}_{\Theta}\notag\\
&\otimes\left(\bra{k}U_j^\dagger\otimes1_B\right)\rho_{AB}\left(U_j\ket{k}\otimes1_B\right)\,.
\end{align}

\begin{example}
For the qubit Pauli measurements~\eqref{eq:h2_bcw} yields:
\begin{align}\label{eq:qubit_h2equal}
H_{\rm coll}(K|B\Theta)=\log3-\log\left(2^{-H_{\rm coll}(A|B)}+1\right)&\notag\\
\text{with}\;\;\Theta\in\{\sigma_{\bX},\sigma_{\bY},\sigma_{\bZ}\}&\,.
\end{align}
\end{example}

Since the collision entropy has an interpretation in terms of the pretty good guessing probability~\eqref{eq:h2_cq},
\begin{align}
H_{\rm coll}(X|B)=-\log p_{\rm guess}^{\rm pg}(X|B)\,,
\end{align}
and the pretty good recovery map~\eqref{eq:h2_qq},
\begin{align}
H_{\rm coll}(A|B)=-\log \big(d_A\cdot F^{\rm pg}(A|B)\big)\,,
\end{align}
the uncertainty equality~\eqref{eq:h2_bcw} can be understood as an entanglement-assisted game of guessing complementary measurement outcomes (as described in Sec.~\ref{sec:bipartiteguessing}). Namely, we can rewrite~\eqref{eq:h2_bcw} as,
\begin{align}
p_{\rm guess}^{\rm pg}(K|B\Theta)=\frac{d\cdot F^{\rm pg}(A|B)+1}{d+1}\,.
\end{align}
This gives a one-to-one relation between uncertainty (certainty) as measured by $p_{\rm guess}^{\rm pg}(K|B\Theta)$ and the absence (presence) of entanglement as measured by $F^{\rm pg}(A|B)$. In contrast, quantum memory assisted uncertainty relations for two measurements, e.g., as in~\eqref{eqnBipartiteMemory19}, only provide us with a connection between uncertainty and entanglement in one direction. Namely, they state that low uncertainty implies the presence of entanglement (cf.~Sec.~\ref{sec:entwit}).

The uncertainty equality~\eqref{eq:h2_bcw} is derived by extending the proof from~\cite{ballester07} that made use of the fact that a full set of mutually unbiased bases generates a complex projective two-design~\cite{klappenecker05}. From this, it is also immediate that an equality as~\eqref{eq:h2_bcw} holds for other complex projective two-designs as well. This includes in particular so-called symmetric informationally complete positive operator valued measures: SIC-POVMs.\footnote{We refer to~\cite{renes04} for a detailed discussion of SIC-POVMs.} More precisely, any SIC-POVM
\begin{align}
\left\{\frac{1}{d}\ket{\psi_k}\bra{\psi_k}\right\}_{k=1}^{d^2}
\end{align}
gives rise to the uncertainty equality
\begin{align}
H_{\rm coll}(K|B\Theta)=\log\Big(d(d+1)\Big)-\log\left(2^{-H_{\rm coll}(A|B)}+1\right)&\notag\\
\text{with}\;\;\Theta\in\{\theta_1,\ldots,\theta_{d+1}\}&\,.
\end{align}
Other examples that generate complex projective two-designs are unitary two-designs.\footnote{We refer to~\cite{dankert09} for a detailed discussion of unitary two-designs.} This includes in particular the Clifford group for $n$ qubit systems.

\textcite{berta14} also showed that the relation~\eqref{eq:h2_bcw} for a full set of $d+1$ MUBs generates the following relation in terms of the von Neumann entropy,
\begin{align}
H(K|B\Theta)\geq\log(d+1)-1+\min\Big\{0,H(A|B)\Big\}&\notag\\
\text{with}\;\;\Theta\in\{\theta_1,\ldots,\theta_{d+1}\}&\,.
\end{align}
This corresponds to the generalization of~\eqref{eq:mub_full} to quantum memory. Note that the entropy dependent term on the right-hand side only makes a contribution if the conditional entropy $H(A|B)$ is negative. This is consistent with~\eqref{eq:mub_full}.

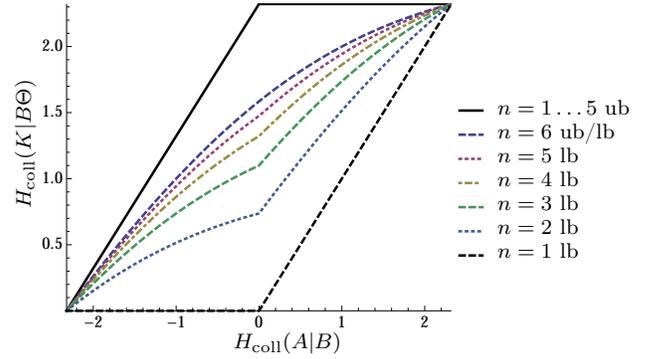
\begin{figure}[tbp]
\begin{center}
\begin{overpic}[width=0.7\columnwidth]{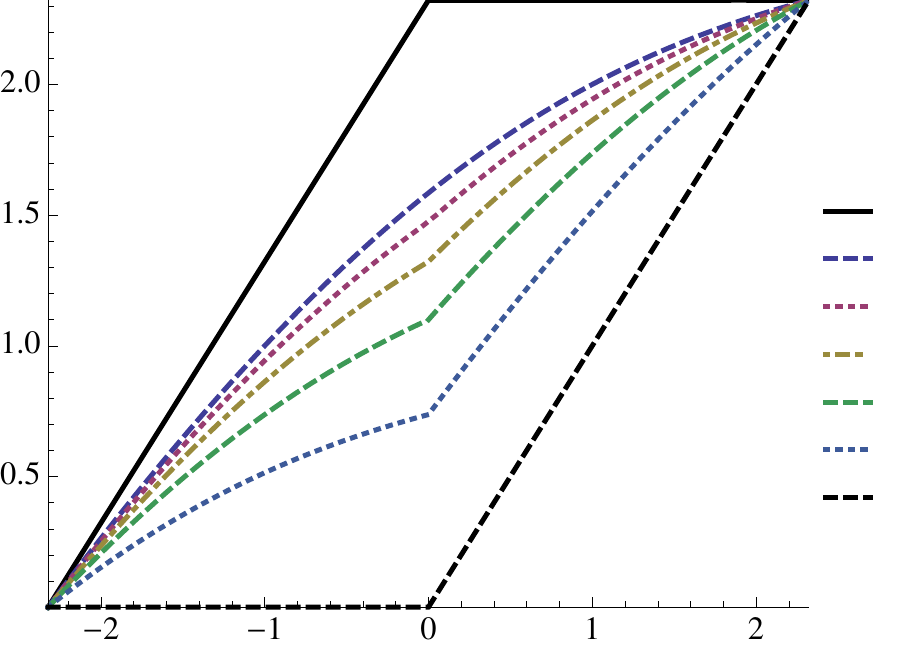}
\put(-6,27){\footnotesize \rotatebox{90}{$H_{\rm coll}(K|B\Theta)$}}
\put(42,-3.8){\footnotesize $H_{\rm coll}(A|B)$}
\put(100,47.5){\footnotesize $n=1\ldots5$ ub}
\put(100,42.3){\footnotesize $n=6$ ub/lb}
\put(100,37.1){\footnotesize $n=5$ lb}
\put(100,31.8){\footnotesize $n=4$ lb}
\put(100,26.5){\footnotesize $n=3$ lb}
\put(100,21.3){\footnotesize $n=2$ lb}
\put(100,16.0){\footnotesize $n=1$ lb}
\end{overpic}
\hspace{1.1cm}
\vspace{0.2cm}
\caption{For $d=5$ and a various number of MUBs $n \leq d+1=6$ we plot the lower bounds (lb) on the entropic uncertainty $H_{\rm coll}(K|\Theta)$ from~\eqref{eq:less_qsi} as a function of $H_{\rm coll}(A|B)$. Moreover, for $n<6$ we only have the trivial upper bound (ub) on $H_{\rm coll}(K|B\Theta)$, whereas for $n=6$ the lower and upper coincide as in~\eqref{eq:h2_bcw}.}
\label{fig:mub}
\end{center}
\end{figure}

For smaller sets of $L<d+1$ MUBs, \textcite{berta13} extended~\eqref{eq:wu_master} to quantum memory,
\begin{align}\label{eq:less_qsi}
&H_{\rm coll}(K|B\Theta)\notag\\
&\geq\left\{\begin{array}{lcl}
-\log\frac{d\cdot2^{-H_{\rm coll}(A|B)}+L-1}{L\cdot d}&\mathrm{for}&H_{\rm coll}(A|B)\geq0\\
-\log\frac{d+(L-1)2^{-H_{\rm coll}(A|B)}}{L\cdot d}&\mathrm{for}&H_{\rm coll}(A|B)<0\end{array}\right.\notag\\
&\qquad\qquad\qquad\qquad\qquad\qquad\quad\text{with}\;\;\Theta\in\{\theta_1,\ldots,\theta_{L}\}\,.
\end{align}
Moreover, for all $d$ and $L$ there exist states that achieve equality. Note that for $L=d+1$ the distinction of cases in~\eqref{eq:less_qsi} collapses and furthermore become an upper bound as shown in~\eqref{eq:h2_bcw}. In Fig.~\ref{fig:mub} we illustrate this by means of an example for $d=5$ (with $L\leq6$).

\subsubsection{Measurements in random bases}\label{sec:randomunitary_qsi}

In the unconditional case we found that measurements in random bases lead to strong uncertainty relations as, e.g., in~\eqref{eq:adamczak14_multiple}. Hence, we might expect that we can generalize this to quantum memory,
\begin{align}\label{eq:random_conjecture}
H(K|B\Theta)\stackrel{?}{\geq}O\left(\log d\cdot\left(1-\frac{1}{L}\right)\right)+\min\Big\{0,H(A|B)\Big\}&\notag\\
\mathrm{with}\;\;\Theta\in\{\theta_1,\dots,\theta_L\}\;\;\text{chosen at random}&\,.
\end{align}
Unfortunately, the previous works~\cite{fawzi11} and~\cite{adamczak14} make use of measure concentration and $\eps$-nets arguments that seem to fail for quantum memory. It is, however, possible to use some of the techniques from~\cite{berta13} based on operator Chernoff bounds to derive relations of the form~\eqref{eq:random_conjecture}. The downside is that we only get strong uncertainty relations for a large number $L$ of measurements,
\begin{align}
L\geq O\Big(d\log(d)\Big)\,.
\end{align}
We conclude that it is an open problem to show the existence of small(er) sets of $L>2$ measurements that generate strong uncertainty relations that hold with quantum memory.


\subsubsection{Product measurements on multiple qubits}\label{sec:bb84six_qm}

Let us now consider uncertainty relations for multiple-qubit systems, which have application in quantum cryptography. For historical reasons we start with the $n$ qubit six-state measurements and only discuss the BB84 measurements afterwards (see Sec.~\ref{sec:product_measurements_no_memory} for definitions of these measurements). For the six-state measurements, \textcite{bertathesis} showed that for any bipartite state $\rho_{A^nB}$ with the $A^n$-system given by $n$ qubits,
\begin{align}\label{eq:sixstate_old}
H_{\rm coll}(K^n|B\Theta^n)\geq n\cdot\log\frac{3}{2}+1-\log\left(2^{-H_{\rm coll}(A^n|B)}+1\right)&\notag\\
\mathrm{with}\;\;\Theta^n\in\{\theta_1,\ldots,\theta_{3^n}\}&\,.
\end{align}
This extends~\eqref{eq:qubit_h2equal} from one to $n$ qubits. The bound~\eqref{eq:sixstate_old} also implies a similar relation in terms of the von Neumann entropy~\cite{berta14}, extending~\eqref{eq:sixstate_uncond} to
\begin{align}\label{eq:sixstate_vNqm}
H(K^n|B\Theta^n)\geq n\cdot\log\frac{3}{2}+\min\Big\{0,H(A^n|B)_\rho\Big\}&\notag\\
\mathrm{with}\;\;\Theta^n\in\{\theta_1,\ldots,\theta_{3^n}\}&\,.
\end{align}
Moreover, \textcite{dupuis15} improved~\eqref{eq:sixstate_old} to the conceptually different bound
\begin{align}\label{eq:sixstate_new}
H_{\rm coll}(K^n|B\Theta^n)\geq n\cdot\gamma_{6s}\left(\frac{H_{\rm coll}(A^n|B)}{n}\right)-1\,,
\end{align}
where
\begin{align}\label{eq:gamma}
&\gamma_{6s}(x):=\left\{\begin{array}{ll}x&\mathrm{if}\;\;x\geq\frac{\log3}{2}\\f^{-1}(x)\log3&\mathrm{if}\;\;0<x<\frac{\log3}{2}\end{array}\right.\notag\\
&\mathrm{with}\quad f(x)=h_{\bin}(x)+x\log3-1\,,
\end{align}
and $h_{\bin}$ denotes the binary entropy. Using the equivalence between the collision entropy and the min-entropy from~\eqref{eq:hmin_h2} this readily implies a relation as~\eqref{eq:sixstate_new}, but with both the collision entropy terms $H_{\rm coll}$ replaced with min-entropy terms $H_{\min}$. Importantly, this variant remains non-trivial for all values of $H_{\rm min}(A^n|B)$. What's more, \textcite{dupuis15} establish a meta theorem that can be used to derive uncertainty relations also for other kinds of measurements.

For the $n$ qubit BB84 measurements~\textcite{dupuis15} found
\begin{align}\label{eq:bb84_h2qm}
H_{\rm coll}(K^n|B\Theta^n)\geq n\cdot\gamma_{BB84}\left(\frac{H_{\rm coll}(A^n|B)}{n}\right)-1&\notag\\
\mathrm{with}\;\;\Theta^n\in\{\theta_1,\ldots,\theta_{2^n}\}&\,.
\end{align}
where
\begin{align}\label{eq:gammabar}
&\gamma_{BB84}(x):=\left\{\begin{array}{ll}x&\mathrm{if}\;\;x\geq\frac{1}{2}\\g^{-1}(x)&\mathrm{if}\;\;0<x<\frac{1}{2}\end{array}\right.\notag\\
&\mathrm{with}\quad g(x)=h_{\bin}(x)+x-1\,.
\end{align}
Again using the equivalence between the collision entropy and the min-entropy from~\eqref{eq:hmin_h2}, we get a relation as~\eqref{eq:bb84_h2qm} but with both the collision entropy terms $H_{\rm coll}$ replaced with min-entropy terms $H_{\min}$. We note that this is also non-trivial for one qubit $(n=1)$ and only the two measurements $\Theta\in\{\sigma_{\bX},\sigma_{\bZ}\}$. The relation~\eqref{eq:bb84_h2qm} and its min-entropy analogue can be understood in terms of the bipartite guessing game with quantum memory as mentioned in Sec.~\ref{sec:bipartiteguessing}.


\subsubsection{General sets of measurements}

Section~\ref{sec:general_multi_measurements} discussed the work of \textcite{liu15} for unipartite systems without memory. Here, we note that they also gave bipartite uncertainty relations with quantum memory. Again for simplicity we only state the case of $L=3$ observables (in any dimension $d\geq2$). We find as the direct extension of~\eqref{eq:liu_multiple},
\begin{align}\label{eq:liu_qsi}
H(K|B\Theta)\geq\frac{1}{3}\log\frac{1}{m}+\frac{2}{3}H(A|B)&\notag\\
\mathrm{with}\;\;\Theta\in\{V^{(1)},V^{(2)},V^{(3)}\}&\,,
\end{align}
where the multiple overlap constant $m$ is defined as in~\eqref{eq:b_constant}. As in the unconditional case, this has to be compared with the bounds implied by two measurement relations as in~\eqref{eq:multiple_qubitqsi}. We refer to~\textcite{liu15} for a fully worked out example where~\eqref{eq:liu_qsi} can become stronger than any bounds implied by two measurement relations.


\subsection{Tripartite quantum memory uncertainty relations}\label{scttripartiteQMemory}

\subsubsection{Tripartite uncertainty relation}\label{sec:tripartiteuncertaintyrelations}

\begin{figure}[tbp]
\begin{center}
\begin{overpic}[width=8.1cm]{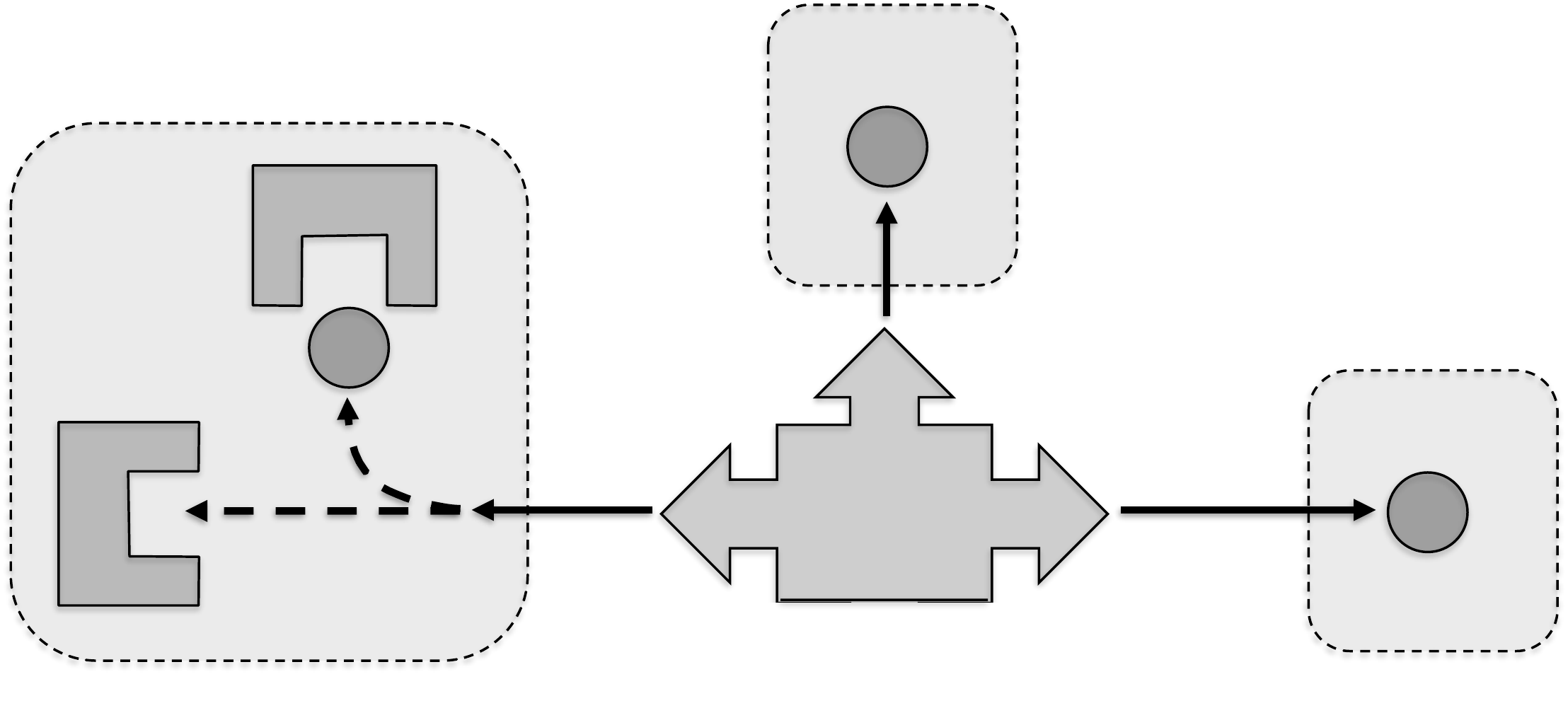}
\put(88.2,18){\footnotesize Bob}
\put(4,30){\footnotesize Alice}
\put(51,42){\footnotesize Charlie}
\put(52.2,13){\footnotesize $\rho_{ABC}$}
\put(28.7,28){\footnotesize $\bZ$}
\put(10,20){\footnotesize $\bX$}
\put(88,6.5){\footnotesize $X$?}
\put(50.3,29.5){\footnotesize  $Z$?}
\end{overpic}
\caption{Diagram showing the tripartite quantum memory setup. First, a source prepares $ABC$ in state $\rho_{ABC}$, and sends $A$ to Alice, $B$ to Bob, and $C$ to Charlie. Second, Alice measures either $\bX$ or $\bZ$ on $A$ and asks: how uncertain is Bob about her $X$ outcome, given $B$, and how uncertain is Charlie about her $Z$ outcome, given $C$? As shown in~\eqref{eqnTripartiteMemory2} there is a trade-off that is quantified by the complementarity of the measurements $X$ and $Z$. We can interpret this scenario as a guessing game, also called a monogamy game. In this game, Bob and Charlie play against Alice. They prepare $\rho_{ABC}$ where they send $A$ to Alice, Bob keeps $B$ and Charlie keeps $C$. Alice then randomly chooses a measurement obtaining the measurement outcome $K$. Afterwards, she sends her choice of basis to Bob and Charlie. They win the game if and only if both output $K$. This game measures the same kind of uncertainty as the relation~\eqref{eqnTripartiteMemory2}, explicitly exploiting the monogamy of entanglement: if Bob produces $K=X$ correctly in case Alice measured $\bX$, then this is a certificate that Charlie cannot produce a good guess of $K=Z$ in case Alice measured $\bZ$.
}
\label{figtripartitememory}
\end{center}
\end{figure}
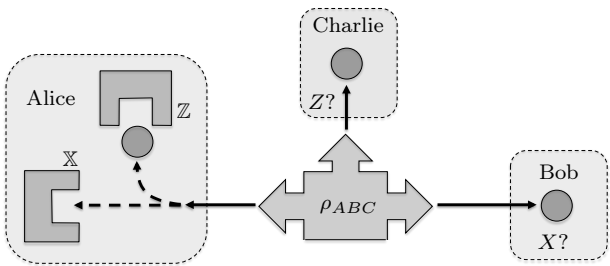

The physical scenario corresponding to tripartite uncertainty relations is shown in Fig.~\ref{figtripartitememory}. Suppose there is a source that outputs the systems $ABC$ in state $\rho_{ABC}$. Systems $A$, $B$, and $C$ are respectively sent to Alice, Bob, and Charlie. Then Alice performs either the $\bX$ or $\bZ$ measurement. If she measures $\bX$, then Bob's goal is to minimize his uncertainty about $X$. If she measure $\bZ$, then Charlie's goal is to minimize his uncertainty about $Z$. \textcite{renes09} considered exactly this scenario but restricted to the case where the $\bX$ and $\bZ$ bases are related by the Fourier matrix $F$,
\begin{align}\label{eqnfourier1}
\ket{\bX^x} = F \ket{\bZ^x}\quad\mathrm{with}\quad F = \sum_{z,z'} \frac{\omega^{-zz'}}{\sqrt{d}} \dyad{\bZ^z}{\bZ^{z'}}&\notag\\
\text{where}\quad \om = e^{2\pi i/d}&\,.
\end{align}
Notice that this makes $\bX$ and $\bZ$ mutually unbiased, although in general not all pairs of MUBs are related by the Fourier matrix. They quantified Bob's and Charlie's uncertainty in terms of the conditional entropies $H(X|B)$ and $H(Z|C)$ respectively, and proved that any tripartite state $\rho_{ABC}$ satisfies the relation
\begin{align}\label{eqnTripartiteMemory1}
H(X|B)+H(Z|C)\geq\log d\,.
\end{align}
Here, $d$ is the dimension of the $A$ system and the classical-quantum states $\rho_{XB}$ and $\rho_{ZC}$ are defined similarly as in~\eqref{eq:cq_postmeas}. \textcite{renes09} also conjectured that this relation generalizes to arbitrary measurements given by bases,
\begin{align}\label{eqnTripartiteMemory2}
H(X|B)+H(Z|C)\geq q_{\rm MU}\,,
\end{align}
with $q_{\rm MU}$ as in~\eqref{eq:shannonUR}. Intuitively, what~\eqref{eqnTripartiteMemory1} says is that the more Bob knows about $Z$, the less Charlie knows about $X$, and vice-versa. This is a signature of the well-known trade-off \textit{monogamy of entanglement}, which roughly says that the more Bob is entangled with Alice, the less he is with Charlie.\footnote{We refer to~\cite{horodecki09} for an in-depth review about entanglement.} The trade-off described by~\eqref{eqnTripartiteMemory1} and~\eqref{eqnTripartiteMemory2} can be viewed as a fine-grained notion of this monogamy. Namely, the monogamy appears at the level of measurement pairs $(\bX, \bZ)$.

Also note that~\eqref{eqnTripartiteMemory2} implies both the Maassen-Uffink relation~\eqref{eq:shannonUR} and its classical memory extension~\eqref{eqnClassicalMem55}, due to the data-processing inequality~\eqref{eq:data_processing}. That is, 
\begin{align}\label{eqnTripartiteMemory45}
H(X|B)\leq H(X|Y)\leq H(X)\,,
\end{align}
for any measurement $\bY$ on $B$. As we will see in Sec.~\ref{sec:qm_tightens} the quantum memory extension~\eqref{eqnTripartiteMemory2} is strictly stronger than the classical memory extension~\eqref{eqnClassicalMem55}.

\subsubsection{Proof of quantum memory uncertainty relations}

The quantum memory uncertainty relation~\eqref{eqnTripartiteMemory2} was first proved by~\textcite{berta10}. Although these authors explicitly stated their relation in the bipartite form~\eqref{eqnBipartiteMemory13}, they noted  that two relations are equivalent.

The equivalence between the bipartite and tripartite relations can be seen as follows. To obtain the bipartite relation~\eqref{eqnBipartiteMemory13} from the tripartite relation~\eqref{eqnTripartiteMemory2}, apply the latter to a purification $\ket{\psi}_{ABC}$ of $\rho_{AB}$. Now for tripartite pure states we have,
\begin{align}\label{eqnTripartiteMemory46}
H(Z|C) =  H(Z|B) -  H(A|B)\,,
\end{align}
and inserting this into~\eqref{eqnTripartiteMemory2} gives~\eqref{eqnBipartiteMemory13}. Conversely we first prove~\eqref{eqnTripartiteMemory2} for tripartite pure states $\ket{\psi}_{ABC}$ by inserting~\eqref{eqnTripartiteMemory46} into \eqref{eqnBipartiteMemory13}. Then, note that the proof for mixed states $\rho_{ABC}$ follows by applying~\eqref{eqnTripartiteMemory2} to a purification $\ket{\psi}_{ABCD}$ of $\rho_{ABC}$, and making use of the data-processing inequality~\eqref{eq:data_processing},
\begin{align}
H(Z|CD)\leq H(Z|C)\,.
\end{align}

The original proof of~\eqref{eqnTripartiteMemory2} was based on so-called smooth entropies.\footnote{We refer to~\textcite{mybook} for an introduction to smooth entropies.} The proof was subsequently simplified by~\textcite{tomamichel11} and \textcite{coles10}, which culminated in the concise proof given by~\textcite{colbeck11}. The latter proof distills the main ideas of the previous proofs: the use of duality relations for entropies as in~\eqref{eqndualentropy1} and the data-processing inequality as in~\eqref{eq:data_processing}. More generally, the proof technique applies to a whole family of entropies satisfying a few axioms (including the R\'enyi entropies). We will present the proof in App.~\ref{sec:memory_proof}. Finally, we note that a direct matrix analysis proof was given by~\textcite{frank12}.


\subsubsection{Quantum memory tightens the bound}\label{sec:qm_tightens}

Here, we argue that the tripartite uncertainty relation in terms of quantum memory~\eqref{eqnTripartiteMemory2} is tighter than the corresponding relation in terms of classical memory~\eqref{eqnClassicalMem55}. We will show that there exist states $\rho_{ABC}$ for which~\eqref{eqnTripartiteMemory2} is an equality but~\eqref{eqnClassicalMem55} is loose, even if one optimizes over all choices of measurements on $B$ and $C$.

Let us introduce some notation. Consider a bipartite state $\rho_{AB}$ and let $\bX_A$ and $\bY_B$ be measurements on systems $A$ and $B$, respectively. Now, how small can we make the uncertainty $\bX_A$ given that we can optimize over all choices of $\bY_B$? That is, consider the quantity
\begin{align}\label{eqnalphadef1}
\al (\bX_A, \rho_{AB}):= \min_{\bY_B} H(X_A | Y_B)\,.
\end{align}
This is to be compared to the classical-quantum conditional entropy
\begin{align}\label{eqnbetadef1}
\bt(\bX_A, \rho_{AB}):= H(X_A | B) \,.
\end{align}
Due to the data-processing inequality~\eqref{eq:data_processing}, we have that
\begin{align}\label{eqnclassicalvsquantummem}
\al(\bX_A, \rho_{AB}) \geq \bt(\bX_A, \rho_{AB})\,,
\end{align}
and naively one might guess that~\eqref{eqnclassicalvsquantummem} is satisfied with equality in general. However, this is false~\cite{hiai91,divincenzo04}. In general there is a non-zero gap: $\al - \bt > 0$. There are many examples to illustrate this, in fact one can argue that \emph{most} states $\rho_{AB}$ exhibit a gap between $\al$ and $\bt$~\cite{hayden13}. This phenomenon is called locking and we discuss it more in Sec.~\ref{sec:locking}. It is closely related to a measure of quantum correlations known as quantum discord \cite{ollivier01,modi12}. Non-zero discord is associated with the potential to have a gap between $\al$ and $\bt$. We discuss discord in more detail in Sec.~\ref{sec:discord}. For now we note that discord is defined as,
\begin{align}\label{eqndiscorddef1}
D(A|B) := \min_{\bY_B} H(A|Y_B)-H(A|B)
\end{align}
where the optimization is over all POVMs $\bY_B$ on $B$.

\begin{example}\label{exQmemlocking}
Let $\bX_A = \{\dya{0},\dya{1}\}$ and consider the bipartite quantum state:
\begin{align}\label{eq:gap_example}
\rho_{AB} = \frac{1}{2}\Big(\dya{0}\ot \dya{0} + \dya{1}\ot \dya{+}\Big)\,.
\end{align}
For this state, the gap between $\al$ and $\bt$ is precisely given by the discord,
\begin{align}
D(A|B) = \al(\bX_A, \rho_{AB}) - \bt(\bX_A, \rho_{AB})\,.
\end{align}
It is known that $D(A|B) = 0$ if and only if system $B$ is classical, i.e., if $\rho_{AB}$ is a quantum-classical state. But the state $\rho_{AB}$ in~\eqref{eq:gap_example} is not quantum-classical. Hence, $D(A|B) >0$ and we have $\al > \bt$.
\end{example}

Now we give an example state for which the quantum memory relation~\eqref{eqnTripartiteMemory2} is an equality but the measured relation~\eqref{eqnClassicalMem55} is loose.

\begin{example}\label{exQmem}
Consider the tripartite pure state $\ket{\psi}_{ABC}= (\ket{000}+\ket{11+})/\sqrt{2}$, with $\bZ$ being the standard basis and $\bX$ being the $\{\ket{+} ,\ket{-}\}$ basis. We have
\begin{align}
H(Z|C) &= 1 - H(\rho_C) \approx 0.4\\
H(X|B) &= H(\rho_C) \approx 0.6\,.
\end{align}
Hence, this state satisfies the quantum memory relation~\eqref{eqnTripartiteMemory1} with equality,
\begin{align}
H(X | B)+H(Z| C)=1\,.
\end{align}
However, the classical memory relation~\eqref{eqnClassicalMem55} is not satisfied with equality. This follows from Ex.~\ref{exQmemlocking}, noting that $\rho_{AC}$ is the same state as in~\eqref{eq:gap_example}.
\end{example}


\subsubsection{Tripartite guessing game}\label{sec:tripartite_guessing}

Tripartite uncertainty relations can be understood in the language of guessing games as outlined in Fig.~\ref{figtripartitememory}. \textcite{tomamichelfehr12} show that there is a fundamental trade-off between Bob's guessing probability $p_{\rm guess}(K|B\Theta)$ and Charlie's guessing probability $p_{\rm guess}(K|C\Theta)$,
\begin{align}\label{eq:marco_guessing1}
p_{\rm guess}(K|B\Theta)+ p_{\rm guess}(K|C\Theta)\leq 2b\,,
\end{align}
with the overlap $b$ as in~\eqref{eqnbdef}. Alternatively, one can rewrite this in terms of the min-entropy using the concavity of the logarithm,
\begin{align}\label{eq:marco_guessing2}
H_{\min}(K|B\Theta)+ H_{\min}(K|C\Theta)\geq 2\log \frac{1}{b}\,.
\end{align}
Note that this relation~\eqref{eq:marco_guessing2} is an extension of~\eqref{eqnguessingprobaverage3} to the tripartite scenario. This relation again shows a trade-off between Bob's and Charlie's winning probability, which is closely connected to the idea of monogamy of entanglement (cf.~Sec.~\ref{sec:tripartiteuncertaintyrelations}).


\subsubsection{Extension to R\'enyi entropies}\label{sec:renyi_muffinkq}

The Maassen-Uffink relation for R\'enyi entropies~\eqref{eq:mu_relation} naturally generalizes to a tripartite uncertainty relation with quantum memory. It is expressed in terms of the conditional R\'enyi entropies, whose definition and properties are discussed in App.~\ref{app:renyi}. For these entropies, the following relation holds~\cite{colbeck11}\footnote{More precisely, the relation follows from the work~\cite{colbeck11} in conjunction with properties of the conditional R\'enyi entropy presented in~\cite{lennert13}. It is thus first mentioned in the later work~\cite{lennert13}. Notably, \textcite{colbeck11} proves a tripartite uncertainty relation for a different definition of the conditional R\'enyi entropy~\cite{tomamichel08}.}
\begin{align}\label{eqnTripartiteMemory435}
H_{\alpha}(X | B) + H_{\beta}( Z | C) \geq q_{\rm MU} \quad \textrm{for} \quad \frac{1}{\alpha} + \frac{1}{\beta} = 2 \,.
\end{align}
Notably, the tripartite uncertainty relation~\eqref{eqnTripartiteMemory2} is the special case where $\alpha = \beta = 1$. Another interesting special case is $\alpha = \infty$ and $\beta = 1/2$, which respectively correspond to the min- and max-entropies that we introduced in \eqref{eqncqminentropydef} and \eqref{eqndualentropy3}. The resulting relation,
\begin{align}\label{eqnminmaxuncertaintyqmem}
H_{\min}(X | B) + H_{\max}( Z | C) \geq q_{\rm MU}\,,
\end{align}
was first proved by \textcite{tomamichel11}, and is fundamental to quantum key distribution (see Sec.~\ref{sec:qkd}).


\subsubsection{Arbitrary measurements}\label{scttripartitepovm}

All of the tripartite uncertainty relations stated above can be generalized to arbitrary POVMs $\bX$ and $\bZ$. \textcite{coles10,tomamichel11} independently noted that~\eqref{eqnTripartiteMemory2} holds for POVMs with the overlap $c$ given by~\eqref{eqncjkPOVM}. This was strengthened by~\textcite{mythesis} to the overlap $c'$ given by~\eqref{eqncPOVMprime}. Further strengthening was given by~\textcite{coles14}. However, their bound is implicit, involving an optimization of a single real-valued parameter over a bounded interval. Namely, they showed a lower bound
\begin{align}
q_{\rm CP2}:= \max_{0\leq p \leq 1} \lambda_{\min}(\Delta(p))\,,
\end{align}
where $\lambda_{\min}[\cdot]$ denotes the minimum eigenvalue and
\begin{align}
\Delta(p) &:= p\hspace{2pt}\delta(\bX , \bZ) + (1-p)\delta(\bZ , \bX)\\
\delta(\bX , \bZ)&:= \sum_x a_x(\bX , \bZ) \cdot \bX^x\\
a_x(\bX , \bZ)&:= - \log \Big\|\sum_z \bZ^z \bX^x \bZ^z   \Big\|\,.
\end{align}
Using the fact that $\delta(\bX , \bZ) \geq \min_x a_x(\bX , \bZ) \cdot \1$, it is straightforward to show that $ q_{\rm CP2} \geq \log (1/c')$. 


\subsection{Mutual information approach}\label{sctmutinfo}

While entropy quantifies the lack of information, it is both intuitive and useful to also consider measures that quantify the \textit{presence} of information or correlation. Consider the \textit{mutual information} $I(X\!:\!Y)$, which quantifies the correlation between random variables $X$ and $Y$, and is given by
\begin{align}\label{eqnMutInfo1}
I(X\!:\!Y) &:= H(X) + H(Y) - H(XY)\\
&= H(X) - H(X|Y)  \,.
\end{align}
It quantifies the information gained\,---\,or equivalently, the reduction of ignorance\,---\,about $X$ when given access to $Y$. It is worth noting that the mutual information is particularly well-suited for applications in information theory. For example, the capacity of a channel can be expressed in terms of its mutual information~\cite{shannon48}, that is, in terms of the correlations between a receiver and a sender. Hence, we will also discuss the application of ``information exclusion relations'' (uncertainty relations expressed via the mutual information) to information transmission over channels.


\subsubsection{Information exclusion principle}

\textcite{hall95,hall97} pioneered an alternative formulation of the uncertainty principle based on the mutual information, which he called the \textit{information exclusion principle}. Information exclusion relations are closely related to entropic uncertainty relations that allow for memory. The idea is that one is interested in the trade-off between a memory system $Y$ being correlated to $\bX$ versus being correlated to $\bZ$ (with $\bX$ and $\bZ$ being two measurements on some quantum system $A$).


\subsubsection{Classical memory}

We show now how information exclusion relations follow directly from entropic uncertainty relations~\cite{hall95}. Consider a generic uncertainty relation involving Shannon entropy terms, of the form $\sum_{n=1}^N H(X_n)_{\rho} \geq q$ as in~\eqref{eq:shannon_generic}. Recall the discussion in Sec.~\ref{sctCMemory} which showed that the uncertainty relation $\sum_{n=1}^N H(X_n | Y) \geq q$ as in~\eqref{eqnClassicalMem3} immediately follows, where $Y$ is some classical memory. Now with the definition of the mutual information~\eqref{eqnMutInfo1} we can rewrite this as
\begin{align}\label{eqnMutInfo123}
\sum_{n=1}^N H(X_n) - I(X_n : Y) \geq q\,.
\end{align}
We have $H(X_n) \leq \log d$ for each $n$ with $d$ the dimension of the quantum system $A$ that is measured. Combining this with \eqref{eqnMutInfo123} gives
\begin{align}\label{eqnMutInfo124}
\sum_{n=1}^N I(X_n :Y) \leq  N\log d - q\,.
\end{align}
For example, if we take the Maassen-Uffink relation~\eqref{eq:shannonUR} as the starting point, we end up with
\begin{align}\label{eqnMutInfo125}
I(X : Y) +I(Z : Y) \leq  \log (d^2 c) =: r_{\rm H}
\end{align}
The information exclusion relation in~\eqref{eqnMutInfo125} was presented by~\textcite{hall95}. Note that we have $\log d \leq r_{\rm H}  \leq 2\log d$, with $r_{\rm H}$ reaching the extreme points respectively for $c = 1/d$ and $c = 1$. Equation~\eqref{eqnMutInfo125} has an intuitive interpretation: any classical memory cannot be highly correlated to two complementary measurement outcomes of a quantum system. In the fully complementary case, the bound becomes $r_{\rm H}=\log d$, implying that if the classical memory is perfectly correlated to $X$, $I(X : Y) = \log d$, then it must be completely uncorrelated to $Z$, $I(Z : Y) =0$.


\subsubsection{Stronger bounds}

Notice that~\eqref{eqnMutInfo125} uses the same overlap $c$ as appearing in the Maassen-Uffink uncertainty relation~\eqref{eq:shannonUR}. However, \textcite{grudka12} realized that this often leads to a fairly weak bound. They noted that the complementarity of the mutual information should depend not only on the maximum element $c$ of overlap matrix $[c_{xz}]$ (see~\eqref{eq:overlap} for its definition), but also on other elements of this matrix. They conjectured a stronger information exclusion relation, of the form $I(X:Y)+I(Z:Y) \leq r_{\rm G}$ with
\begin{align}\label{eqn23782bb}
r_{\rm G} = \log_2 \left(d \cdot \sum_{\text{d largest}} c_{xz}\right),
\end{align}
with the sum over the largest $d$ terms of the matrix $[c_{xz}]$. This conjecture was proved by~\textcite{coles14}, where the bound was further strengthened to
\begin{align}\label{eqn23785}
I(X\colo Y)+I(Z\colo Y) \leq r_{\rm CP},
\end{align}
with
\begin{subequations}\label{eqn237431}
\begin{align}
r_{\rm CP}&:= \min \Big\{r(\bX,\bZ), r(\bZ,\bX)\Big\}\\
r(\bX,\bZ)&:= \log\left(\ddd \sum_x \max_z c_{xz}\right)\\
r(\bZ,\bX)&:= \log\left(\ddd \sum_z \max_x c_{xz}\right)\,.
\end{align}
\end{subequations}
One can easily verify that $r_{\rm CP} \leq r_{\rm G} \leq r_{\rm H}$.

\begin{example}
The unitary in~\eqref{eqnU3930} from Ex.~\ref{exQutrit} provides a simple example where all three bounds are different, namely $r_{\rm H} = \log 6$, $r_{\rm G} = \log 5$, and $r_{\rm CP} = \log (9/2)$. 
\end{example}

Notice that the behavior of the bounds $r_{\rm H}$ and $r_{\rm CP}$ are qualitatively different in that they become trivial under different conditions. The former is trivial if at least one row or column of $[c_{xz}]$ is trivial (i.e., composed of all zeros except for one element being one), whereas the latter is trivial only if all rows and columns $[c_{xz}]$ are trivial. Hence, the latter gives a non-trivial bound for a much larger range of scenarios.


\subsubsection{Quantum memory}\label{sec:mutual_quantum}

It is natural to ask whether system $Y$ can be generalized to a quantum memory $B$. \textcite{coles14} showed that \eqref{eqn23785} indeed extends to
\begin{equation}\label{eqn23785quantum}
I(X\colo B)+I(Z\colo B) \leq r_{\rm CP} - H(A|B)\,.
\end{equation}
Here, the quantum mutual information of a bipartite quantum state $\rho_{AB}$ is defined as
\begin{align}
I(A\!:\!B) &:= H(\rho_A) + H(\rho_B) - H(\rho_{AB})\\
&= H(\rho_A) - H(A|B)  \,,
\end{align}
and evaluated on the classical-quantum state $\rho_{XB}$ as in~\eqref{eq:cq_postmeas}. Notice that if we specialize to the case where $B=Y$ is classical, then $H(A|Y) \geq 0$ and hence~\eqref{eqn23785quantum} also implies~\eqref{eqn23785}. 

\begin{example}
Consider a maximally entangled state $\rho_{AB}$ for which both $I(X\colo B)$ and $I(Z\colo B)$ become equal to $\log d$. Hence, the upper bound $r_{\rm CP}$ must be weakened in such a way that it becomes trivial, and indeed the term $-H(A|B)$ accomplishes this. Namely, we have $-H(A|B) = \log d$ for the maximally entangled state.
\end{example}

In general, a negative value of $H(A|B)$ implies that $\rho_{AB}$ has distillable entanglement \cite{devetak05}, and this results in a bound in~\eqref{eqn23785quantum} that is larger than $r_{\rm CP}$. In the other extreme, when $H(A|B)$ is positive, which intuitively means that the correlations between Alice and Bob are weak, \eqref{eqn23785quantum} strengthens the bound in~\eqref{eqn23785}.


\subsubsection{A conjecture}\label{sec:mutual_conjecture}

Following the resolved conjectures by~\textcite{kraus87,renes09,grudka12}, we point to a recent open conjecture by~\textcite{schneeloch14}. They ask if for any bipartite quantum state $\rho_{AB}$,
\begin{equation}\label{eqninfoconjecture}
I(X_A \colo X_B)+I(Z_A \colo Z_B) \stackrel{?}{\leq} I(A \colo B)\,,
\end{equation}
where $X_A$ and $Z_A$ are the registers associated with measuring two MUBs $\bX_A$ and $\bZ_A$ on system $A$, and likewise for $X_B$ and $Z_B$ on system $B$. The relation~\eqref{eqninfoconjecture} would say that the quantum mutual information is lower bounded by the sum of the classical mutual informations in two mutually unbiased bases. We note that a stronger conjecture, in which $X_B$ and $Z_B$ are replaced by the quantum memory $B$, is violated in general.


\subsection{Quantum channel formulation}\label{sctQchannelmemory}

\subsubsection{Bipartite formulation}

\textcite{christandl05} considered the question of how well information can be transmitted over a quantum channel. A quantum channel is the general form for quantum dynamics~\cite{davies76} (more general than unitary evolution). Mathematically a quantum channel $\cE$ is a completely positive trace preserving map, and can be represented in its Kraus form,
\begin{align}
\cE(\cdot) = \sum_j K_j (\cdot) K_j\ad \,, \quad\text{where }\sum_j K_j\ad K_j = \1 \,.
\end{align}
\textcite{christandl05} addressed the topic of sending classical information over a quantum channel, or more specifically, sending two complementary types of classical information over a quantum channel. They consider a scenario where Alice chooses a state, with probability $1/d$, from a set of $d$ orthonormal states, which we label as $\bZ = \{\dya{\bZ^z}\}$. She then sends the state over the channel $\cE$ to Bob, and Bob tries to distinguish which state she sent. Likewise Alice and Bob may play the same game but with the $\bX = \{\dya{\bX^x}\}$ states instead, where the $\bX$ and $\bZ$ states are related by the Fourier matrix $F$, given by~\eqref{eqnfourier1}. Bob's distinguishability for the $\bZ$ states can be quantified by the so-called Holevo quantity~\cite{holevo73b},
\begin{align}\label{eqnholevo}
\chi(\cE, \bZ) = &H\left(\sum_z \frac{1}{d}\cE\Big(\dya{\bZ^z}\Big)\right)\notag\\
&- \sum_z \frac{1}{d}H\Big(\cE\big(\dya{\bZ^z}\big)\Big)\,.
\end{align}
Likewise, $\chi(\cE, \bX)$ is a measure of Bob's distinguishability for the $\bX$ states. \textcite{christandl05} proved that
\begin{align}\label{eqnchristandlwinter}
\chi(\cE, \bX) + \chi(\cE, \bZ) \leq \log d + I_{\rm coh}\left(\frac{\1}{d},\cE\right)\,,
\end{align}
where the coherent information $I_{\rm coh}(\rho,\cE)$ is a measure of the quality of a quantum channel $\cE$ introduced by \textcite{schumacher96}. For the maximally-mixed input state $\1/d$ it is given by
\begin{align}\label{eqncoherentinfo}
I_{\rm coh}\left(\frac{\1}{d},\cE \right) = H\Big(\cE(\1/d)\Big) - H\Big(\big(\id \ot \cE\big)\big(\dya{\Phi}\big)\Big)\,,
\end{align}
where $\ket{\Phi} = \sum_j (1/\sqrt{d})\ket{j}\ket{j}$ is a maximally entangled state. \textcite{coles10} noted that \eqref{eqnchristandlwinter} holds for arbitrary MUBs, and that it naturally generalizes to arbitrary orthonormal bases $\bX$ and $\bZ$ with the right-hand side of \eqref{eqnchristandlwinter} replaced by
\begin{align}
\log \left(d^2 c\right) + I_{\rm coh}\left(\frac{\1}{d},\cE\right)\,.
\end{align}
Later this bound was improved by~\textcite{coles14} to
\begin{align}
r_{\rm CP} + I_{\rm coh}\left(\frac{\1}{d},\cE\right)\,.
\end{align}
While~\eqref{eqnchristandlwinter} may look similar to some uncertainty relations discussed in this section, especially~\eqref{eqn23785quantum}, it is important to note the conceptual difference. The relations discussed previously were from a static perspective, whereas~\eqref{eqnholevo} refers to a dynamic perspective involving a sender and a receiver. Intuitively, what~\eqref{eqnchristandlwinter} says is that if Alice can transmit both the $\bZ$ states and the $\bX$ states well to Bob, then $\cE$ is a noiseless quantum channel, i.e., it is close to a perfect channel (as quantified by the coherent information).


\subsubsection{Static-dynamic isomorphism}\label{sctstaticdynamiciso}

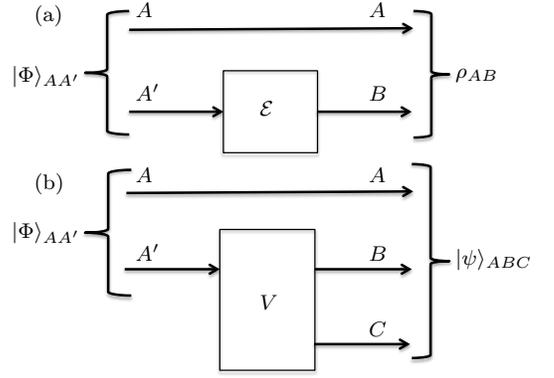
\begin{figure}[tbp]
\begin{center}
\begin{overpic}[width=5cm]{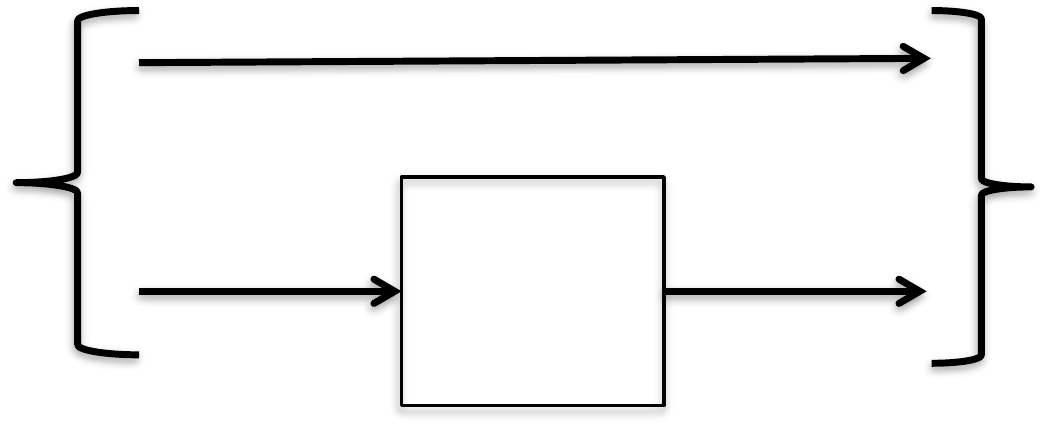}
\put(-12,37){\footnotesize (a)}
\put(15,38){\footnotesize $A$}
\put(15,16){\footnotesize $A'$}
\put(-18,21){\footnotesize $\ket{\Phi}_{AA'}$}
\put(48,12){\footnotesize $\cE$}
\put(77,38){\footnotesize $A$}
\put(77,16){\footnotesize $B$}
\put(100,22){\footnotesize $\rho_{AB}$}
\end{overpic}
\vspace{0.1cm}
\begin{overpic}[width=5cm]{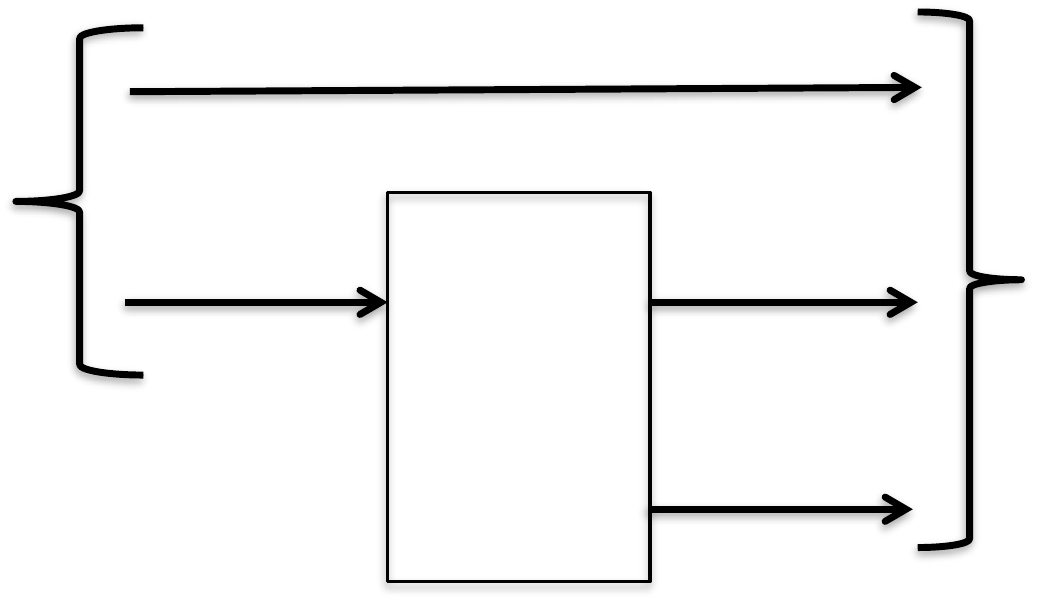}
\put(-12,50){\footnotesize (b)}
\put(-18,37){\footnotesize $\ket{\Phi}_{AA'}$}
\put(15,52){\footnotesize $A$}
\put(15,31){\footnotesize $A'$}
\put(77,52){\footnotesize $A$}
\put(77,31){\footnotesize $B$}
\put(77,11){\footnotesize $C$}
\put(48,18){\footnotesize $V$}
\put(100,30){\footnotesize $\ket{\psi}_{ABC}$}
\end{overpic}
\caption{How to convert the dynamical evolution of a system into (a) a bipartite mixed state or (b) a tripartite pure state.}
\label{fgrchannel}
\end{center}
\end{figure}

With that said, there is a close, mathematical relationship between the static and dynamic perspectives. In fact, there is an isomorphism, known as the Choi-Jamio\l{}kowski isomorphism~\cite{choi75,jamiliolkowski72}, that relates the two perspectives (e.g., see \textcite{zyczkowski04}). Every quantum channel $\cE$ corresponds to a bipartite mixed state defined by
\begin{align}\label{eqnstatechannel}
\rho_{AB} = (\id \ot \cE)(\dya{\Phi})\,,
\end{align}
where $\ket{\Phi} = \sum_j (1/\sqrt{d})\ket{j}\ket{j}$ is maximally entangled (see Fig.~\ref{fgrchannel}(a)). Note that $\rho_{AB}$ here has the property that $\rho_A =\tr_B(\rho_{AB}) = \1 / d_A$ is maximally mixed. Likewise, every bipartite mixed state $\rho_{AB}$ with marginal $\rho_A = \1 / d_A$ corresponds to a quantum channel whose action on some operator $O$ is defined as,
\begin{align}\label{eqnstatechannel2}
\cE(O) = d_A \tr_A \Big[\left(O^T \ot \1\right) \rho_{AB}\Big]\,,
\end{align}
where the transpose denoted by $(\cdot)^T$ is taken in the standard basis. One can easily verify that the condition that $\rho_A = \1 / d_A$ is connected to the fact that $\cE$ is trace-preserving.

This isomorphism can be exploited to derive uncertainty relations for quantum channels as corollaries from uncertainty relations for states, and vice versa. This point was emphasized, e.g., by~\textcite{coles10}. For example, if one has an uncertainty relation for bipartite states $\rho_{AB}$, such as \eqref{eqnBipartiteMemory13}, then one can apply this relation to the state in \eqref{eqnstatechannel} in order to obtain an uncertainty relation for channels.

Specifically, notice that if Alice measures observable $\bZ $ on system $A$ in Fig.~\ref{fgrchannel}(a) and obtains outcome $\dya{\bZ^z}$, then the state corresponding to the transpose, $\dya{\bZ^z}^T$, will be sent through the channel $\cE$. In other words,
\begin{align}\label{eqnricochet}
\frac{1}{d}\dya{\bZ^z}^T = \tr_A \Big[(\dya{\bZ^z}\ot \1)\dya{\Phi}\Big]\,.
\end{align}
This implies that the Holevo quantity $\chi(\cE, \bZ^T)$ can be thought of as a classical-quantum mutual information as,
\begin{align}\label{eqnholevo2}
\chi(\cE, \bZ^T) = I(Z:B) = \log d - H(Z|B)&\notag\\
\mathrm{where}\quad\bZ^T = \left\{\dya{\bZ^z}^T\right\}&\,,
\end{align}
and the right-hand side is evaluated for the state 
\begin{align}\label{eqnrhozbholevo}
\rho_{ZB} &= \sum_z \dya{z}\ot \tr_A\Big[(\dya{\bZ^z}\ot \1_B)\rho_{AB}\Big]\\
& = \sum_z \frac{1}{d} \dya{z} \ot \cE\left(\dya{\bZ^z}^T\right)\,.
\end{align}
Using \eqref{eqnholevo2}, one can verify that the channel uncertainty relation \eqref{eqnchristandlwinter} is a corollary of the bipartite state uncertainty relation, either \eqref{eqnBipartiteMemory13} or \eqref{eqn23785quantum}.


\subsubsection{Tripartite formulation}

One can formulate uncertainty relations for a dynamic tripartite scenario where Alice sends the $\bZ$ states over quantum channel $\cE$ to Bob or the $\bX$ states over the \textit{complementary} quantum channel $\cF$ to Charlie. The relationship between a channel and its complementary channel can be seen via the Stinespring dilation~\cite{stinespring54}, in which one writes the channel in terms of an isometry $V$ that maps $A\to BC$, namely
\begin{align}
\cE(O) &= \tr_C[VOV\ad],\label{eqnstatechannel5}\\
\cF(O) &= \tr_B[VOV\ad]\,.\label{eqnstatechannel50}
\end{align}
Analogous to~\eqref{eqnstatechannel}, we consider the tripartite pure state defined by
\begin{align}\label{eqnstatechannel6}
\ket{\psi}_{ABC} = (\1 \ot V)\ket{\Phi}\,.
\end{align}
This mapping is depicted in Fig.~\ref{fgrchannel}(b). The tripartite uncertainty relations presented in Section~\ref{scttripartiteQMemory} can then be applied to the state $\ket{\psi}_{ABC}$ in~\eqref{eqnstatechannel6} in order to derive uncertainty relations for complementary quantum channels. For example, \textcite{coles10} read~\eqref{eqnTripartiteMemory2} in this way to obtain
\begin{align}\label{eqntripartitequantumchannel}
\chi(\cE, \bX) + \chi(\cF, \bZ) \leq \log\left(d^2c\right)\,,
\end{align}
for two orthonormal bases $\bX$ and $\bZ$. This relation implies that if Alice can send the $\bZ$ states well to Charlie over the channel $\cF$, then Bob cannot distinguish very well the outputs of the channel $\cE$ associated with Alice sending a complementary set of states $\bX$.



\section{Position-Momentum Uncertainty Relations}\label{sec:infinite-dim}

As discussed in Sec.~\ref{sec:introduction}, the first precise statement of the uncertainty principle was formulated for position and momentum measurements. Namely, \textcite{kennard27} showed that for all states (with $\hbar=1$),
\begin{align}\label{eq:Kennard}
\sigma(Q)\cdot\sigma(P)\geq\frac{1}{2}\,,
\end{align}
where $\sigma(Q)$ denotes the standard deviation of the probability density $\Gamma_{Q}(q)$ when measuring the position $Q$, and similarly for $\sigma(P)$ when measuring the momentum $P$. 

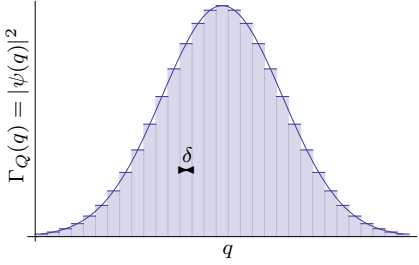
\begin{figure}[tbp]
\begin{center}
\begin{overpic}[width=0.6\columnwidth]{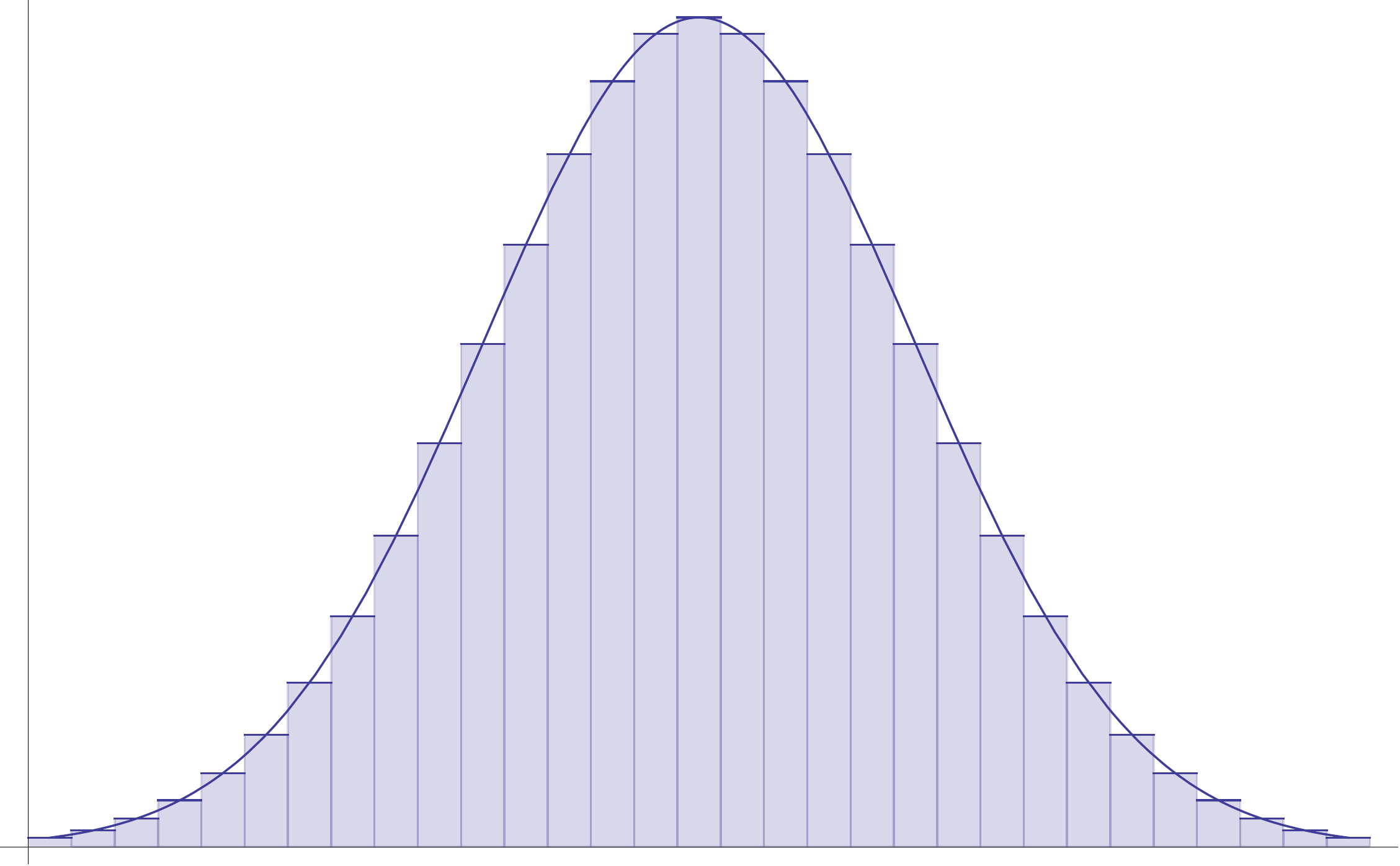}
  \put(-5.5,15){\rotatebox{90}{\footnotesize $\Gamma_{Q}(q)=|\psi(q)|^2$}}
  \put(50,-3){\footnotesize $q$}
  \put(39.7,20.5){\footnotesize $\delta$}
  \put(39.4,18.5){\vector(1,0){2}}
  \put(42,18.5){\vector(-1,0){2}}
\end{overpic}
\caption{Gaussian wave packet in position space with $\Gamma_{Q}(q)$ as in~\eqref{eq:Gauss_Q}, as well as the finite resolution discretization from~\eqref{eq:discrete_distribution} in intervals of size $\delta$.}
\label{fig:gaussian}
\end{center}
\end{figure}

\begin{example}
Consider Gaussian wave packets (see Fig.~\ref{fig:gaussian}) with position probability density\footnote{In all of Sec.~\ref{sec:infinite-dim}, we use the letter $\Gamma$ instead of $P$ for probability distributions since the momentum operator is already denoted by the letter $P$.}
\begin{align}\label{eq:Gauss_Q}
\Gamma_{Q}(q)=\frac{1}{\sqrt{2\pi\sigma^{2}}}\cdot\exp\left(-q^{2}\cdot\frac{1}{2\sigma^{2}}\right)\,,
\end{align}
and corresponding momentum probability density
\begin{align}\label{eq:Gauss_P}
\Gamma_{P}(p)=\sqrt{\frac{2\sigma^{2}}{\pi}}\cdot\exp\left(-p^{2}\cdot2\sigma^{2}\right)\,.
\end{align}
It is then straightforward to check that these achieve equality in~\eqref{eq:Kennard} and hence minimize the uncertainty in terms of the product of the two standard deviations.
\end{example}

In contrast to Kennard's formulation~\eqref{eq:Kennard}, the relations developed in Sec.~\ref{sec:finite_dimensional}--\ref{sctMemory} are phrased in terms of entropy measures and apply to finite-dimensional systems (whereas position and momentum measurements can only be modeled on infinite-dimensional spaces). In this section we review entropic uncertainty relations with and without a memory system for position and momentum measurements.\footnote{Entropic uncertainty relations for completely general quantum systems described by von Neumann algebras and measurements described by measure spaces are also studied in the literature~\cite{furrer13,frank12}.} We discuss applications to continuous variable quantum cryptography later in Sec.~\ref{sec:cvqkd}.


\subsection{Entropy for infinite-dimensional systems}

On a technical level, the position operator $Q$ and the momentum operator $P$ with the canonical commutation relation
\begin{align}
[P,Q]=i\1
\end{align}
can only be represented as unbounded operators on infinite-dimensional spaces. Hence, we need to extend our setup from finite-dimensional Hilbert spaces to separable Hilbert spaces $A$ with $\mathrm{dim}(A)=\infty$. However, quantum states can still be represented as linear, positive semi-definite operators. Hence, we just keep the notation the same as for finite-dimensional spaces without going into any mathematical details. We start with describing how to define entropy for infinite-dimensional systems.


\subsubsection{Shannon entropy for discrete distributions}

Imagine a finite resolution detector that measures the position $Q$ by indicating in which interval 
\begin{align}\label{eq:intervals}
\mathcal{I}_{k;\delta}:=\big(k\delta,(k+1)\delta\big] \qquad (k\in\mathbb{Z})\,,
\end{align}
of size $\delta > 0$ the value $q$ falls. This defines a discrete probability distribution $\Gamma_{Q_{\delta}}$ with infinitely many elements. If the initial state is described by a pure state wave function $\ket{\psi(q)}_{Q}$ we get $\{\Gamma_{Q_{\delta}}(k)\}_{k\in\mathbb{Z}}$ with
\begin{align}\label{eq:discrete_distribution}
\Gamma_{Q_{\delta}}(k)=\int_{k\delta}^{(k+1)\delta} \big|\psi(q)\big|^2\,\mathrm{d}q\,.
\end{align}
We then define the Shannon entropy of $\Gamma_{Q_{\delta}}$ in the usual way as
\begin{align}\label{eq:discrete_Shannon}
H(Q_{\delta}):=-\sum_{k=-\infty}^{\infty}\Gamma_{Q_{\delta}}(k)\log\Gamma_{Q_{\delta}}(k)\,.
\end{align}
Despite the fact that there are now infinitely many terms in the sum, $H(Q_{\delta})$ keeps many of the properties of its finite-dimensional analogue. In particular, $H(Q_{\delta})\geq0$ and the Shannon entropy can still be thought of as an information measure.

\subsubsection{Shannon entropy for continuous distributions}

The differential Shannon entropy is defined in the limit of infinitely small interval size $\delta\to0$,
\begin{align}\label{eq:diff_limit}
h(Q):&=\lim_{\delta\to0}\Big(H(Q_{\delta})+\log\delta\Big)\\
&=\lim_{\delta\to0}\left(-\sum_{k=-\infty}^{\infty}\Gamma_{Q_{\delta}}(k)\log\frac{\Gamma_{Q_{\delta}}(k)}{\delta}\right)\,.
\end{align}
The term $H(Q_{\delta})$ scales with the interval $\delta\to0$ and hence the normalization in~\eqref{eq:diff_limit}. This makes the differential Shannon entropy an entropy density. There is also a closed formula for the differential Shannon entropy (at least when $\Gamma_{Q}(q)$ is continuous),
\begin{align}\label{eq:differential}
h(Q)=-\int\mathrm{d}q\;\Gamma_{Q}(q)\log\Gamma_{Q}(q)\,,
\end{align}
where $\Gamma_{Q}(q)$ denotes the probability density when measuring the position $Q$. For the momentum probability density $\Gamma_{P}(p)$ we define the discrete and differential Shannon entropy in the same way. Since probability densities can be larger than one, not all of the properties of discrete Shannon entropy carry over. For example the differential Shannon entropy can be negative. 

\begin{example}
For Gaussian wave packets as in~\eqref{eq:Gauss_Q} and~\eqref{eq:Gauss_P} we have,
\begin{align}
h(Q)=\frac{1}{2}\log\left(2\pi e\sigma^{2}\right)\quad\mathrm{and}\quad h(P)=\frac{1}{2}\log\frac{\pi e}{2\sigma^{2}}\,.
\end{align}
By inspection we find that $h(Q)<0$ for $\sigma$ sufficiently small and $h(P)<0$ for $\sigma$ sufficiently large. 
\end{example}

Nevertheless the uncertainty principle can still be expressed in term of differential Shannon entropies.


\subsection{Differential relations}

Extending the work of~\textcite{everett57,hirschman57}, \textcite{biaynicki75} and independently~\textcite{beckner75} showed for position and momentum measurements $Q$ and $P$, respectively, that
\begin{align}\label{eq:cont_pq}
h(P)+h(Q)\geq\log(e\pi)\,.
\end{align}
We emphasize that~\eqref{eq:cont_pq} holds even though either one of the two differential Shannon entropies on the left-hand side can become negative. As in Kennard's relation~\eqref{eq:Kennard} Gaussian wave packets again minimize the uncertainty and lead to equality in~\eqref{eq:cont_pq}. This shows that the relation is tight. It is shown in Sec.~\ref{sec:stddev} the entropic relation~\eqref{eq:cont_pq} also implies Kennard's relation~\eqref{eq:Kennard} and is therefore stronger.

Recently alternative bounds were shown by~\textcite{frank12-2,rumin12,hall12}. In particular, extending the work of~\textcite{beckner75,hall99,rumin11}, \textcite{frank12-2} showed that
\begin{align}\label{eq:cont_mixed}
h(Q)+h(P)\geq\log(2\pi)+H(\rho_A)\,,
\end{align}
where
\begin{align}
H(\rho_A):=-\tr\big[\rho_{A}\log\rho_{A}\big]
\end{align}
denotes the von Neumann entropy of the infinite-dimensional input state before any measurement was performed. We note that in contrast to the differential Shannon entropy, the von Neumann entropy is always non-negative since there is no regularization in its definition (even for infinite-dimensional systems). In~\eqref{eq:cont_mixed} the state independent bound $\log(2\pi)\leq\log(e\pi)$ is worse than in~\eqref{eq:cont_pq}, but interestingly~\eqref{eq:cont_mixed} becomes an equality for a thermal state in the infinite temperature limit~\cite{hall99,frank12-2}. Hence, the relation~\eqref{eq:cont_mixed} is also tight if we insist on having the von Neumann entropy $H(\rho_A)$ on the right-hand side.


\subsection{Finite-spacing relations}

It has been argued in the literature that ideal position and momentum measurements can effectively never be performed because every detector has a finite accuracy. We can then ask: in what other than a purely mathematical sense does~\eqref{eq:cont_pq} and~\eqref{eq:cont_mixed} express the uncertainty principle?\footnote{This criticism also applies to Kennard's relation~\eqref{eq:Kennard} and a finite spacing version thereof has been derived by~\textcite{rudnicki12}.} Certainly a more operational way to express uncertainty is in terms of the discrete Shannon entropy as defined in~\eqref{eq:discrete_Shannon}. A series of works~\cite{partovi83,birula84,gonzales95,rudnicki11,rudnicki12} established that for measurements with finite spacing $\delta_q$ for the position and finite spacing $\delta_p$ for the momentum we have that
\begin{align}\label{eq:finite_pq}
&H(Q_\delta)+H(P_\delta)\notag\\
&\geq\log(2\pi)-\log\left(\delta_{q}\delta_{p}\cdot S_{0}^{(1)}\left(1,\frac{\delta_q\delta_p}{4}\right)^{2}\right)\,,
\end{align}
where $S_{0}^{(1)}(\cdot,\cdot)$ denotes the $0$th radial prolate spheroidal wave function of the first kind~\cite{slepian64}. This way of expressing the uncertainty principle has the advantage that the discrete Shannon entropy is always non-negative and has a clear information theoretic interpretation. As we will see later, it is the discrete formulation of the uncertainty principle that becomes relevant for applications in continuous variable quantum cryptography (see Sec.~\ref{sec:diff_guess} and~\ref{sec:cvqkd}). 

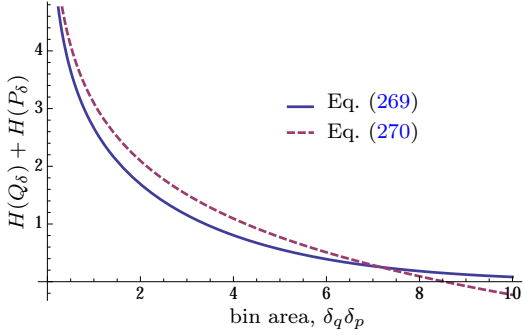
\begin{figure}[tbp]
\begin{center}
\begin{overpic}[width=0.75\columnwidth]{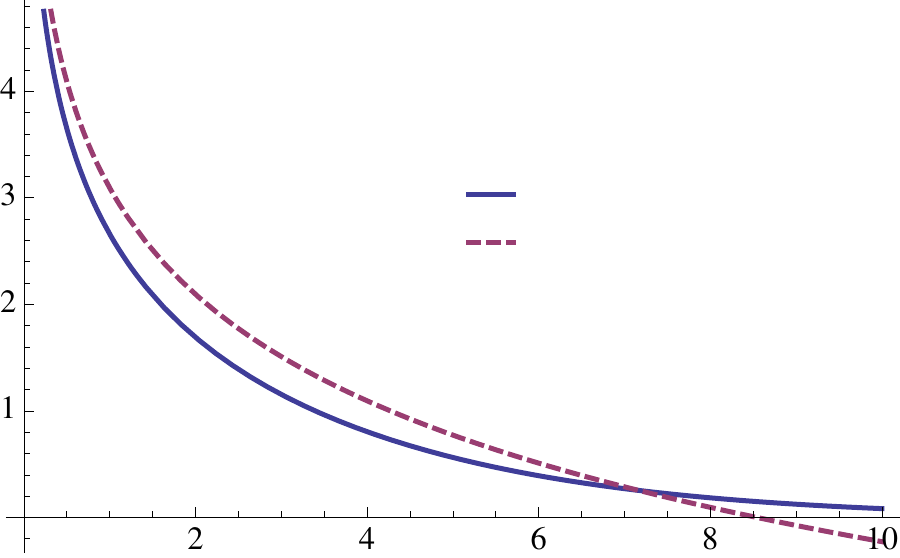}
\put(-5.5,15){\rotatebox{90}{\footnotesize $H(Q_\delta)+H(P_\delta)$}}
\put(40,-4){\footnotesize bin area, $\delta_{q}\delta_{p}$}
\put(60,40){\footnotesize Eq.~\eqref{eq:finite_pq}}
\put(60,34){\footnotesize Eq.~\eqref{eq:finite_pq2}}
\end{overpic}
\vspace{0.3cm}
\caption{Comparison of the lower bounds in~\eqref{eq:finite_pq} and \eqref{eq:finite_pq2} on the uncertainty generated by the finite-spacing position and momentum measurements $Q_\delta,P_\delta$ as in~\eqref{eq:discrete_distribution}. Note that the latter bound becomes negative and hence trivial for larger spacings $\delta_q\delta_p\gtrsim8.5$. 
}
\label{fig:pq_bounds}
\end{center}
\end{figure}

Interestingly~\eqref{eq:finite_pq} is not tight for general $\delta>0$ since we also know that~\cite{birula84}
\begin{align}\label{eq:finite_pq2}
H(Q_{\delta q})+H(P_{\delta p})\geq\log(e\pi)-\log\left(\delta_{q}\delta_{p}\right)\,,
\end{align}
which becomes tighter for $\delta\to0$ (see Fig.~\ref{fig:pq_bounds}). \textcite{rudnicki15} employs a majorization-based approach along the lines of Sec.~\ref{sec:major} to improve on~\eqref{eq:finite_pq} and~\eqref{eq:finite_pq2} for large spacing. However, this does not yield a closed formula and we refer to~\textcite{rudnicki11,rudnicki15} for a discussion of tightness and a more detailed comparison. We will further comment on this issue in the next section (Sec.~\ref{sec:diff_cond}) after extending~\eqref{eq:cont_pq} and~\eqref{eq:finite_pq} to a quantum memory system.

\subsection{Uncertainty given a memory system}\label{sec:diff_cond}

For finite-dimensional systems we can write the conditional von Neumann entropy of bipartite quantum states $\rho_{AB}$ as $H(A|B)=H(AB)-H(B)$. However, for infinite-dimensional systems this is in general not a sensible notion of conditional entropy. This is because for some states both terms $H(AB)$ and $H(B)$ can become infinite even though the entropy of $A$ is finite and hence the conditional entropy should also remain finite.

\begin{example}
Consider a bipartite system with $A$ one qubit and $B$ composed of infinitely many qubits indexed by $k \in \mathbb{N}$. Let $\ket{\psi}_{AB_k}$ be maximally entangled between $A$ and the $k^{\mathrm{th}}$ qubit on $B$, and let $\ket{\phi^k}_{B/B_k}$ be some pure states on $B$ (except $B_k$) such that $\avg{\phi^k|\phi^{k'}}=\delta_{kk'}$. Now, for a probability distribution $p_k\propto\frac{1}{k(\log k)^2}$ for $k>2$~\cite{wehrl78}, the bipartite quantum state
\begin{align}\label{eq:example_state}
\rho_{AB}=\sum_kp_k\ket{\psi}\bra{\psi}_{AB_k}\otimes\ket{\phi^k}\bra{\phi^k}_{B/B_k}\text{ has}&\\
\text{$H(AB)=\infty$ and $H(B)=\infty$.}&
\end{align}
However, any sensible definition of conditional entropy for this state $\rho_{AB}$ should give $H(A|B)=-1$.\footnote{We refer to \textcite{kuznetsova11} for an extended discussion.}
\end{example}

Observe that the conditional entropy of finite-dimensional classical-quantum states $\rho_{XB}$ as in~\eqref{eq:cq} can be rewritten in terms of the relative entropy~\cite{umegaki62},
\begin{align}\label{eq:umegaki}
D(\rho\|\sigma):= \tr[\rho(\log\rho-\log\sigma)]&,\\
\mathrm{as}\quad H(X|B)=-\sum_x D(P_X(x)\rho_B^x\|\rho_B)&\,.\label{eq:condentrpy_discrete}
\end{align}
\textcite{furrer13} pointed out that~\eqref{eq:condentrpy_discrete} can be lifted~to
\begin{align}\label{eq:conditional_umegaki}
H(Q_{\delta}|B):=-\sum_{k=-\infty}^{\infty} D \big(\rho_{B}^{k;\delta} \big\|\rho_{B} \big)\,,
\end{align}
where $\rho_B^{k;\delta}$ denotes the (sub-normalized) marginal state on $B$ when the position $Q$ is measured in $\mathcal{I}_{k;\delta}$, i.e.,
$P_{Q_\delta}(k):=\tr\!\big[\rho_B^{k;\delta}\big]$ 
is the probability to measure in $\mathcal{I}_{k;\delta}$.


\subsubsection{Tripartite quantum memory uncertainty relations}

With~\eqref{eq:conditional_umegaki} as the definition for classical-quantum entropy~\textcite{furrer13} find,
\begin{align}\label{eq:discrete_qm}
&H(Q_{\delta q}|B)+H(P_{\delta p}|C)\notag\\
&\geq\log(2\pi)-\log\left(\delta_{q}\delta_{p}\cdot S_{0}^{(1)}\left(1,\frac{\delta_q\delta_p}{4}\right)^{2}\right)\;.
\end{align}
This is the extension of~\eqref{eq:finite_pq} to quantum memories and likewise not tight. By taking the limit $\delta\to0$ we find the differential quantum conditional entropy
\begin{align}\label{eq:umegaki_limit}
h(Q|B):&=\lim_{\delta\to0}\Big(H(Q_{\delta}|B)+\log\delta\Big)\\
&=\int\mathrm{d}q\;D(\rho_{B}^{q}\|\rho_{B})\,,
\end{align}
where the second equality holds under a particular finiteness assumption~\cite{furrer13}. With~\eqref{eq:discrete_qm} we then immediately find the extension of~\eqref{eq:cont_pq} to quantum memories,
\begin{align}\label{eq:cont_qm}
h(Q|B)+h(P|C)\geq\log(2\pi)\,.
\end{align}

\begin{figure}[tbp]
\begin{center}
\begin{overpic}[width=0.8\columnwidth]{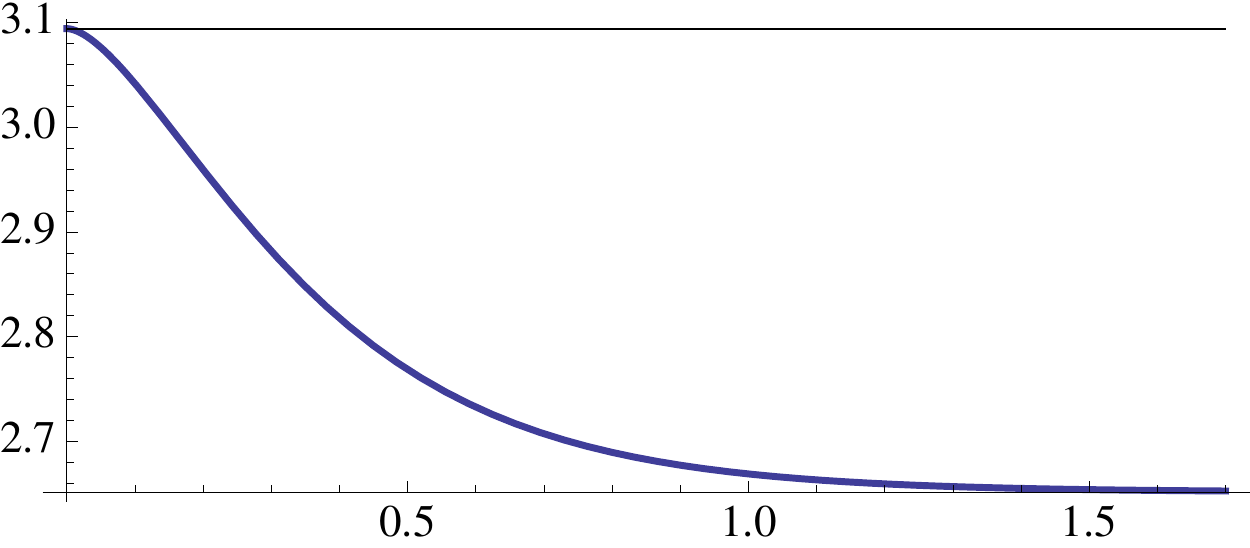}
\put(-6,9){\rotatebox{90}{\footnotesize $h(Q|B)+h(P)$}}
\put(35,-4){\footnotesize squeezing strength, $r$}
\put(50,28){\footnotesize $r = \frac12 \arccosh(\nu)$}
\end{overpic}
\end{center}
\caption{The uncertainty $h(Q|B)+h(P)$ of the EPR state from Ex.~\ref{ex:epr_state} in terms of the squeezing strength $r$.}
\label{fig:squeezing}
\end{figure}

\begin{example}\label{ex:epr_state}
For the EPR state on $AB$ (or likewise $AC$) in the limit of perfect correlations~\eqref{eq:cont_qm} becomes an equality. For finite squeezing strength $r=\arccosh(\nu)/2$ the EPR state is a Gaussian state with covariance matrix
\begin{align}\label{eq:epr_state}
&\Gamma_{AB}(\nu)=\frac{1}{2}\begin{pmatrix}\nu\1_2&\sqrt{\nu^2-1}Z_2\\ \sqrt{\nu^2-1}Z_2&\nu\1_2\end{pmatrix}\\
&\mathrm{with}\;\;\1_2=\begin{pmatrix}1&0\\0&1\end{pmatrix}\;\;\mathrm{and}\;\;Z_2=\begin{pmatrix}1&0\\0&-1\end{pmatrix}\,.
\end{align}
We refer to the review article~\cite{weedbrook11} for more details about Gaussian quantum information theory. The left-hand side of~\eqref{eq:cont_qm} for this state generated by $\Gamma_{AB}(\nu)$ is then calculated to be~\cite{furrer13}
\begin{align}\label{eq:epr_bound}
&h(Q|B)+h(P)\notag\\
&=\log(e\pi\nu)-\frac{\nu+1}{2}\log\left(\frac{\nu+1}{2}\right)+\frac{\nu-1}{2}\log\left(\frac{\nu-1}{2}\right)\,,
\end{align}
which converges to $\log(2\pi)$ for $\nu\to\infty$. In Fig.~\ref{fig:squeezing} we plot~\eqref{eq:epr_bound} as a function of the squeezing strength $r=\frac12 \arccosh(\nu)$:
\begin{enumerate}
\item For $r=0$ the system $B$ is uncorrelated and we have the lower bound $h(Q)+h(P)\geq\log(e\pi)$ as in~\eqref{eq:cont_pq}.
\item For $r>0$ we have to take the quantum memory $B$ into account and only the lower bound $h(Q|B)+h(P)\geq\log(2\pi)$ from~\eqref{eq:cont_qm} holds.
\item For $r\to\infty$ we get maximal correlations and the bound~\eqref{eq:cont_qm} becomes an equality.
\end{enumerate}
We note that in typical experiments for applications (see Sec.~\ref{sec:cvqkd}) a squeezing strength of $r\approx1.5$ is achievable~\cite{schnabel13}. For this the lower bound~\eqref{eq:cont_qm} is already very tight.
\end{example}

The state independent bound in~\eqref{eq:cont_qm} is $\log(2\pi)$ whereas it is $\log(e\pi)$ for the case without quantum memory in~\eqref{eq:cont_pq}. Hence, in contrast to the finite-dimensional case, a quantum memory reduces the state independent uncertainty limit. This is because for the approximate EPR state there exists a gap between the accessible classical correlation and the classical-quantum correlation. That is, even when minimized over all measurements $Q_B$ on $B$, we have
$h(Q|Q_B)-h(Q|B)\approx\log\big(\frac{e}2\big)$.

\subsubsection{Bipartite quantum memory uncertainty relations}

Similarly as for finite-dimensional systems it is possible to formulate uncertainty relations with quantum memory in a bipartite form. For continuous position and momentum measurements~\textcite{frank12} showed that,
\begin{align}\label{eq:bipartite_pq}
h(Q|B)+h(P|B)\geq\log(2\pi)+H(A|B)_{\rho}\,.
\end{align}
This is the extension of~\eqref{eq:cont_mixed} to a quantum memory system. However, we note that~\eqref{eq:bipartite_pq} only holds if all the terms appearing in $H(A|B)=H(AB)-H(B)$ are finite (which is in general too restrictive).\footnote{This restriction is connected with the question about a sensible notion of conditional entropy for fully quantum states~\cite{kuznetsova11}.}


\subsubsection{Mutual information approach}

A conceptually different approach was taken by~\textcite{hall95} where the uncertainty relative to a memory system is quantified in terms mutual information instead of conditional entropy (see Sec.~\ref{sctmutinfo} for a general discussion). Similarly as for the conditional entropy in~\eqref{eq:conditional_umegaki}, mutual information for classical-quantum states is most generally defined in terms of the relative entropy in~\eqref{eq:umegaki},
\begin{align}
I(Q_\delta:B):=\sum_{k=-\infty}^{\infty}\left(D\big(\rho_{B}^{k;\delta}\big\|\rho_{B}\big)+H(B)_{\rho^{k;\delta}}\right)\,.
\end{align}
In contrast to entropy, however, the mutual information stays finite when taking the limit $\delta\to0$,
\begin{align}
I(Q:B):=\lim_{\delta\to0}I(Q_\delta:B)\,.
\end{align}
Hence, no regularization in terms of the interval size $\delta$ is taken. For classical memories $M$ it was shown that~\cite{hall95},
\begin{align}
I(Q:M)+I(P:M)\leq1+\log\sigma (Q)+\log\sigma (P)\,.
\end{align}
It is an open question to find a generalization that also holds for quantum memories. This would be in analogy to what is known for finite-dimensional systems (see Sec.~\ref{sec:mutual_quantum}).


\subsection{Extension to min- and max-entropy}\label{sec:diff_guess}

As for finite-dimensional systems, entropic uncertainty relations like~\eqref{eq:cont_pq} and~\eqref{eq:finite_pq} cannot only be shown for the Shannon entropy, but also more generally for pairs of R\'enyi entropies~\cite{bialynicki06,birula07,rudnicki12,rastegin15b}. Here, we focus on a special case that is important for applications in continuous variable quantum cryptography (see Sec.~\ref{sec:cvqkd}). We study relations in terms of the R\'enyi entropy of order $\infty$ and its dual quantity the R\'enyi entropy of order $1/2$. These are exactly the min- and max-entropy, respectively.


\subsubsection{Finite-spacing relations}

Following the finite resolution detector picture as in~\eqref{eq:intervals} and~\eqref{eq:discrete_distribution}, the conditional min-entropy is defined as
\begin{align}\label{eq:hmin_pq}
H_{\min}(Q_\delta|B):=-\log p_{\rm guess}(Q_\delta|B)\,.
\end{align}
Here, we have the optimal guessing probability as in~\eqref{eq:guess},
\begin{align}
&p_{\rm guess}(X|B)\notag\\
&:=\sup_{\bX_B} \left\{ \sum_{k=-\infty}^{\infty} \Gamma_{Q_\delta}(k) \tr\left[ \bX_B^k \rho_B^{k;\delta} \right]:\, \textrm{$\bX_B$ POVM on $B$}\right\}.
\end{align}
In analogy to the finite-dimensional case, the min-entropy quantifies the uncertainty of the classical register $Q_\delta$ from the perspective of an observer with access to the quantum memory $B$. The conditional max-entropy is given by
\begin{align}\label{eq:maxpq_discrete}
H_{\max}(Q_\delta|B):=\log F_{\rm dec}(Q_\delta|B)\,,
\end{align}
where we have the optimal decoupling fidelity
\begin{align}
&F_{\rm dec}(Q_\delta|B)\notag\\
&:=\sup\Bigg\{\Bigg(\sum_{k=-\infty}^{\infty}\sqrt{F\big(\rho_B^{k;\delta},\sigma_B\big)}\Bigg)^2:\textrm{$\sigma_B$ state on $B$}\Bigg\}.
\end{align}
The decoupling fidelity is a measure of how much information the quantum memory $B$ contains about the classical register $Q_\delta$.\footnote{For finite-dimensional systems the expression~\eqref{eq:maxpq_discrete} is equivalent to the max-entropy as defined in~\eqref{eqndualentropy3}, see~\cite{koenig08,furrer13}.} For these definitions \textcite{furrer13} show
\begin{align}\label{eq:cv/marco-needs-you}
&H_{\min}(Q_\delta|B)+H_{\max}(P_\delta|C)\notag\\
&\geq\log(2\pi)-\log\left(\delta_{q}\delta_{p}\cdot S_{0}^{(1)}\left(1,\frac{\delta_q\delta_p}{4}\right)^{2}\right)\,,
\end{align}
as well as the same relation with $Q_\delta$ and $P_{\delta}$ interchanged. We note that the special case with trivial quantum memories $B,C$ was already shown by~\textcite{rudnicki12}. \textcite{furrer13} show that the relation~\eqref{eq:cv/marco-needs-you} is tight for any spacing $\delta>0$ even in the absence of any correlations (i.e., there exist states for which the relation becomes an equality). Note that this is in contrast to the situation for the Shannon entropy, where neither~\eqref{eq:finite_pq} and~\eqref{eq:finite_pq2}, nor~\eqref{eq:discrete_qm} are tight.


\subsubsection{Differential relations}

For the differential version we take the limit $\delta\to0$,
\begin{align}
h_{\min}(Q|B):=\lim_{\delta\to0}\Big(H_{\min}(Q_{\delta}|B)+\log\delta\Big)\,,
\end{align}
and similarly for $h_{\max}(Q|B)$.\footnote{Under some finiteness assumptions we have
$h_{\max}(Q|B)=2 \log\sup\left\{\int\mathrm{d}q\; \sqrt{F(\rho_B^q,\sigma_B)}:\;\textrm{$\sigma_B$ state on $B$}\right\}$ as well as
$h_{\min}(Q|B)=-\log\sup\left\{\int\mathrm{d}q\;\rho_{B}^{q}(\bX_B^q):\;q\mapsto\textrm{$\bX_B^q$ POVM on $B$}\right\}$.}
We then find the uncertainty relation~\cite{furrer13},
\begin{align}\label{eq:pq_contminmax}
h_{\min}(Q|B)+h_{\max}(P|C)\geq\log(2\pi)
\end{align}
as well as the same relation with $Q$ and $P$ interchanged. \textcite{bialynicki06} shows that~\eqref{eq:pq_contminmax} becomes an equality for pure Gaussian states as in~\eqref{eq:Gauss_Q} and~\eqref{eq:Gauss_P}. Note that this implies in particular that the unconditional special case
\begin{align}
h_{\min}(Q)+h_{\max}(P)\geq\log(2\pi)
\end{align}
is tight. Hence, the optimal state-independent constant is $\log(2\pi)$ for the min- and max-entropy, whereas the optimal constant for the Shannon entropy in~\eqref{eq:cont_pq} is $\log(e\pi)$.


\subsection{Other infinite-dimensional measurements}\label{sec:infinte}

As a multidimensional extension of~\eqref{eq:cont_pq}, \textcite{huang11} shows that for any measurements of the form
\begin{align}\label{eq:ab_meas}
A=\sum_{i=1}^{n}a_iQ_i+a'_iP_i,\;\;B=\sum_{i=1}^{n}b_iQ_i+b'_iP_i&\notag\\
\mathrm{with}\quad a_i,a'_i,b_i,b'_i\in\mathbb{R}&\,,
\end{align}
we have that
\begin{align}\label{eq:huang}
h(A)+h(B)\geq\log(e\pi)+\log|[A,B]|\,.
\end{align}
\textcite{huang11} also shows that for any measurement pair $A,B$ as in~\eqref{eq:ab_meas} there exist states for which~\eqref{eq:huang} becomes an equality.

Moreover, the techniques for deriving position-momentum uncertainty relations can also be applied to other complementary observable pairs that are modeled on infinite-dimensional Hilbert spaces. For example, for a particle on a circle we have the position angle $\varphi$ and the conjugate angular momentum observable $L_z$. Consider a measurement device that either tells in which of
\begin{align}
M:=2\pi/\delta_\varphi\quad\text{bins of size $\delta_\varphi$}
\end{align}
the particle is in or the exact value of the angular momentum $L_z$. We get a discrete probability distribution $P_{\varphi_\delta}$ for the angle defined similarly as in~\eqref{eq:discrete_distribution}, as well as a discrete probability distribution $P_{L_z}$ over the $L_z$ eigenstates. Improving on the earlier work of~\textcite{partovi83}, \textcite{birula84} showed that
\begin{align}\label{eq:angle_momentum}
H(\varphi_\delta)+H(L_z)\geq\log M\,.
\end{align}
By inspection~\eqref{eq:angle_momentum} becomes an equality for any eigenstate of the $L_z$ observable. The relation was also extended to two angles $\varphi$ and $\theta$ and the corresponding pair of observables $L_z$ and $L^2$~\cite{birula85}.

Another example are the number $N$ and the phase $\Phi$ for the harmonic oscillator. \textcite{hall92} showed that
\begin{align}\label{eq:number_phase}
H(N)+h(\Phi)\geq\log2\pi\,,
\end{align}
where $P_N(n)$ represents the probability distribution in the number basis $\{\ket{n}\}$, and the probability density in the phase basis is
\begin{align}\label{eq:number_phase2}
P_\Phi(\phi):=\frac{|\langle e^{i\phi} | \psi \rangle|^{2}}{2\pi}\quad\mathrm{with}\quad |e^{i \phi}\rangle:=\sum_n e^{i n \phi}|n\rangle
\end{align}
the Susskind-Glogower phase kets (which are not normalized).\footnote{Due to the non-orthogonality of the phase kets $|e^{i \phi}\rangle$ there is no observable corresponding to the phase distribution $P_\Phi(\phi)$. This, however, will not concern us further since $P_\Phi(\phi)$ is well-defined.} This can also be seen as a special case of the results in~\cite{biaynicki75}. Equation~\eqref{eq:number_phase} becomes an equality for number states. Furthermore, \textcite{hall94} also extends~\eqref{eq:number_phase} to noisy harmonic oscillators degraded by Gaussian noise.

Finally, time-energy entropic uncertainty relations for systems with discrete energy spectra were discussed by~\textcite{hall08}.



\section{Applications}\label{sec:app}

\subsection{Quantum randomness}\label{sec:randomness}

Randomness is a crucial resource for many everyday information processing tasks, ranging from online gambling to scientific simulations and cryptography. Randomness is a scarce resource since computers are designed to perform deterministic operations. Even more importantly classical physics is deterministic, meaning that every outcome of an experiment can in principle be predicted by an observer who has full knowledge of the initial state of the physical system and the operations that are performed on it. The study of pseudorandomness tries to circumvent this problem~\cite{vadhan12}.

Quantum mechanics with its inherent nondeterminism allows us to consider a stronger notion of randomness, namely randomness that is information-theoretically secure. Formally, we want to generate a random variable $L$ that is uniformly distributed over all bit strings $\{0,1\}^{\ell}$ of a given length $\ell$. Moreover, we want that this random variable is independent of any side information an observer might have, including information about the process that is used to calculate $L$ and any random seeds that are used to prepare $L$.  A classical-quantum product state 
\begin{align}
\pi_{LE} = \frac{1}{2^{\ell}} \sum_{i = 1}^{2^{\ell}} \proj{i}_L \otimes \pi_E
\end{align}
describes $\ell$ bits of uniform randomness that is independent of its environment, or side information, $E$. Often, the best we can hope for is to approximate such a state. Namely, we say that $\rho_{ZE}$ describes a state where $L$ is $\delta$-close to uniform on $\ell$ bits and independent of $E$ if
\begin{align}
\bigg\| \rho_{LE} - \frac{1}{2^{\ell}} \sum_{i = 1}^{2^{\ell}} \proj{i}_L \otimes \rho_E \bigg\|_{\rm tr} \leq \delta\,,
\end{align}
where $\|\cdot\|_{\rm tr}$ denotes the trace norm.
This bound implies that
$L$ cannot be distinguished from a uniform and independent random variable with probability 
more than $\frac{1}{2} (1 + \delta)$. This viewpoint is at the core
of universally composable security frameworks~\cite{canetti01,unruh10}, which ensure that a secret key satisfying the above property can safely be employed in any cryptographic protocol requiring a secret key.

Entropic uncertainty relations can help us since they certify that the random variables resulting from a quantum measurement are uncertain and thus contain randomness. However, in order to extract approximately uniform and independent randomness we will need an additional step, which we describe next.


\subsubsection{The operational significance of conditional min-entropy}\label{sec:rand_min}

The importance of the min-entropy in cryptography is partly due to the following lemma, called the \emph{Leftover Hashing Lemma}~\cite{mcinnes87,impagliazzo89,zuckerman89}. Informally, it states that there exists a family of functions $\{ f_s \}_{s}$, where $f_s : \mathcal{X} \to [2^{\ell}]$, called hash functions, such that the random variable $L = f_S(X)$, which results by applying the function $f_S$ with $S$ a seed chosen uniformly at random, is close to uniform and independent of $S$ if the initial min-entropy is sufficiently large. 

More formally, \textcite{rennerkoenig05,renner05} show the following result for the quantum case. There exists a family $\{ f_s \}_{s}$ as described above such that for any classical-quantum state 
\begin{align}
\rho_{XE} = \sum_x P_X(x) \proj{x}_X \otimes \rho_E^x
\end{align}
with $H_{\min}(X|E) \geq k$, the classical-quantum-classical state $\rho_{LES}$ after applying $f_S$, namely
\begin{align}
\rho_{LES} = \sum_{s,x} \frac{P_X(x)}{|S|} \proj{f_s(x)}_L \otimes \rho_E^x \otimes \proj{s}_S \,,
\end{align}
describes a state where $L$ is $\delta$-close to uniform on $\ell$ bits and independent of $E$ and $S$ with $\delta = 2^{\frac12 ( \ell - k ) }$.

The special case where the environment $E$ is trivial has been discussed extensively in the computer science literature~\cite{vadhan12}. Since hashing is a classical process, one might expect that the physical nature of the side information is irrelevant and that a classical treatment is sufficient.
However, this is not true in general. For example, the output of certain extractors may be partially known if side information about their input is stored in a quantum memory, while the same output is
almost uniform conditioned on any classical side information.\footnote{See~\textcite{gavinsky07} for a
concrete example and~\textcite{koenig07} for a more general discussion of this topic.}

A generalization of this result is possible by considering a variation of the min-entropy, which is called $\eps$-\emph{smooth} min-entropy, and denoted $H_{\min}^{\eps}(X|E)$, for a small $\eps > 0$. This is defined by maximizing the min-entropy over states that are in a ball of radius $\eps$ around the state $\rho$.\footnote{See \textcite{tomamichel09} for a precise definition of smooth min-entropy.}

The generalized Leftover Hashing Lemma~\cite{renner05,tomamichel10} asserts that there exists a family $\{ f_s \}_{s}$ such that for any state $\rho_{XE}$ with $H_{\min}^{\eps}(X|E) \geq k$, we find that $L = f_S(X)$ is $(\eps+\delta)$-close to uniform and independent of $E$ and $S$, with $\delta$ as defined above.

The latter result is tight in the following sense. If $L = f_S(X)$ is $\eps$-close to uniform and independent from $E$ and $S$ for any family of functions $\{ f_s \}_s$, then we must have $H_{\min}^{\eps'}(X|E) \geq \ell$ with $\eps' = \sqrt{2\eps}$. 

Due to this tightness result we are justified to say that the smooth min-entropy characterizes (at least approximately) how much uniform randomness can be extracted from a random source $X$ that is correlated with its environment $E$.


\subsubsection{Certifying quantum randomness}\label{sec:certifyingrandomness}

Note that we can certify randomness, if we can somehow conclude that $H_{\min}(X|E)$ is large. In principle, all entropic uncertainty relations that involve a quantum memory are suitable for this task, whenever
we can verify the terms lower bounding the entropy. Tripartite uncertainty relations are especially suitable to this task, and the security of quantum key distribution below rests on our ability to make such estimates.
For example, \textcite{vallone14} specialize the uncertainty relation for min- and max-entropy in~\eqref{eqnminmaxuncertaintyqmem} to assert that
\begin{align}
H_{\min}(X|E)_{\rho} \geq \log d - H_{\max}(Z)\,,
\end{align}
where $\mathbb{X}$ and $\mathbb{Z}$ are mutually unbiased measurements on a $d$-dimensional Hilbert space. Here, $E$ is the environment of the measured system and the max-entropy $H_{\max}(Z) = H_{\nicefrac12}(Z)$ can be estimated using statistical tests, resulting in confidence about $H_{\min}(X|E)$. 
As discussed above, the Leftover Hashing Lemma now allows to extract uniform randomness from $X$.

\textcite{miller13} derive a lower bound on an entropy difference instead of a conditional entropy. Assume that $\mathbb{X}$ and $\mathbb{Z}$ are complementary binary measurements on a qubit.
Then, the following relation holds,
\begin{align}
H_{\alpha}(XB) - H_{\alpha}(B) &\geq q(\alpha,\delta)  \quad \textrm{for} \quad \alpha \in (1, 2]\,,
\end{align}
where $\delta$ is determined by the equality
\begin{align}
\tr\big[ \bra{\mathbb{Z}^0} \rho_{AB} \ket{\mathbb{Z}^0}^{\alpha} \big] &= \delta \tr\big[\rho_B^{\alpha}\big]\,,
\end{align}
and $q$ is a function satisfying $\lim_{\alpha \to 1} q(\alpha,\delta) = 1 - 2h(\delta)$. The authors then proceed to use this result to bound the smooth min-entropy and apply the generalized Leftover Hashing Lemma.


\subsection{Quantum key distribution (QKD)}\label{sec:qkd}

The goal of a key distribution scheme is for two honest parties to agree on a shared key by communicating over a public channel in such a way that the key is secret from any potential adversary eavesdropping on the channel. Traditionally the two honest parties trying to share a key are called Alice and Bob and the eavesdropper is called Eve. By a simple symmetry argument it is evident that key distribution is impossible if only classical information is considered: Since Eve will hear all communication from Alice to Bob, at any point in the protocol she will have at least as much information about Alice's key as Bob\,---\,in particular, if Bob knows Alice's key then so does Eve.

Quantum key distribution (QKD) was first proposed by~\textcite{bb84} and~\textcite{ekert91} to get out of this impasse.\footnote{We refer to~\cite{scarani09a} for a recent review.} Since quantum information cannot be copied or cloned~\cite{wootters82}, the above impossibility argument no longer applies when Alice and Bob are allowed to communicate over a quantum channel. Roughly speaking, the main idea is that whenever the eavesdropper interacts with the  channel (for example by performing a measurement on a particle) she will necessarily introduce noise in the quantum communication between Alice and Bob, which they can then detect and subsequently abort the protocol.


\subsubsection{A simple protocol}\label{sec:qkd/prot}

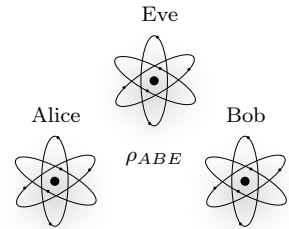
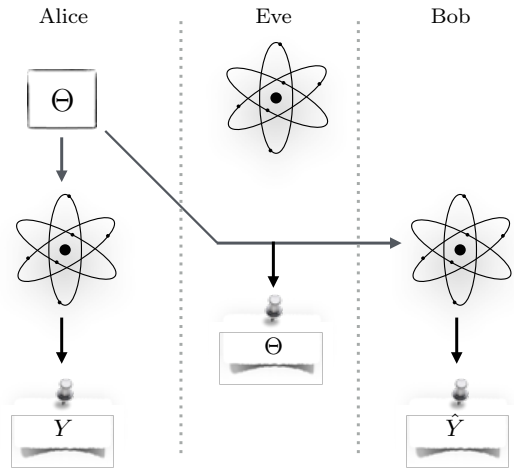
\begin{figure}
\subfigure[\,{Preparation phase}]{
\begin{overpic}[width=0.5\columnwidth]{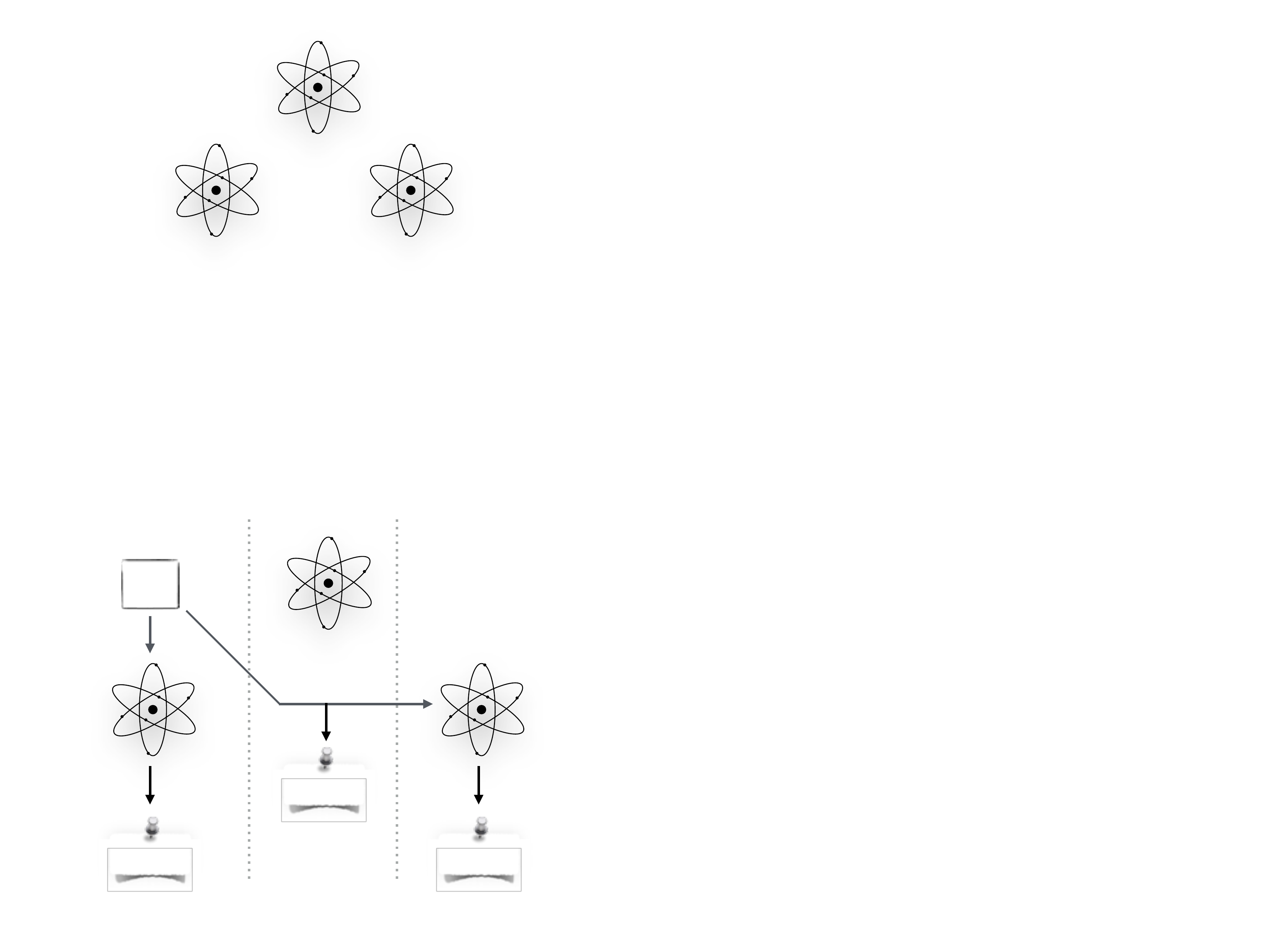}
\put(42,30){\footnotesize $\rho_{ABE}$}
\put(13,42){\footnotesize Alice}
\put(73,42){\footnotesize Bob}
\put(47,73){\footnotesize Eve}
\end{overpic}
\label{fig:qkd_prep}
}
\subfigure[\,{Measurement phase:} Alice and Bob measure their quantum system in the basis indicated by $\Theta$ to recover $Y$ and $\hat{Y}$, respectively. Eve stores $\Theta$ and keeps her quantum memory intact. The uncertainty relation is applied to the resulting state $\rho_{Y\hat{Y}E\Theta}$.]{
\begin{overpic}[width=0.9\columnwidth]{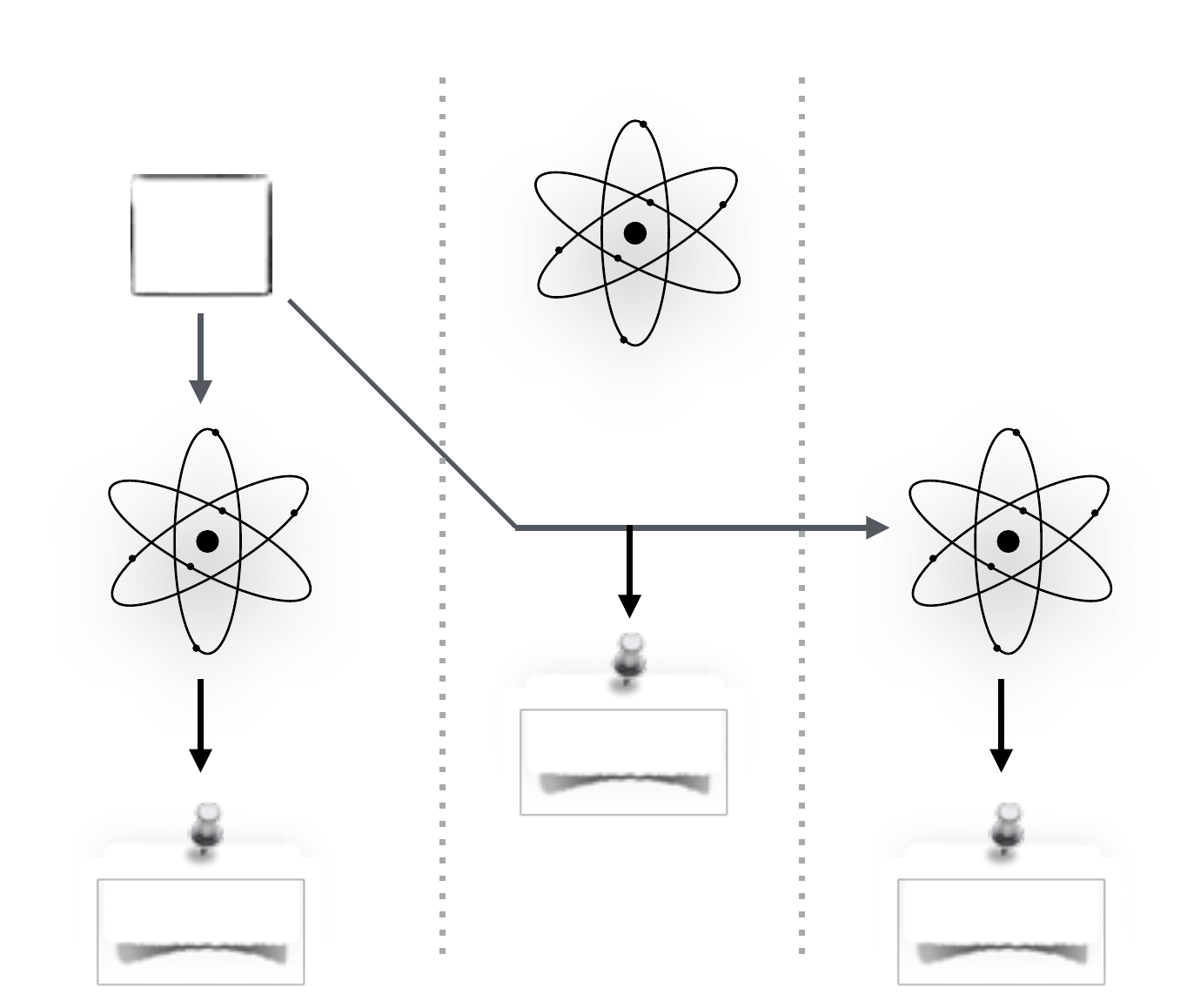}
\put(13,78){\footnotesize Alice}
\put(50,78){\footnotesize Eve}
\put(80,78){\footnotesize Bob}
\put(15,63){\large $\Theta$}
\put(15.5,7.5){$Y$}
\put(82.5,7.5){$\hat{Y}$}
\put(51.5,21.5){$\Theta$}
\end{overpic}
\label{fig:qkd_meas}
}
\caption{Preparation and measurement phase of the QKD protocol described in Sec.~\ref{sec:qkd/prot}.}
\label{fig:qkd}
\end{figure}

We will focus on a truncated version of Ekert's protocol~\cite{ekert91}, which proceeds as follows: 
\begin{description}
\item[Preparation] Alice and Bob share a maximally entangled two-qubit state using the public channel. Eve can coherently interact with the channel.
\item[Measurement] They randomly agree (using the public channel) on either the basis $\bZ = \{ \proj{0}, \proj{1} \}$ or $\bX = \{ \proj{+}, \proj{-} \}$, and measure their respective qubits in this basis. (These two steps are repeated many times.)
\item[Parameter estimation] Alice announces her measurement results on a random subset of these rounds. If their measurement results agree on most rounds, they conclude that their correlations contain some secrecy and proceed to correct their errors and extract a secret key (we will not discuss this further here). If not, they abort the protocol.
\end{description}


\subsubsection{Security criterion for QKD}

To show security of QKD we thus need to show that the following two statements are mutually exclusive: a) Alice's and Bob's measurement results agree in most rounds, and b) Eve has a lot of information about Alice's or Bob's measurement outcomes.

Security of quantum key distribution against general attacks was first formally established by~\textcite{mayers96,mayers01} as well as~\textcite{biham00stoc,biham06} and~\textcite{SP00}. In all these security arguments, the complementarity or uncertainty principle is invoked in some form to argue that if Alice and Bob have large agreement on the qubits measured in one basis, then necessarily Eve's information about the bits measured in the complementary basis must be low.

In Sec.~\ref{sec:qkd-intuitive} we attempt to present the security argument in a concise and intuitive way, and for this purpose we adopt a notion of security---certifying that the raw key has high Shannon entropy---that has proven to be insufficient in practice. However, our ultimate goal is to extract a secret key, and not to have a bit string with high Shannon entropy. This ultimately requires the use of different entropies and a post-processing step in the protocol to distill a secret key. A discussion of these issues follows in Sec.~\ref{sec:qkd-min-entropy}.

Entropic uncertainty relations were first used in this context by \textcite{cerf02} and \textcite{grosshans04}. In particular, \textcite{koashi06} established security by leveraging the Maassen-Uffink relation~\eqref{eq:shannonUR}. However, entropic uncertainty relations with quantum memory provide a more direct avenue to formalize security arguments for QKD, as we will see in the following.


\subsubsection{Proof of security via an entropic uncertainty relation}\label{sec:qkd-intuitive}

\paragraph{Single round.}
We will here broadly follow an argument outlined by~\textcite{berta10}. First, note that during the preparation step (as described above) the eavesdropper might interfere and we can thus not know if Alice and Bob will indeed share a maximally entangled state after the preparation step is complete. However, without loss of generality we may assume that Alice (A), Bob (B), and Eve (E) share an arbitrary state $\rho_{ABE}$ after the preparation step, where $A$ and $B$ are qubits and $E$ is any quantum system held by Eve (see~Fig.~\ref{fig:qkd_prep}).

Let $\Theta$ be a binary register in a fully mixed state that determines if the qubits are to be measured in the basis $\bX$ or $\bZ$ and let $Y$ denote the output of Alice's measurement. Then we can write $H(Y|B\Theta) = \frac{1}{2} H(X|B) + \frac12 H(Z|B)$ and similarly $H(Y|E\Theta) = \frac{1}{2} H(X|E) + \frac12 H(Z|E)$. Thus, the tripartite entropic uncertainty principle with quantum memory~\eqref{eqnTripartiteMemory2} can be cast into the form
\begin{align}
H(Y|E \Theta) + H(Y| B \Theta) \geq q_{\rm MU} = 1, 
\end{align}
where we used that $q_{\rm MU}=1$ for the measurements $\bX$ and $\bZ$. The entropies are evaluated for the state $\rho_{Y\Theta BE}$ after the measurement on Alice's qubit is performed.

Next we perform Bob's measurement, which yields an estimate $\hat{Y}$ of $Y$. The data-processing inequality~\eqref{eq:data-proc} implies that $H(Y|B\Theta) \leq H(Y|\hat{Y})$, and thus we conclude that $H(Y|E \Theta) \geq 1 - H(Y|\hat{Y})$. This ensures that Eve's uncertainty---in terms of von Neumann entropy---of Alice's measurement outcome is large as long as the conditional entropy $H(Y|\hat{Y})$ is small (see Fig.~\ref{fig:qkd_meas}). This is a quantitative expression of the above-mentioned security criterion.\footnote{Note that in practice we need a stronger statement, namely a bound on the min-entropy. This is discussed in Sec.~\ref{sec:qkd-min-entropy}.}

\begin{example}
If Alice and Bob's measurement outcomes agree with high probability, let us say with probability $1-\delta$, then $H(Y|\hat{Y})$ evaluates to $h_{\bin}(\delta) = \delta \log \frac{1}{\delta} + (1-\delta) \log \frac{1}{1-\delta}$. Hence, we find that
\begin{align}
H(Y|E \Theta) \geq 1 - h_{\bin}(\delta)\,,
\end{align}
which is positive as long as $\delta$ is strictly less than $50\%$. 
\end{example}

\paragraph{Multiple rounds.}
The protocol extends over multiple rounds and we can repeat the above argument for each round individually and then attempt to add up the resulting entropies\,---\,but it is much more convenient to use a stronger uncertainty relation that describes the situation for multiple rounds directly.

For this purpose, let us model the situation after Alice and Bob have exchanged $n$ qubits but before they measure them. {This is a hypothetical situation since in the actual protocol Alice and Bob measure their qubits after every round. However, we can always imagine that Alice and Bob delay their measurement since Eve's strategy cannot depend on the timing of their measurement.} 
After the exchange Alice and Bob each hold $n$ qubits in systems $A^n = A_1 A_2 \ldots A_n$ and $B^n = B_1 B_2 \ldots B_n$, respectively. This is described by an arbitrary state $\rho_{A^n B^n E}$ where $E$ is any quantum system held by the eavesdropper. Again, we model the random measurement choice using a register, a bit string $\Theta^n = (\Theta_1, \Theta_2, \ldots, \Theta_n)$ of length $n$ in a fully mixed state, where $\Theta_i$ determines the choice of measurement on the systems indexed by $i$. Analogously, we store the measurement outcomes on Alice's system in a bit string $Y^n = (Y_1, Y_2, \ldots, Y_n)$ and on Bob's system in a bit string $\hat{Y}^n = (\hat{Y}_1, \hat{Y}_2, \ldots, \hat{Y}_n)$.

The crucial observation is that the tripartite uncertainty principle in~\eqref{eqnTripartiteMemory2} implies that 
\begin{align}
&H(X_1 X_2 Z_3 X_4 \ldots X_{n-1} Z_n | E) \nonumber\\
&\qquad + H(Z_1 Z_2 X_3 Z_4 \ldots Z_{n-1} X_n | B) \geq n \label{eq:vn-qkd0}\,,
\end{align}
where we made sure that all $n$ systems are measured in the opposite basis in the two terms, and used that $\log \frac{1}{c^n} = n$. A similar averaging argument as for the one round case and the data-processing inequality~\eqref{eq:data-proc} then reveal the bounds
\begin{align}
&H(Y^n|E \Theta^n) + H(Y^n|\hat{Y}^n) \nonumber\\
&\qquad \geq H(Y^n|E \Theta^n) + H(Y^n|B^n \Theta^n) \geq n \,. \label{eq:vn-qkd}
\end{align}
Hence, Eve's uncertainty (in terms of von Neumann entropy) of the measurement outcome $Y^n$ increases linearly in the number of rounds. Notably, this is true without assuming anything about the attack. In particular, the state $\rho_{A^nB^nE}$ after preparation but before the uncertainty principle is applied does not need to have any particular structure and is assumed to be arbitrary.


\subsubsection{Finite size effects and min-entropy}\label{sec:qkd-min-entropy}

So far we have argued that security of QKD is ensured if Eve's uncertainty of the key expressed in terms of the von Neumann entropy is large. This might be a reasonable \emph{ad-hoc} criterion\,---\,but more operationally what we want to say is that a key is secure if it can be safely used in any other protocol, for example one-time pad encryption, that requires a secret key. This leads to the notion of composable security, first studied by~\textcite{renner05} in the context of QKD. It turns out that in order to achieve composable security for a key of finite length, it is not sufficient to consider Eve's uncertainty in terms of the von Neumann entropy. Instead, it is necessary to ensure that the smooth min-entropy of the measurement results is large~\cite{rennerkoenig05}, so that we can extract a secret key, i.e., uniform randomness that is independent of the eavesdropper's memory. (Recall the discussion of randomness in Sec.~\ref{sec:randomness}.) Thus, instead of the inequality~\eqref{eq:vn-qkd0} involving von Neumann entropies, we want to apply a generalization of the Maassen-Uffink uncertainty relation with quantum memory~\eqref{eqnTripartiteMemory435}. This leads to the following relation~\cite{tomamichel11},
\begin{align}
H_{\min}^{\eps}(Y^n|E\Theta^n)+H_{\max}^{\eps}(Y^n|\hat{Y}^n)\geq n\,,
\end{align}
where $H_{\min}^{\eps}$ and $H_{\max}^{\eps}$ denote the smooth min- and max-entropies, variations of the min- and max-entropy (that we will not discuss further here). Hence, in order to ensure security it is sufficient to estimate the smooth max-entropy $H_{\max}^{\eps}(Y^n|\hat{Y}^n)$. This can be done by a suitable parameter estimation procedure, as is shown by~\textcite{tomamichellim11}.


\subsubsection{Continuous variable QKD}\label{sec:cvqkd}

Quantum information processing with continuous variables~\cite{weedbrook11} offers an interesting and practical alternative to finite-dimensional systems. Here, we discuss a particular variation of the above QKD protocol where Alice and Bob measure the quadrature components of an electromagnetic field, and then extract a secret key from the correlations contained in the resulting continuous variables.

If Alice and Bob share a squeezed Gaussian state, \textcite{furrer12} show that the security of such protocols can be shown rigorously using entropic uncertainty relations, including finite size effects. For this purpose, it is convenient to employ a smoothed extension of~\eqref{eq:cv/marco-needs-you} as first shown by~\textcite{furrer10}. This yields
\begin{align}
&H_{\min}^{\eps}(Y^n|E\Theta^n) + H_{\max}^{\eps}(Y^n|\hat{Y}^n) \nonumber\\
&\qquad \quad \geq n \log\left( \frac{2\pi}{\delta^2} \cdot S_{0}^{(1)}\bigg(1,\frac{\delta^2}{4}\bigg)^{-2}\right) \,,
\end{align}
where $Y_i$ is the outcome of the quadrature measurement in the basis (position or momentum) specified by $\Theta_i$ discretized with bin size $\delta$. We point to~\textcite{gehring15} for an implementation.


\subsection{Two-party cryptography}\label{sec:bounded_noisy}

In this section we discuss applications of entropic uncertainty relations to cryptographic tasks between two mutually distrustful parties (traditionally called Alice and Bob). This setup is in contrast to quantum key distribution where Alice and Bob do trust each other and only a third party is eavesdropping. Typical tasks for two-party cryptography are bit commitment, oblivious transfer or secure identification. 

It turns out, however, that even using quantum communication it is only possible to obtain security if we make some assumptions about the adversary~\cite{lo97-2,lo97-1,mayers97}. What makes this problem harder is that unlike in QKD where Alice and Bob trust each other to check on any eavesdropping activity, here every party has to fend for himself. Nevertheless, since tasks like secure identification are of great practical importance, one is willing to make such assumptions in practice.

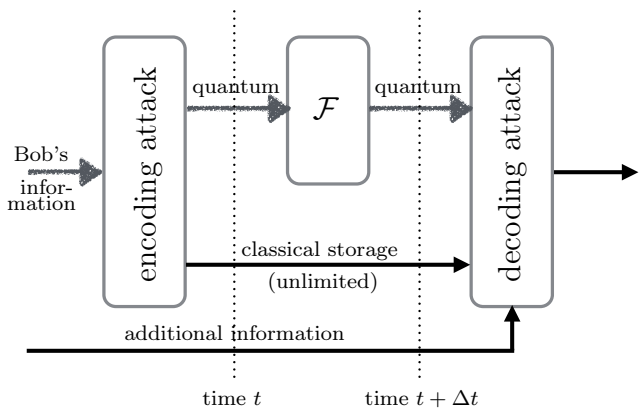
\begin{figure}
\begin{overpic}[width=1\columnwidth]{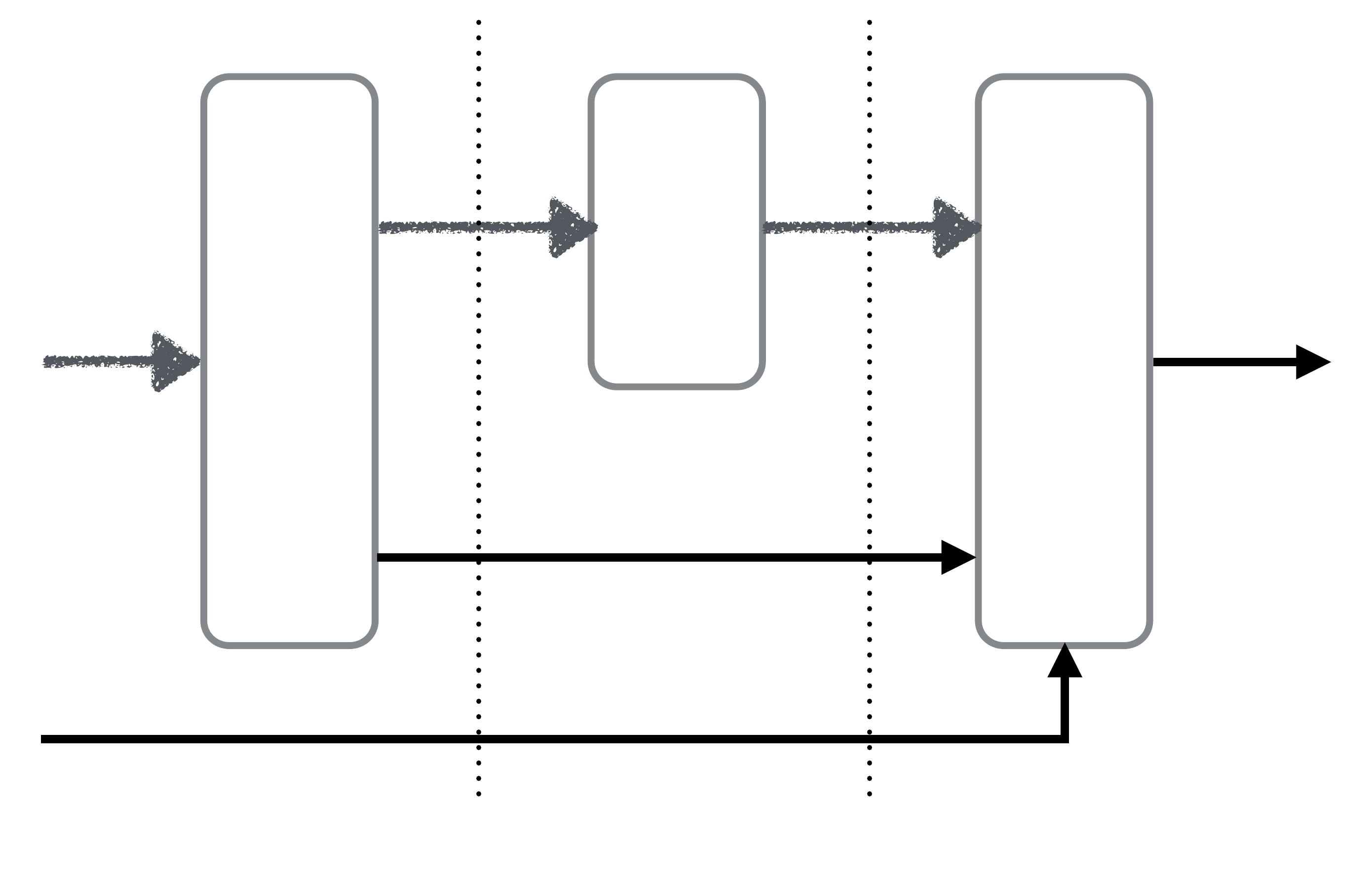}
\put(30,2){\footnotesize time $t$}
\put(55,2){\footnotesize time $t+\Delta t$}
\put(18,12){\footnotesize additional information}
\put(36,25){\footnotesize classical storage}
\put(40,20){\footnotesize (unlimited)}
\put(28.5,50){\footnotesize quantum}
\put(56.5,50){\footnotesize quantum}
\put(47,46){\large $\mathcal{F}$}
\put(19.5,21){\rotatebox{90}{\large encoding attack}}
\put(75.5,21){\rotatebox{90}{\large decoding attack}}
\put(1,39.5){\footnotesize Bob's}
\put(2,34.5){\footnotesize infor-}
\put(0,32){\footnotesize mation}
\end{overpic}
\caption{The noisy-storage model: \textcite{wehner08,koenig09} assume that during waiting times $\Delta t$ in the protocol, the adversary can only keep quantum information in an imperfect and limited storage device described by a quantum channel $\mathcal{F}$. This is the only restriction and the adversary is otherwise arbitrarily powerful. In particular, he can first store all incoming qubits, and has a quantum computer to encode them into an arbitrary quantum error-correcting code to protect them against the noise of the channel $\mathcal{F}$. He can also keep an unlimited amount of classical memory, and perform any operation instantaneously.}
\label{fig:noisyCrypto}
\end{figure}

Classically, such assumptions are typically computational assumptions. We assume a particular problem such as factoring is difficult to solve, and in addition that the adversary has limited computational resources, in particular not enough to solve the difficult problem. On the other hand, it is also possible to obtain security based on \emph{physical} assumptions, where we will first consider assuming that the adversary's memory resources are limited. Even a limit on classical memory can lead to security~\cite{cachin97,maurer92}. However, classical memory is typically cheap and plentiful. 
More significantly, however,~\textcite{dziembowski04} have shown that any classical protocol in which the honest players need to store $n$ bits to execute the protocol can be broken by an adversary who is able to store more than $O(n^2)$ bits. Motivated by this unsatisfactory gap it is an evident question to ask if quantum communication can be of any help. The situation is rather different if we allow quantum communication. We can have quantum protocols (see below) that require \emph{no} quantum memory to be executed, but that are secure as long as the adversary's quantum memory is not larger than $n-O(\log^2 n)$ qubits~\cite{dupuis15}, where $n$ is the number of qubits sent during the protocol. This is essentially optimal, since any protocol that allows the adversary to store $n$ qubits is known to be insecure~\cite{lo97-2,lo97-1,mayers97}. The assumption of a memory limitation is known as the bounded~\cite{damgaard08}, or more generally, noisy-storage model~\cite{wehner08}, as 
illustrated in Fig.~\ref{fig:noisyCrypto}. 

Security proofs in this model are intimately connected to entropic uncertainty relations. What's more, the uncertainty relations of~\textcite{dupuis15} together with the work of~\textcite{koenig09} demonstrate
that any physical assumption that limits the adversary's entanglement leads to security. 

\subsubsection{Weak string erasure}\label{sec:wse}

The relation between cryptographic security and entropic uncertainty relations can easily be understood by looking at a simple cryptographic building block known as weak string erasure (WSE)~\cite{koenig09}. Weak string erasure is universal for two-party secure computation, in the sense that any other protocol can be obtained by repeated executions of weak string erasure, following by additional quantum or classical communication~\cite{kilian88}. Importantly, the storage assumption only needs to hold during some time $\Delta t$ during the execution of weak string erasure. 

Weak string erasure generates the following outputs if both Alice and Bob are honest: Alice obtains an $n$-bit string $K^n$, and Bob obtains a random subset $I\subseteq[n]$, and the bits $K_I\subseteq K^n$ as indexed by the subset $I$. In addition, the following demands are made for security. If Bob is honest, then Alice does not know anything about $I$. In turn, if Alice is honest, then Bob should not know too much about $K^n$ (except for $K_I$). More precisely, Bob should not be able to guess $K^n$ too well, that is with~\eqref{eqncqminentropydef},
\begin{align}\label{eq:wse}
H_{\min}(K^n|B)\geq\lambda\cdot n\quad\text{for some}\quad\lambda\in[0,1]\,,
\end{align}
where $B$ denotes all of Bob's knowledge. We refer to~\textcite{koenig09} for a more detailed definition. A simple protocol for implementing weak string erasure is as follows:
\begin{enumerate}
\item Alice prepares a random $n$ bit string $K^n$, encodes each bit $K_i$ in one of the BB84 bases $\Theta\in\{\sigma_{\bX},\sigma_{\bZ}\}$ at random, and sends these $n$ qubits to Bob.
\item Bob measures the $n$ qubits in randomly chosen bases $\Theta'\in\{\sigma_{\bX},\sigma_{\bZ}\}$.
\item After the waiting time $\Delta t$, Alice sends the classical $n$ bit string $\Theta^n$ to Bob and outputs $K^n$.
\item Bob computes $I=\{i:\theta_i=\theta_i'\}$ and outputs $I$ and~$K_I$.
\end{enumerate}
Note that if both parties are honest, then the protocol is correct in the sense that Alice outputs $K^n$ and Bob $I$ with $K_I\subseteq K^n$. Moreover, when Alice is dishonest, it is intuitively obvious that she is unable to gain any information about the index set $I$ (even with an arbitrary quantum memory), since she never receives any information from Bob during the protocol. A precise argument for this can be found in, e.g., \textcite{koenig09}. On the other hand, note that a dishonest Bob with a quantum memory can easily cheat by just keeping the $n$ qubits he gets from Alice and wait until he receives the $n$ bit string $\Theta^n$ from Alice as well. Namely, he can then measure the $n$ qubits in the same basis $\Theta^n$ as Alice and get the full $n$ bit string $K^n$ (that is, $H_{\min}(K^n|B\Theta^n)=0$). However, if Bob only has a limited quantum memory, then he could not keep a perfect copy of the $n$ qubits he gets from Alice. 

The security analysis is linked immediately to a guessing game, whenever we consider a purified version of the protocol in which Alice does not prepare BB84 states herself, but instead makes EPR pairs $\ket{\psi}_{AB} = \left(\ket{00}_{AB} + \ket{11}_{AB}\right)/\sqrt{2}$ and sends $B$ to Bob, while measuring $A$ in a randomly chosen BB84 basis. In the analysis, one can indeed give even more power to Bob, in which we imagine that he prepares a state $\rho_{AB}$ in each round of the protocol and Alice measures $A$ in randomly chosen BB84 basis. Alice then sends him the basis choice. Recall that $H_{\rm min}(K^n|B\Theta^n) = - \log p_{\rm guess}(K^n|B\Theta^n)$, that is, the min-entropy security guarantee that WSE demands is precisely related to Bob's ability to win the guessing game~\cite{ballester08}. The storage assumption translates into one particular example of how the entanglement in $\rho_{AB}$ is limited, putting a limit on $H_{\rm min}(A|B)$ of the states that Bob can prepare.


\subsubsection{Bounded-storage model}\label{sec:bounded}

To illustrate further how a bound on entropic uncertainty leads to security, let us first consider a special case of the noisy-storage model, also known as the bounded-storage model. Here, the channel
$\mathcal{F} = \id_2^{\otimes q}$ in Fig.~\ref{fig:noisyCrypto} is just the identity on $q$ qubits. 
This \emph{bounded-storage model} was introduced and first studied by~\textcite{schaffner07,damgaard07,damgaard08}. 

While more refined bounds are known~\cite{dupuis15}, let us first explain how entropic uncertainty relations for a classical memory system can be used to obtain weak security statements in this setting. To this end, we differentiate Bob's knowledge into $B=QM\Theta^n$, where $Q$ denotes the $q$ qubits of quantum memory, $M$ denotes (unbounded) classical information, and $\Theta^n$ is the $n$ bit basis information string Alice sent to Bob. Since the conditional min-entropy obeys a chain rule~\cite{renner05}, we can separate the quantum memory as
\begin{align}
H_{\min}(K^n|B)&=H_{\min}(K^n|QM\Theta^n)\label{eq:min_lock1}\\
&\geq H_{\min}(K^n|M\Theta^n)-q\,.\label{eq:min_lock2}
\end{align}
Analyzing $H_{\min}(K^n|M\Theta^n)$ is then directly determined by Bob's ability to win the guessing game, in which he only has classical information $M$. Using the min-entropy uncertainty relation~\eqref{eq:schaffner_min} for the $n$ qubit BB84 measurements (with an extension to classical side information $M$ as sketched in Sec.~\ref{sctCMemory}), we get
\begin{align}
H_{\min}(K^n|M\Theta^n)\geq-n\cdot\log\left(\frac{1}{2}+\frac{1}{2\sqrt{2}}\right)\,.
\end{align}
Hence, we find a non-trivial lower bound
\begin{align}
H_{\min}(K^n|B)>0\quad\text{as long as}\quad q\lesssim n\cdot0.22\,.
\end{align}
This security analysis can be refined and improving on the work of~\textcite{damgaard07}, \textcite{nelly12} make use of the following stronger smooth min-entropy uncertainty relation which is based on~\eqref{eq:bb84_halpha},
\begin{align}\label{eq:minsmooth_boundedstorage}
&H_{\min}^{\eps}(K^n|M\Theta^n)\notag\\
&\geq n\cdot\sup_{s\in(0,1]}\left(\frac{1}{s}\big(1+s-\log\left(1+2^s\right)\big)-\frac{1}{sn}\log\frac{2}{\eps^2}\right)\,.
\end{align}
One can use this uncertainty relation together with the more refined analysis of~\textcite{koenig09} instead of~\eqref{eq:min_lock2}, 
to obtain perfect security $(\lambda\to1)$ against quantum memory of size
\begin{align}
q\leq \frac{n}{2}\ .
\end{align}
for $n\to\infty$. 
Ultimately, \textcite{dupuis15} show by deriving strong entropic uncertainty relations that the protocol from Sec.~\ref{sec:wse} implements a WSE scheme against $q$ qubits of quantum memory for
\begin{align}
\lambda=\frac{1}{2}\left(\gamma_{BB84}\left(-\frac{q}{n}\right)-\frac{1}{n}\right)\,,
\end{align}
where the function $\gamma_{BB84}(\cdot)$ is as in~\eqref{eq:gammabar}. Asymptotically $(n\to\infty)$, this provides perfect security $(\lambda\to1)$ against quantum memories of size
\begin{align}
q\leq n-O\left(\log^2n\right)\,.
\end{align}
This is basically optimal, since no protocol can be secure if $q=n$. Finally, we mention that alternatively we could also use a six-state encoding $\{\sigma_{\bX},\sigma_{\bY},\sigma_{\bZ}\}$ for the weak string erasure protocol described in Sec.~\ref{sec:wse}. We refer to~\textcite{mandayam11,nelly12,dupuis15} for a security analysis.


\subsubsection{Noisy-storage model}\label{sec:noisy}

Let us now consider the general case of arbitrary storage devices $\mathcal{F}$ in Fig.~\ref{fig:noisyCrypto}~\cite{wehner08}. This model is motivated by the fact that counting qubits is generally a significant overestimate of the storage capabilities of a quantum memory, and indeed for example for continous variable systems there is no dimension bound to which to apply the bounded-storage analysis. The first general security analysis was given by~\textcite{koenig09}, which was then refined significantly by~\textcite{christandl13,berta14}, leading to the asymptotically tight security analysis by~\textcite{dupuis15}. Here, one cannot just use the chain rule to separate the quantum memory as in~\eqref{eq:min_lock1} -- \eqref{eq:min_lock2}. Such a separation is only possible when relating the security to the classical capacity of the storage channel $\mathcal{F}$~\cite{koenig09}. Instead, we have to apply a min-entropy uncertainty relation with quantum memory to directly lower bound
\begin{align}
H_{\min}(K^n|B)=H_{\min}(K^n|QM\Theta^n)\,.
\end{align}
We use a variant of the relation~\eqref{eq:bb84_h2qm} for the $n$ qubit BB84 measurements to bound~\cite{dupuis15},
\begin{align}\label{eq:noisy_storage}
H_{\min}^{\eps}(K^n|QM\Theta^n)\geq&\;n\cdot\gamma_{BB84}\left(\frac{H_{\min}(A^n|QM)}{n}\right)\notag\\
&-1-\log\left(\frac{2}{\eps^2}\right)\,,
\end{align}
where the function $\gamma_{BB84}(\cdot)$ is as in~\eqref{eq:gammabar}. In order to get an idea how to lower bound the right-hand side of~\eqref{eq:noisy_storage} under a noisy quantum memory $Q$ assumption, recall that $H_{\min}(A^n|QM)$ is a measure of entanglement between $A^n$ and $B = QM$. In particular, one can relate this amount of entanglement to Bob's ability to store the $n$ EPR pairs that Alice sends in the purified version of the protocol, that is, the quantum capacity of the storage channel $\mathcal{F}$. If $\mathcal{F}$ cannot preserve said entanglement, then $H_{\min}(K^n|QM\Theta^n)$ in~\eqref{eq:noisy_storage} will be lower bounded non-trivially leading to a secure WSE scheme for some trade-off between the security parameter $\lambda$ from~\eqref{eq:wse}, the number $n$ of qubits sent, and the noisiness of the quantum memory $Q$. We refer to~\textcite{dupuis15} for any details.

Again we could also use a six-state encoding $\{\sigma_{\bX},\sigma_{\bY},\sigma_{\bZ}\}$ for the weak string erasure protocol described in Sec.~\ref{sec:wse}. We refer to~\textcite{berta14,dupuis15} for a security analysis. 


\subsubsection{Uncertainty in other protocols}

Entropic uncertainty relations feature in many other quantum cryptographic protocols~\cite{broadbent15}. The entropic relation for channels~\eqref{eqnchristandlwinter} was used in~\textcite{buhrman08} to obtain cheat-sensitivity for a quantum string commitment protocol. The same relations as relevant for the noisy-storage model, have also been used to prove security in the isolated qubit model~\textcite{liu14,liu15b}. In this model, the adversary is given a quantum memory of potentially long-lived qubits, but they are isolated in the sense that he is unable to perform coherent operations on 
many qubits simultaneously. In particular, the uncertainty relation of~\textcite{damgaard07} was used in~\textcite{liu14} to obtain security. It would possible to use the relation~\eqref{eq:bb84_halpha} from~\textcite{nelly12} to obtain improved security parameters. Furthermore, tripartite~\cite{tomamichelfehr12} 
uncertainty relations have been used to ensure the security of position-based cryptography. Finally, in relativistic cryptography, security of two-party protocols is possible under the assumptions that each player is split into several non-communicating agents. Tripartite uncertainty relations have been used to establish security in this setting~\cite{kaniewski13}.


\subsection{Entanglement witnessing}\label{sec:entwit}

Entanglement is a central resource in quantum information processing. Hence, methods for detecting entanglement are crucial for quantum information technologies. Entanglement witnessing refers to the process of verifying that a source is producing entangled particles. Entangled states are defined as those states that are non-separable, i.e., they cannot be written as a convex combination of product states. A common theme in entanglement witnessing is to prove a mathematical identity that all separable states must satisfy; let us refer to such an identity as an entanglement witness. Experimentally demonstrating that one's source violates this identity will then guarantee that the source produces entangled particles. 

Entanglement witnessing is a well-developed field (e.g., see the review articles by~\textcite{horodecki09, guhne09}), and there are many types of entanglement witnesses. Here, we focus mostly on entanglement witnesses that follow from entropic uncertainty relations. 

In what follows, we restrict the discussion to bipartite entanglement. We note that entanglement witnessing typically occurs in the distant-laboratories paradigm, where two parties - Alice and Bob - can each perform local measurements on their respective systems, but neither party can perform a global measurement on the bipartite system.

For introductory purposes, let us mention a simple, well-known bipartite entanglement witness for two qubits. Although it is non-entropic, it is based on complementary observables, and so it can be directly compared to the entropic witnesses discussed below. Namely, consider the operator
\begin{align}
E_{XZ}&:= E_X + E_Z, \text{  where}\\
E_X &:= \dya{+} \ot \dya{-} \hspace{4pt}+\hspace{4pt}\dya{-} \ot \dya{+},\\ 
E_Z &:= \dya{0} \ot \dya{1} +\dya{1} \ot \dya{0}\,.
\end{align}
Note that $E_X$ and $E_Z$ are ``error operators'' in that they project onto the subspaces where Alice's and Bob's measurement outcomes are different. For a maximally entangled state of the form $\ket{\psi} = (\ket{00}+\ket{11})/\sqrt{2}$, there is no probability for error in either basis, so we have $\expct{\psi}{E_{XZ}} = 0$. On the other hand, for any separable state $\rho_{AB}$, we have that (e.g., see \textcite{namiki12})
\begin{align}\label{eqnlinearentanglementwitness}
\tr[\rho_{AB}E_{XZ}] \geq\frac{1}{2} \,.
\end{align}
Hence, if $\avg{E_X} + \avg{E_Z} < 1/2$, where $\avg{O}:=\tr [O \rho_{AB}]$, then $\rho_{AB}$ is entangled. This witness is depicted as the solid line in Fig.~\ref{figEntanglement}.


\subsubsection{Shannon entropic witness}

Some early work on entanglement witnessing using entropic uncertainty relations was done by \textcite{giovannetti04, guhne04}, and further improvements were later made by \textcite{huang10}. The following discussion focuses primarily on more recent developments, e.g., where entanglement witnessing is based on the bipartite uncertainty relation with quantum memory in \eqref{eqnBipartiteMemory13}. \textcite{berta10} discussed how this can be used for entanglement witnessing, and the approach was implemented by~\textcite{prevedel11,li11}. Specifically, from \eqref{eqnBipartiteMemory13}, one finds that all separable states satisfy
\begin{align}\label{eqnBipartiteMemoryseparable}
H(X_A | X_B) + H( Z_A | Z_B) \geq q_{\rm MU}\,,
\end{align}
where the $q_{\rm MU}$ parameter refers to Alice's observables, and Bob's observables $\bX_B$ and $\bZ_B$ are arbitrary. One can see this by noting that $H(A|B)\geq 0$ for any separable state, and furthermore that measuring Bob's system in some basis $\bX_B$ cannot reduce his uncertainty about Alice's measurement, i.e., $H(X_A | X_B) \geq H(X_A | B)$. 

One can use \eqref{eqnBipartiteMemoryseparable} for entanglement witnessing, using a protocol where Alice and Bob have many copies of $\rho_{AB}$ and they both measure on each copy either their $\bX$ or $\bZ$ observable. The quantities $H(X_A | X_B)$ and $H( Z_A | Z_B)$ can then be calculated from their joint probability distributions $\Pr(X_A=x_A, X_B = x_B)$ and $\Pr(Z_A=z_A, Z_B=z_B)$, and if \eqref{eqnBipartiteMemoryseparable} is violated, then $\rho_{AB}$ must be entangled. 

Fig.~\ref{figEntanglement} depicts this entanglement witness (long-dashed curve) for the case of qubits and mutually unbiased bases. A comparison of this curve to the black line shows that \eqref{eqnlinearentanglementwitness} detects more entangled states than~\eqref{eqnBipartiteMemoryseparable}. However, the ``quality'' of entanglement that~\eqref{eqnBipartiteMemoryseparable} detects is higher. This is because \eqref{eqnBipartiteMemoryseparable} holds for all non-distillable states, i.e., states from which Alice and Bob cannot distill any EPR (maximally-entangled) states using local operations and classical communication (see, e.g., \cite{horodecki09} for a discussion of local operations and classical communication). In this sense, \eqref{eqnBipartiteMemoryseparable} detects \textit{distillable} entanglement whereas~\eqref{eqnlinearentanglementwitness} detects all forms of entanglement.

One can make this quantitative using a result by \textcite{devetak05} that the coherent information (i.e., minus the conditional entropy) lower bounds the distillable entanglement $E_D$, i.e., the optimal asymptotic rate for distilling EPR states using LOCC:
\begin{align}\label{eqnDevWin}
E_D \geq -H(A|B) \,.
\end{align}
Combining this with~\eqref{eqnBipartiteMemory13} gives
\begin{align}\label{eqnBipartiteMemoryRearranged}
E_D \geq q_{\rm MU} -H(X_A | X_B) - H( Z_A | Z_B)\,.
\end{align}
This reveals an advantage of the entropic uncertainty approach to entanglement witnessing. Namely, that it can give quantitative lower bounds, in contrast to witnesses like that in~\eqref{eqnlinearentanglementwitness} that only answer a ``yes or no'' question.

Another advantage of the entropic uncertainty approach is that it requires no structure on Bob's side. While~\eqref{eqnlinearentanglementwitness} requires both Alice's and Bob's measurements to be mutually unbiased, the entropic uncertainty approach allows for arbitrary measurements on Bob's system.


\subsubsection{Other entropic witnesses}

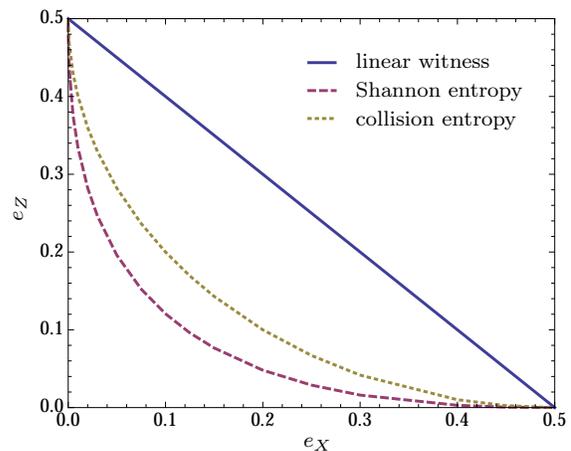
\begin{figure}[tbp]
\begin{center}
\begin{overpic}[width=7cm]{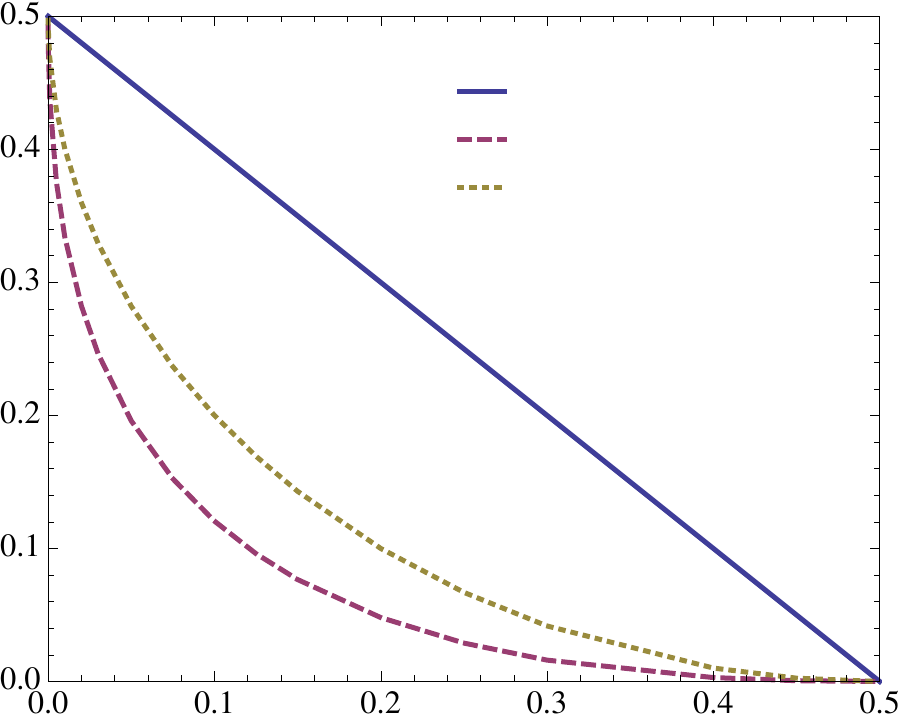}
\put(50,-3){\footnotesize $e_X$}
\put(-5,40){\rotatebox{90}{\footnotesize $e_Z$}}
\put(60,69){\footnotesize linear witness}
\put(60,63.5){\footnotesize Shannon entropy}
\put(60,58){\footnotesize collision entropy}
\end{overpic}
\vspace{0.1cm}
\caption{Entanglement witnessing for a bipartite two-qubit state using mutually unbiased observables. Suppose Alice and Bob observe $\Pr[X_A\!=\!X_B\!=\!0] = \Pr[X_A\!=\!X_B\!=\!1] = \frac{1-e_X}{2}$ and $\Pr[X_A\!=\!0, X_B\!=\!1] = \Pr[X_A\!=1\!, X_B\!=\!0] = \frac{e_X}2$, and analogously for $Z$ and $e_Z$. The region below the curve in the plot indicates the region for which one can guarantee entanglement for the respective witnesses.}
\label{figEntanglement}
\end{center}
\end{figure}

Bipartite quantum memory uncertainty relations generally lead to entanglement witnesses. For example, \textcite{berta13} discuss how the uncertainty relation in \eqref{eq:h2_bcw} allows for entanglement witnessing using a set of $n$ MUBs on Alice's system (more precisely, a subset of size $n$ of MUBs chosen from a set of $d_A+1$ MUBs, where $d_A$ is a prime power and $2\leq n\leq d_A +1$). Consider such a set $\{\bX_j\}$ of $n$ MUBs on Alice's system, and consider a set of $n$ arbitrary POVMs $\{\bY_j\}$ on Bob's system. \textcite{berta13} show that all separable states must satisfy
\begin{equation}\label{eq:entwitnessh2}
\sum_{j=1}^n 2^{-H_2(X_{j}|Y_{j})} \leq 1+\frac{n-1}{d_A}\,.
\end{equation}
Fig.~\ref{figEntanglement} compares this entanglement witness (short-dashed curve) to the previously discussed ones, in the qubit case with $n=2$. Notice that \eqref{eq:entwitnessh2} detects more entangled states than \eqref{eqnBipartiteMemoryseparable}, but not as much as \eqref{eqnlinearentanglementwitness}.

Similar to the Shannon entropy case in \eqref{eqnBipartiteMemoryRearranged}, the uncertainty relation \eqref{eq:h2_bcw} actually allows one to give a quantitative lower bound on an entanglement-like measure. Namely, \eqref{eq:h2_bcw} allows one to lower bound $-H_{\rm coll}(A|B)$.
  
\subsubsection{Continuous variable witnesses}

The method of witnessing entanglement through entropic uncertainty relations was also extended to continuous variable systems by~\textcite{walborn09}, and further studied by~\textcite{saboia11,huang13}.


\subsection{Steering inequalities}\label{sec:steering}

First highlighted by \textcite{schroedinger35}, steering is a phenomenon for bipartite quantum systems that is related to entanglement (although not precisely the same). Like the previous subsection, we consider the distant-laboratories paradigm involving two parties, Alice and Bob, where Alice (Bob) has access to system $A$ ($B$). Steering corresponds to one party's (say Alice's) measurement choice giving rise to different ensembles of states on the other party's (Bob's) system. Not all quantum states exhibit steering, e.g., separable states are non-steerable. At the other extreme, all states that violate a Bell inequality are steerable. While Bell inequalities are derived for states that admit a local hidden variable (LHV) model, \textcite{wiseman07} formalized the notion of steerability as those states $\rho_{AB}$ that do not admit a local hidden state (LHS) model. An LHS model is a model where, say, system $B$ has a local quantum state that is classically correlated to arbitrary observables on system $A$.  This formalization has led researchers to derive steering inequalities~\cite{cavalcanti09}, in analogy to Bell inequalities.

\textcite{walborn11,schneeloch13} show how entropic uncertainty relations can be used to derive steering inequalities. The idea is that if $B$ has a local hidden state, then its measurement probabilities must obey a single system uncertainty relation, even if they are conditioned on the measurement outcomes on $A$. More precisely, an LHS model implies that the joint probability distribution for discrete observables $\bX_A$ on $A$ and $\bX_B$ on $B$ has the form
\begin{align}\label{eqnLHS}
P (\bX_A , \bX_B) = \sum_{\lambda} P (\Lambda = \lambda) P (\bX_A | \Lambda = \lambda) P_Q (\bX_B | \Lambda = \lambda) \,.
\end{align}
Here, $\Lambda$ is the hidden variable that determines Bob's local state, $\lambda$ is a particular value that this variable may take, and the subscript $Q$ on $P_Q (\bX_B | \Lambda = \lambda)$ emphasizes that the probability distribution arises from a single quantum state. Next, we have that 
\begin{align}\label{eqnLHS2}
H(X_B | X_A)& \geq H(X_B | X_A\Lambda)\\
& = \sum_{\lambda} P (\Lambda = \lambda) H(X_B | X_A\Lambda = \lambda)\\
& = \sum_{\lambda} P (\Lambda = \lambda) H(X_B |\Lambda = \lambda )\,,
\end{align}
where the notation $H(X_B | X_A\Lambda = \lambda)$ should be read as the entropy of $X_B$ conditioned on $X_A$ and conditioned on the event that $\Lambda = \lambda$. Hence, for two observables $\bX_B$ and $\bZ_B$ on $B$, and some other observables $\bX_A$ and $\bZ_A$ on $A$, we have 
\begin{align}
&H(X_B | X_A)+ H(Z_B | Z_A)\notag\\
&\geq \sum_{\lambda} P (\Lambda = \lambda) [ H(X_B |\Lambda = \lambda ) +H(Z_B |\Lambda = \lambda )]
\end{align}
Combining this with, say, Maassen-Uffink's uncertainty relation \eqref{eq:shannonUR} gives the following steering inequality \cite{schneeloch13},
\begin{align}\label{eqnLHS3}
H(X_B | X_A)+ H(Z_B | Z_A) \geq q_{\rm MU}\,,
\end{align}
where $q_{\rm MU}$ refers to Bob's observables. Any state $\rho_{AB}$ that admits an LHS model must satisfy \eqref{eqnLHS3}. Hence, an experimental violation of \eqref{eqnLHS3} would constitute a demonstration of steering. Similar steering inequalities can be derived for continuous variables \cite{walborn11}.


\subsection{Wave-particle duality}\label{sec:waveparticle}

\begin{figure}[tbp]
\begin{center}
\includegraphics{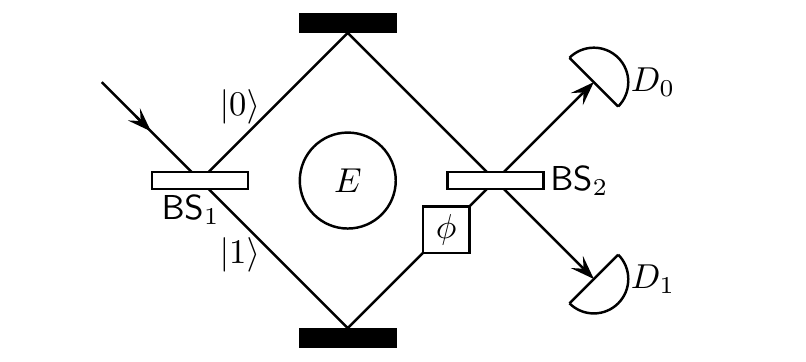}
\caption{Mach-Zehnder interferometer for single photons. A photon impinges on a beam splitter, after which we label the two possible paths by the $\bZ$ basis states $\ket{0}, \ket{1}$. The photon may interact with some environment $E$ inside the interferometer. Then a phase $\phi$ is applied to the lower path, and the two paths are recombined on a second beam splitter. Finally the photon is detected at either $D_0$ or $D_1$.}
\label{figMZ}
\end{center}
\end{figure}

Wave-particle duality is the fundamental concept that a single quantum system can exhibit either wave behavior or particle behavior: one cannot design an interferometer that can simultaneously show both behaviors. This idea was qualitatively discussed, e.g., by Feynman, and was subsequently put on quantitative grounds by \textcite{wootters79}, \textcite{jaeger95}, \textcite{englert96}, \textcite{englert00}, and others, who proved inequalities known as wave-particle duality relations (WPDRs). Many such relations consider the Mach-Zehnder interferometer for single photons, shown in Fig.~\ref{figMZ}. In this case, particle behavior is associated with knowing the path that the photon travels through the interferometer. Wave behavior on the other hand is associated with seeing oscillations in the probability to detect the photon in a given output mode as one varies the relative phase $\phi$ between the two interferometer arms. Denoting the which-path observable as $\bZ = \{\dya{0}, \dya{1}\}$, particle behavior can be quantified by the path predictability $\cP = 2 p_{\guess}(Z)-1$ (which is related to the probability $p_{\guess}(Z)$ of guessing the path correctly). The wave behavior is quantified by the fringe visibility
\begin{align}
\cV =  \frac{p^{\max}_0 - p^{\min}_0}{p^{\max}_0 + p^{\min}_0}\quad\mathrm{with}\quad p^{\max}_0&:= \max_{\phi}p_0\notag\\
p^{\min}_0&:= \min_{\phi} p_0\,,
\end{align}
where $p_0$ is the probability for the photon to be detected by $D_0$ (see Fig.~\ref{figMZ}). \textcite{wootters79} prove that
\begin{align}\label{eqnWPDR1}
\cP^2 + \cV ^2 \leq 1\,,
\end{align}
which implies $\cV = 0$ when $\cP = 1$ (full particle behavior means no wave behavior) and vice-versa.

More generally, suppose the photon may interact with some environment system $E$ inside the interferometer. Measuring $E$ might reveal, e.g., some information about which path the photon took, so it is natural to consider the path distinguishability
\begin{align}
\cD  = 2 p_{\guess}(Z|E)-1\,.
\end{align}
\textcite{englert96,jaeger95} prove a stronger version of \eqref{eqnWPDR1}, namely 
\begin{align}\label{eqnWPDR2}
\cD^2 + \cV ^2 \leq 1\,.
\end{align}

WPDRs such as~\eqref{eqnWPDR1} and~\eqref{eqnWPDR2} have often been thought to be conceptually different from uncertainty relations, although this has been debated. For example, \textcite{durr00} and \textcite{busch06} found connections between certain WPDRs and Robertson's uncertainty relation involving the standard deviation. More recently, \textcite{coles14d} showed that~\eqref{eqnWPDR1},~\eqref{eqnWPDR2}, and some other WPDRs are actually entropic uncertainty relations in disguise. In particular, they correspond to the uncertainty relation for the min- and max-entropy in~\eqref{eqnminmaxuncertaintyqmem}, applied to complementary qubit observables. Namely, \eqref{eqnWPDR1} is equivalent to the uncertainty relation,
\begin{align}\label{eqnWPDR3}
H_{\min}(Z)+ \min_{W\in XY}H_{\max}(W)\geq 1\,,
\end{align}
where the $\min_{W\in XY}$ corresponds to minimizing over all observables in the $xy$ plane of the Bloch sphere. Likewise \eqref{eqnWPDR2} is equivalent to the uncertainty relation
\begin{align}\label{eqnWPDR4}
H_{\min}(Z|E)+ \min_{W\in XY}H_{\max}(W)\geq 1\,.
\end{align}
This unifies the wave-particle duality principle with the entropic uncertainty principle, showing that the former is a special case of the latter. 

Naturally, other entropies could be used in place of the min- and max-entropy, and although one might not obtain a precise equivalence to the WPDRs above, the conceptual meaning may be similar. \textcite{bosyk13} took this approach using uncertainty relations involving R\'enyi entropies. \textcite{vaccaro11} employed the Shannon entropy to formulate a WPDR in terms of the mutual information. Moreover, they added the conceptual insight that wave and particle behavior are related to symmetry and asymmetry, respectively. Finally, \textcite{englert08} considered entropic measures of wave and particle behavior for interferometers with more than two paths.


\subsection{Quantum metrology}\label{sec:metrology}

Quantum metrology deals with the physical limits on the accuracy of measurements \cite{giovannetti11}. The uncertainty principle plays an important role in establishing such physical limits. Typically in quantum metrology one is interested in estimating an optical phase, e.g., the phase shift in an interferometer (as in Fig.~\ref{figMZ}). Hence, uncertainty relations involving the phase have applications here. Recall that we briefly discussed an entropic uncertainty relation for the number and phase in Sec.~\ref{sec:infinte}, specifically in~\eqref{eq:number_phase}. While quantum metrology is a broad field (see, e.g., \textcite{giovannetti11} for a review), we mention here a few works that exploit entropic uncertainty relations.

The Heisenberg limit is a well-known limit in quantum metrology stating that the uncertainty in the phase estimation scales as $1/\avg{N}$. Here, $\avg{N}$ is the mean photon number of the light that is used to probe the phase. \textcite{hall12b} note that the Heisenberg limit is heuristic, and put it on rigorous footing by proving the following bound,
\begin{align}\label{eqnheisenberglimit1}
\delta \hat{\Phi} \geq k / \avg{N+1}\,,
\end{align}
where $\delta \hat{\Phi}$ is the root-mean-square deviation of the phase estimate $\hat{\Phi}$ from the actual phase $\Phi$, and $k:= \sqrt{2\pi / e^2}$. To prove~\eqref{eqnheisenberglimit1}, \textcite{hall12b} define the random variable $\Theta:= \hat{\Phi} - \Phi$ and apply the entropic uncertainty relation in~\eqref{eq:number_phase}, giving 
\begin{align}\label{eqnheisenberglimit2}
H(N)+h(\Theta)\geq\log2\pi \,.
\end{align}
Then they combine \eqref{eqnheisenberglimit2} with some identities that relate $h(\Theta)$ to $\delta \hat{\Phi}$ and $H(N)$ to $\avg{N+1}$.

\textcite{hall12} consider a more general scenario where one may have some prior information about the phase, and they likewise use the entropic uncertainty relation in~\eqref{eq:number_phase} to obtain a rigorous statement of the Heisenberg limit.


\subsection{Other applications in quantum information theory}

Recent efforts to understand the classical-quantum boundary, in the context of both physics and information-processing, have led to quantitative measures of ``quantumness'' like coherence and discord, which are discussed in Sec.~\ref{sec:coherence} and~\ref{sec:discord}, respectively. We further discuss information locking in Sec.~\ref{sec:locking} and touch on quantum coding in Sec.~\ref{sec:coding}.


\subsubsection{Coherence}\label{sec:coherence}

\textcite{baumgratz14} introduced a framework for quantifying coherence, which is a measure that does not increase under incoherent operations. There are a variety of coherence measures, but one in particular has an operational meaning in terms of the number of distillable maximally coherent states~\cite{winter15},
\begin{align}\label{eqncoherence1}
\Phi (\bZ, \rho) :=D\left(\rho\middle\|\sum_z \proj{\bZ^{z}}\rho \proj{\bZ^{z}}\right)\,,
\end{align}
the relative entropy of coherence. Note that the coherence is a function of the state $\rho$ as well as an orthonormal basis $\bZ=\{\proj{\bZ^{z}}\}$.

The following connection between coherence and entropic uncertainty was established in \cite{coles10, coles12b}. Let $\rho_S$ be any state for system $S$ and let $\bZ$ be a projective measurement on $S$. Then, we have
\begin{align}\label{eqncoherence2}
\Phi (\bZ, \rho_S) = H(Z|E)\,,
\end{align}
where $E$ is a purifying system for $\rho_S$. This states that the relative entropy of coherence for a projective measurement is equivalent to the uncertainty of that measurement given the purifying system, or in other words, given access to the environment $E$. The right-hand-side of \eqref{eqncoherence2} quantifies uncertainty in the presence of quantum memory, and uncertainty relations for such measures have been discussed in Sec.~\ref{sctMemory}. Hence, one can reinterpret such uncertainty relations as, e.g., in~\eqref{eqnBipartiteMemory13}, as lower bounds on the \textit{coherence} of $\rho_S$ for different measurements. This idea was discussed by~\textcite{korzekwa14}, although they focused more on the perspective of~\textcite{luo05} of separating total uncertainty into a ``classical'' and ``quantum'' part.  In particular, for a rank-one projective measurement $\bZ=\{\proj{\bZ^{z}}\}$ and a quantum state $\rho$, they defined the classical uncertainty as the entropy of the state, $C(\bZ, \rho) := H(\rho)$, and the quantum uncertainty as the relative entropy of coherence,
\begin{align}
Q(\bZ, \rho) := D\left(\rho\middle\|\sum_z \proj{\bZ^{z}}\rho \proj{\bZ^{z}}\right)\,.
\end{align}
It is straightforward to show that overall uncertainty is the sum of the classical and quantum parts
\begin{align}
H(Z) = Q(\bZ, \rho) + C(\bZ, \rho)\,.
\end{align}
\textcite{korzekwa14} derive several uncertainty relations for the quantum uncertainty $Q(\bZ, \rho)$. However, using~\eqref{eqncoherence2}, one can reinterpret their relations as entropic uncertainty relations in the presence of quantum memory. In particular, their uncertainty relations follow directly from combining~\eqref{eqnBipartiteMemory13} with~\eqref{eqncoherence2}.


\subsubsection{Discord}\label{sec:discord}

\textcite{ollivier01} quantified quantum correlations by discord,
\begin{align}\label{eqndiscorddef}
D(B|A) := I(A:B) - J(B|A)\,,
\end{align}
which is the difference between the quantum mutual information $I(A:B)$ and the classical correlations,
\begin{align}
J(B|A) :=  \max_{\bX} I(X:B)\,,
\end{align}
where the optimization is over all POVMs $\bX$ acting on system $A$. In Sec.~\ref{scttripartiteQMemory}, Ex.~\ref{exQmemlocking}, we discussed how discord quantifies the gap between conditioning on classical versus quantum memory. Another connection to discord is the following. In an effort to strengthen the uncertainty relation with quantum memory in~\eqref{eqnBipartiteMemory13}, \textcite{pati12} introduced an additional term that depends on the discord of the state $\rho_{AB}$. Namely, they proved the inequality
\begin{align}\label{eqndiscordUR}
H(X|B)+H(Z|B)\geq\;&q_{\rm MU} + H(A|B)\notag\\
&+ \max \Big\{0, D(B|A) - J(B|A)\Big\}\,.
\end{align}
Clearly this strengthens the bound in~\eqref{eqnBipartiteMemory13} for states $\rho_{AB}$ whose discord exceeds their classical correlations: $D(B|A) > J(B|A)$. Indeed, \textcite{pati12} showed that this is true for Werner states, for which~\eqref{eqndiscordUR} becomes an equality.

In turn, this result was used by~\textcite{hu13} to obtain a strong upper bound on discord. That is, the uncertainty relation \eqref{eqndiscordUR} allows one to bound the discord by
\begin{align}\label{eqndiscordupperbound}
D(B|A)\leq \frac{1}{2}\Big(I(A:B) + \delta_T\Big)\,,
\end{align}
where
\begin{align}\label{eqndiscordupperbound2}
\delta_T:=H(X|B)+H(Z|B)-q_{\rm MU}-H(A|B)\,.
\end{align}
Here, $\delta_T$ is the gap between the left and right hand sides in the uncertainty relation~\eqref{eqnBipartiteMemory13}.

Further connections between quantum correlations and entropic uncertainty relations have been elucidated in the context of non-Markovian dynamics \cite{karpat15}, entanglement creation \cite{coles12c}, teleportation \cite{hu12}, and monogamy \cite{hu13b}.


\subsubsection{Locking of classical correlations}\label{sec:locking}

One operational way of understanding entropic uncertainty relations is in terms of information locking~\cite{divincenzo04}. In the following we present a cryptographic view on information locking as discussed by~\textcite{fawzi11}.

A locking scheme is a protocol that encodes a classical message into a quantum state using a classical key of size smaller than the message. The goal is that without knowing the key the message is locked in the quantum state such that any possible measurement only reveals a negligible amount of information about the message. Furthermore, knowing the key it is possible to unlock and completely recover the message. The connection of information locking to entropic uncertainty is best presented by means of a simple example based on the Maassen-Uffink bound for the $n$ qubit BB84 measurements~\eqref{eq:bb84_h1},
\begin{align}\label{eq:bb84_h11}
H(K^n|\Theta^n)\geq n\cdot\frac{1}{2}\;\;\mathrm{with}\;\;\Theta^n\in\{\theta_1,\ldots,\theta_{2^n}\}\,.
\end{align}
In order to encode a uniformly random $n$ bit string $X$ we choose at random an $n$ qubit BB84 basis $\theta_i$ (the key) and encode the message in this basis. Based on~\eqref{eq:bb84_h11}, \textcite{divincenzo04} show that for any measurement on this quantum state the mutual information (accessible information) between the outcome of that measurement and the original classical message $X$ is at most $n/2$. That is, $n/2$ bits are locked in the quantum state and are not accessible without knowing the basis choice (the key). This is remarkable because any non-trivial purely classical encryption of an $n$ bit string message requires a key of size at least $n$. Of course, this then raises the question about the optimal trade-off between the number of lockable bits and the key size. For that purpose~\textcite{fawzi11} make use of the uncertainty relation~\eqref{eq:locking_ex},
\begin{align}
H(K|\Theta)\geq n\cdot(1-2\eps)-h_{\bin}(\eps)\;\;\mathrm{with}\;\;\Theta=\{\theta_1,\dots,\theta_L\}\,.
\end{align}
Based on this they show that a key size of $L=O(\log(n/\eps))$ allows for locking an $n$ bit string up to a mutual information smaller than $\eps>0$. State-of-the-art results use stronger definitions for information locking in terms of the trace norm instead of the mutual information and are based on so-called metric uncertainty relations~\cite{hayden13,fawzi11}.\footnote{We emphasize that the security definitions for information locking are not composable (see, e.g., \textcite{renner05} for a discussion).} Finally, we mention that~\textcite{guha14} initiated the study of the information locking capacity of quantum channels, which is also intimately related to uncertainty.


\subsubsection{Quantum Shannon theory}\label{sec:coding}

The original partial results and conjectures for entropic uncertainty relations with quantum memory by \textcite{christandl05,renes08,renes09} were inspired by applications in quantum Shannon theory. 
More recently, entropic uncertainty relations and in particular their equality conditions have been used to analyze the performance of quantum Polar codes~\cite{renes14,renes15}.



\section{Miscellaneous Topics}\label{sctMisc}

\subsection{Tsallis and other entropy functions}\label{sec:other-entropies}

From a mathematical perspective it is insightful to consider uncertainty relations for various generalizations of the Shannon entropy. While the R\'enyi entropies were discussed above, the Tsallis entropies are another family of interest. The Tsallis entropy of order $\alpha$ is defined as
\begin{align}\label{eqntsallis}
H_{\alpha}^T(X) :=& \left( \frac{\log e}{1-\alpha}\right)  \left(\sum_x P_X(x)^{\alpha} -1 \right) \notag\\
& \textrm{for} \quad \alpha \in (0, 1) \cup (1,\infty) \,,
\end{align}
and as the corresponding limit for $\alpha \in \{0, 1, \infty\}$. Similar to the R\'enyi entropies, the $\alpha = 1$ Tsallis entropy corresponds to the Shannon entropy. Note that for $x\approx 1$ we have $\log x \approx\log e\cdot (x -1)$, so when $\sum_x P_X(x)^{\alpha} \approx 1$ the Tsallis entropy approximates the R\'enyi entropy.

Rastegin has studied uncertainty relations in terms of the Tsallis entropy. For example, \textcite{rastegin12b} proved the following uncertainty relation for Tsallis entropies, for a set of three MUBs $\{\bX,\bY,\bZ\}$ on a qubit. For $\alpha \in (0,1]$ and for integers $\alpha \geq 2$, we have
\begin{align}\label{eqnrastegin1}
&H_{\alpha}^T (X) +H_{\alpha}^T (Y) +H_{\alpha}^T (Z) \geq 2\log e \cdot f_{\alpha}(2)\,,\\
& \textrm{where } \quad f_{\alpha}(x) := \left(\frac{1 - x^{1-\alpha}}{\alpha - 1} \right)\,.
\end{align}
This generalizes the result in~\eqref{eq:mub_qubitpauli}, which is recovered by taking the limit $\alpha \to 1$, noting that $\lim_{\alpha \to 1}f_{\alpha}(x) = \log x/\log e$.

A more general scenario was considered in~\cite{rastegin13}, where system $A$ has dimension $d$, and the measurements under consideration form a set of $n$ MUBs, $\{\bX_j\}$. For $\alpha \in (0,2]$, \textcite{rastegin13} shows that
\begin{align}\label{eqnrastegin2}
\frac{1}{n}\sum_{j=1}^n H_{\alpha}^T (X_j)\geq 2\log e \cdot f_{\alpha}\left(\frac{nd}{n+d-1} \right)\,.
\end{align}
This result is quite general in that it holds for any $n$ and $d$. Furthermore, in the case of $n = d+1$ and $\alpha \to 1$, one recovers the result presented in \eqref{eq:mub_full}. \textcite{rastegin13} also tightened~\eqref{eqnrastegin2} for mixed states:
\begin{align}\label{eqnrastegin3}
\frac{1}{n}\sum_{j=1}^n H_{\alpha}^T (X_j)  \geq 2\log e \cdot f_{\alpha}\left(\frac{nd}{n+d\tr(\rho^2)-1} \right)\,.
\end{align}
Other entropy families are also discussed in the literature. For example, \textcite{zozor14} consider a broad class of entropies defined as
\begin{align}\label{eqnzozor1}
H_{(\eta,\phi)}(X):= \eta\left( \sum_{x} \phi (P_X(x)) \right)\,.
\end{align}
Here, $\eta: \bbR\to\bbR$ and $\phi:[0;1]\to \bbR$ are generic continuous functions such that either $\phi$ is strictly concave and $\eta$ is strictly increasing, or $\phi$ is strictly convex and $\eta$ is strictly decreasing. Additionally, they imposed $\phi(0)=0$ and $\eta(\phi(1))=0$. This family includes as special cases both the R\'enyi and Tsallis families and hence also the Shannon entropy. In addition to giving an overview of the literature on entropic uncertainty relations, \textcite{zozor14} derived a new uncertainty relation for the $H_{(\eta,\phi)}$ entropies. For any two POVMs $\bX$ and $\bZ$, and for any two pairs of functionals $(\eta_1,\phi_1)$ and $(\eta_2,\phi_2)$, their relation takes the form,
\begin{align}\label{eqnzozor2}
H_{(\eta_1,\phi_1)}(X)+H_{(\eta_2,\phi_2)}(Z) \geq B_{(\eta_1,\phi_1), (\eta_2,\phi_2)}(t)\,,
\end{align}
where the right-hand side is a function of the triplet
\begin{align}\label{eqnzozor3}
t := \{c_{\bX}, c_{\bZ}, c \}, \quad c_{\bX}:= \max_x \| \bX^x \|, \quad c_{\bZ}:= \max_z \| \bZ^z \|\,,
\end{align}
and $c$ is defined in \eqref{eqncjkPOVM}. The reader is referred to \textcite{zozor14} for the explicit form of $B_{(\eta_1,\phi_1), (\eta_2,\phi_2)}(t)$. In general, this bound can be computed, since it only involves a one-parameter optimization over a bounded interval. Note that the functionals associated with the two terms in \eqref{eqnzozor2} may be different. This gives a very general result allowing the authors to consider, e.g., R\'enyi entropy uncertainty relations that go beyond the usual conjugacy curve, defined by $(1/\alpha)+(1/\beta) = 2$.


\subsection{Certainty relations}\label{sec:certainty}

Instead of lower bounding sums of entropies for different observables, one can also ask whether there exist non-trivial \textit{upper} bounds on such sums. These bounds are called certainty relations. Of course, one would not expect to find non-trivial upper bounds for, say, the maximally mixed state $\rho_A = \1 /d$. However, one might, e.g., restrict to pure states $\ket{\psi}_A$.

For some sets of observables, even restricting to pure states is not enough to get a certainty relation. For example, consider the Pauli $\sg_X$ and $\sg_Z$ observables for one qubit. One cannot find a certainty relation for these two observables because there exist states, namely the eigenstates of $\sg_Y$, that are unbiased with respect to the eigenbases of $\sg_X$ and $\sg_Z$, and hence lead to maximum uncertainty in these two bases: $H(X)+H(Z)=2$.

Recently~\textcite{korzekwa14b} proved a general result that non-trivial certainty relations are not possible for two arbitrary orthonormal bases $\bX$ and $\bZ$, in any finite dimension $d$. This follows from the fact that one can always find a pure state $\ket{\psi}_A$ that is unbiased with respect to both $\bX$ and $\bZ$.

However, there do exist non-trivial certainty relation, e.g., for a $d+1$ set of MUBs. This is connected to the fact that there are no states that are unbiased to all bases in a $d+1$ set of MUBs. Consider a result of~\textcite{sanchez93}, which deals with three MUBs ($\bX, \bY, \bZ$) on a qubit system in a pure state:
\begin{align}\label{eqnsanchezcertainty1}
H(X)+H(Y)+H(Z)\leq \frac{3}{2}\log 6 -\frac{\sqrt{3}}{2}\log (2+\sqrt{3})\,. 
\end{align}
The right-hand side of \eqref{eqnsanchezcertainty1} is $\approx 2.23$. Comparing this to the lower bound of 2, from \eqref{eq:mub_qubitpauli}, one sees that the allowable range for $H(X)+H(Y)+H(Z)$ is quite small. \textcite{sanchez93} noted that \eqref{eqnsanchezcertainty1} is in fact the optimal certainty relation for these observables. More generally, considering a $d+1$ set of MUBs $\{\bX_j\}$, \textcite{sanchez93} showed that
\begin{align}\label{eqnsanchezcertainty2}
\sum_{j=1}^{n} H(X_j) &\leq  n \log (n+\sqrt{n})\notag\\
&\hspace{9pt}- \frac{1}{d} (n+(n-2)\sqrt{n})\log (2+\sqrt{n}) \,. 
\end{align}
where $n= d+1$. Note that \eqref{eqnsanchezcertainty1} is a special case of \eqref{eqnsanchezcertainty2} corresponding to $d=2$.

Rastegin obtained some generalizations of~\eqref{eqnsanchezcertainty1} to the R\'enyi and Tsallis entropy families. In the R\'enyi case~\textcite{rastegin14} found, for all $\alpha \in (0,1]$,
\begin{align}\label{eqnrastegin123}
H_{\alpha}(X)+H_{\alpha}(Y)+H_{\alpha}(Z)\leq 3 R_{\alpha}\,,
\end{align}
where 
\begin{align}\label{eqnrastegin124}
R_{\alpha} := \frac{1}{1-\alpha } \log \left(\left(\frac{1+1/\sqrt{3}}{2}\right)^{\alpha} +\left(\frac{1-1/\sqrt{3}}{2}\right)^{\alpha}  \right)\,.
\end{align}
Likewise~\textcite{rastegin12b} found a similar sort of bound for the Tsallis entropies, but with $\log (x)$ in~\eqref{eqnrastegin124} replaced by $x-1$.

While the above certainty relations are for MUBs, very recently~\textcite{puchala15} studied a more general situation with sets of $n>2$ orthonormal bases in dimension $d$. Their certainty relations are upper bounds on the sum of Shannon entropies, similar to \eqref{eqnsanchezcertainty2}, but are not restricted to MUBs. Certainty relations for unitary $k$-designs with $k=2,4$ in terms of the mutual information were also covered by~\textcite{matthews09}.

Finally, it is worth reminding the reader that for the collision entropy one can obtain an \textit{equality}, as in~\eqref{eq:h_2meta_noqsi}. An equation of this sort is both an uncertainty and a certainty relation. Stated another way, an equation implies that the strongest uncertainty relation coincides with the strongest certainty relation, leaving no gap between the two bounds. Equations such as~\eqref{eq:h_2meta_noqsi} can, in turn, be used to derive certainty relations for other entropies, such as the min-entropy, due to the fact that $H_{\min} \leq H_{2}$.

The generalization of~\eqref{eq:h_2meta_noqsi} to bipartite states $\rho_{AB}$ was given in \eqref{eq:h2_bcw}. Equation~\eqref{eq:h2_bcw} is a certainty relation in the presence of quantum memory. It relates the amount of uncertainty to the amount of entanglement, as quantified by the conditional entropy $H_{2}(A|B)$. Similar to the unipartite case, \eqref{eq:h2_bcw} can be used to derive certainty relations (in the presence of quantum memory) for other entropies, such as the min-entropy, as discussed by~\textcite{berta13}.

Studying bipartite certainty relations in the presence of quantum memory is largely an open problem. For example, one could ask whether~\eqref{eqnsanchezcertainty1} or~\eqref{eqnsanchezcertainty2} can be appropriately generalized to the quantum memory case.


\subsection{Measurement uncertainty}\label{sctmeasurementuncertainty}

This review has focused on preparation uncertainty relations. Two other aspects of the uncertainty principle are (1) the joint measurability of observable pairs and (2) the disturbance of one observable caused by the measurement of another observable. Joint measurability and measurement-disturbance are two aspects of measurement uncertainty, which deals with fundamental restrictions on one's ability to measure things. For a detailed discussion of measurement uncertainty, we refer the reader to \textcite{ozawa03,hall04,busch07,busch14b}. It is important, though, that we briefly mention measurement uncertainty here because the topic has seen significant debate recently (see, e.g., \textcite{busch13, busch14, busch14b}).
Rather than delve into the conceptual issues of measurement uncertainty, we will simply give a taste here of a few recent works that have taken an entropic approach, in particular, to measurement-disturbance.


\subsubsection{State-independent measurement-disturbance relations}

One approach to measurement uncertainty is to ask: how well can a measurement device perform on particular idealized sets of input states, e.g., the basis states associated with two complementary observables $\bX$ and $\bZ$? This is often called a state-independent approach, although it could also be called a calibration approach, since one is calibrating a device's performance based on idealized input states. For example, this approach was discussed by~\textcite{busch13} for the position and momentum observables. However, the quantities in their relation were not entropic so we will not discuss it further.

\begin{figure}[tbp]
\begin{center}
\begin{overpic}[width=6.1cm]{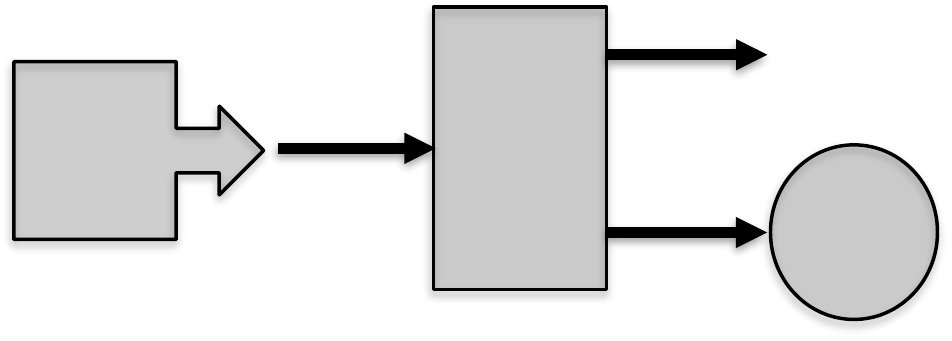}
\put(-14.7,28.7){\footnotesize (a)}
\put(2.8,18.7){\footnotesize $X=x$}
\put(51,18.7){\footnotesize $\cM$}
\put(34,21.6){\footnotesize $A$}
\put(32.3,15.2){\footnotesize $\ket{\bX^x}$}
\put(68.8,31.2){\footnotesize $A'$}
\put(68.5,12.5){\footnotesize $M$}
\put(83.5,12){\footnotesize Guess}
\put(85,7){\footnotesize for $x$}
\end{overpic}
\begin{overpic}[width=8.1cm]{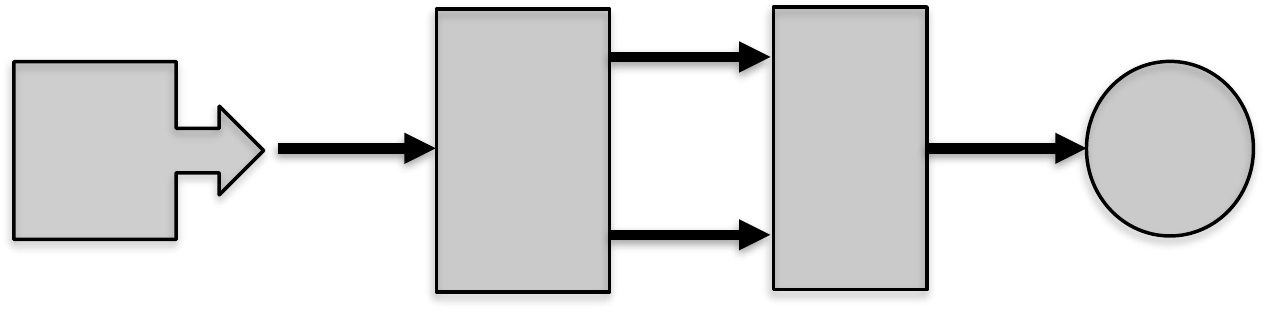}
\put(2,22.7){\footnotesize (b)}
\put(2.4,12){\footnotesize $Z=z$}
\put(24.3,9.4){\footnotesize $\ket{\bZ^z}$}
\put(26,14){\footnotesize $A$}
\put(51.8,21.6){\footnotesize $A'$}
\put(51.5,7.7){\footnotesize $M$}
\put(38.5,12){\footnotesize $\cM$}
\put(65,12){\footnotesize $\cR$}
\put(87.4,13.6){\footnotesize Guess}
\put(88.4,9.5){\footnotesize for $z$}
\put(77.6,14.2){\footnotesize $\hat{Z}$}
\end{overpic}
\vspace{0.1cm}
\caption{Two scenarios considered by \textcite{buscemi14}, which capture (a) the noise of an attempted $\bX$ measurement, and (b) the disturbance of the $\bZ$ observable.}
\label{figmeasdisturbance}
\end{center}
\end{figure}
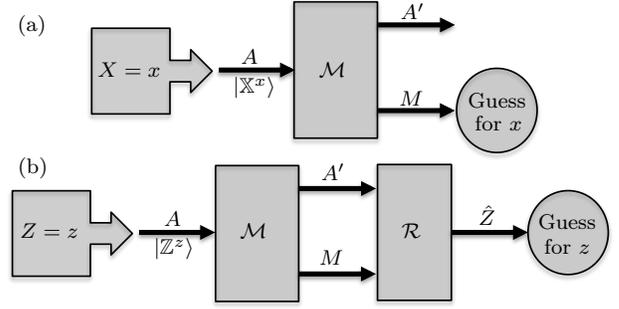

More recently the calibration approach was taken by~\textcite{buscemi14} using entropic quantities. Consider a measurement apparatus represented by a quantum channel $\cM$ acting on system $A$, and two counterfactual preparation schemes which will be fed into this apparatus, as shown in Fig.~\ref{figmeasdisturbance}. In one scheme, $A$ is prepared in a basis state of $\bX$, say $\ket{\bX^x}$, where the index $x$ is chosen with uniformly random probability. The output of $\cM$ consists of a classical system $M$ as well as a ``disturbed'' version of the original quantum system $A'$. The classical output $M$ represents an attempted measurement of the $\bX$ observable, and it provides a guess for the index $x$. The measurement noise is then quantified by $\sN(\cM , \bX) := H(X | M)$, where $X$ is the random variable associated with the $\bX$ observable on the input system, i.e., associated with the index $x$. In the other scheme, Fig.~\ref{figmeasdisturbance}(b), $A$ is prepared in a basis state of $\bZ$, say $\ket{\bZ^z}$, again with uniform probability. Now the question is: can one recover a good guess of $z$ from the outputs of $\cM$? If not, then the interpretation is that the attempted measurement of $\bX$ ``disturbs'' the $\bZ$ observable. To quantify this, \textcite{buscemi14} defined the disturbance of $\bZ$ by $\sD(\cM , \bZ) := \min_{\cR} H(Z| \hat{Z})$. Here, $Z$ is the random variable associated with the observable $\bZ$ on the input system, and $\cR$ is a recovery map, i.e., a quantum channel that maps $A'M$ to a classical system $\hat{Z}$ that provides a guess for $z$. Their measurement-disturbance relation states that
\begin{align}\label{eqMDRbuscemi}
\sN(\cM , \bX) + \sD(\cM , \bZ) \geq q_{\rm MU}\,,
\end{align}
with $q_{\rm MU}$ as in~\eqref{eq:shannonUR}. This shows a trade-off between the ability to measure the $\bX$ states versus the ability to leave the $\bZ$ states undisturbed.

Fig.~\ref{figmeasdisturbance} is a dynamic scenario, similar to the scenario in Sec.~\ref{sctQchannelmemory}. Hence, to derive \eqref{eqMDRbuscemi}, \textcite{buscemi14} started with a ``static'' uncertainty relation (namely the Maassen-Uffink relation) and then applied the static-dynamic isomorphism from Sec.~\ref{sctstaticdynamiciso}. In particular they employed the property in \eqref{eqnricochet}.


\subsubsection{State-dependent measurement-disturbance relations}

Now let us consider a sequential measurement scenario where system $A$ is prepared in an arbitrary state $\rho_A$ and fed into the measurement apparatus.

For simplicity, consider the sequential measurement of orthonormal bases, $\bX$ followed by $\bZ$, where the first measurement is a von Neumann measurement, i.e., it projects the system onto an $\bX$-basis state. One can apply Maassen-Uffink's uncertainty relation to each outcome of the $\bX$ measurement, i.e., to each state $\ket{\bX^x}$, giving 
\begin{align}\label{eqnsequential1}
H(Z)_{\ket{\bX^x}} = H(X)_{\ket{\bX^x}}+H(Z)_{\ket{\bX^x}}  \geq q_{\rm MU}\,.
\end{align}
Multiplying this by the probability $p^x = \expct{\bX^x}{\rho_A} $ for outcome $x$, and summing over $x$ gives
\begin{align}\label{eqnsequential2}
H(Z|X) \geq q_{\rm MU}\,,
\end{align}
where $H(Z|X)$ denotes the uncertainty for a future $\bZ$ measurement given the outcome of the previous $\bX$ measurement. Equation~\eqref{eqnsequential2} was discussed in detail by \textcite{baek14}, and was also briefly mentioned by \textcite{coles14c}. Note that~\eqref{eqnsequential2} holds for any fixed input state $\rho_A$, it is a state-dependent relation.

While \eqref{eqnsequential2} assumes the $\bX$ measurement is an ideal von Neumann measurement, it is interesting to ask what happens if the first measurement is non-ideal, i.e., a noisy measurement. There are various ways to address this. One approach, given by~\textcite{coles14b}, quantified the imperfection of the $\bX$ measurement by the predictive error,
\begin{align}\label{eq:predictive error}
\sE(\rho_A, \bX, \cE) := H_{\max}(X | M_X)\,.
\end{align}
That is, the max-entropy of a future (perfect) $\bX$ measurement given the register $M_X$ that stores the outcome of the previous (imperfect) measurement of $\bX$. Here, $\cE$, which maps $A \to AM_X$, is the channel that performs this imperfect $\bX$ measurement. One is interested in the disturbance of the $\bZ$ observables caused by the imperfect $\bX$ measurement. \textcite{coles14b} quantified the disturbance of $\bZ$ using the R\'enyi relative entropies for $\alpha\in[1/2,\infty]$,
\begin{align}\label{eq:DistMeas}
\sD_\alpha (\rho_A, \bZ, \cE) :=  D_\alpha( P_Z || P_Z^{\cE}) \,.
\end{align}
Here, $P_Z$ is the initial probability distribution for the $\bZ$ measurement and $P_Z^\cE$ is the final probability distribution for $\bZ$, i.e., after the imperfect $\bX$ measurement. With these definitions, they found the measurement-disturbance relation
\begin{align}\label{eq1DiscURclas}
\sD_\alpha(\rho_A, \bZ, \cE)+\sE(\rho_A, \bX, \cE) + H_{\alpha}(Z)_{P} \geq q_{\rm MU}\,. 
\end{align}
On the one hand this gives a trade-off between measuring $\bX$ well and causing large $\bZ$ disturbance. On the other hand, the trade-off gets weaker as more initial uncertainty is contained in $P_Z$, as quantified by the term $H_{\alpha}(Z)_{P}$. So there is an interplay between initial uncertainty, measurement error, and disturbance.



\section{Perspectives}\label{sec:perspectives}

We have discussed modern formulations of Heisenberg's uncertainty principle where uncertainty is quantified by entropy. Such formulations are directly relevant to quantum information processing tasks as discussed in Sec.~\ref{sec:app}. 

Technological applications such as QKD (Sec.~\ref{sec:qkd}) provide the driving force for obtaining more refined entropic uncertainty relations. For example, to prove security of QKD protocols involving more than two measurements, new entropic uncertainty relations are needed\,---\,namely ones that allow for quantum memory and for multiple measurements. This is an important frontier that requires more research. Device-independent randomness, i.e., certifying randomness obtained from untrusted devices (Sec.~\ref{sec:certifyingrandomness}), is another emerging application for which entropic uncertainty relations appear to be useful but more research is needed to find uncertainty relations that are specifically tailored to this application.

Aside from their technological applications, we believe that entropic uncertainty relations have a beauty to them. They give insight into the structure of quantum theory, and for that reason alone they are worth pursuing. For example, Sec.~\ref{sec:mutual_conjecture} noted a simple conjecture\,---\,that the sum of the mutual informations for two MUBs lower bounds the quantum mutual information.

New tools are being developed to prove entropic uncertainty relations. For example, the majorization approach (Sec.~\ref{sec:major}) is promising. The relation between the majorization approach and the relative entropy approach (see App.~\ref{app:shannon_proof}) remains to be clarified, and a unified framework would be insightful. For uncertainty relations with memory, \textcite{dupuis15} established a meta theorem to derive uncertainty relations. Yet, it is known that the resulting relations are not tight in all regimes, calling for further improvements.


One of the most exciting things about entropic uncertainty relations is that they give insight into basic physics. For example Sec~\ref{sec:waveparticle} discussed how entropic uncertainty relations allow one to unify the uncertainty principle with the wave-particle duality principle. A natural framework for quantifying wave-particle duality will likely come from applying entropic uncertainty relations to interferometers. Likewise, a hot topic in quantum foundations is measurement uncertainty. Sec.~\ref{sctmeasurementuncertainty} noted that entropic uncertainty relations may play an important role in obtaining conceptually clear formulations of measurement uncertainty. In that respect, very recently the notion of preparation uncertainty was combined with measurement reversibility~\cite{berta16} and the corresponding entropic uncertainty relations were successfully tested on the IBM Quantum Experience~\cite{IBM_Quantum}.

Furthermore, entropic uncertainty relations will continue to help researchers characterize the boundary between separable vs.\ entangled states (Sec.~\ref{sec:entwit}), as well as steerable vs.\ non-steerable states (Sec.~\ref{sec:steering}).

Entropic uncertainty relations may play a role in the study of phase transitions in condensed matter physics~\cite{romera15}. Entropic uncertainty relations are also studied in the context of special and general relativity~\cite{jia15,feng13}. Given that quantum information is playing an increasing role in cosmology~\cite{hayden07}, it would not be surprising to see future work on entropic uncertainty relations in the context of black hole physics.


\begin{acknowledgments}

We thank Kais Abdelkhalek, Iwo Bia\l{}ynicki-Birula, Matthias Christandl, Rupert L.~Frank, Gilad Gour, Michael J.~W.~Hall, Hans Maassen, Joseph M.~Renes, Renato Renner, Lukasz Rudnicki, Christian Schaffner, Reinhard F.~Werner, Mark M.~Wilde, and Karol Zyczkowski for feedback. PJC acknowledges support from Industry Canada, Sandia National Laboratories, NSERC Discovery Grant, and Ontario Research Fund (ORF). MB acknowledges funding provided by the Institute for Quantum Information and Matter, an NSF Physics Frontiers Center (NFS Grant PHY-1125565) with support of the Gordon and Betty Moore Foundation (GBMF-12500028). Additional funding support was provided by the ARO grant for Research on Quantum Algorithms at the IQIM (W911NF-12-1-0521). MT is funded by an University of Sydney Postdoctoral Fellowship and acknowledges support from the ARC Centre of Excellence for Engineered Quantum Systems (EQUS). SW is supported by STW, Netherlands and an NWO VIDI grant.
\end{acknowledgments}


\appendix

\section{Mutually unbiased bases}\label{app:mub}

Sec.~\ref{sctMUBdefinition} defined MUBs, and sets of $n$ MUBs. The study of MUBs is closely related to the study of entropic uncertainty. Strong entropic uncertainty relations have been derived generically for sets of MUBs (particularly for $d+1$ sets of MUBs). Hence, constructing a new set of MUBs immediately yields a new entropic uncertainty relation. On the other hand, there is the interesting open question whether a set of $n$ MUBs $\{\bX_j\}$ yields the strongest bound $b$ in a generic uncertainty relation of the form
\begin{align}
\sum_{j=1}^{n} H(X_j) \geq b\,.
\end{align}
A review of MUBs can be found in \cite{durt10}. Here, we discuss the connection of MUBs to Hadamard matrices, as well as the existence and construction of MUBs.


\subsection{Connection to Hadamard matrices}

Any two orthonormal bases are related by a unitary, and in the case of MUBs, that unitary is called a Hadamard matrix $H$. The general form of such matrices is
\begin{align}\label{eqnHaddef}
H = \sum_{j,k} \frac{e^{i \phi_{jk}}}{\sqrt{d}} \dyad{j}{k}\,,
\end{align}
where the phase factors $\phi_{jk}$ must be appropriately chosen so that $H$ is unitary. Notice that each matrix element has a magnitude of $1/\sqrt{d}$, which is the defining property of Hadamard unitaries. The best known Hadamard is the Fourier matrix, defined in \eqref{eqnfourier1},
\begin{align}\label{eqnFordef}
F = \sum_{j,k} \frac{\om^{-jk}}{\sqrt{d}} \dyad{j}{k} \quad \text{with}\quad\om = e^{2\pi i/d}\,,
\end{align}
which relates the generalized Pauli operators
\begin{align}\label{eqnPaulis}
\sg_{\bZ} = \sum_j \om^j \dya{j}, \quad \sg_{\bX} = F\sg_{\bZ} F\ad = \sum_j \dyad{j+1}{j}\,.
\end{align}
For $d=2$ these are just the usual Pauli matrices from Ex.~\ref{exPauli}.

It should be clear that the problem of finding MUBs is equivalent to the problem of finding Hadamard matrices. We note that Hadamard matrices can be categorized into equivalence classes, based on whether there exists a diagonal unitary or permutation that that maps one Hadamard to another. A detailed catalog of Hadamard matrices can be found online~\cite{hadamard_matrix_weblink}.


\subsection{Existence}

That there exist MUB pairs in any finite dimension follows, e.g., from the fact that we can write down the Fourier matrix in \eqref{eqnFordef} for any $d$. In fact, for any $d$ there exists a set of 3 MUBs, e.g., formed from the eigenvectors of $\sg_{\bX}$, $\sg_{\bZ}$, and $\sg_{\bX}\sg_{\bZ}$. It is also known that a set of MUBs can at most be of size $d+1$ \cite{bandyopadhyay02}. Such $d+1$ sets are called complete sets of MUBs. Complete sets play a role in tomography since they are informationally complete, and they have the useful property of forming a complex projective two-design~\cite{klappenecker05}. Complete sets of MUBs are known to exist in prime power dimensions, i.e., $d = p^m$ where $p$ is a prime and $m$ is a positive integer~\cite{bandyopadhyay02}. However, even for the smallest number that is not a prime power, namely 6, the existence problem remains unsolved.


\subsection{Simple constructions}

When $d$ is a prime, a simple construction~\cite{wootters89,bandyopadhyay02} of a complete set of MUBs is to consider the eigenvectors of the $d+1$ products of the form
\begin{align}\label{eqnMUBprime}
\{\sg_{\bZ}, \sg_{\bX}, \sg_{\bX}\sg_{\bZ}, \sg_{\bX}\sg_{\bZ}^2, ..., \sg_{\bX}\sg_{\bZ}^{d-1}\}.
\end{align}
More generally for $d=p^m$, a construction is known where each basis $B_i$ comes from the common eigenvectors of a corresponding set $C_i$ of commuting matrices~\cite{bandyopadhyay02}. The elements of $C_i$ are a subset of size $|C_i | = d-1$ of the $d^2-1$ Pauli products $\sg_{\bX}^j\sg_{\bZ}^k$ (excluding the identity). The subset is chosen such that all the elements of $C_i$ commute and $C_i \bigcap C_j = \{\1 \}$ for $i \neq j$.


\section{Proof of Maassen-Uffink's relation}\label{app:shannon_proof}

Here, we give a proof of Maassen-Uffink's uncertainty relation for the Shannon entropy~\eqref{eq:shannonUR}. Our proof closely follows the ideas in~\cite{colbeck11} and makes use of the data-processing inequality for the relative entropy~\cite{lieb73,uhlmann77,lindblad75}. In fact, we will prove the slightly stronger relation stated in~\eqref{eq:shannonURmixed}:
\begin{align}\label{eq:mu_stronger}
H(X)+H(Z)\geq\log\frac{1}{c}+H(\rho_A)\,.
\end{align}

\begin{proof}
For the proof of~\eqref{eq:mu_stronger} we consider the classical state $\rho_{X}=\cX_{A\to X}(\rho_{A})$ generated by applying the measurement map
\begin{align}\label{eq:mu_2}
\cX_{A\to X}(\cdot)=\sum_{x}\dya{\bX^{x}}\cdot \dya{\bX^{x}}\,,
\end{align}
where the auxiliary Hilbert space $X$ allows us to represent the classical random variable $X$ in the quantum formalism.

It is easy to verify that the Shannon entropy of the distribution $P_{X}$ is equal to the von Neumann entropy of the state $\rho_{X}$. From this we get
\begin{align}\label{eq:mu_1}
H(X) =-\tr\left[\rho_{X}\log\rho_{X}\right]&=-\tr\left[\cX(\rho_{A})\log\cX(\rho_{A})\right]\\
&=-\tr\left[\rho_{A}\log\cX(\rho_{A})\right]\,,
\end{align}
where the last equality is straightforward to check by writing out the trace and the measurement map $\cX_{A\to X}$. By phrasing the right-hand side of~\eqref{eq:mu_1} in terms relative entropy,
$D(\rho\|\sigma) = \tr[\rho (\log \rho - \log \sigma)]$,
we arrive at
\begin{align}
H(X)=D(\rho_{A}\|\cX(\rho_{A}))+H(\rho_A ) \,.
\end{align}
We then apply the measurement map
\begin{align}\label{eq:mu_3}
\cZ_{A\to Z}(\cdot)=\sum_{z}\dya{\bZ^{z}}\cdot \dya{\bZ^{z}}
\end{align}
to both arguments of the relative entropy, and find by the data-processing inequality for the relative entropy that
\begin{align}\label{eq:mu_4}
D(\rho_{A}\|\cX(\rho_{A})) &\geq D(\cZ(\rho_{A})\|\cZ\circ\cX(\rho_{A}))\\
&=D(\rho_{Z}\|\cZ\circ\cX(\rho_{A}))\,,
\end{align}
where $\rho_{Z}=\cZ_{A\to Z}(\rho_{A})$. By writing out both measurement maps we find the classical state
\begin{align}
\cZ\circ\cX(\rho_{A})=\sum_{z}\dya{\bZ^{z}} \cdot\sum_x |\ip{\bX^{x}}{\bZ^{z}}|^2\expct{\bX^{x}}{\rho_{A}}\,,
\end{align}
and the right-hand side of~\eqref{eq:mu_4} becomes
\begin{align}
&D(\rho_{Z} \|\cZ\circ\cX(\rho_{A}))= -H(\rho_Z ) \notag \\ 
&\quad - \sum_z \expct{\bZ^{z}}{\rho_{A}}\log \bigg(\sum_x |\ip{\bX^{x}}{\bZ^{z}}|^2\expct{\bX^{x}}{\rho_{A}}\bigg).
\end{align}
Now, the logarithm is a monotonic function and hence we find
\begin{align}
&-\sum_z \expct{\bZ^{z}}{\rho_{A}}\log\bigg(\sum_x |\ip{\bX^{x}}{\bZ^{z}}|^2\expct{\bX^{x}}{\rho_{A}}\bigg)\notag\\
\geq &-\sum_z \expct{\bZ^{z}}{\rho_{A}}\log\bigg(\max_{x',z'} \big|\ip{\bX^{x'}}{\bZ^{z'}} \big|^2 \sum_x\expct{\bX^{x}}{\rho_{A}}\bigg)\\
= &-\log\max_{x',z'} \big|\ip{\bX^{x'}}{\bZ^{z'}}\big|^2\,.\label{eq:mu_5}
\end{align}
By combining~\eqref{eq:mu_1}--\eqref{eq:mu_5} and noting that $H(Z)$ equals the von Neumann entropy of $\rho_Z$, we arrive at the claim~\eqref{eq:mu_stronger}.
\end{proof}


\section{R\'enyi entropies for joint quantum systems}\label{app:renyi}

Here, we define general conditional R\'enyi entropies. This allows us to exhibit their intuitive properties in a general setting without having to discuss various special cases individually. We will exhibit these properties to show a generalization of the Maassen-Uffink relation to the tripartite quantum memory setting.


\subsection{Definitions}

For any bipartite quantum state $\rho_{AB}$ and $\alpha \in [\frac12, \infty]$, we define
the \emph{quantum conditional R\'enyi entropy} as 
\begin{align}\label{eq:alpha-conditional}
H_{\alpha}(A|B):=& - \min_{\sigma_B} D_{\alpha}(\rho_{AB} \| \1_A \otimes \sigma_B)\,,\notag\\
&\textrm{where $\sigma_B$ is a quantum state on $B$}\,.
\end{align}
Here, $D_{\alpha}$ is the R\'enyi divergence of order $\alpha$~\cite{lennert13,wilde13}, namely\footnote{This quantum generalization is not unique\,---\,in fact other generalizations based on Petz's notion of R\'enyi divergence~\cite{ohya93} have also been explored, for example by~\textcite{tomamichel13}. However, for the purpose of the present review it is convenient to stick with the proposed definition in~\eqref{eq:alpha-conditional} and~\eqref{eq:alpha-divergence} as it entails the most important special cases encountered here and in the literature.}
\begin{align}\label{eq:alpha-divergence}
D_{\alpha}(\rho\|\sigma) :=& \frac{1}{\alpha-1} \log \tr \bigg[\Big( \sigma^{\frac{1-\alpha}{2\alpha}} \rho  \sigma^{\frac{1-\alpha}{2\alpha}} \Big)^{\alpha} \bigg] \notag\\
& \textrm{for}\ \alpha \in \bigg[\frac12, 1\bigg) \cup (1,\infty)
\end{align}
and as the corresponding limit for $\alpha \in \{1,\infty\}$. These divergences are measures of distinguishability between quantum states and some of their properties will be discussed in App.~\ref{pre:prop}. Note the following special cases that we have encountered previously. First, the conditional min- and max-entropy are simply recovered as $H_{\min} \equiv H_{\infty}$ and $H_{\max} \equiv H_{\nicefrac12}$. The conditional von Neumann entropy is recovered as $H \equiv H_1$. Finally, the conditional collision entropy can be expressed as
\begin{align}
H_{\rm coll}(A|B) = - D_2(\rho_{AB} \| \1_A \otimes \rho_B) \,.
\end{align}
Note that $H_{2}(A|B) \leq H_{\rm coll}(A|B)$ since the former involves a minimization over marginal states $\sigma_B$. The two expressions are not equal in general and we want to mostly work with $H_{\rm coll}(A|B)$ because it has the operational interpretation as in~\eqref{eq:h2_cq} and~\eqref{eq:h2_qq}.


\subsection{Entropic properties}\label{pre:prop}

We present the properties for the whole family of R\'enyi divergences and entropies, but recall that the properties also apply to the relative entropy and the von Neumann entropy as special cases. Most properties of the conditional R\'enyi entropy can be derived from properties of the underlying R\'enyi divergence.\footnote{These divergences have been investigated in a series of recent works~\cite{lennert13,wilde13,beigi13,frank13,mosonyiogawa13} and proofs of the properties discussed here can be found in these references.}


\subsubsection{Positivity and monotonicity}

First, we remark that $D_{\alpha}(\rho\|\sigma)$ is guaranteed to be non-negative when the arguments $\rho$ and $\sg$ are normalized, and $D_{\alpha}(\rho\|\sigma)=0$ when $\rho = \sg$. Also, $\alpha \mapsto D_{\alpha}(\rho\|\sigma)$ is monotonically increasing in $\alpha$. Thus, for any $\beta \geq \alpha$, we have
\begin{align}
0 &\leq D_{\alpha}(\rho\|\sigma) \leq D_{\beta}(\rho\|\sigma)\,, \quad \textrm{and} \\
\log d_A &\geq H_{\alpha}(A|B) \geq H_{\beta}(A|B) \geq -\log \min \{ d_A, d_B\} \, .
\end{align}
This means that the conditional R\'enyi entropies, in particular also the conditional von Neumann entropy, can be negative. However, this can only happen in the presence of quantum entanglement and the conditional entropies are thus always positive when one of the two systems is classical. The maximum $\log d_A$ is achieved for a state of the form $\rho_{AB} = \frac{\1_A}{d_A} \otimes \rho_B$. On the other hand, the minimum $-\log d_A$ is achieved for the maximally entangled pure state $\ket{\psi}_{AB} = \frac1{\sqrt{d_A}} \sum_x \ket{x}_A \otimes \ket{x}_B$.


\subsubsection{Data-processing inequalities}

Any quantum channel is described by a completely-positive and trace-preserving (CPTP) map. The R\'enyi divergences satisfy a data-processing inequality. Namely, for all $\alpha \geq \frac12$ and any CPTP map $\mathcal{E}$, we find the following relation~\cite{frank13}:
\begin{align}\label{eq:data-proc}
D_{\alpha}\big( \cE(\rho) \big\| \cE(\sigma) \big) \leq D_{\alpha}(\rho\|\sigma)\,.
\end{align}
This is an expression of the intuitive property that it is easier to distinguish between the inputs rather than the outputs of any quantum channel. In fact, this property holds more generally for any completely positive trace non-increasing map~$\cE$ which satisfies $\tr[\cE(\rho)] = 1$. This has two important implications for conditional entropies. First, consider an arbitrary CPTP map $\mathcal{E}_{B\to B'}$ acting on the side information that takes $\rho_{AB}$ to $\tau_{AB'} = \id_A \otimes \cE_{B \to B'}(\rho_{AB})$. Then we have $H_{\alpha}(A|B) \leq H_{\alpha}(A|B')$. This tells us that any physically allowed information processing of the side information $B$ may only increase the uncertainty we have about $A$. 

\begin{example}
An often encountered special case of the data-processing inequality is that $H_{\alpha}(A|BC)\leq H_{\alpha}(A|B)$ for any tripartite state $\rho_{ABC}$. This expresses the fact that throwing away part of the side information can only increase the uncertainty about $A$.
\end{example}

The second application concerns rank-$1$ projective measurements on the $A$ system. More precisely, we consider any rank-$1$ projective measurement $\cX_{A \to X}$ that takes $\rho_{AB}$ to
\begin{align}
\rho_{XB} &= \cX_{A \to X} \otimes \id_B(\rho_{AB})\\
&= \sum_x  \Big(\dya{\bX^x}_A \otimes \1_B \Big) \rho_{AB} \Big(\dya{\bX^x}_A \otimes \1_B \Big)\,.
\end{align}
Then, we find that $H_{\alpha}(A|B)\leq H_{\alpha}(X|B)$, which reveals that measuring out system $A$ completely can only increase the uncertainty we have about it.\footnote{The above inequality holds more generally for all CPTP maps on $\cE_{A \to A'}$ that satisfy $\cE_{A \to A'}(\1_A) = \1_{A'}$ (unital maps).}


\subsubsection{Duality and additivity}

We will see that the following property is essential for deriving uncertainty relations with quantum side information. For any tripartite state $\rho_{ABC}$, the conditional R\'enyi entropies satisfy the following \emph{duality relation}. For $\alpha, \beta \in [\frac12, \infty]$ such that $\frac{1}{\alpha} + \frac{1}{\beta} = 2$, we have~\cite{lennert13,beigi13}
\begin{align}\label{eq:dual}
H_{\alpha}(A|B) + H_{\beta}(A|C) \geq 0\,,
\end{align}
with equality if $\rho_{ABC}$ is pure. 

This is a quantitative manifestation of the monogamy of quantum correlations. For example, if system $A$ is highly entangled with system $B$ we find that the conditional von Neumann entropy $H(A|B)$ is negative. However, the duality relation~\eqref{eq:dual} now shows that for any third system $C$ correlated with $A$ and $B$, it holds that $H(A|C) \geq -H(A|B)$, that is, the uncertainty of $A$ from an observer with access to $C$ is necessarily large in this case.

The R\'enyi entropies are additive. Namely, given a product state of the form $\rho_{ABCD} = \rho_{AC} \otimes \rho_{BD}$, they satisfy $H_{\alpha}(AB|CD)= H_{\alpha}(A|C)+H_{\alpha}(B|D)$. This is in fact a consequence of the above duality relation.\footnote{Recall that by definition~\eqref{eq:alpha-conditional}, we have 
\begin{align}
H_{\alpha}(AB|CD)&= - \min_{\sigma_{CD}} D_{\alpha}(\rho_{ABCD}\|\1_{AB} \otimes \sigma_{CD}) \notag\\
&\geq - \min_{\sigma_{C}, \sigma_{D}} D_{\alpha}(\rho_{ABCD}\|\1_{AB} \otimes \sigma_{C} \otimes \sigma_D) \notag\\
&= H_{\alpha}(A|C)+ H_{\alpha}(B|D) \notag\,.
\end{align}
The reverse inequality then follows due to the duality relation.
}


\subsection{Axiomatic proof of uncertainty relation with quantum memory}\label{sec:memory_proof}

Here, we give a concise proof of the generalized Maassen-Uffink relation~\eqref{eqnTripartiteMemory435},
\begin{align}
\label{eqnTripartiteMemory435new}
H_{\alpha}(X|B)_{\rho} + H_{\beta}( Z | C) \geq q_{\rm MU}  \,,
\end{align}
where $\frac{1}{\alpha} + \frac{1}{\beta} = 2$. Let us note that the proof applies to a general class of entropic quantities that satisfy certain properties, but we will specialize it here to conditional R\'enyi entropies.

Let us consider measurements $\bX = \{\bX_A^x\}$ and $\bZ=\{\bZ_A^z\}$ in two orthonormal bases such that $\bX_A^x$ and $\bZ_A^z$ are rank-one projectors. The proof for POVMs follows essentially the same steps, as detailed in~\textcite{colbeck11} (based on ideas of~\textcite{coles10} and~\textcite{tomamichel11}).

\begin{proof}[Proof of~\eqref{eqnTripartiteMemory435new}]
First, let us define the isometry $V := \sum_z \ket{z}_Z \ot \bZ_A^z$ associated with the $\bZ$ measurement on system $A$, and the state $\tilde{\rho}_{ZABC}:= V \rho_{ABC}V\ad$. We find the following sequence of inequalities (which will be explained in detail below),
\begin{align}
&H_{\beta }(Z|C)\notag\\
&\geq -H_{\alpha}(Z|AB)\label{eq:ap-step1} \\
&=\min_{\sigma_{AB}} D_{\alpha}\big(\tilde{\rho}_{ZAB} \big\| \1_Z \ot \sigma_{AB}\big) \label{eq:ap-step2} \\
&\geq \min_{\sigma_{AB}} D_{\alpha}\bigg(\rho_{AB} \bigg\| \sum_z \bZ_A^z \sigma_{AB} \bZ_A^z \bigg) \label{eq:ap-step3}\\
&\geq \min_{\sigma_{AB}} D_{\alpha}\bigg(\overline{\rho}_{XB} \bigg\| \sum_{x,z} \big|\ip{\bX_A^x}{\bZ_A^z}\big|^2 \bX_A^x \ot \tr_A \big[ \bZ_A^z\sigma_{AB} \big] \bigg) \label{eq:ap-step4} \,.
\end{align}
where we have used $\overline{\rho}_{XB}:= \sum_k \bX_A^x \rho_{AB} \bX_A^x$. To establish~\eqref{eq:ap-step1}, we apply the duality relation~\eqref{eq:dual} to the state $\tilde{\rho}_{ZABC}$. Equation~\eqref{eq:ap-step2} is simply the definition of the conditional entropy as in~\eqref{eq:alpha-conditional}. To find~\eqref{eq:ap-step3}, we apply the data-processing inequality for the partial isometry $V\ad$ as a trace non-increasing map, and note that $V\ad (\1_Z \otimes \sigma_{AB}) V = \sum_z \bZ_A^z \sigma_{AB} \bZ_A^z$. Next, \eqref{eq:ap-step4} follows by applying the data-processing inequality for the measurement cptp map $\mathcal{X}(\cdot ) = \sum_x \bX^x \cdot \bX^x$.

Next we observe that
\begin{align}
&\sum_{x,z} \big|\ip{\bX_A^x}{\bZ_A^z}\big|^2 \bX_A^x \ot \tr_A \big[ \bZ_A^z\sigma_{AB} \big]\notag\\
&\leq c \sum_{x,z} \bX_A^x \ot \tr_A \big[ \bZ_A^z\sigma_{AB} \big] = c\, \1_A \otimes \sigma_B ,
\end{align}
where we recall that $c = \max_{x,z} \big|\ip{\bX_A^x}{\bZ_A^z}\big|^2$ as defined in~\eqref{eq:overlap}. Moreover, we need that for any $\sigma'$ and positive $\lambda$ such that $\sigma \leq \lambda \sigma'$, we have
$D_{\alpha}(\rho\|\sigma) \geq D_{\alpha}(\rho\|\sigma') + \log\frac{1}{\lambda}$.\footnote{For a proof of this property, see~\cite[Prop.~4]{lennert13}.}
Continuing from~\eqref{eq:ap-step4}, we thus find that
\begin{align}
H_{\beta }(Z|C)
&\geq \min_{\sigma_B} D_{\alpha}\big(\overline{\rho}_{XB}  \big\| \1_X \ot \sigma_{B}\big) + q_{\rm MU} \label{eq:ap-step5}\\
&= - H_{\alpha}(X|B)+ q_{\rm MU} \label{eq:ap-step6}\,,
\end{align}
where~\eqref{eq:ap-step6} again follows by the definition of the conditional entropy.
\end{proof}



%


\begin{thebibliography}{268}%
\makeatletter
\providecommand \@ifxundefined [1]{%
 \@ifx{#1\undefined}
}%
\providecommand \@ifnum [1]{%
 \ifnum #1\expandafter \@firstoftwo
 \else \expandafter \@secondoftwo
 \fi
}%
\providecommand \@ifx [1]{%
 \ifx #1\expandafter \@firstoftwo
 \else \expandafter \@secondoftwo
 \fi
}%
\providecommand \natexlab [1]{#1}%
\providecommand \enquote  [1]{``#1''}%
\providecommand \bibnamefont  [1]{#1}%
\providecommand \bibfnamefont [1]{#1}%
\providecommand \citenamefont [1]{#1}%
\providecommand \href@noop [0]{\@secondoftwo}%
\providecommand \href [0]{\begingroup \@sanitize@url \@href}%
\providecommand \@href[1]{\@@startlink{#1}\@@href}%
\providecommand \@@href[1]{\endgroup#1\@@endlink}%
\providecommand \@sanitize@url [0]{\catcode `\\12\catcode `\$12\catcode
  `\&12\catcode `\#12\catcode `\^12\catcode `\_12\catcode `\%12\relax}%
\providecommand \@@startlink[1]{}%
\providecommand \@@endlink[0]{}%
\providecommand \url  [0]{\begingroup\@sanitize@url \@url }%
\providecommand \@url [1]{\endgroup\@href {#1}{\urlprefix }}%
\providecommand \urlprefix  [0]{URL }%
\providecommand \Eprint [0]{\href }%
\providecommand \doibase [0]{http://dx.doi.org/}%
\providecommand \selectlanguage [0]{\@gobble}%
\providecommand \bibinfo  [0]{\@secondoftwo}%
\providecommand \bibfield  [0]{\@secondoftwo}%
\providecommand \translation [1]{[#1]}%
\providecommand \BibitemOpen [0]{}%
\providecommand \bibitemStop [0]{}%
\providecommand \bibitemNoStop [0]{.\EOS\space}%
\providecommand \EOS [0]{\spacefactor3000\relax}%
\providecommand \BibitemShut  [1]{\csname bibitem#1\endcsname}%
\let\auto@bib@innerbib\@empty
\bibitem [{\citenamefont {Abdelkhalek}\ \emph {et~al.}(2015)\citenamefont
  {Abdelkhalek}, \citenamefont {Schwonnek}, \citenamefont {Maassen},
  \citenamefont {Furrer}, \citenamefont {Duhme}, \citenamefont {Raynal},
  \citenamefont {Englert},\ and\ \citenamefont {Werner}}]{abdelkhalek15}%
  \BibitemOpen
  \bibfield  {author} {\bibinfo {author} {\bibnamefont {Abdelkhalek},
  \bibfnamefont {K.}}, \bibinfo {author} {\bibfnamefont {R.}~\bibnamefont
  {Schwonnek}}, \bibinfo {author} {\bibfnamefont {H.}~\bibnamefont {Maassen}},
  \bibinfo {author} {\bibfnamefont {F.}~\bibnamefont {Furrer}}, \bibinfo
  {author} {\bibfnamefont {J.}~\bibnamefont {Duhme}}, \bibinfo {author}
  {\bibfnamefont {P.}~\bibnamefont {Raynal}}, \bibinfo {author} {\bibfnamefont
  {B.-G.}\ \bibnamefont {Englert}}, \ and\ \bibinfo {author} {\bibfnamefont
  {R.~F.}\ \bibnamefont {Werner}}} (\bibinfo {year} {2015}),\ \href {\doibase
  10.1142/S0219749915500458} {\bibfield  {journal} {\bibinfo  {journal}
  {International Journal of Quantum Information}\ }\textbf {\bibinfo {volume}
  {13}}~(\bibinfo {number} {06}),\ \bibinfo {pages} {1550045}}\BibitemShut
  {NoStop}%
\bibitem [{\citenamefont {Adamczak}\ \emph {et~al.}(2016)\citenamefont
  {Adamczak}, \citenamefont {Lata{\l}a}, \citenamefont {Pucha{\l}a},\ and\
  \citenamefont {{\.{Z}}yczkowski}}]{adamczak14}%
  \BibitemOpen
  \bibfield  {author} {\bibinfo {author} {\bibnamefont {Adamczak},
  \bibfnamefont {R.}}, \bibinfo {author} {\bibfnamefont {R.}~\bibnamefont
  {Lata{\l}a}}, \bibinfo {author} {\bibfnamefont {Z.}~\bibnamefont
  {Pucha{\l}a}}, \ and\ \bibinfo {author} {\bibfnamefont {K.}~\bibnamefont
  {{\.{Z}}yczkowski}}} (\bibinfo {year} {2016}),\ \href {\doibase
  10.1063/1.4944425} {\bibfield  {journal} {\bibinfo  {journal} {Journal of
  Mathematical Physics}\ }\textbf {\bibinfo {volume} {57}}~(\bibinfo {number}
  {3}),\ \bibinfo {pages} {032204}}\BibitemShut {NoStop}%
\bibitem [{\citenamefont {Ambainis}(2010)}]{ambainis10}%
  \BibitemOpen
  \bibfield  {author} {\bibinfo {author} {\bibnamefont {Ambainis},
  \bibfnamefont {A.}}} (\bibinfo {year} {2010}),\ \href@noop {} {\bibfield
  {journal} {\bibinfo  {journal} {Quantum Information and Computation}\
  }\textbf {\bibinfo {volume} {10}}~(\bibinfo {number} {9{\&}10}),\ \bibinfo
  {pages} {0848}}\BibitemShut {NoStop}%
\bibitem [{\citenamefont {Azarchs}(2004)}]{azarchs04}%
  \BibitemOpen
  \bibfield  {author} {\bibinfo {author} {\bibnamefont {Azarchs}, \bibfnamefont
  {A.}}} (\bibinfo {year} {2004}),\ \emph {\bibinfo {title} {{Entropic
  Uncertainty Relations for Incomplete Sets of Mutually Unbiased
  Observables}}},\ \href {http://arxiv.org/abs/quant-ph/0412083} {\bibinfo
  {type} {Semester thesis}}\ (\bibinfo  {school} {California Institute of
  Technology})\BibitemShut {NoStop}%
\bibitem [{\citenamefont {Baek}\ \emph {et~al.}(2014)\citenamefont {Baek},
  \citenamefont {Farrow},\ and\ \citenamefont {Son}}]{baek14}%
  \BibitemOpen
  \bibfield  {author} {\bibinfo {author} {\bibnamefont {Baek}, \bibfnamefont
  {K.}}, \bibinfo {author} {\bibfnamefont {T.}~\bibnamefont {Farrow}}, \ and\
  \bibinfo {author} {\bibfnamefont {W.}~\bibnamefont {Son}}} (\bibinfo {year}
  {2014}),\ \href {\doibase 10.1103/PhysRevA.89.032108} {\bibfield  {journal}
  {\bibinfo  {journal} {Physical Review A}\ }\textbf {\bibinfo {volume}
  {89}}~(\bibinfo {number} {3}),\ \bibinfo {pages} {032108}}\BibitemShut
  {NoStop}%
\bibitem [{\citenamefont {Ballester}\ and\ \citenamefont
  {Wehner}(2007)}]{ballester07}%
  \BibitemOpen
  \bibfield  {author} {\bibinfo {author} {\bibnamefont {Ballester},
  \bibfnamefont {M.}}, \ and\ \bibinfo {author} {\bibfnamefont
  {S.}~\bibnamefont {Wehner}}} (\bibinfo {year} {2007}),\ \href {\doibase
  10.1103/PhysRevA.75.022319} {\bibfield  {journal} {\bibinfo  {journal}
  {Physical Review A}\ }\textbf {\bibinfo {volume} {75}}~(\bibinfo {number}
  {2}),\ \bibinfo {pages} {022319}}\BibitemShut {NoStop}%
\bibitem [{\citenamefont {Ballester}\ \emph {et~al.}(2008)\citenamefont
  {Ballester}, \citenamefont {Wehner},\ and\ \citenamefont
  {Winter}}]{ballester08}%
  \BibitemOpen
  \bibfield  {author} {\bibinfo {author} {\bibnamefont {Ballester},
  \bibfnamefont {M.~A.}}, \bibinfo {author} {\bibfnamefont {S.}~\bibnamefont
  {Wehner}}, \ and\ \bibinfo {author} {\bibfnamefont {A.}~\bibnamefont
  {Winter}}} (\bibinfo {year} {2008}),\ \href {\doibase
  10.1109/TIT.2008.928276} {\bibfield  {journal} {\bibinfo  {journal} {IEEE
  Transactions on Information Theory}\ }\textbf {\bibinfo {volume}
  {54}}~(\bibinfo {number} {9}),\ \bibinfo {pages} {4183}}\BibitemShut
  {NoStop}%
\bibitem [{\citenamefont {Bandyopadhyay}\ \emph {et~al.}(2002)\citenamefont
  {Bandyopadhyay}, \citenamefont {Boykin}, \citenamefont {Roychowdhury},\ and\
  \citenamefont {Vatan}}]{bandyopadhyay02}%
  \BibitemOpen
  \bibfield  {author} {\bibinfo {author} {\bibnamefont {Bandyopadhyay},},
  \bibinfo {author} {\bibnamefont {Boykin}}, \bibinfo {author} {\bibfnamefont
  {V.}~\bibnamefont {Roychowdhury}}, \ and\ \bibinfo {author} {\bibnamefont
  {Vatan}}} (\bibinfo {year} {2002}),\ \href {\doibase
  10.1007/s00453-002-0980-7} {\bibfield  {journal} {\bibinfo  {journal}
  {Algorithmica}\ }\textbf {\bibinfo {volume} {34}}~(\bibinfo {number} {4}),\
  \bibinfo {pages} {512}}\BibitemShut {NoStop}%
\bibitem [{\citenamefont {Barnum}\ and\ \citenamefont
  {Knill}(2002)}]{barnum02}%
  \BibitemOpen
  \bibfield  {author} {\bibinfo {author} {\bibnamefont {Barnum}, \bibfnamefont
  {H.}}, \ and\ \bibinfo {author} {\bibfnamefont {E.}~\bibnamefont {Knill}}}
  (\bibinfo {year} {2002}),\ \href {\doibase 10.1063/1.1459754} {\bibfield
  {journal} {\bibinfo  {journal} {Journal of Mathematical Physics}\ }\textbf
  {\bibinfo {volume} {43}}~(\bibinfo {number} {5}),\ \bibinfo {pages}
  {2097}}\BibitemShut {NoStop}%
\bibitem [{\citenamefont {Baumgratz}\ \emph {et~al.}(2014)\citenamefont
  {Baumgratz}, \citenamefont {Cramer},\ and\ \citenamefont
  {Plenio}}]{baumgratz14}%
  \BibitemOpen
  \bibfield  {author} {\bibinfo {author} {\bibnamefont {Baumgratz},
  \bibfnamefont {T.}}, \bibinfo {author} {\bibfnamefont {M.}~\bibnamefont
  {Cramer}}, \ and\ \bibinfo {author} {\bibfnamefont {M.~B.}\ \bibnamefont
  {Plenio}}} (\bibinfo {year} {2014}),\ \href {\doibase
  10.1103/PhysRevLett.113.140401} {\bibfield  {journal} {\bibinfo  {journal}
  {Physical Review Letters}\ }\textbf {\bibinfo {volume} {113}}~(\bibinfo
  {number} {14}),\ \bibinfo {pages} {140401}}\BibitemShut {NoStop}%
\bibitem [{\citenamefont {Beckner}(1975)}]{beckner75}%
  \BibitemOpen
  \bibfield  {author} {\bibinfo {author} {\bibnamefont {Beckner}, \bibfnamefont
  {W.}}} (\bibinfo {year} {1975}),\ \href {\doibase 10.2307/1970980} {\bibfield
   {journal} {\bibinfo  {journal} {Annals of Mathematics}\ }\textbf {\bibinfo
  {volume} {102}}~(\bibinfo {number} {1}),\ \bibinfo {pages} {159}}\BibitemShut
  {NoStop}%
\bibitem [{\citenamefont {Beigi}(2013)}]{beigi13}%
  \BibitemOpen
  \bibfield  {author} {\bibinfo {author} {\bibnamefont {Beigi}, \bibfnamefont
  {S.}}} (\bibinfo {year} {2013}),\ \href {\doibase 10.1063/1.4838855}
  {\bibfield  {journal} {\bibinfo  {journal} {Journal of Mathematical Physics}\
  }\textbf {\bibinfo {volume} {54}}~(\bibinfo {number} {12}),\ \bibinfo {pages}
  {122202}}\BibitemShut {NoStop}%
\bibitem [{\citenamefont {Bennett}\ and\ \citenamefont
  {Brassard}(1984)}]{bb84}%
  \BibitemOpen
  \bibfield  {author} {\bibinfo {author} {\bibnamefont {Bennett}, \bibfnamefont
  {C.~H.}}, \ and\ \bibinfo {author} {\bibfnamefont {G.}~\bibnamefont
  {Brassard}}} (\bibinfo {year} {1984}),\ in\ \href@noop {} {\emph {\bibinfo
  {booktitle} {Proc. IEEE International Conference on Computers, Systems and
  Signal Processing 1984}}},\ Vol.~\bibinfo {volume} {1}\ (\bibinfo
  {publisher} {IEEE},\ \bibinfo {address} {Bangalore})\ pp.\ \bibinfo {pages}
  {175--179}\BibitemShut {NoStop}%
\bibitem [{\citenamefont {Bergh}\ and\ \citenamefont
  {L{\"{o}}fstr{\"{o}}m}(1976)}]{bergh76}%
  \BibitemOpen
  \bibfield  {author} {\bibinfo {author} {\bibnamefont {Bergh}, \bibfnamefont
  {J.}}, \ and\ \bibinfo {author} {\bibfnamefont {J.}~\bibnamefont
  {L{\"{o}}fstr{\"{o}}m}}} (\bibinfo {year} {1976}),\ \href {\doibase
  10.1007/978-3-642-66451-9} {\emph {\bibinfo {title} {{Interpolation
  Spaces}}}},\ \bibinfo {series} {Grundlehren der mathematischen
  Wissenschaften}, Vol.\ \bibinfo {volume} {223}\ (\bibinfo  {publisher}
  {Springer Berlin Heidelberg},\ \bibinfo {address} {Berlin,
  Heidelberg})\BibitemShut {NoStop}%
\bibitem [{\citenamefont {Berta}(2013)}]{bertathesis}%
  \BibitemOpen
  \bibfield  {author} {\bibinfo {author} {\bibnamefont {Berta}, \bibfnamefont
  {M.}}} (\bibinfo {year} {2013}),\ \emph {\bibinfo {title} {{Quantum Side
  Information: Uncertainty Relations, Extractors, Channel Simulations}}},\
  \href {http://arxiv.org/abs/1310.4581} {Ph.D. thesis}\ (\bibinfo  {school}
  {ETH Zurich})\BibitemShut {NoStop}%
\bibitem [{\citenamefont {Berta}\ \emph {et~al.}(2013)\citenamefont {Berta},
  \citenamefont {Brandao}, \citenamefont {Christandl},\ and\ \citenamefont
  {Wehner}}]{christandl13}%
  \BibitemOpen
  \bibfield  {author} {\bibinfo {author} {\bibnamefont {Berta}, \bibfnamefont
  {M.}}, \bibinfo {author} {\bibfnamefont {F.~G. S.~L.}\ \bibnamefont
  {Brandao}}, \bibinfo {author} {\bibfnamefont {M.}~\bibnamefont {Christandl}},
  \ and\ \bibinfo {author} {\bibfnamefont {S.}~\bibnamefont {Wehner}}}
  (\bibinfo {year} {2013}),\ \href {\doibase 10.1109/TIT.2013.2268533}
  {\bibfield  {journal} {\bibinfo  {journal} {IEEE Transactions on Information
  Theory}\ }\textbf {\bibinfo {volume} {59}}~(\bibinfo {number} {10}),\
  \bibinfo {pages} {6779}}\BibitemShut {NoStop}%
\bibitem [{\citenamefont {Berta}\ \emph {et~al.}(2010)\citenamefont {Berta},
  \citenamefont {Christandl}, \citenamefont {Colbeck}, \citenamefont {Renes},\
  and\ \citenamefont {Renner}}]{berta10}%
  \BibitemOpen
  \bibfield  {author} {\bibinfo {author} {\bibnamefont {Berta}, \bibfnamefont
  {M.}}, \bibinfo {author} {\bibfnamefont {M.}~\bibnamefont {Christandl}},
  \bibinfo {author} {\bibfnamefont {R.}~\bibnamefont {Colbeck}}, \bibinfo
  {author} {\bibfnamefont {J.~M.}\ \bibnamefont {Renes}}, \ and\ \bibinfo
  {author} {\bibfnamefont {R.}~\bibnamefont {Renner}}} (\bibinfo {year}
  {2010}),\ \href {\doibase 10.1038/nphys1734} {\bibfield  {journal} {\bibinfo
  {journal} {Nature Physics}\ }\textbf {\bibinfo {volume} {6}}~(\bibinfo
  {number} {9}),\ \bibinfo {pages} {659}}\BibitemShut {NoStop}%
\bibitem [{\citenamefont {Berta}\ \emph
  {et~al.}(2014{\natexlab{a}})\citenamefont {Berta}, \citenamefont {Coles},\
  and\ \citenamefont {Wehner}}]{berta13}%
  \BibitemOpen
  \bibfield  {author} {\bibinfo {author} {\bibnamefont {Berta}, \bibfnamefont
  {M.}}, \bibinfo {author} {\bibfnamefont {P.~J.}\ \bibnamefont {Coles}}, \
  and\ \bibinfo {author} {\bibfnamefont {S.}~\bibnamefont {Wehner}}} (\bibinfo
  {year} {2014}{\natexlab{a}}),\ \href {\doibase 10.1103/PhysRevA.90.062127}
  {\bibfield  {journal} {\bibinfo  {journal} {Physical Review A}\ }\textbf
  {\bibinfo {volume} {90}}~(\bibinfo {number} {6}),\ \bibinfo {pages}
  {062127}}\BibitemShut {NoStop}%
\bibitem [{\citenamefont {Berta}\ \emph
  {et~al.}(2014{\natexlab{b}})\citenamefont {Berta}, \citenamefont {Fawzi},\
  and\ \citenamefont {Wehner}}]{berta14}%
  \BibitemOpen
  \bibfield  {author} {\bibinfo {author} {\bibnamefont {Berta}, \bibfnamefont
  {M.}}, \bibinfo {author} {\bibfnamefont {O.}~\bibnamefont {Fawzi}}, \ and\
  \bibinfo {author} {\bibfnamefont {S.}~\bibnamefont {Wehner}}} (\bibinfo
  {year} {2014}{\natexlab{b}}),\ \href {\doibase 10.1109/TIT.2013.2291780}
  {\bibfield  {journal} {\bibinfo  {journal} {IEEE Transactions on Information
  Theory}\ }\textbf {\bibinfo {volume} {60}}~(\bibinfo {number} {2}),\ \bibinfo
  {pages} {1168}}\BibitemShut {NoStop}%
\bibitem [{\citenamefont {Berta}\ \emph {et~al.}(2016)\citenamefont {Berta},
  \citenamefont {Wehner},\ and\ \citenamefont {Wilde}}]{berta16}%
  \BibitemOpen
  \bibfield  {author} {\bibinfo {author} {\bibnamefont {Berta}, \bibfnamefont
  {M.}}, \bibinfo {author} {\bibfnamefont {S.}~\bibnamefont {Wehner}}, \ and\
  \bibinfo {author} {\bibfnamefont {M.~M.}\ \bibnamefont {Wilde}}} (\bibinfo
  {year} {2016}),\ \href {\doibase 10.1088/1367-2630/18/7/073004} {\bibfield
  {journal} {\bibinfo  {journal} {New Journal of Physics}\ }\textbf {\bibinfo
  {volume} {18}}~(\bibinfo {number} {7}),\ \bibinfo {pages}
  {073004}}\BibitemShut {NoStop}%
\bibitem [{\citenamefont {Bia{\l}ynicki-Birula}(1984)}]{birula84}%
  \BibitemOpen
  \bibfield  {author} {\bibinfo {author} {\bibnamefont {Bia{\l}ynicki-Birula},
  \bibfnamefont {I.}}} (\bibinfo {year} {1984}),\ \href {\doibase
  10.1016/0375-9601(84)90118-X} {\bibfield  {journal} {\bibinfo  {journal}
  {Physics Letters A}\ }\textbf {\bibinfo {volume} {103}}~(\bibinfo {number}
  {5}),\ \bibinfo {pages} {253}}\BibitemShut {NoStop}%
\bibitem [{\citenamefont {Bialynicki-Birula}(2006)}]{bialynicki06}%
  \BibitemOpen
  \bibfield  {author} {\bibinfo {author} {\bibnamefont {Bialynicki-Birula},
  \bibfnamefont {I.}}} (\bibinfo {year} {2006}),\ \href {\doibase
  10.1103/PhysRevA.74.052101} {\bibfield  {journal} {\bibinfo  {journal}
  {Physical Review A}\ }\textbf {\bibinfo {volume} {74}}~(\bibinfo {number}
  {5}),\ \bibinfo {pages} {052101}}\BibitemShut {NoStop}%
\bibitem [{\citenamefont {Bia{\l}ynicki-Birula}(2007)}]{birula07}%
  \BibitemOpen
  \bibfield  {author} {\bibinfo {author} {\bibnamefont {Bia{\l}ynicki-Birula},
  \bibfnamefont {I.}}} (\bibinfo {year} {2007}),\ in\ \href {\doibase
  10.1063/1.2713446} {\emph {\bibinfo {booktitle} {AIP Conference
  Proceedings}}},\ Vol.\ \bibinfo {volume} {889}\ (\bibinfo  {publisher}
  {AIP},\ \bibinfo {address} {Vaxjo})\ pp.\ \bibinfo {pages}
  {52--61}\BibitemShut {NoStop}%
\bibitem [{\citenamefont {Bia{\l}ynicki-Birula}\ and\ \citenamefont
  {Madajczyk}(1985)}]{birula85}%
  \BibitemOpen
  \bibfield  {author} {\bibinfo {author} {\bibnamefont {Bia{\l}ynicki-Birula},
  \bibfnamefont {I.}}, \ and\ \bibinfo {author} {\bibfnamefont {J.~L.}\
  \bibnamefont {Madajczyk}}} (\bibinfo {year} {1985}),\ \href {\doibase
  10.1016/0375-9601(85)90277-4} {\bibfield  {journal} {\bibinfo  {journal}
  {Physics Letters A}\ }\textbf {\bibinfo {volume} {108}}~(\bibinfo {number}
  {8}),\ \bibinfo {pages} {384}}\BibitemShut {NoStop}%
\bibitem [{\citenamefont {Bia{\l}ynicki-Birula}\ and\ \citenamefont
  {Mycielski}(1975)}]{biaynicki75}%
  \BibitemOpen
  \bibfield  {author} {\bibinfo {author} {\bibnamefont {Bia{\l}ynicki-Birula},
  \bibfnamefont {I.}}, \ and\ \bibinfo {author} {\bibfnamefont
  {J.}~\bibnamefont {Mycielski}}} (\bibinfo {year} {1975}),\ \href {\doibase
  10.1007/BF01608825} {\bibfield  {journal} {\bibinfo  {journal}
  {Communications in Mathematical Physics}\ }\textbf {\bibinfo {volume}
  {44}}~(\bibinfo {number} {2}),\ \bibinfo {pages} {129}}\BibitemShut {NoStop}%
\bibitem [{\citenamefont {Bia{\l}ynicki-Birula}\ and\ \citenamefont
  {Rudnicki}(2011)}]{birula10}%
  \BibitemOpen
  \bibfield  {author} {\bibinfo {author} {\bibnamefont {Bia{\l}ynicki-Birula},
  \bibfnamefont {I.}}, \ and\ \bibinfo {author} {\bibfnamefont
  {{\L}.}~\bibnamefont {Rudnicki}}} (\bibinfo {year} {2011}),\ in\ \href
  {\doibase 10.1007/978-90-481-3890-6} {\emph {\bibinfo {booktitle}
  {Statistical Complexity}}},\ \bibinfo {editor} {edited by\ \bibinfo {editor}
  {\bibfnamefont {K.}~\bibnamefont {Sen}}}\ (\bibinfo  {publisher} {Springer
  Netherlands},\ \bibinfo {address} {Dordrecht})\ pp.\ \bibinfo {pages}
  {1--34}\BibitemShut {NoStop}%
\bibitem [{\citenamefont {Biham}\ \emph {et~al.}(2000)\citenamefont {Biham},
  \citenamefont {Boyer}, \citenamefont {Boykin}, \citenamefont {Mor},\ and\
  \citenamefont {Roychowdhury}}]{biham00stoc}%
  \BibitemOpen
  \bibfield  {author} {\bibinfo {author} {\bibnamefont {Biham}, \bibfnamefont
  {E.}}, \bibinfo {author} {\bibfnamefont {M.}~\bibnamefont {Boyer}}, \bibinfo
  {author} {\bibfnamefont {P.~O.}\ \bibnamefont {Boykin}}, \bibinfo {author}
  {\bibfnamefont {T.}~\bibnamefont {Mor}}, \ and\ \bibinfo {author}
  {\bibfnamefont {V.}~\bibnamefont {Roychowdhury}}} (\bibinfo {year} {2000}),\
  in\ \href {\doibase 10.1145/335305.335406} {\emph {\bibinfo {booktitle}
  {Proc. ACM STOC 2000}}}\ (\bibinfo  {publisher} {ACM Press},\ \bibinfo
  {address} {New York, NY})\ pp.\ \bibinfo {pages} {715--724}\BibitemShut
  {NoStop}%
\bibitem [{\citenamefont {Biham}\ \emph {et~al.}(2006)\citenamefont {Biham},
  \citenamefont {Boyer}, \citenamefont {Boykin}, \citenamefont {Mor},\ and\
  \citenamefont {Roychowdhury}}]{biham06}%
  \BibitemOpen
  \bibfield  {author} {\bibinfo {author} {\bibnamefont {Biham}, \bibfnamefont
  {E.}}, \bibinfo {author} {\bibfnamefont {M.}~\bibnamefont {Boyer}}, \bibinfo
  {author} {\bibfnamefont {P.~O.}\ \bibnamefont {Boykin}}, \bibinfo {author}
  {\bibfnamefont {T.}~\bibnamefont {Mor}}, \ and\ \bibinfo {author}
  {\bibfnamefont {V.}~\bibnamefont {Roychowdhury}}} (\bibinfo {year} {2006}),\
  \href {\doibase 10.1007/s00145-005-0011-3} {\bibfield  {journal} {\bibinfo
  {journal} {Journal of Cryptology}\ }\textbf {\bibinfo {volume}
  {19}}~(\bibinfo {number} {4}),\ \bibinfo {pages} {381}}\BibitemShut {NoStop}%
\bibitem [{\citenamefont {Boltzmann}(1872)}]{boltzmann1872}%
  \BibitemOpen
  \bibfield  {author} {\bibinfo {author} {\bibnamefont {Boltzmann},
  \bibfnamefont {L.}}} (\bibinfo {year} {1872}),\ in\ \href@noop {} {\emph
  {\bibinfo {booktitle} {Sitzungsberichte der Akademie der Wissenschaften zu
  Wien}}},\ Vol.~\bibinfo {volume} {66},\ pp.\ \bibinfo {pages}
  {275--370}\BibitemShut {NoStop}%
\bibitem [{\citenamefont {Bosyk}\ \emph {et~al.}(2013)\citenamefont {Bosyk},
  \citenamefont {Portesi}, \citenamefont {Holik},\ and\ \citenamefont
  {Plastino}}]{bosyk13}%
  \BibitemOpen
  \bibfield  {author} {\bibinfo {author} {\bibnamefont {Bosyk}, \bibfnamefont
  {G.~M.}}, \bibinfo {author} {\bibfnamefont {M.}~\bibnamefont {Portesi}},
  \bibinfo {author} {\bibfnamefont {F.}~\bibnamefont {Holik}}, \ and\ \bibinfo
  {author} {\bibfnamefont {A.}~\bibnamefont {Plastino}}} (\bibinfo {year}
  {2013}),\ \href {\doibase 10.1088/0031-8949/87/06/065002} {\bibfield
  {journal} {\bibinfo  {journal} {Physica Scripta}\ }\textbf {\bibinfo {volume}
  {87}}~(\bibinfo {number} {6}),\ \bibinfo {pages} {065002}}\BibitemShut
  {NoStop}%
\bibitem [{\citenamefont {Broadbent}\ and\ \citenamefont
  {Schaffner}(2016)}]{broadbent15}%
  \BibitemOpen
  \bibfield  {author} {\bibinfo {author} {\bibnamefont {Broadbent},
  \bibfnamefont {A.}}, \ and\ \bibinfo {author} {\bibfnamefont
  {C.}~\bibnamefont {Schaffner}}} (\bibinfo {year} {2016}),\ \href {\doibase
  10.1007/s10623-015-0157-4} {\bibfield  {journal} {\bibinfo  {journal}
  {Designs, Codes and Cryptography}\ }\textbf {\bibinfo {volume}
  {78}}~(\bibinfo {number} {1}),\ \bibinfo {pages} {351}}\BibitemShut {NoStop}%
\bibitem [{\citenamefont {Brukner}\ and\ \citenamefont
  {Zeilinger}(1999)}]{brukner99}%
  \BibitemOpen
  \bibfield  {author} {\bibinfo {author} {\bibnamefont {Brukner}, \bibfnamefont
  {{\v{C}}.}}, \ and\ \bibinfo {author} {\bibfnamefont {A.}~\bibnamefont
  {Zeilinger}}} (\bibinfo {year} {1999}),\ \href {\doibase
  10.1103/PhysRevLett.83.3354} {\bibfield  {journal} {\bibinfo  {journal}
  {Physical Review Letters}\ }\textbf {\bibinfo {volume} {83}}~(\bibinfo
  {number} {17}),\ \bibinfo {pages} {3354}}\BibitemShut {NoStop}%
\bibitem [{\citenamefont {Bru{\ss}}(1998)}]{bruss98}%
  \BibitemOpen
  \bibfield  {author} {\bibinfo {author} {\bibnamefont {Bru{\ss}},
  \bibfnamefont {D.}}} (\bibinfo {year} {1998}),\ \href {\doibase
  10.1103/PhysRevLett.81.3018} {\bibfield  {journal} {\bibinfo  {journal}
  {Physical Review Letters}\ }\textbf {\bibinfo {volume} {81}}~(\bibinfo
  {number} {14}),\ \bibinfo {pages} {3018}}\BibitemShut {NoStop}%
\bibitem [{\citenamefont {Bruzda}\ \emph {et~al.}(2015)\citenamefont {Bruzda},
  \citenamefont {Tadej},\ and\ \citenamefont
  {{\.{Z}}yczkowski}}]{hadamard_matrix_weblink}%
  \BibitemOpen
  \bibfield  {author} {\bibinfo {author} {\bibnamefont {Bruzda}, \bibfnamefont
  {W.}}, \bibinfo {author} {\bibfnamefont {W.}~\bibnamefont {Tadej}}, \ and\
  \bibinfo {author} {\bibfnamefont {K.}~\bibnamefont {{\.{Z}}yczkowski}}}
  (\bibinfo {year} {2015}),\ \href
  {http://chaos.if.uj.edu.pl/{~}karol/hadamard/} {\enquote {\bibinfo {title}
  {{Complex Hadamard Matrices}},}\ }\BibitemShut {NoStop}%
\bibitem [{\citenamefont {Buhrman}\ \emph {et~al.}(2008)\citenamefont
  {Buhrman}, \citenamefont {Christandl}, \citenamefont {Hayden}, \citenamefont
  {Lo},\ and\ \citenamefont {Wehner}}]{buhrman08}%
  \BibitemOpen
  \bibfield  {author} {\bibinfo {author} {\bibnamefont {Buhrman}, \bibfnamefont
  {H.}}, \bibinfo {author} {\bibfnamefont {M.}~\bibnamefont {Christandl}},
  \bibinfo {author} {\bibfnamefont {P.}~\bibnamefont {Hayden}}, \bibinfo
  {author} {\bibfnamefont {H.-K.}\ \bibnamefont {Lo}}, \ and\ \bibinfo {author}
  {\bibfnamefont {S.}~\bibnamefont {Wehner}}} (\bibinfo {year} {2008}),\ \href
  {\doibase 10.1103/PhysRevA.78.022316} {\bibfield  {journal} {\bibinfo
  {journal} {Physical Review A}\ }\textbf {\bibinfo {volume} {78}}~(\bibinfo
  {number} {2}),\ \bibinfo {pages} {022316}}\BibitemShut {NoStop}%
\bibitem [{\citenamefont {Buscemi}\ \emph {et~al.}(2014)\citenamefont
  {Buscemi}, \citenamefont {Hall}, \citenamefont {Ozawa},\ and\ \citenamefont
  {Wilde}}]{buscemi14}%
  \BibitemOpen
  \bibfield  {author} {\bibinfo {author} {\bibnamefont {Buscemi}, \bibfnamefont
  {F.}}, \bibinfo {author} {\bibfnamefont {M.~J.~W.}\ \bibnamefont {Hall}},
  \bibinfo {author} {\bibfnamefont {M.}~\bibnamefont {Ozawa}}, \ and\ \bibinfo
  {author} {\bibfnamefont {M.~M.}\ \bibnamefont {Wilde}}} (\bibinfo {year}
  {2014}),\ \href {\doibase 10.1103/PhysRevLett.112.050401} {\bibfield
  {journal} {\bibinfo  {journal} {Physical Review Letters}\ }\textbf {\bibinfo
  {volume} {112}}~(\bibinfo {number} {5}),\ \bibinfo {pages}
  {050401}}\BibitemShut {NoStop}%
\bibitem [{\citenamefont {Busch}\ \emph {et~al.}(2007)\citenamefont {Busch},
  \citenamefont {Heinonen},\ and\ \citenamefont {Lahti}}]{busch07}%
  \BibitemOpen
  \bibfield  {author} {\bibinfo {author} {\bibnamefont {Busch}, \bibfnamefont
  {P.}}, \bibinfo {author} {\bibfnamefont {T.}~\bibnamefont {Heinonen}}, \ and\
  \bibinfo {author} {\bibfnamefont {P.}~\bibnamefont {Lahti}}} (\bibinfo {year}
  {2007}),\ \href {\doibase 10.1016/j.physrep.2007.05.006} {\bibfield
  {journal} {\bibinfo  {journal} {Physics Reports}\ }\textbf {\bibinfo {volume}
  {452}}~(\bibinfo {number} {6}),\ \bibinfo {pages} {155}}\BibitemShut
  {NoStop}%
\bibitem [{\citenamefont {Busch}\ \emph {et~al.}(2013)\citenamefont {Busch},
  \citenamefont {Lahti},\ and\ \citenamefont {Werner}}]{busch13}%
  \BibitemOpen
  \bibfield  {author} {\bibinfo {author} {\bibnamefont {Busch}, \bibfnamefont
  {P.}}, \bibinfo {author} {\bibfnamefont {P.}~\bibnamefont {Lahti}}, \ and\
  \bibinfo {author} {\bibfnamefont {R.~F.}\ \bibnamefont {Werner}}} (\bibinfo
  {year} {2013}),\ \href {\doibase 10.1103/PhysRevLett.111.160405} {\bibfield
  {journal} {\bibinfo  {journal} {Physical Review Letters}\ }\textbf {\bibinfo
  {volume} {111}}~(\bibinfo {number} {16}),\ \bibinfo {pages}
  {160405}}\BibitemShut {NoStop}%
\bibitem [{\citenamefont {Busch}\ \emph
  {et~al.}(2014{\natexlab{a}})\citenamefont {Busch}, \citenamefont {Lahti},\
  and\ \citenamefont {Werner}}]{busch14b}%
  \BibitemOpen
  \bibfield  {author} {\bibinfo {author} {\bibnamefont {Busch}, \bibfnamefont
  {P.}}, \bibinfo {author} {\bibfnamefont {P.}~\bibnamefont {Lahti}}, \ and\
  \bibinfo {author} {\bibfnamefont {R.~F.}\ \bibnamefont {Werner}}} (\bibinfo
  {year} {2014}{\natexlab{a}}),\ \href {\doibase 10.1103/RevModPhys.86.1261}
  {\bibfield  {journal} {\bibinfo  {journal} {Review of Modern Physics}\
  }\textbf {\bibinfo {volume} {86}}~(\bibinfo {number} {4}),\ \bibinfo {pages}
  {1261}}\BibitemShut {NoStop}%
\bibitem [{\citenamefont {Busch}\ \emph
  {et~al.}(2014{\natexlab{b}})\citenamefont {Busch}, \citenamefont {Lahti},\
  and\ \citenamefont {Werner}}]{busch14}%
  \BibitemOpen
  \bibfield  {author} {\bibinfo {author} {\bibnamefont {Busch}, \bibfnamefont
  {P.}}, \bibinfo {author} {\bibfnamefont {P.}~\bibnamefont {Lahti}}, \ and\
  \bibinfo {author} {\bibfnamefont {R.~F.}\ \bibnamefont {Werner}}} (\bibinfo
  {year} {2014}{\natexlab{b}}),\ \href {\doibase 10.1103/PhysRevA.89.012129}
  {\bibfield  {journal} {\bibinfo  {journal} {Physical Review A}\ }\textbf
  {\bibinfo {volume} {89}}~(\bibinfo {number} {1}),\ \bibinfo {pages}
  {012129}}\BibitemShut {NoStop}%
\bibitem [{\citenamefont {Busch}\ and\ \citenamefont
  {Shilladay}(2006)}]{busch06}%
  \BibitemOpen
  \bibfield  {author} {\bibinfo {author} {\bibnamefont {Busch}, \bibfnamefont
  {P.}}, \ and\ \bibinfo {author} {\bibfnamefont {C.}~\bibnamefont
  {Shilladay}}} (\bibinfo {year} {2006}),\ \href {\doibase
  10.1016/j.physrep.2006.09.001} {\bibfield  {journal} {\bibinfo  {journal}
  {Physics Reports}\ }\textbf {\bibinfo {volume} {435}}~(\bibinfo {number}
  {1}),\ \bibinfo {pages} {1}}\BibitemShut {NoStop}%
\bibitem [{\citenamefont {Cachin}\ and\ \citenamefont
  {Maurer}(1997)}]{cachin97}%
  \BibitemOpen
  \bibfield  {author} {\bibinfo {author} {\bibnamefont {Cachin}, \bibfnamefont
  {C.}}, \ and\ \bibinfo {author} {\bibfnamefont {U.}~\bibnamefont {Maurer}}}
  (\bibinfo {year} {1997}),\ in\ \href {\doibase 10.1007/BFb0052243} {\emph
  {\bibinfo {booktitle} {Proc. CRYPTO 1997}}},\ \bibinfo {series} {LNCS}, Vol.\
  \bibinfo {volume} {1294}\ (\bibinfo  {publisher} {Springer})\ pp.\ \bibinfo
  {pages} {292--306}\BibitemShut {NoStop}%
\bibitem [{\citenamefont {Canetti}(2001)}]{canetti01}%
  \BibitemOpen
  \bibfield  {author} {\bibinfo {author} {\bibnamefont {Canetti}, \bibfnamefont
  {R.}}} (\bibinfo {year} {2001}),\ in\ \href {\doibase
  10.1109/SFCS.2001.959888} {\emph {\bibinfo {booktitle} {Proc. IEEE FOCS
  2001}}}\ (\bibinfo  {publisher} {IEEE})\ pp.\ \bibinfo {pages}
  {136--145}\BibitemShut {NoStop}%
\bibitem [{\citenamefont {Cavalcanti}\ \emph {et~al.}(2009)\citenamefont
  {Cavalcanti}, \citenamefont {Jones}, \citenamefont {Wiseman},\ and\
  \citenamefont {Reid}}]{cavalcanti09}%
  \BibitemOpen
  \bibfield  {author} {\bibinfo {author} {\bibnamefont {Cavalcanti},
  \bibfnamefont {E.~G.}}, \bibinfo {author} {\bibfnamefont {S.~J.}\
  \bibnamefont {Jones}}, \bibinfo {author} {\bibfnamefont {H.~M.}\ \bibnamefont
  {Wiseman}}, \ and\ \bibinfo {author} {\bibfnamefont {M.~D.}\ \bibnamefont
  {Reid}}} (\bibinfo {year} {2009}),\ \href {\doibase
  10.1103/PhysRevA.80.032112} {\bibfield  {journal} {\bibinfo  {journal}
  {Physical Review A}\ }\textbf {\bibinfo {volume} {80}}~(\bibinfo {number}
  {3}),\ \bibinfo {pages} {032112}}\BibitemShut {NoStop}%
\bibitem [{\citenamefont {Cerf}\ \emph {et~al.}(2002)\citenamefont {Cerf},
  \citenamefont {Bourennane}, \citenamefont {Karlsson},\ and\ \citenamefont
  {Gisin}}]{cerf02}%
  \BibitemOpen
  \bibfield  {author} {\bibinfo {author} {\bibnamefont {Cerf}, \bibfnamefont
  {N.~J.}}, \bibinfo {author} {\bibfnamefont {M.}~\bibnamefont {Bourennane}},
  \bibinfo {author} {\bibfnamefont {A.}~\bibnamefont {Karlsson}}, \ and\
  \bibinfo {author} {\bibfnamefont {N.}~\bibnamefont {Gisin}}} (\bibinfo {year}
  {2002}),\ \href {\doibase 10.1103/PhysRevLett.88.127902} {\bibfield
  {journal} {\bibinfo  {journal} {Physical Review Letters}\ }\textbf {\bibinfo
  {volume} {88}}~(\bibinfo {number} {12}),\ \bibinfo {pages}
  {127902}}\BibitemShut {NoStop}%
\bibitem [{\citenamefont {Chen}\ and\ \citenamefont {Fei}(2015)}]{chen15}%
  \BibitemOpen
  \bibfield  {author} {\bibinfo {author} {\bibnamefont {Chen}, \bibfnamefont
  {B.}}, \ and\ \bibinfo {author} {\bibfnamefont {S.-M.}\ \bibnamefont {Fei}}}
  (\bibinfo {year} {2015}),\ \href {\doibase 10.1007/s11128-015-0949-5}
  {\bibfield  {journal} {\bibinfo  {journal} {Quantum Information Processing}\
  }\textbf {\bibinfo {volume} {14}}~(\bibinfo {number} {6}),\ \bibinfo {pages}
  {2227}}\BibitemShut {NoStop}%
\bibitem [{\citenamefont {Choi}(1975)}]{choi75}%
  \BibitemOpen
  \bibfield  {author} {\bibinfo {author} {\bibnamefont {Choi}, \bibfnamefont
  {M.}}} (\bibinfo {year} {1975}),\ \href {\doibase
  10.1016/0024-3795(75)90075-0} {\bibfield  {journal} {\bibinfo  {journal}
  {Linear Algebra Appl.}\ }\textbf {\bibinfo {volume} {10}}~(\bibinfo {number}
  {3}),\ \bibinfo {pages} {285}}\BibitemShut {NoStop}%
\bibitem [{\citenamefont {Christandl}\ and\ \citenamefont
  {Winter}(2005)}]{christandl05}%
  \BibitemOpen
  \bibfield  {author} {\bibinfo {author} {\bibnamefont {Christandl},
  \bibfnamefont {M.}}, \ and\ \bibinfo {author} {\bibfnamefont
  {A.}~\bibnamefont {Winter}}} (\bibinfo {year} {2005}),\ \href {\doibase
  10.1109/TIT.2005.853338} {\bibfield  {journal} {\bibinfo  {journal} {IEEE
  Transactions on Information Theory}\ }\textbf {\bibinfo {volume}
  {51}}~(\bibinfo {number} {9}),\ \bibinfo {pages} {3159}}\BibitemShut
  {NoStop}%
\bibitem [{\citenamefont {Coles}(2012{\natexlab{a}})}]{coles12c}%
  \BibitemOpen
  \bibfield  {author} {\bibinfo {author} {\bibnamefont {Coles}, \bibfnamefont
  {P.~J.}}} (\bibinfo {year} {2012}{\natexlab{a}}),\ \href {\doibase
  10.1103/PhysRevA.86.062334} {\bibfield  {journal} {\bibinfo  {journal}
  {Physical Review A}\ }\textbf {\bibinfo {volume} {86}}~(\bibinfo {number}
  {6}),\ \bibinfo {pages} {062334}}\BibitemShut {NoStop}%
\bibitem [{\citenamefont {Coles}(2012{\natexlab{b}})}]{coles12b}%
  \BibitemOpen
  \bibfield  {author} {\bibinfo {author} {\bibnamefont {Coles}, \bibfnamefont
  {P.~J.}}} (\bibinfo {year} {2012}{\natexlab{b}}),\ \href {\doibase
  10.1103/PhysRevA.85.042103} {\bibfield  {journal} {\bibinfo  {journal}
  {Physical Review A}\ }\textbf {\bibinfo {volume} {85}}~(\bibinfo {number}
  {4}),\ \bibinfo {pages} {042103}}\BibitemShut {NoStop}%
\bibitem [{\citenamefont {Coles}\ \emph {et~al.}(2012)\citenamefont {Coles},
  \citenamefont {Colbeck}, \citenamefont {Yu},\ and\ \citenamefont
  {Zwolak}}]{colbeck11}%
  \BibitemOpen
  \bibfield  {author} {\bibinfo {author} {\bibnamefont {Coles}, \bibfnamefont
  {P.~J.}}, \bibinfo {author} {\bibfnamefont {R.}~\bibnamefont {Colbeck}},
  \bibinfo {author} {\bibfnamefont {L.}~\bibnamefont {Yu}}, \ and\ \bibinfo
  {author} {\bibfnamefont {M.}~\bibnamefont {Zwolak}}} (\bibinfo {year}
  {2012}),\ \href {\doibase 10.1103/PhysRevLett.108.210405} {\bibfield
  {journal} {\bibinfo  {journal} {Physical Review Letters}\ }\textbf {\bibinfo
  {volume} {108}}~(\bibinfo {number} {21}),\ \bibinfo {pages}
  {210405}}\BibitemShut {NoStop}%
\bibitem [{\citenamefont {Coles}\ and\ \citenamefont
  {Furrer}(2015)}]{coles14b}%
  \BibitemOpen
  \bibfield  {author} {\bibinfo {author} {\bibnamefont {Coles}, \bibfnamefont
  {P.~J.}}, \ and\ \bibinfo {author} {\bibfnamefont {F.}~\bibnamefont
  {Furrer}}} (\bibinfo {year} {2015}),\ \href {\doibase
  10.1016/j.physleta.2014.11.002} {\bibfield  {journal} {\bibinfo  {journal}
  {Physics Letters A}\ }\textbf {\bibinfo {volume} {379}}~(\bibinfo {number}
  {3}),\ \bibinfo {pages} {105}}\BibitemShut {NoStop}%
\bibitem [{\citenamefont {Coles}\ \emph {et~al.}(2014)\citenamefont {Coles},
  \citenamefont {Kaniewski},\ and\ \citenamefont {Wehner}}]{coles14d}%
  \BibitemOpen
  \bibfield  {author} {\bibinfo {author} {\bibnamefont {Coles}, \bibfnamefont
  {P.~J.}}, \bibinfo {author} {\bibfnamefont {J.}~\bibnamefont {Kaniewski}}, \
  and\ \bibinfo {author} {\bibfnamefont {S.}~\bibnamefont {Wehner}}} (\bibinfo
  {year} {2014}),\ \href {\doibase 10.1038/ncomms6814} {\bibfield  {journal}
  {\bibinfo  {journal} {Nature Communications}\ }\textbf {\bibinfo {volume}
  {5}},\ \bibinfo {pages} {5814}}\BibitemShut {NoStop}%
\bibitem [{\citenamefont {Coles}\ and\ \citenamefont
  {Piani}(2014{\natexlab{a}})}]{coles14c}%
  \BibitemOpen
  \bibfield  {author} {\bibinfo {author} {\bibnamefont {Coles}, \bibfnamefont
  {P.~J.}}, \ and\ \bibinfo {author} {\bibfnamefont {M.}~\bibnamefont {Piani}}}
  (\bibinfo {year} {2014}{\natexlab{a}}),\ \href {\doibase
  10.1103/PhysRevA.89.010302} {\bibfield  {journal} {\bibinfo  {journal}
  {Physical Review A}\ }\textbf {\bibinfo {volume} {89}}~(\bibinfo {number}
  {1}),\ \bibinfo {pages} {010302}}\BibitemShut {NoStop}%
\bibitem [{\citenamefont {Coles}\ and\ \citenamefont
  {Piani}(2014{\natexlab{b}})}]{coles14}%
  \BibitemOpen
  \bibfield  {author} {\bibinfo {author} {\bibnamefont {Coles}, \bibfnamefont
  {P.~J.}}, \ and\ \bibinfo {author} {\bibfnamefont {M.}~\bibnamefont {Piani}}}
  (\bibinfo {year} {2014}{\natexlab{b}}),\ \href {\doibase
  10.1103/PhysRevA.89.022112} {\bibfield  {journal} {\bibinfo  {journal}
  {Physical Review A}\ }\textbf {\bibinfo {volume} {89}}~(\bibinfo {number}
  {2}),\ \bibinfo {pages} {022112}}\BibitemShut {NoStop}%
\bibitem [{\citenamefont {Coles}\ \emph {et~al.}(2011)\citenamefont {Coles},
  \citenamefont {Yu}, \citenamefont {Gheorghiu},\ and\ \citenamefont
  {Griffiths}}]{coles10}%
  \BibitemOpen
  \bibfield  {author} {\bibinfo {author} {\bibnamefont {Coles}, \bibfnamefont
  {P.~J.}}, \bibinfo {author} {\bibfnamefont {L.}~\bibnamefont {Yu}}, \bibinfo
  {author} {\bibfnamefont {V.}~\bibnamefont {Gheorghiu}}, \ and\ \bibinfo
  {author} {\bibfnamefont {R.}~\bibnamefont {Griffiths}}} (\bibinfo {year}
  {2011}),\ \href {\doibase 10.1103/PhysRevA.83.062338} {\bibfield  {journal}
  {\bibinfo  {journal} {Physical Review A}\ }\textbf {\bibinfo {volume}
  {83}}~(\bibinfo {number} {6}),\ \bibinfo {pages} {062338}}\BibitemShut
  {NoStop}%
\bibitem [{\citenamefont {Damgaard}\ \emph {et~al.}(2007)\citenamefont
  {Damgaard}, \citenamefont {Fehr}, \citenamefont {Renner}, \citenamefont
  {Salvail},\ and\ \citenamefont {Schaffner}}]{damgaard07}%
  \BibitemOpen
  \bibfield  {author} {\bibinfo {author} {\bibnamefont {Damgaard},
  \bibfnamefont {I.~B.}}, \bibinfo {author} {\bibfnamefont {S.}~\bibnamefont
  {Fehr}}, \bibinfo {author} {\bibfnamefont {R.}~\bibnamefont {Renner}},
  \bibinfo {author} {\bibfnamefont {L.}~\bibnamefont {Salvail}}, \ and\
  \bibinfo {author} {\bibfnamefont {C.}~\bibnamefont {Schaffner}}} (\bibinfo
  {year} {2007}),\ in\ \href {\doibase 10.1007/978-3-540-74143-5_20} {\emph
  {\bibinfo {booktitle} {Proc. CRYPTO 2007}}},\ \bibinfo {series} {LNCS}, Vol.\
  \bibinfo {volume} {4622}\ (\bibinfo  {publisher} {Springer})\ pp.\ \bibinfo
  {pages} {360--378}\BibitemShut {NoStop}%
\bibitem [{\citenamefont {Damgaard}\ \emph {et~al.}(2008)\citenamefont
  {Damgaard}, \citenamefont {Fehr}, \citenamefont {Salvail},\ and\
  \citenamefont {Schaffner}}]{damgaard08}%
  \BibitemOpen
  \bibfield  {author} {\bibinfo {author} {\bibnamefont {Damgaard},
  \bibfnamefont {I.~B.}}, \bibinfo {author} {\bibfnamefont {S.}~\bibnamefont
  {Fehr}}, \bibinfo {author} {\bibfnamefont {L.}~\bibnamefont {Salvail}}, \
  and\ \bibinfo {author} {\bibfnamefont {C.}~\bibnamefont {Schaffner}}}
  (\bibinfo {year} {2008}),\ \href {\doibase 10.1137/060651343} {\bibfield
  {journal} {\bibinfo  {journal} {SIAM Journal of Computing}\ }\textbf
  {\bibinfo {volume} {37}}~(\bibinfo {number} {6}),\ \bibinfo {pages}
  {1865}}\BibitemShut {NoStop}%
\bibitem [{\citenamefont {Dammeier}\ \emph {et~al.}(2015)\citenamefont
  {Dammeier}, \citenamefont {Schwonnek},\ and\ \citenamefont
  {Werner}}]{dammeier15}%
  \BibitemOpen
  \bibfield  {author} {\bibinfo {author} {\bibnamefont {Dammeier},
  \bibfnamefont {L.}}, \bibinfo {author} {\bibfnamefont {R.}~\bibnamefont
  {Schwonnek}}, \ and\ \bibinfo {author} {\bibfnamefont {R.~F.}\ \bibnamefont
  {Werner}}} (\bibinfo {year} {2015}),\ \href {\doibase
  10.1088/1367-2630/17/9/093046} {\bibfield  {journal} {\bibinfo  {journal}
  {New Journal of Physics}\ }\textbf {\bibinfo {volume} {17}}~(\bibinfo
  {number} {9}),\ \bibinfo {pages} {093046}}\BibitemShut {NoStop}%
\bibitem [{\citenamefont {Dankert}\ \emph {et~al.}(2009)\citenamefont
  {Dankert}, \citenamefont {Cleve}, \citenamefont {Emerson},\ and\
  \citenamefont {Livine}}]{dankert09}%
  \BibitemOpen
  \bibfield  {author} {\bibinfo {author} {\bibnamefont {Dankert}, \bibfnamefont
  {C.}}, \bibinfo {author} {\bibfnamefont {R.}~\bibnamefont {Cleve}}, \bibinfo
  {author} {\bibfnamefont {J.}~\bibnamefont {Emerson}}, \ and\ \bibinfo
  {author} {\bibfnamefont {E.}~\bibnamefont {Livine}}} (\bibinfo {year}
  {2009}),\ \href {\doibase 10.1103/PhysRevA.80.012304} {\bibfield  {journal}
  {\bibinfo  {journal} {Physical Review A}\ }\textbf {\bibinfo {volume}
  {80}}~(\bibinfo {number} {1}),\ \bibinfo {pages} {012304}}\BibitemShut
  {NoStop}%
\bibitem [{\citenamefont {Davies}(1976)}]{davies76}%
  \BibitemOpen
  \bibfield  {author} {\bibinfo {author} {\bibnamefont {Davies}, \bibfnamefont
  {E.~B.}}} (\bibinfo {year} {1976}),\ \href
  {https://books.google.com.au/books?id=I5kuAAAAIAAJ} {\emph {\bibinfo {title}
  {{Quantum Theory of Open Systems}}}}\ (\bibinfo  {publisher} {Academic
  Press})\BibitemShut {NoStop}%
\bibitem [{\citenamefont {Deutsch}(1983)}]{deutsch83}%
  \BibitemOpen
  \bibfield  {author} {\bibinfo {author} {\bibnamefont {Deutsch}, \bibfnamefont
  {D.}}} (\bibinfo {year} {1983}),\ \href {\doibase 10.1103/PhysRevLett.50.631}
  {\bibfield  {journal} {\bibinfo  {journal} {Physical Review Letters}\
  }\textbf {\bibinfo {volume} {50}}~(\bibinfo {number} {9}),\ \bibinfo {pages}
  {631}}\BibitemShut {NoStop}%
\bibitem [{\citenamefont {Devetak}\ and\ \citenamefont
  {Winter}(2003)}]{devetak03}%
  \BibitemOpen
  \bibfield  {author} {\bibinfo {author} {\bibnamefont {Devetak}, \bibfnamefont
  {I.}}, \ and\ \bibinfo {author} {\bibfnamefont {A.}~\bibnamefont {Winter}}}
  (\bibinfo {year} {2003}),\ \href {\doibase 10.1103/PhysRevA.68.042301}
  {\bibfield  {journal} {\bibinfo  {journal} {Physical Review A}\ }\textbf
  {\bibinfo {volume} {68}}~(\bibinfo {number} {4}),\ \bibinfo {pages}
  {042301}}\BibitemShut {NoStop}%
\bibitem [{\citenamefont {Devetak}\ and\ \citenamefont
  {Winter}(2005)}]{devetak05}%
  \BibitemOpen
  \bibfield  {author} {\bibinfo {author} {\bibnamefont {Devetak}, \bibfnamefont
  {I.}}, \ and\ \bibinfo {author} {\bibfnamefont {A.}~\bibnamefont {Winter}}}
  (\bibinfo {year} {2005}),\ \href {\doibase 10.1098/rspa.2004.1372} {\bibfield
   {journal} {\bibinfo  {journal} {Proceedings of the Royal Society A}\
  }\textbf {\bibinfo {volume} {461}}~(\bibinfo {number} {2053}),\ \bibinfo
  {pages} {207}}\BibitemShut {NoStop}%
\bibitem [{\citenamefont {Dey}\ \emph {et~al.}(2013)\citenamefont {Dey},
  \citenamefont {Pramanik},\ and\ \citenamefont {Majumdar}}]{dey13}%
  \BibitemOpen
  \bibfield  {author} {\bibinfo {author} {\bibnamefont {Dey}, \bibfnamefont
  {A.}}, \bibinfo {author} {\bibfnamefont {T.}~\bibnamefont {Pramanik}}, \ and\
  \bibinfo {author} {\bibfnamefont {A.~S.}\ \bibnamefont {Majumdar}}} (\bibinfo
  {year} {2013}),\ \href {\doibase 10.1103/PhysRevA.87.012120} {\bibfield
  {journal} {\bibinfo  {journal} {Physical Review A}\ }\textbf {\bibinfo
  {volume} {87}}~(\bibinfo {number} {1}),\ \bibinfo {pages}
  {012120}}\BibitemShut {NoStop}%
\bibitem [{\citenamefont {Dietz}(2006)}]{dietz06}%
  \BibitemOpen
  \bibfield  {author} {\bibinfo {author} {\bibnamefont {Dietz}, \bibfnamefont
  {K.}}} (\bibinfo {year} {2006}),\ \href {\doibase 10.1088/0305-4470/39/6/016}
  {\bibfield  {journal} {\bibinfo  {journal} {Journal of Physics A:
  Mathematical and Theoretical}\ }\textbf {\bibinfo {volume} {39}}~(\bibinfo
  {number} {6}),\ \bibinfo {pages} {1433}}\BibitemShut {NoStop}%
\bibitem [{\citenamefont {DiVincenzo}\ \emph {et~al.}(2004)\citenamefont
  {DiVincenzo}, \citenamefont {Horodecki}, \citenamefont {Leung}, \citenamefont
  {Smolin},\ and\ \citenamefont {Terhal}}]{divincenzo04}%
  \BibitemOpen
  \bibfield  {author} {\bibinfo {author} {\bibnamefont {DiVincenzo},
  \bibfnamefont {D.}}, \bibinfo {author} {\bibfnamefont {M.}~\bibnamefont
  {Horodecki}}, \bibinfo {author} {\bibfnamefont {D.}~\bibnamefont {Leung}},
  \bibinfo {author} {\bibfnamefont {J.}~\bibnamefont {Smolin}}, \ and\ \bibinfo
  {author} {\bibfnamefont {B.}~\bibnamefont {Terhal}}} (\bibinfo {year}
  {2004}),\ \href {\doibase 10.1103/PhysRevLett.92.067902} {\bibfield
  {journal} {\bibinfo  {journal} {Physical Review Letters}\ }\textbf {\bibinfo
  {volume} {92}}~(\bibinfo {number} {6}),\ \bibinfo {pages}
  {067902}}\BibitemShut {NoStop}%
\bibitem [{\citenamefont {Dupuis}\ \emph {et~al.}(2015)\citenamefont {Dupuis},
  \citenamefont {Fawzi},\ and\ \citenamefont {Wehner}}]{dupuis15}%
  \BibitemOpen
  \bibfield  {author} {\bibinfo {author} {\bibnamefont {Dupuis}, \bibfnamefont
  {F.}}, \bibinfo {author} {\bibfnamefont {O.}~\bibnamefont {Fawzi}}, \ and\
  \bibinfo {author} {\bibfnamefont {S.}~\bibnamefont {Wehner}}} (\bibinfo
  {year} {2015}),\ \href {\doibase 10.1109/TIT.2014.2371464} {\bibfield
  {journal} {\bibinfo  {journal} {IEEE Transactions on Information Theory}\
  }\textbf {\bibinfo {volume} {61}}~(\bibinfo {number} {2}),\ \bibinfo {pages}
  {1093}}\BibitemShut {NoStop}%
\bibitem [{\citenamefont {Dupuis}\ \emph {et~al.}(2013)\citenamefont {Dupuis},
  \citenamefont {Florjanczyk}, \citenamefont {Hayden},\ and\ \citenamefont
  {Leung}}]{hayden13}%
  \BibitemOpen
  \bibfield  {author} {\bibinfo {author} {\bibnamefont {Dupuis}, \bibfnamefont
  {F.}}, \bibinfo {author} {\bibfnamefont {J.}~\bibnamefont {Florjanczyk}},
  \bibinfo {author} {\bibfnamefont {P.}~\bibnamefont {Hayden}}, \ and\ \bibinfo
  {author} {\bibfnamefont {D.}~\bibnamefont {Leung}}} (\bibinfo {year}
  {2013}),\ \href {\doibase 10.1098/rspa.2013.0289} {\bibfield  {journal}
  {\bibinfo  {journal} {Proceedings of the Royal Society A}\ }\textbf {\bibinfo
  {volume} {469}}~(\bibinfo {number} {2159}),\ \bibinfo {pages}
  {20130289}}\BibitemShut {NoStop}%
\bibitem [{\citenamefont {D{\"{u}}rr}\ and\ \citenamefont
  {Rempe}(2000)}]{durr00}%
  \BibitemOpen
  \bibfield  {author} {\bibinfo {author} {\bibnamefont {D{\"{u}}rr},
  \bibfnamefont {S.}}, \ and\ \bibinfo {author} {\bibfnamefont
  {G.}~\bibnamefont {Rempe}}} (\bibinfo {year} {2000}),\ \href {\doibase
  10.1119/1.1285869} {\bibfield  {journal} {\bibinfo  {journal} {American
  Journal of Physics}\ }\textbf {\bibinfo {volume} {68}}~(\bibinfo {number}
  {11}),\ \bibinfo {pages} {1021}}\BibitemShut {NoStop}%
\bibitem [{\citenamefont {Durt}\ \emph {et~al.}(2010)\citenamefont {Durt},
  \citenamefont {Englert}, \citenamefont {Bengtsson},\ and\ \citenamefont
  {{\.{Z}}yczkowski}}]{durt10}%
  \BibitemOpen
  \bibfield  {author} {\bibinfo {author} {\bibnamefont {Durt}, \bibfnamefont
  {T.}}, \bibinfo {author} {\bibfnamefont {B.-G.}\ \bibnamefont {Englert}},
  \bibinfo {author} {\bibfnamefont {I.}~\bibnamefont {Bengtsson}}, \ and\
  \bibinfo {author} {\bibfnamefont {K.}~\bibnamefont {{\.{Z}}yczkowski}}}
  (\bibinfo {year} {2010}),\ \href {\doibase 10.1142/S0219749910006502}
  {\bibfield  {journal} {\bibinfo  {journal} {International Journal of Quantum
  Information}\ }\textbf {\bibinfo {volume} {08}}~(\bibinfo {number} {04}),\
  \bibinfo {pages} {535}}\BibitemShut {NoStop}%
\bibitem [{\citenamefont {Dziembowski}\ and\ \citenamefont
  {Maurer}(2004)}]{dziembowski04}%
  \BibitemOpen
  \bibfield  {author} {\bibinfo {author} {\bibnamefont {Dziembowski},
  \bibfnamefont {S.}}, \ and\ \bibinfo {author} {\bibfnamefont
  {U.}~\bibnamefont {Maurer}}} (\bibinfo {year} {2004}),\ in\ \href {\doibase
  10.1007/978-3-540-24676-3_8} {\emph {\bibinfo {booktitle} {Proc. EUROCRYPT
  2004}}},\ \bibinfo {series} {LNCS}, Vol.\ \bibinfo {volume} {3027}\ (\bibinfo
   {publisher} {Springer})\ pp.\ \bibinfo {pages} {126--137}\BibitemShut
  {NoStop}%
\bibitem [{\citenamefont {Eberle}\ \emph {et~al.}(2013)\citenamefont {Eberle},
  \citenamefont {H{\"{a}}ndchen},\ and\ \citenamefont {Schnabel}}]{schnabel13}%
  \BibitemOpen
  \bibfield  {author} {\bibinfo {author} {\bibnamefont {Eberle}, \bibfnamefont
  {T.}}, \bibinfo {author} {\bibfnamefont {V.}~\bibnamefont {H{\"{a}}ndchen}},
  \ and\ \bibinfo {author} {\bibfnamefont {R.}~\bibnamefont {Schnabel}}}
  (\bibinfo {year} {2013}),\ \href {\doibase 10.1364/OE.21.011546} {\bibfield
  {journal} {\bibinfo  {journal} {Optics Express}\ }\textbf {\bibinfo {volume}
  {21}}~(\bibinfo {number} {9}),\ \bibinfo {pages} {11546}}\BibitemShut
  {NoStop}%
\bibitem [{\citenamefont {Einstein}\ \emph {et~al.}(1935)\citenamefont
  {Einstein}, \citenamefont {Podolsky},\ and\ \citenamefont {Rosen}}]{epr35}%
  \BibitemOpen
  \bibfield  {author} {\bibinfo {author} {\bibnamefont {Einstein},
  \bibfnamefont {A.}}, \bibinfo {author} {\bibfnamefont {B.}~\bibnamefont
  {Podolsky}}, \ and\ \bibinfo {author} {\bibfnamefont {N.}~\bibnamefont
  {Rosen}}} (\bibinfo {year} {1935}),\ \href {\doibase 10.1103/PhysRev.47.777}
  {\bibfield  {journal} {\bibinfo  {journal} {Physical Review}\ }\textbf
  {\bibinfo {volume} {47}}~(\bibinfo {number} {10}),\ \bibinfo {pages}
  {777}}\BibitemShut {NoStop}%
\bibitem [{\citenamefont {Ekert}(1991)}]{ekert91}%
  \BibitemOpen
  \bibfield  {author} {\bibinfo {author} {\bibnamefont {Ekert}, \bibfnamefont
  {A.~K.}}} (\bibinfo {year} {1991}),\ \href {\doibase
  10.1103/PhysRevLett.67.661} {\bibfield  {journal} {\bibinfo  {journal}
  {Physical Review Letters}\ }\textbf {\bibinfo {volume} {67}}~(\bibinfo
  {number} {6}),\ \bibinfo {pages} {661}}\BibitemShut {NoStop}%
\bibitem [{\citenamefont {Englert}(1996)}]{englert96}%
  \BibitemOpen
  \bibfield  {author} {\bibinfo {author} {\bibnamefont {Englert}, \bibfnamefont
  {B.-G.}}} (\bibinfo {year} {1996}),\ \href {\doibase
  10.1103/PhysRevLett.77.2154} {\bibfield  {journal} {\bibinfo  {journal}
  {Physical Review Letters}\ }\textbf {\bibinfo {volume} {77}}~(\bibinfo
  {number} {11}),\ \bibinfo {pages} {2154}}\BibitemShut {NoStop}%
\bibitem [{\citenamefont {Englert}\ and\ \citenamefont
  {Bergou}(2000)}]{englert00}%
  \BibitemOpen
  \bibfield  {author} {\bibinfo {author} {\bibnamefont {Englert}, \bibfnamefont
  {B.-G.}}, \ and\ \bibinfo {author} {\bibfnamefont {J.~A.}\ \bibnamefont
  {Bergou}}} (\bibinfo {year} {2000}),\ \href {\doibase
  10.1016/S0030-4018(99)00718-X} {\bibfield  {journal} {\bibinfo  {journal}
  {Optics Communications}\ }\textbf {\bibinfo {volume} {179}}~(\bibinfo
  {number} {1-6}),\ \bibinfo {pages} {337}}\BibitemShut {NoStop}%
\bibitem [{\citenamefont {Englert}\ \emph {et~al.}(2008)\citenamefont
  {Englert}, \citenamefont {Kaszlikowski}, \citenamefont {Kwek},\ and\
  \citenamefont {Chee}}]{englert08}%
  \BibitemOpen
  \bibfield  {author} {\bibinfo {author} {\bibnamefont {Englert}, \bibfnamefont
  {B.-G.}}, \bibinfo {author} {\bibfnamefont {D.}~\bibnamefont {Kaszlikowski}},
  \bibinfo {author} {\bibfnamefont {L.~C.}\ \bibnamefont {Kwek}}, \ and\
  \bibinfo {author} {\bibfnamefont {W.~H.}\ \bibnamefont {Chee}}} (\bibinfo
  {year} {2008}),\ \href {\doibase 10.1142/S0219749908003220} {\bibfield
  {journal} {\bibinfo  {journal} {International Journal of Quantum
  Information}\ }\textbf {\bibinfo {volume} {06}}~(\bibinfo {number} {01}),\
  \bibinfo {pages} {129}}\BibitemShut {NoStop}%
\bibitem [{\citenamefont {Everett}(1957)}]{everett57}%
  \BibitemOpen
  \bibfield  {author} {\bibinfo {author} {\bibnamefont {Everett}, \bibfnamefont
  {H.}}} (\bibinfo {year} {1957}),\ \href {\doibase 10.1103/RevModPhys.29.454}
  {\bibfield  {journal} {\bibinfo  {journal} {Review of Modern Physics}\
  }\textbf {\bibinfo {volume} {29}}~(\bibinfo {number} {3}),\ \bibinfo {pages}
  {454}}\BibitemShut {NoStop}%
\bibitem [{\citenamefont {Fawzi}\ \emph {et~al.}(2011)\citenamefont {Fawzi},
  \citenamefont {Hayden},\ and\ \citenamefont {Sen}}]{fawzi11}%
  \BibitemOpen
  \bibfield  {author} {\bibinfo {author} {\bibnamefont {Fawzi}, \bibfnamefont
  {O.}}, \bibinfo {author} {\bibfnamefont {P.}~\bibnamefont {Hayden}}, \ and\
  \bibinfo {author} {\bibfnamefont {P.}~\bibnamefont {Sen}}} (\bibinfo {year}
  {2011}),\ in\ \href {\doibase 10.1145/1993636.1993738} {\emph {\bibinfo
  {booktitle} {Proc. ACM STOC 2011}}}\ (\bibinfo  {publisher} {ACM Press},\
  \bibinfo {address} {New York, NY})\ pp.\ \bibinfo {pages}
  {773--782}\BibitemShut {NoStop}%
\bibitem [{\citenamefont {Feng}\ \emph {et~al.}(2013)\citenamefont {Feng},
  \citenamefont {Zhang}, \citenamefont {Gould},\ and\ \citenamefont
  {Fan}}]{feng13}%
  \BibitemOpen
  \bibfield  {author} {\bibinfo {author} {\bibnamefont {Feng}, \bibfnamefont
  {J.}}, \bibinfo {author} {\bibfnamefont {Y.-Z.}\ \bibnamefont {Zhang}},
  \bibinfo {author} {\bibfnamefont {M.~D.}\ \bibnamefont {Gould}}, \ and\
  \bibinfo {author} {\bibfnamefont {H.}~\bibnamefont {Fan}}} (\bibinfo {year}
  {2013}),\ \href {\doibase 10.1016/j.physletb.2013.08.069} {\bibfield
  {journal} {\bibinfo  {journal} {Physics Letters B}\ }\textbf {\bibinfo
  {volume} {726}}~(\bibinfo {number} {1-3}),\ \bibinfo {pages}
  {527}}\BibitemShut {NoStop}%
\bibitem [{\citenamefont {Frank}\ and\ \citenamefont {Lieb}(2012)}]{frank12-2}%
  \BibitemOpen
  \bibfield  {author} {\bibinfo {author} {\bibnamefont {Frank}, \bibfnamefont
  {R.~L.}}, \ and\ \bibinfo {author} {\bibfnamefont {E.~H.}\ \bibnamefont
  {Lieb}}} (\bibinfo {year} {2012}),\ \href {\doibase
  10.1007/s00023-012-0175-y} {\bibfield  {journal} {\bibinfo  {journal} {Annals
  Henri Poincar{\'{e}}}\ }\textbf {\bibinfo {volume} {13}}~(\bibinfo {number}
  {8}),\ \bibinfo {pages} {1711}}\BibitemShut {NoStop}%
\bibitem [{\citenamefont {Frank}\ and\ \citenamefont
  {Lieb}(2013{\natexlab{a}})}]{frank12}%
  \BibitemOpen
  \bibfield  {author} {\bibinfo {author} {\bibnamefont {Frank}, \bibfnamefont
  {R.~L.}}, \ and\ \bibinfo {author} {\bibfnamefont {E.~H.}\ \bibnamefont
  {Lieb}}} (\bibinfo {year} {2013}{\natexlab{a}}),\ \href {\doibase
  10.1007/s00220-013-1775-1} {\bibfield  {journal} {\bibinfo  {journal}
  {Communications in Mathematical Physics}\ }\textbf {\bibinfo {volume}
  {323}}~(\bibinfo {number} {2}),\ \bibinfo {pages} {487}}\BibitemShut
  {NoStop}%
\bibitem [{\citenamefont {Frank}\ and\ \citenamefont
  {Lieb}(2013{\natexlab{b}})}]{frank13}%
  \BibitemOpen
  \bibfield  {author} {\bibinfo {author} {\bibnamefont {Frank}, \bibfnamefont
  {R.~L.}}, \ and\ \bibinfo {author} {\bibfnamefont {E.~H.}\ \bibnamefont
  {Lieb}}} (\bibinfo {year} {2013}{\natexlab{b}}),\ \href {\doibase
  10.1063/1.4838835} {\bibfield  {journal} {\bibinfo  {journal} {Journal of
  Mathematical Physics}\ }\textbf {\bibinfo {volume} {54}}~(\bibinfo {number}
  {12}),\ \bibinfo {pages} {122201}}\BibitemShut {NoStop}%
\bibitem [{\citenamefont {Friedland}\ \emph {et~al.}(2013)\citenamefont
  {Friedland}, \citenamefont {Gheorghiu},\ and\ \citenamefont
  {Gour}}]{friedland13}%
  \BibitemOpen
  \bibfield  {author} {\bibinfo {author} {\bibnamefont {Friedland},
  \bibfnamefont {S.}}, \bibinfo {author} {\bibfnamefont {V.}~\bibnamefont
  {Gheorghiu}}, \ and\ \bibinfo {author} {\bibfnamefont {G.}~\bibnamefont
  {Gour}}} (\bibinfo {year} {2013}),\ \href {\doibase
  10.1103/PhysRevLett.111.230401} {\bibfield  {journal} {\bibinfo  {journal}
  {Physical Review Letters}\ }\textbf {\bibinfo {volume} {111}}~(\bibinfo
  {number} {23}),\ \bibinfo {pages} {230401}}\BibitemShut {NoStop}%
\bibitem [{\citenamefont {Furrer}\ \emph {et~al.}(2011)\citenamefont {Furrer},
  \citenamefont {Aberg},\ and\ \citenamefont {Renner}}]{furrer10}%
  \BibitemOpen
  \bibfield  {author} {\bibinfo {author} {\bibnamefont {Furrer}, \bibfnamefont
  {F.}}, \bibinfo {author} {\bibfnamefont {J.}~\bibnamefont {Aberg}}, \ and\
  \bibinfo {author} {\bibfnamefont {R.}~\bibnamefont {Renner}}} (\bibinfo
  {year} {2011}),\ \href {\doibase 10.1007/s00220-011-1282-1} {\bibfield
  {journal} {\bibinfo  {journal} {Communications in Mathematical Physics}\
  }\textbf {\bibinfo {volume} {306}}~(\bibinfo {number} {1}),\ \bibinfo {pages}
  {165}}\BibitemShut {NoStop}%
\bibitem [{\citenamefont {Furrer}\ \emph {et~al.}(2014)\citenamefont {Furrer},
  \citenamefont {Berta}, \citenamefont {Tomamichel}, \citenamefont {Scholz},\
  and\ \citenamefont {Christandl}}]{furrer13}%
  \BibitemOpen
  \bibfield  {author} {\bibinfo {author} {\bibnamefont {Furrer}, \bibfnamefont
  {F.}}, \bibinfo {author} {\bibfnamefont {M.}~\bibnamefont {Berta}}, \bibinfo
  {author} {\bibfnamefont {M.}~\bibnamefont {Tomamichel}}, \bibinfo {author}
  {\bibfnamefont {V.~B.}\ \bibnamefont {Scholz}}, \ and\ \bibinfo {author}
  {\bibfnamefont {M.}~\bibnamefont {Christandl}}} (\bibinfo {year} {2014}),\
  \href {\doibase 10.1063/1.4903989} {\bibfield  {journal} {\bibinfo  {journal}
  {Journal of Mathematical Physics}\ }\textbf {\bibinfo {volume} {55}},\
  \bibinfo {pages} {122205}}\BibitemShut {NoStop}%
\bibitem [{\citenamefont {Furrer}\ \emph {et~al.}(2012)\citenamefont {Furrer},
  \citenamefont {Franz}, \citenamefont {Berta}, \citenamefont {Leverrier},
  \citenamefont {Scholz}, \citenamefont {Tomamichel},\ and\ \citenamefont
  {Werner}}]{furrer12}%
  \BibitemOpen
  \bibfield  {author} {\bibinfo {author} {\bibnamefont {Furrer}, \bibfnamefont
  {F.}}, \bibinfo {author} {\bibfnamefont {T.}~\bibnamefont {Franz}}, \bibinfo
  {author} {\bibfnamefont {M.}~\bibnamefont {Berta}}, \bibinfo {author}
  {\bibfnamefont {A.}~\bibnamefont {Leverrier}}, \bibinfo {author}
  {\bibfnamefont {V.~B.}\ \bibnamefont {Scholz}}, \bibinfo {author}
  {\bibfnamefont {M.}~\bibnamefont {Tomamichel}}, \ and\ \bibinfo {author}
  {\bibfnamefont {R.~F.}\ \bibnamefont {Werner}}} (\bibinfo {year} {2012}),\
  \href {\doibase 10.1103/PhysRevLett.109.100502} {\bibfield  {journal}
  {\bibinfo  {journal} {Physical Review Letters}\ }\textbf {\bibinfo {volume}
  {109}}~(\bibinfo {number} {10}),\ \bibinfo {pages} {100502}}\BibitemShut
  {NoStop}%
\bibitem [{\citenamefont {Gavinsky}\ \emph {et~al.}(2009)\citenamefont
  {Gavinsky}, \citenamefont {Kempe}, \citenamefont {Kerenidis}, \citenamefont
  {Raz},\ and\ \citenamefont {de~Wolf}}]{gavinsky07}%
  \BibitemOpen
  \bibfield  {author} {\bibinfo {author} {\bibnamefont {Gavinsky},
  \bibfnamefont {D.}}, \bibinfo {author} {\bibfnamefont {J.}~\bibnamefont
  {Kempe}}, \bibinfo {author} {\bibfnamefont {I.}~\bibnamefont {Kerenidis}},
  \bibinfo {author} {\bibfnamefont {R.}~\bibnamefont {Raz}}, \ and\ \bibinfo
  {author} {\bibfnamefont {R.}~\bibnamefont {de~Wolf}}} (\bibinfo {year}
  {2009}),\ \href {\doibase 10.1137/070706550} {\bibfield  {journal} {\bibinfo
  {journal} {SIAM Journal of Computing}\ }\textbf {\bibinfo {volume}
  {38}}~(\bibinfo {number} {5}),\ \bibinfo {pages} {1695}}\BibitemShut
  {NoStop}%
\bibitem [{\citenamefont {Gehring}\ \emph {et~al.}(2015)\citenamefont
  {Gehring}, \citenamefont {H{\"{a}}ndchen}, \citenamefont {Duhme},
  \citenamefont {Furrer}, \citenamefont {Franz}, \citenamefont {Pacher},
  \citenamefont {Werner},\ and\ \citenamefont {Schnabel}}]{gehring15}%
  \BibitemOpen
  \bibfield  {author} {\bibinfo {author} {\bibnamefont {Gehring}, \bibfnamefont
  {T.}}, \bibinfo {author} {\bibfnamefont {V.}~\bibnamefont {H{\"{a}}ndchen}},
  \bibinfo {author} {\bibfnamefont {J.}~\bibnamefont {Duhme}}, \bibinfo
  {author} {\bibfnamefont {F.}~\bibnamefont {Furrer}}, \bibinfo {author}
  {\bibfnamefont {T.}~\bibnamefont {Franz}}, \bibinfo {author} {\bibfnamefont
  {C.}~\bibnamefont {Pacher}}, \bibinfo {author} {\bibfnamefont {R.~F.}\
  \bibnamefont {Werner}}, \ and\ \bibinfo {author} {\bibfnamefont
  {R.}~\bibnamefont {Schnabel}}} (\bibinfo {year} {2015}),\ \href {\doibase
  10.1038/ncomms9795} {\bibfield  {journal} {\bibinfo  {journal} {Nature
  Communications}\ }\textbf {\bibinfo {volume} {6}},\ \bibinfo {pages}
  {8795}}\BibitemShut {NoStop}%
\bibitem [{\citenamefont {Ghirardi}\ \emph {et~al.}(2003)\citenamefont
  {Ghirardi}, \citenamefont {Marinatto},\ and\ \citenamefont
  {Romano}}]{ghirardi03}%
  \BibitemOpen
  \bibfield  {author} {\bibinfo {author} {\bibnamefont {Ghirardi},
  \bibfnamefont {G.}}, \bibinfo {author} {\bibfnamefont {L.}~\bibnamefont
  {Marinatto}}, \ and\ \bibinfo {author} {\bibfnamefont {R.}~\bibnamefont
  {Romano}}} (\bibinfo {year} {2003}),\ \href {\doibase
  10.1016/j.physleta.2003.08.029} {\bibfield  {journal} {\bibinfo  {journal}
  {Physics Letters A}\ }\textbf {\bibinfo {volume} {317}}~(\bibinfo {number}
  {1-2}),\ \bibinfo {pages} {32}}\BibitemShut {NoStop}%
\bibitem [{\citenamefont {Gibbs}(1876)}]{gibbs1876}%
  \BibitemOpen
  \bibfield  {author} {\bibinfo {author} {\bibnamefont {Gibbs}, \bibfnamefont
  {J.~W.}}} (\bibinfo {year} {1876}),\ \href@noop {} {\bibfield  {journal}
  {\bibinfo  {journal} {Transactions of the Connecticut Academy of Arts and
  Sciences}\ }\textbf {\bibinfo {volume} {III}},\ \bibinfo {pages}
  {108}}\BibitemShut {NoStop}%
\bibitem [{\citenamefont {Giovannetti}(2004)}]{giovannetti04}%
  \BibitemOpen
  \bibfield  {author} {\bibinfo {author} {\bibnamefont {Giovannetti},
  \bibfnamefont {V.}}} (\bibinfo {year} {2004}),\ \href {\doibase
  10.1103/PhysRevA.70.012102} {\bibfield  {journal} {\bibinfo  {journal}
  {Physical Review A}\ }\textbf {\bibinfo {volume} {70}}~(\bibinfo {number}
  {1}),\ \bibinfo {pages} {012102}}\BibitemShut {NoStop}%
\bibitem [{\citenamefont {Giovannetti}\ \emph {et~al.}(2011)\citenamefont
  {Giovannetti}, \citenamefont {Lloyd},\ and\ \citenamefont
  {Maccone}}]{giovannetti11}%
  \BibitemOpen
  \bibfield  {author} {\bibinfo {author} {\bibnamefont {Giovannetti},
  \bibfnamefont {V.}}, \bibinfo {author} {\bibfnamefont {S.}~\bibnamefont
  {Lloyd}}, \ and\ \bibinfo {author} {\bibfnamefont {L.}~\bibnamefont
  {Maccone}}} (\bibinfo {year} {2011}),\ \href {\doibase
  10.1038/nphoton.2011.35} {\bibfield  {journal} {\bibinfo  {journal} {Nature
  Photonics}\ }\textbf {\bibinfo {volume} {5}}~(\bibinfo {number} {4}),\
  \bibinfo {pages} {222}}\BibitemShut {NoStop}%
\bibitem [{\citenamefont {Grosshans}\ and\ \citenamefont
  {Cerf}(2004)}]{grosshans04}%
  \BibitemOpen
  \bibfield  {author} {\bibinfo {author} {\bibnamefont {Grosshans},
  \bibfnamefont {F.}}, \ and\ \bibinfo {author} {\bibfnamefont {N.~J.}\
  \bibnamefont {Cerf}}} (\bibinfo {year} {2004}),\ \href {\doibase
  10.1103/PhysRevLett.92.047905} {\bibfield  {journal} {\bibinfo  {journal}
  {Physical Review Letters}\ }\textbf {\bibinfo {volume} {92}}~(\bibinfo
  {number} {4}),\ \bibinfo {pages} {047905}}\BibitemShut {NoStop}%
\bibitem [{\citenamefont {Grudka}\ \emph {et~al.}(2013)\citenamefont {Grudka},
  \citenamefont {Horodecki}, \citenamefont {Horodecki}, \citenamefont
  {Horodecki}, \citenamefont {Klobus},\ and\ \citenamefont
  {Pankowski}}]{grudka12}%
  \BibitemOpen
  \bibfield  {author} {\bibinfo {author} {\bibnamefont {Grudka}, \bibfnamefont
  {A.}}, \bibinfo {author} {\bibfnamefont {M.}~\bibnamefont {Horodecki}},
  \bibinfo {author} {\bibfnamefont {P.}~\bibnamefont {Horodecki}}, \bibinfo
  {author} {\bibfnamefont {R.}~\bibnamefont {Horodecki}}, \bibinfo {author}
  {\bibfnamefont {W.}~\bibnamefont {Klobus}}, \ and\ \bibinfo {author}
  {\bibfnamefont {L.}~\bibnamefont {Pankowski}}} (\bibinfo {year} {2013}),\
  \href {\doibase 10.1103/PhysRevA.88.032106} {\bibfield  {journal} {\bibinfo
  {journal} {Physical Review A}\ }\textbf {\bibinfo {volume} {88}}~(\bibinfo
  {number} {3}),\ \bibinfo {pages} {032106}}\BibitemShut {NoStop}%
\bibitem [{\citenamefont {Guha}\ \emph {et~al.}(2014)\citenamefont {Guha},
  \citenamefont {Hayden}, \citenamefont {Krovi}, \citenamefont {Lloyd},
  \citenamefont {Lupo}, \citenamefont {Shapiro}, \citenamefont {Takeoka},\ and\
  \citenamefont {Wilde}}]{guha14}%
  \BibitemOpen
  \bibfield  {author} {\bibinfo {author} {\bibnamefont {Guha}, \bibfnamefont
  {S.}}, \bibinfo {author} {\bibfnamefont {P.}~\bibnamefont {Hayden}}, \bibinfo
  {author} {\bibfnamefont {H.}~\bibnamefont {Krovi}}, \bibinfo {author}
  {\bibfnamefont {S.}~\bibnamefont {Lloyd}}, \bibinfo {author} {\bibfnamefont
  {C.}~\bibnamefont {Lupo}}, \bibinfo {author} {\bibfnamefont {J.~H.}\
  \bibnamefont {Shapiro}}, \bibinfo {author} {\bibfnamefont {M.}~\bibnamefont
  {Takeoka}}, \ and\ \bibinfo {author} {\bibfnamefont {M.~M.}\ \bibnamefont
  {Wilde}}} (\bibinfo {year} {2014}),\ \href {\doibase
  10.1103/PhysRevX.4.011016} {\bibfield  {journal} {\bibinfo  {journal}
  {Physical Review X}\ }\textbf {\bibinfo {volume} {4}}~(\bibinfo {number}
  {1}),\ \bibinfo {pages} {011016}}\BibitemShut {NoStop}%
\bibitem [{\citenamefont {G{\"{u}}hne}\ and\ \citenamefont
  {Lewenstein}(2004)}]{guhne04}%
  \BibitemOpen
  \bibfield  {author} {\bibinfo {author} {\bibnamefont {G{\"{u}}hne},
  \bibfnamefont {O.}}, \ and\ \bibinfo {author} {\bibfnamefont
  {M.}~\bibnamefont {Lewenstein}}} (\bibinfo {year} {2004}),\ \href {\doibase
  10.1103/PhysRevA.70.022316} {\bibfield  {journal} {\bibinfo  {journal}
  {Physical Review A}\ }\textbf {\bibinfo {volume} {70}}~(\bibinfo {number}
  {2}),\ \bibinfo {pages} {022316}}\BibitemShut {NoStop}%
\bibitem [{\citenamefont {G{\"{u}}hne}\ and\ \citenamefont
  {T{\'{o}}th}(2009)}]{guhne09}%
  \BibitemOpen
  \bibfield  {author} {\bibinfo {author} {\bibnamefont {G{\"{u}}hne},
  \bibfnamefont {O.}}, \ and\ \bibinfo {author} {\bibfnamefont
  {G.}~\bibnamefont {T{\'{o}}th}}} (\bibinfo {year} {2009}),\ \href {\doibase
  10.1016/j.physrep.2009.02.004} {\bibfield  {journal} {\bibinfo  {journal}
  {Physics Reports}\ }\textbf {\bibinfo {volume} {474}}~(\bibinfo {number}
  {1-6}),\ \bibinfo {pages} {1}}\BibitemShut {NoStop}%
\bibitem [{\citenamefont {Hall}(1993)}]{hall92}%
  \BibitemOpen
  \bibfield  {author} {\bibinfo {author} {\bibnamefont {Hall}, \bibfnamefont
  {M.~J.~W.}}} (\bibinfo {year} {1993}),\ \href {\doibase
  10.1080/09500349314550841} {\bibfield  {journal} {\bibinfo  {journal}
  {Journal of Modern Optics}\ }\textbf {\bibinfo {volume} {40}}~(\bibinfo
  {number} {5}),\ \bibinfo {pages} {809}}\BibitemShut {NoStop}%
\bibitem [{\citenamefont {Hall}(1994)}]{hall94}%
  \BibitemOpen
  \bibfield  {author} {\bibinfo {author} {\bibnamefont {Hall}, \bibfnamefont
  {M.~J.~W.}}} (\bibinfo {year} {1994}),\ \href {\doibase
  10.1103/PhysRevA.49.42} {\bibfield  {journal} {\bibinfo  {journal} {Physical
  Review A}\ }\textbf {\bibinfo {volume} {49}}~(\bibinfo {number} {1}),\
  \bibinfo {pages} {42}}\BibitemShut {NoStop}%
\bibitem [{\citenamefont {Hall}(1995)}]{hall95}%
  \BibitemOpen
  \bibfield  {author} {\bibinfo {author} {\bibnamefont {Hall}, \bibfnamefont
  {M.~J.~W.}}} (\bibinfo {year} {1995}),\ \href {\doibase
  10.1103/PhysRevLett.74.3307} {\bibfield  {journal} {\bibinfo  {journal}
  {Physical Review Letters}\ }\textbf {\bibinfo {volume} {74}}~(\bibinfo
  {number} {17}),\ \bibinfo {pages} {3307}}\BibitemShut {NoStop}%
\bibitem [{\citenamefont {Hall}(1997)}]{hall97}%
  \BibitemOpen
  \bibfield  {author} {\bibinfo {author} {\bibnamefont {Hall}, \bibfnamefont
  {M.~J.~W.}}} (\bibinfo {year} {1997}),\ \href {\doibase
  10.1103/PhysRevA.55.100} {\bibfield  {journal} {\bibinfo  {journal} {Physical
  Review A}\ }\textbf {\bibinfo {volume} {55}}~(\bibinfo {number} {1}),\
  \bibinfo {pages} {100}}\BibitemShut {NoStop}%
\bibitem [{\citenamefont {Hall}(1999)}]{hall99}%
  \BibitemOpen
  \bibfield  {author} {\bibinfo {author} {\bibnamefont {Hall}, \bibfnamefont
  {M.~J.~W.}}} (\bibinfo {year} {1999}),\ \href {\doibase
  10.1103/PhysRevA.59.2602} {\bibfield  {journal} {\bibinfo  {journal}
  {Physical Review A}\ }\textbf {\bibinfo {volume} {59}}~(\bibinfo {number}
  {4}),\ \bibinfo {pages} {2602}}\BibitemShut {NoStop}%
\bibitem [{\citenamefont {Hall}(2004)}]{hall04}%
  \BibitemOpen
  \bibfield  {author} {\bibinfo {author} {\bibnamefont {Hall}, \bibfnamefont
  {M.~J.~W.}}} (\bibinfo {year} {2004}),\ \href {\doibase
  10.1103/PhysRevA.69.052113} {\bibfield  {journal} {\bibinfo  {journal}
  {Physical Review A}\ }\textbf {\bibinfo {volume} {69}}~(\bibinfo {number}
  {5}),\ \bibinfo {pages} {052113}}\BibitemShut {NoStop}%
\bibitem [{\citenamefont {Hall}(2008)}]{hall08}%
  \BibitemOpen
  \bibfield  {author} {\bibinfo {author} {\bibnamefont {Hall}, \bibfnamefont
  {M.~J.~W.}}} (\bibinfo {year} {2008}),\ \href {\doibase
  10.1088/1751-8113/41/25/255301} {\bibfield  {journal} {\bibinfo  {journal}
  {Journal of Physics A}\ }\textbf {\bibinfo {volume} {41}}~(\bibinfo {number}
  {25}),\ \bibinfo {pages} {255301}}\BibitemShut {NoStop}%
\bibitem [{\citenamefont {Hall}\ \emph {et~al.}(2012)\citenamefont {Hall},
  \citenamefont {Berry}, \citenamefont {Zwierz},\ and\ \citenamefont
  {Wiseman}}]{hall12b}%
  \BibitemOpen
  \bibfield  {author} {\bibinfo {author} {\bibnamefont {Hall}, \bibfnamefont
  {M.~J.~W.}}, \bibinfo {author} {\bibfnamefont {D.~W.}\ \bibnamefont {Berry}},
  \bibinfo {author} {\bibfnamefont {M.}~\bibnamefont {Zwierz}}, \ and\ \bibinfo
  {author} {\bibfnamefont {H.~M.}\ \bibnamefont {Wiseman}}} (\bibinfo {year}
  {2012}),\ \href {\doibase 10.1103/PhysRevA.85.041802} {\bibfield  {journal}
  {\bibinfo  {journal} {Physical Review A}\ }\textbf {\bibinfo {volume}
  {85}}~(\bibinfo {number} {4}),\ \bibinfo {pages} {041802}}\BibitemShut
  {NoStop}%
\bibitem [{\citenamefont {Hall}\ and\ \citenamefont {Wiseman}(2012)}]{hall12}%
  \BibitemOpen
  \bibfield  {author} {\bibinfo {author} {\bibnamefont {Hall}, \bibfnamefont
  {M.~J.~W.}}, \ and\ \bibinfo {author} {\bibfnamefont {H.~M.}\ \bibnamefont
  {Wiseman}}} (\bibinfo {year} {2012}),\ \href {\doibase
  10.1088/1367-2630/14/3/033040} {\bibfield  {journal} {\bibinfo  {journal}
  {New Journal of Physics}\ }\textbf {\bibinfo {volume} {14}}~(\bibinfo
  {number} {3}),\ \bibinfo {pages} {033040}}\BibitemShut {NoStop}%
\bibitem [{\citenamefont {Hausladen}\ and\ \citenamefont
  {Wootters}(1994)}]{hausladen94}%
  \BibitemOpen
  \bibfield  {author} {\bibinfo {author} {\bibnamefont {Hausladen},
  \bibfnamefont {P.}}, \ and\ \bibinfo {author} {\bibfnamefont {W.~K.}\
  \bibnamefont {Wootters}}} (\bibinfo {year} {1994}),\ \href {\doibase
  10.1080/09500349414552221} {\bibfield  {journal} {\bibinfo  {journal}
  {Journal of Modern Optics}\ }\textbf {\bibinfo {volume} {41}}~(\bibinfo
  {number} {12}),\ \bibinfo {pages} {2385}}\BibitemShut {NoStop}%
\bibitem [{\citenamefont {Hayden}\ \emph {et~al.}(2004)\citenamefont {Hayden},
  \citenamefont {Leung}, \citenamefont {Shor},\ and\ \citenamefont
  {Winter}}]{hayden04b}%
  \BibitemOpen
  \bibfield  {author} {\bibinfo {author} {\bibnamefont {Hayden}, \bibfnamefont
  {P.}}, \bibinfo {author} {\bibfnamefont {D.}~\bibnamefont {Leung}}, \bibinfo
  {author} {\bibfnamefont {P.~W.}\ \bibnamefont {Shor}}, \ and\ \bibinfo
  {author} {\bibfnamefont {A.}~\bibnamefont {Winter}}} (\bibinfo {year}
  {2004}),\ \href {\doibase 10.1007/s00220-004-1087-6} {\bibfield  {journal}
  {\bibinfo  {journal} {Communications in Mathematical Physics}\ }\textbf
  {\bibinfo {volume} {250}}~(\bibinfo {number} {2}),\ \bibinfo {pages}
  {1}}\BibitemShut {NoStop}%
\bibitem [{\citenamefont {Hayden}\ and\ \citenamefont
  {Preskill}(2007)}]{hayden07}%
  \BibitemOpen
  \bibfield  {author} {\bibinfo {author} {\bibnamefont {Hayden}, \bibfnamefont
  {P.}}, \ and\ \bibinfo {author} {\bibfnamefont {J.}~\bibnamefont {Preskill}}}
  (\bibinfo {year} {2007}),\ \href {\doibase 10.1088/1126-6708/2007/09/120}
  {\bibfield  {journal} {\bibinfo  {journal} {Journal of High Energy Physics}\
  }\textbf {\bibinfo {volume} {2007}}~(\bibinfo {number} {09}),\ \bibinfo
  {pages} {120}}\BibitemShut {NoStop}%
\bibitem [{\citenamefont {Heisenberg}(1927)}]{heisenberg27}%
  \BibitemOpen
  \bibfield  {author} {\bibinfo {author} {\bibnamefont {Heisenberg},
  \bibfnamefont {W.}}} (\bibinfo {year} {1927}),\ \href@noop {} {\bibfield
  {journal} {\bibinfo  {journal} {Zeitschrift f{\"{u}}r Physik}\ }\textbf
  {\bibinfo {volume} {43}}~(\bibinfo {number} {3-4}),\ \bibinfo {pages}
  {172}}\BibitemShut {NoStop}%
\bibitem [{\citenamefont {Hiai}\ and\ \citenamefont {Petz}(1991)}]{hiai91}%
  \BibitemOpen
  \bibfield  {author} {\bibinfo {author} {\bibnamefont {Hiai}, \bibfnamefont
  {F.}}, \ and\ \bibinfo {author} {\bibfnamefont {D.}~\bibnamefont {Petz}}}
  (\bibinfo {year} {1991}),\ \href {\doibase 10.1007/BF02100287} {\bibfield
  {journal} {\bibinfo  {journal} {Communications in Mathematical Physics}\
  }\textbf {\bibinfo {volume} {143}}~(\bibinfo {number} {1}),\ \bibinfo {pages}
  {99}}\BibitemShut {NoStop}%
\bibitem [{\citenamefont {Hirschman}(1957)}]{hirschman57}%
  \BibitemOpen
  \bibfield  {author} {\bibinfo {author} {\bibnamefont {Hirschman},
  \bibfnamefont {I.~I.}}} (\bibinfo {year} {1957}),\ \href@noop {} {\bibfield
  {journal} {\bibinfo  {journal} {American Journal of Mathematics}\ }\textbf
  {\bibinfo {volume} {79}}~(\bibinfo {number} {1}),\ \bibinfo {pages}
  {152}}\BibitemShut {NoStop}%
\bibitem [{\citenamefont {Holevo}(1973)}]{holevo73b}%
  \BibitemOpen
  \bibfield  {author} {\bibinfo {author} {\bibnamefont {Holevo}, \bibfnamefont
  {A.~S.}}} (\bibinfo {year} {1973}),\ \href@noop {} {\bibfield  {journal}
  {\bibinfo  {journal} {Problems of Information Transmission}\ }\textbf
  {\bibinfo {volume} {9}}~(\bibinfo {number} {3}),\ \bibinfo {pages}
  {177}}\BibitemShut {NoStop}%
\bibitem [{\citenamefont {Horodecki}\ \emph {et~al.}(2006)\citenamefont
  {Horodecki}, \citenamefont {Oppenheim},\ and\ \citenamefont
  {Winter}}]{horodecki06}%
  \BibitemOpen
  \bibfield  {author} {\bibinfo {author} {\bibnamefont {Horodecki},
  \bibfnamefont {M.}}, \bibinfo {author} {\bibfnamefont {J.}~\bibnamefont
  {Oppenheim}}, \ and\ \bibinfo {author} {\bibfnamefont {A.}~\bibnamefont
  {Winter}}} (\bibinfo {year} {2006}),\ \href {\doibase
  10.1007/s00220-006-0118-x} {\bibfield  {journal} {\bibinfo  {journal}
  {Communications in Mathematical Physics}\ }\textbf {\bibinfo {volume}
  {269}}~(\bibinfo {number} {1}),\ \bibinfo {pages} {107}}\BibitemShut
  {NoStop}%
\bibitem [{\citenamefont {Horodecki}\ \emph {et~al.}(2009)\citenamefont
  {Horodecki}, \citenamefont {Horodecki}, \citenamefont {Horodecki},\ and\
  \citenamefont {Horodecki}}]{horodecki09}%
  \BibitemOpen
  \bibfield  {author} {\bibinfo {author} {\bibnamefont {Horodecki},
  \bibfnamefont {R.}}, \bibinfo {author} {\bibfnamefont {P.}~\bibnamefont
  {Horodecki}}, \bibinfo {author} {\bibfnamefont {M.}~\bibnamefont
  {Horodecki}}, \ and\ \bibinfo {author} {\bibfnamefont {K.}~\bibnamefont
  {Horodecki}}} (\bibinfo {year} {2009}),\ \href {\doibase
  10.1103/RevModPhys.81.865} {\bibfield  {journal} {\bibinfo  {journal} {Review
  of Modern Physics}\ }\textbf {\bibinfo {volume} {81}}~(\bibinfo {number}
  {2}),\ \bibinfo {pages} {865}}\BibitemShut {NoStop}%
\bibitem [{\citenamefont {Hu}\ and\ \citenamefont {Fan}(2012)}]{hu12}%
  \BibitemOpen
  \bibfield  {author} {\bibinfo {author} {\bibnamefont {Hu}, \bibfnamefont
  {M.-L.}}, \ and\ \bibinfo {author} {\bibfnamefont {H.}~\bibnamefont {Fan}}}
  (\bibinfo {year} {2012}),\ \href {\doibase 10.1103/PhysRevA.86.032338}
  {\bibfield  {journal} {\bibinfo  {journal} {Physical Review A}\ }\textbf
  {\bibinfo {volume} {86}}~(\bibinfo {number} {3}),\ \bibinfo {pages}
  {032338}}\BibitemShut {NoStop}%
\bibitem [{\citenamefont {Hu}\ and\ \citenamefont
  {Fan}(2013{\natexlab{a}})}]{hu13b}%
  \BibitemOpen
  \bibfield  {author} {\bibinfo {author} {\bibnamefont {Hu}, \bibfnamefont
  {M.-L.}}, \ and\ \bibinfo {author} {\bibfnamefont {H.}~\bibnamefont {Fan}}}
  (\bibinfo {year} {2013}{\natexlab{a}}),\ \href {\doibase
  10.1103/PhysRevA.87.022314} {\bibfield  {journal} {\bibinfo  {journal}
  {Physical Review A}\ }\textbf {\bibinfo {volume} {87}}~(\bibinfo {number}
  {2}),\ \bibinfo {pages} {022314}}\BibitemShut {NoStop}%
\bibitem [{\citenamefont {Hu}\ and\ \citenamefont
  {Fan}(2013{\natexlab{b}})}]{hu13}%
  \BibitemOpen
  \bibfield  {author} {\bibinfo {author} {\bibnamefont {Hu}, \bibfnamefont
  {M.-L.}}, \ and\ \bibinfo {author} {\bibfnamefont {H.}~\bibnamefont {Fan}}}
  (\bibinfo {year} {2013}{\natexlab{b}}),\ \href {\doibase
  10.1103/PhysRevA.88.014105} {\bibfield  {journal} {\bibinfo  {journal}
  {Physical Review A}\ }\textbf {\bibinfo {volume} {88}}~(\bibinfo {number}
  {1}),\ \bibinfo {pages} {014105}}\BibitemShut {NoStop}%
\bibitem [{\citenamefont {Huang}(2010)}]{huang10}%
  \BibitemOpen
  \bibfield  {author} {\bibinfo {author} {\bibnamefont {Huang}, \bibfnamefont
  {Y.}}} (\bibinfo {year} {2010}),\ \href {\doibase 10.1103/PhysRevA.82.012335}
  {\bibfield  {journal} {\bibinfo  {journal} {Physical Review A}\ }\textbf
  {\bibinfo {volume} {82}}~(\bibinfo {number} {1}),\ \bibinfo {pages}
  {012335}}\BibitemShut {NoStop}%
\bibitem [{\citenamefont {Huang}(2011)}]{huang11}%
  \BibitemOpen
  \bibfield  {author} {\bibinfo {author} {\bibnamefont {Huang}, \bibfnamefont
  {Y.}}} (\bibinfo {year} {2011}),\ \href {\doibase 10.1103/PhysRevA.83.052124}
  {\bibfield  {journal} {\bibinfo  {journal} {Physical Review A}\ }\textbf
  {\bibinfo {volume} {83}}~(\bibinfo {number} {5}),\ \bibinfo {pages}
  {052124}}\BibitemShut {NoStop}%
\bibitem [{\citenamefont {Huang}(2013)}]{huang13}%
  \BibitemOpen
  \bibfield  {author} {\bibinfo {author} {\bibnamefont {Huang}, \bibfnamefont
  {Y.}}} (\bibinfo {year} {2013}),\ \href {\doibase 10.1109/TIT.2013.2257936}
  {\bibfield  {journal} {\bibinfo  {journal} {IEEE Transactions on Information
  Theory}\ }\textbf {\bibinfo {volume} {59}}~(\bibinfo {number} {10}),\
  \bibinfo {pages} {6774}}\BibitemShut {NoStop}%
\bibitem [{\citenamefont {{IBM}}(2016)}]{IBM_Quantum}%
  \BibitemOpen
  \bibfield  {author} {\bibinfo {author} {\bibnamefont {{IBM}},}} (\bibinfo
  {year} {2016}),\ \href {http://www.research.ibm.com/quantum/} {\enquote
  {\bibinfo {title} {{IBM} quantum experience},}\ }\BibitemShut {NoStop}%
\bibitem [{\citenamefont {Impagliazzo}\ \emph {et~al.}(1989)\citenamefont
  {Impagliazzo}, \citenamefont {Levin},\ and\ \citenamefont
  {Luby}}]{impagliazzo89}%
  \BibitemOpen
  \bibfield  {author} {\bibinfo {author} {\bibnamefont {Impagliazzo},
  \bibfnamefont {R.}}, \bibinfo {author} {\bibfnamefont {L.~A.}\ \bibnamefont
  {Levin}}, \ and\ \bibinfo {author} {\bibfnamefont {M.}~\bibnamefont {Luby}}}
  (\bibinfo {year} {1989}),\ in\ \href {\doibase 10.1145/73007.73009} {\emph
  {\bibinfo {booktitle} {Proc. ACM STOC 1989}}}\ (\bibinfo  {publisher} {ACM
  Press})\ pp.\ \bibinfo {pages} {12--24}\BibitemShut {NoStop}%
\bibitem [{\citenamefont {Impagliazzo}\ and\ \citenamefont
  {Zuckerman}(1989)}]{zuckerman89}%
  \BibitemOpen
  \bibfield  {author} {\bibinfo {author} {\bibnamefont {Impagliazzo},
  \bibfnamefont {R.}}, \ and\ \bibinfo {author} {\bibfnamefont
  {D.}~\bibnamefont {Zuckerman}}} (\bibinfo {year} {1989}),\ in\ \href
  {\doibase 10.1109/SFCS.1989.63486} {\emph {\bibinfo {booktitle} {Proc. IEEE
  FOCS 1989}}},\ pp.\ \bibinfo {pages} {248--253}\BibitemShut {NoStop}%
\bibitem [{\citenamefont {Ivanovic}(1992)}]{ivanovic92}%
  \BibitemOpen
  \bibfield  {author} {\bibinfo {author} {\bibnamefont {Ivanovic},
  \bibfnamefont {I.~D.}}} (\bibinfo {year} {1992}),\ \href {\doibase
  10.1088/0305-4470/25/7/014} {\bibfield  {journal} {\bibinfo  {journal}
  {Journal of Physics A: Mathematical and Theoretical}\ }\textbf {\bibinfo
  {volume} {25}}~(\bibinfo {number} {7}),\ \bibinfo {pages} {L363}}\BibitemShut
  {NoStop}%
\bibitem [{\citenamefont {Jaeger}\ \emph {et~al.}(1995)\citenamefont {Jaeger},
  \citenamefont {Shimony},\ and\ \citenamefont {Vaidman}}]{jaeger95}%
  \BibitemOpen
  \bibfield  {author} {\bibinfo {author} {\bibnamefont {Jaeger}, \bibfnamefont
  {G.}}, \bibinfo {author} {\bibfnamefont {A.}~\bibnamefont {Shimony}}, \ and\
  \bibinfo {author} {\bibfnamefont {L.}~\bibnamefont {Vaidman}}} (\bibinfo
  {year} {1995}),\ \href {\doibase 10.1103/PhysRevA.51.54} {\bibfield
  {journal} {\bibinfo  {journal} {Physical Review A}\ }\textbf {\bibinfo
  {volume} {51}}~(\bibinfo {number} {1}),\ \bibinfo {pages} {54}}\BibitemShut
  {NoStop}%
\bibitem [{\citenamefont {Jamio{\l}kowski}(1972)}]{jamiliolkowski72}%
  \BibitemOpen
  \bibfield  {author} {\bibinfo {author} {\bibnamefont {Jamio{\l}kowski},
  \bibfnamefont {A.}}} (\bibinfo {year} {1972}),\ \href {\doibase
  10.1016/0034-4877(72)90011-0} {\bibfield  {journal} {\bibinfo  {journal}
  {Reports on Mathematical Physics}\ }\textbf {\bibinfo {volume} {3}}~(\bibinfo
  {number} {4}),\ \bibinfo {pages} {275}}\BibitemShut {NoStop}%
\bibitem [{\citenamefont {Jia}\ \emph {et~al.}(2015)\citenamefont {Jia},
  \citenamefont {Tian},\ and\ \citenamefont {Jing}}]{jia15}%
  \BibitemOpen
  \bibfield  {author} {\bibinfo {author} {\bibnamefont {Jia}, \bibfnamefont
  {L.}}, \bibinfo {author} {\bibfnamefont {Z.}~\bibnamefont {Tian}}, \ and\
  \bibinfo {author} {\bibfnamefont {J.}~\bibnamefont {Jing}}} (\bibinfo {year}
  {2015}),\ \href {\doibase 10.1016/j.aop.2014.10.019} {\bibfield  {journal}
  {\bibinfo  {journal} {Annals of Physics}\ }\textbf {\bibinfo {volume}
  {353}},\ \bibinfo {pages} {37}}\BibitemShut {NoStop}%
\bibitem [{\citenamefont {Julsgaard}\ \emph {et~al.}(2004)\citenamefont
  {Julsgaard}, \citenamefont {Sherson}, \citenamefont {Cirac}, \citenamefont
  {Fiur{\'{a}}{\v{s}}ek},\ and\ \citenamefont {Polzik}}]{julsgaard04}%
  \BibitemOpen
  \bibfield  {author} {\bibinfo {author} {\bibnamefont {Julsgaard},
  \bibfnamefont {B.}}, \bibinfo {author} {\bibfnamefont {J.}~\bibnamefont
  {Sherson}}, \bibinfo {author} {\bibfnamefont {J.~I.}\ \bibnamefont {Cirac}},
  \bibinfo {author} {\bibfnamefont {J.}~\bibnamefont {Fiur{\'{a}}{\v{s}}ek}}, \
  and\ \bibinfo {author} {\bibfnamefont {E.~S.}\ \bibnamefont {Polzik}}}
  (\bibinfo {year} {2004}),\ \href {\doibase 10.1038/nature03064} {\bibfield
  {journal} {\bibinfo  {journal} {Nature}\ }\textbf {\bibinfo {volume}
  {432}}~(\bibinfo {number} {7016}),\ \bibinfo {pages} {482}}\BibitemShut
  {NoStop}%
\bibitem [{\citenamefont {Kalev}\ and\ \citenamefont {Gour}(2014)}]{kalev07}%
  \BibitemOpen
  \bibfield  {author} {\bibinfo {author} {\bibnamefont {Kalev}, \bibfnamefont
  {A.}}, \ and\ \bibinfo {author} {\bibfnamefont {G.}~\bibnamefont {Gour}}}
  (\bibinfo {year} {2014}),\ \href {\doibase 10.1088/1367-2630/16/5/053038}
  {\bibfield  {journal} {\bibinfo  {journal} {New Journal of Physics}\ }\textbf
  {\bibinfo {volume} {16}}~(\bibinfo {number} {5}),\ \bibinfo {pages}
  {053038}}\BibitemShut {NoStop}%
\bibitem [{\citenamefont {Kaniewski}\ \emph {et~al.}(2013)\citenamefont
  {Kaniewski}, \citenamefont {Tomamichel}, \citenamefont {Hanggi},\ and\
  \citenamefont {Wehner}}]{kaniewski13}%
  \BibitemOpen
  \bibfield  {author} {\bibinfo {author} {\bibnamefont {Kaniewski},
  \bibfnamefont {J.}}, \bibinfo {author} {\bibfnamefont {M.}~\bibnamefont
  {Tomamichel}}, \bibinfo {author} {\bibfnamefont {E.}~\bibnamefont {Hanggi}},
  \ and\ \bibinfo {author} {\bibfnamefont {S.}~\bibnamefont {Wehner}}}
  (\bibinfo {year} {2013}),\ \href {\doibase 10.1109/TIT.2013.2247463}
  {\bibfield  {journal} {\bibinfo  {journal} {IEEE Transactions on Information
  Theory}\ }\textbf {\bibinfo {volume} {59}}~(\bibinfo {number} {7}),\ \bibinfo
  {pages} {4687}}\BibitemShut {NoStop}%
\bibitem [{\citenamefont {Kaniewski}\ \emph {et~al.}(2014)\citenamefont
  {Kaniewski}, \citenamefont {Tomamichel},\ and\ \citenamefont
  {Wehner}}]{kaniewski14}%
  \BibitemOpen
  \bibfield  {author} {\bibinfo {author} {\bibnamefont {Kaniewski},
  \bibfnamefont {J.}}, \bibinfo {author} {\bibfnamefont {M.}~\bibnamefont
  {Tomamichel}}, \ and\ \bibinfo {author} {\bibfnamefont {S.}~\bibnamefont
  {Wehner}}} (\bibinfo {year} {2014}),\ \href {\doibase
  10.1103/PhysRevA.90.012332} {\bibfield  {journal} {\bibinfo  {journal}
  {Physical Review A}\ }\textbf {\bibinfo {volume} {90}}~(\bibinfo {number}
  {1}),\ \bibinfo {pages} {012332}}\BibitemShut {NoStop}%
\bibitem [{\citenamefont {Karpat}\ \emph {et~al.}(2015)\citenamefont {Karpat},
  \citenamefont {Piilo},\ and\ \citenamefont {Maniscalco}}]{karpat15}%
  \BibitemOpen
  \bibfield  {author} {\bibinfo {author} {\bibnamefont {Karpat}, \bibfnamefont
  {G.}}, \bibinfo {author} {\bibfnamefont {J.}~\bibnamefont {Piilo}}, \ and\
  \bibinfo {author} {\bibfnamefont {S.}~\bibnamefont {Maniscalco}}} (\bibinfo
  {year} {2015}),\ \href {\doibase 10.1209/0295-5075/111/50006} {\bibfield
  {journal} {\bibinfo  {journal} {EPL (Europhysics Letters)}\ }\textbf
  {\bibinfo {volume} {111}}~(\bibinfo {number} {5}),\ \bibinfo {pages}
  {50006}}\BibitemShut {NoStop}%
\bibitem [{\citenamefont {Kennard}(1927)}]{kennard27}%
  \BibitemOpen
  \bibfield  {author} {\bibinfo {author} {\bibnamefont {Kennard}, \bibfnamefont
  {E.~H.}}} (\bibinfo {year} {1927}),\ \href {\doibase 10.1007/BF01391200}
  {\bibfield  {journal} {\bibinfo  {journal} {Zeitschrift f{\"{u}}r Physik}\
  }\textbf {\bibinfo {volume} {44}}~(\bibinfo {number} {4-5}),\ \bibinfo
  {pages} {326}}\BibitemShut {NoStop}%
\bibitem [{\citenamefont {Kilian}(1988)}]{kilian88}%
  \BibitemOpen
  \bibfield  {author} {\bibinfo {author} {\bibnamefont {Kilian}, \bibfnamefont
  {J.}}} (\bibinfo {year} {1988}),\ in\ \href {\doibase 10.1145/62212.62215}
  {\emph {\bibinfo {booktitle} {Proc. ACM STOC 1988}}}\ (\bibinfo  {publisher}
  {ACM Press},\ \bibinfo {address} {New York, NY})\ pp.\ \bibinfo {pages}
  {20--31}\BibitemShut {NoStop}%
\bibitem [{\citenamefont {Klappenecker}\ and\ \citenamefont
  {Rotteler}(2005)}]{klappenecker05}%
  \BibitemOpen
  \bibfield  {author} {\bibinfo {author} {\bibnamefont {Klappenecker},
  \bibfnamefont {A.}}, \ and\ \bibinfo {author} {\bibfnamefont
  {M.}~\bibnamefont {Rotteler}}} (\bibinfo {year} {2005}),\ in\ \href {\doibase
  10.1109/ISIT.2005.1523643} {\emph {\bibinfo {booktitle} {Proc. IEEE ISIT
  2005}}}\ (\bibinfo  {publisher} {IEEE})\ pp.\ \bibinfo {pages}
  {1740--1744}\BibitemShut {NoStop}%
\bibitem [{\citenamefont {Koashi}(2006)}]{koashi06}%
  \BibitemOpen
  \bibfield  {author} {\bibinfo {author} {\bibnamefont {Koashi}, \bibfnamefont
  {M.}}} (\bibinfo {year} {2006}),\ \href {\doibase 10.1088/1742-6596/36/1/016}
  {\bibfield  {journal} {\bibinfo  {journal} {Journal of Physics: Conference
  Series}\ }\textbf {\bibinfo {volume} {36}}~(\bibinfo {number} {1}),\ \bibinfo
  {pages} {98}}\BibitemShut {NoStop}%
\bibitem [{\citenamefont {K{\"{o}}nig}\ and\ \citenamefont
  {Renner}(2011)}]{koenig07}%
  \BibitemOpen
  \bibfield  {author} {\bibinfo {author} {\bibnamefont {K{\"{o}}nig},
  \bibfnamefont {R.}}, \ and\ \bibinfo {author} {\bibfnamefont
  {R.}~\bibnamefont {Renner}}} (\bibinfo {year} {2011}),\ \href {\doibase
  10.1109/TIT.2011.2146730} {\bibfield  {journal} {\bibinfo  {journal} {IEEE
  Transactions on Information Theory}\ }\textbf {\bibinfo {volume}
  {57}}~(\bibinfo {number} {7}),\ \bibinfo {pages} {4760}}\BibitemShut
  {NoStop}%
\bibitem [{\citenamefont {K{\"{o}}nig}\ \emph {et~al.}(2009)\citenamefont
  {K{\"{o}}nig}, \citenamefont {Renner},\ and\ \citenamefont
  {Schaffner}}]{koenig08}%
  \BibitemOpen
  \bibfield  {author} {\bibinfo {author} {\bibnamefont {K{\"{o}}nig},
  \bibfnamefont {R.}}, \bibinfo {author} {\bibfnamefont {R.}~\bibnamefont
  {Renner}}, \ and\ \bibinfo {author} {\bibfnamefont {C.}~\bibnamefont
  {Schaffner}}} (\bibinfo {year} {2009}),\ \href {\doibase
  10.1109/TIT.2009.2025545} {\bibfield  {journal} {\bibinfo  {journal} {IEEE
  Transactions on Information Theory}\ }\textbf {\bibinfo {volume}
  {55}}~(\bibinfo {number} {9}),\ \bibinfo {pages} {4337}}\BibitemShut
  {NoStop}%
\bibitem [{\citenamefont {K{\"{o}}nig}\ \emph {et~al.}(2012)\citenamefont
  {K{\"{o}}nig}, \citenamefont {Wehner},\ and\ \citenamefont
  {Wullschleger}}]{koenig09}%
  \BibitemOpen
  \bibfield  {author} {\bibinfo {author} {\bibnamefont {K{\"{o}}nig},
  \bibfnamefont {R.}}, \bibinfo {author} {\bibfnamefont {S.}~\bibnamefont
  {Wehner}}, \ and\ \bibinfo {author} {\bibfnamefont {J.}~\bibnamefont
  {Wullschleger}}} (\bibinfo {year} {2012}),\ \href {\doibase
  10.1109/TIT.2011.2177772} {\bibfield  {journal} {\bibinfo  {journal} {IEEE
  Transactions on Information Theory}\ }\textbf {\bibinfo {volume}
  {58}}~(\bibinfo {number} {3}),\ \bibinfo {pages} {1962}}\BibitemShut
  {NoStop}%
\bibitem [{\citenamefont {Korzekwa}\ \emph
  {et~al.}(2014{\natexlab{a}})\citenamefont {Korzekwa}, \citenamefont
  {Jennings},\ and\ \citenamefont {Rudolph}}]{korzekwa14b}%
  \BibitemOpen
  \bibfield  {author} {\bibinfo {author} {\bibnamefont {Korzekwa},
  \bibfnamefont {K.}}, \bibinfo {author} {\bibfnamefont {D.}~\bibnamefont
  {Jennings}}, \ and\ \bibinfo {author} {\bibfnamefont {T.}~\bibnamefont
  {Rudolph}}} (\bibinfo {year} {2014}{\natexlab{a}}),\ \href {\doibase
  10.1103/PhysRevA.89.052108} {\bibfield  {journal} {\bibinfo  {journal}
  {Physical Review A}\ }\textbf {\bibinfo {volume} {89}}~(\bibinfo {number}
  {5}),\ \bibinfo {pages} {052108}}\BibitemShut {NoStop}%
\bibitem [{\citenamefont {Korzekwa}\ \emph
  {et~al.}(2014{\natexlab{b}})\citenamefont {Korzekwa}, \citenamefont
  {Lostaglio}, \citenamefont {Jennings},\ and\ \citenamefont
  {Rudolph}}]{korzekwa14}%
  \BibitemOpen
  \bibfield  {author} {\bibinfo {author} {\bibnamefont {Korzekwa},
  \bibfnamefont {K.}}, \bibinfo {author} {\bibfnamefont {M.}~\bibnamefont
  {Lostaglio}}, \bibinfo {author} {\bibfnamefont {D.}~\bibnamefont {Jennings}},
  \ and\ \bibinfo {author} {\bibfnamefont {T.}~\bibnamefont {Rudolph}}}
  (\bibinfo {year} {2014}{\natexlab{b}}),\ \href {\doibase
  10.1103/PhysRevA.89.042122} {\bibfield  {journal} {\bibinfo  {journal}
  {Physical Review A}\ }\textbf {\bibinfo {volume} {89}}~(\bibinfo {number}
  {4}),\ \bibinfo {pages} {042122}}\BibitemShut {NoStop}%
\bibitem [{\citenamefont {Kraus}(1987)}]{kraus87}%
  \BibitemOpen
  \bibfield  {author} {\bibinfo {author} {\bibnamefont {Kraus}, \bibfnamefont
  {K.}}} (\bibinfo {year} {1987}),\ \href {\doibase 10.1103/PhysRevD.35.3070}
  {\bibfield  {journal} {\bibinfo  {journal} {Physical Review D}\ }\textbf
  {\bibinfo {volume} {35}}~(\bibinfo {number} {10}),\ \bibinfo {pages}
  {3070}}\BibitemShut {NoStop}%
\bibitem [{\citenamefont {Krishna}\ and\ \citenamefont
  {Parthasarathy}(2002)}]{krishna01}%
  \BibitemOpen
  \bibfield  {author} {\bibinfo {author} {\bibnamefont {Krishna}, \bibfnamefont
  {M.}}, \ and\ \bibinfo {author} {\bibfnamefont {K.~R.}\ \bibnamefont
  {Parthasarathy}}} (\bibinfo {year} {2002}),\ \href
  {http://www.jstor.org/stable/25051432} {\bibfield  {journal} {\bibinfo
  {journal} {Indian Journal of Statistics}\ }\textbf {\bibinfo {volume}
  {64}}~(\bibinfo {number} {3}),\ \bibinfo {pages} {842}}\BibitemShut {NoStop}%
\bibitem [{\citenamefont {Kuznetsova}(2011)}]{kuznetsova11}%
  \BibitemOpen
  \bibfield  {author} {\bibinfo {author} {\bibnamefont {Kuznetsova},
  \bibfnamefont {A.~A.}}} (\bibinfo {year} {2011}),\ \href {\doibase
  10.1137/S0040585X97985121} {\bibfield  {journal} {\bibinfo  {journal} {Theory
  of Probability and its Applications}\ }\textbf {\bibinfo {volume}
  {55}}~(\bibinfo {number} {4}),\ \bibinfo {pages} {709}}\BibitemShut {NoStop}%
\bibitem [{\citenamefont {Larsen}(1990)}]{larsen90}%
  \BibitemOpen
  \bibfield  {author} {\bibinfo {author} {\bibnamefont {Larsen}, \bibfnamefont
  {U.}}} (\bibinfo {year} {1990}),\ \href {\doibase 10.1088/0305-4470/23/7/013}
  {\bibfield  {journal} {\bibinfo  {journal} {Journal of Physics A:
  Mathematical and Theoretical}\ }\textbf {\bibinfo {volume} {23}}~(\bibinfo
  {number} {7}),\ \bibinfo {pages} {1041}}\BibitemShut {NoStop}%
\bibitem [{\citenamefont {Li}\ \emph {et~al.}(2011)\citenamefont {Li},
  \citenamefont {Xu}, \citenamefont {Xu}, \citenamefont {Li},\ and\
  \citenamefont {Guo}}]{li11}%
  \BibitemOpen
  \bibfield  {author} {\bibinfo {author} {\bibnamefont {Li}, \bibfnamefont
  {C.-F.}}, \bibinfo {author} {\bibfnamefont {J.-S.}\ \bibnamefont {Xu}},
  \bibinfo {author} {\bibfnamefont {X.-Y.}\ \bibnamefont {Xu}}, \bibinfo
  {author} {\bibfnamefont {K.}~\bibnamefont {Li}}, \ and\ \bibinfo {author}
  {\bibfnamefont {G.-C.}\ \bibnamefont {Guo}}} (\bibinfo {year} {2011}),\ \href
  {\doibase 10.1038/nphys2047} {\bibfield  {journal} {\bibinfo  {journal}
  {Nature Physics}\ }\textbf {\bibinfo {volume} {7}}~(\bibinfo {number} {10}),\
  \bibinfo {pages} {752}}\BibitemShut {NoStop}%
\bibitem [{\citenamefont {Lieb}\ and\ \citenamefont {Ruskai}(1973)}]{lieb73}%
  \BibitemOpen
  \bibfield  {author} {\bibinfo {author} {\bibnamefont {Lieb}, \bibfnamefont
  {E.~H.}}, \ and\ \bibinfo {author} {\bibfnamefont {M.~B.}\ \bibnamefont
  {Ruskai}}} (\bibinfo {year} {1973}),\ \href {\doibase 10.1063/1.1666274}
  {\bibfield  {journal} {\bibinfo  {journal} {Journal of Mathematical Physics}\
  }\textbf {\bibinfo {volume} {14}}~(\bibinfo {number} {12}),\ \bibinfo {pages}
  {1938}}\BibitemShut {NoStop}%
\bibitem [{\citenamefont {Lindblad}(1975)}]{lindblad75}%
  \BibitemOpen
  \bibfield  {author} {\bibinfo {author} {\bibnamefont {Lindblad},
  \bibfnamefont {G.}}} (\bibinfo {year} {1975}),\ \href {\doibase
  10.1007/BF01609396} {\bibfield  {journal} {\bibinfo  {journal}
  {Communications in Mathematical Physics}\ }\textbf {\bibinfo {volume}
  {40}}~(\bibinfo {number} {2}),\ \bibinfo {pages} {147}}\BibitemShut {NoStop}%
\bibitem [{\citenamefont {Liu}\ \emph {et~al.}(2015)\citenamefont {Liu},
  \citenamefont {Mu},\ and\ \citenamefont {Fan}}]{liu15}%
  \BibitemOpen
  \bibfield  {author} {\bibinfo {author} {\bibnamefont {Liu}, \bibfnamefont
  {S.}}, \bibinfo {author} {\bibfnamefont {L.-Z.}\ \bibnamefont {Mu}}, \ and\
  \bibinfo {author} {\bibfnamefont {H.}~\bibnamefont {Fan}}} (\bibinfo {year}
  {2015}),\ \href {\doibase 10.1103/PhysRevA.91.042133} {\bibfield  {journal}
  {\bibinfo  {journal} {Physical Review A}\ }\textbf {\bibinfo {volume}
  {91}}~(\bibinfo {number} {4}),\ \bibinfo {pages} {042133}}\BibitemShut
  {NoStop}%
\bibitem [{\citenamefont {Liu}(2014)}]{liu14}%
  \BibitemOpen
  \bibfield  {author} {\bibinfo {author} {\bibnamefont {Liu}, \bibfnamefont
  {Y.-K.}}} (\bibinfo {year} {2014}),\ in\ \href {\doibase
  10.1007/978-3-662-44381-1_2} {\emph {\bibinfo {booktitle} {Proc. CRYPTO
  2014}}},\ \bibinfo {series} {LNCS}, Vol.\ \bibinfo {volume} {8617},\ \bibinfo
  {editor} {edited by\ \bibinfo {editor} {\bibfnamefont {J.~A.}\ \bibnamefont
  {Garay}}\ and\ \bibinfo {editor} {\bibfnamefont {R.}~\bibnamefont
  {Gennaro}}}\ (\bibinfo  {publisher} {Springer},\ \bibinfo {address} {Santa
  Barbara, CA})\ pp.\ \bibinfo {pages} {19--36}\BibitemShut {NoStop}%
\bibitem [{\citenamefont {Liu}(2015)}]{liu15b}%
  \BibitemOpen
  \bibfield  {author} {\bibinfo {author} {\bibnamefont {Liu}, \bibfnamefont
  {Y.-K.}}} (\bibinfo {year} {2015}),\ in\ \href {\doibase
  10.1007/978-3-662-46803-6_26} {\emph {\bibinfo {booktitle} {Proc. EUROCRYPT
  2015}}},\ \bibinfo {series} {LNCS}, Vol.\ \bibinfo {volume} {9057}\ (\bibinfo
   {publisher} {Springer})\ pp.\ \bibinfo {pages} {785--814}\BibitemShut
  {NoStop}%
\bibitem [{\citenamefont {Lo}(1997)}]{lo97-2}%
  \BibitemOpen
  \bibfield  {author} {\bibinfo {author} {\bibnamefont {Lo}, \bibfnamefont
  {H.-k.}}} (\bibinfo {year} {1997}),\ \href {\doibase
  10.1103/PhysRevA.56.1154} {\bibfield  {journal} {\bibinfo  {journal}
  {Physical Review A}\ }\textbf {\bibinfo {volume} {56}}~(\bibinfo {number}
  {2}),\ \bibinfo {pages} {1154}}\BibitemShut {NoStop}%
\bibitem [{\citenamefont {Lo}\ and\ \citenamefont {Chau}(1997)}]{lo97-1}%
  \BibitemOpen
  \bibfield  {author} {\bibinfo {author} {\bibnamefont {Lo}, \bibfnamefont
  {H.-k.}}, \ and\ \bibinfo {author} {\bibfnamefont {H.~F.}\ \bibnamefont
  {Chau}}} (\bibinfo {year} {1997}),\ \href {\doibase
  10.1103/PhysRevLett.78.3410} {\bibfield  {journal} {\bibinfo  {journal}
  {Physical Review Letters}\ }\textbf {\bibinfo {volume} {78}}~(\bibinfo
  {number} {17}),\ \bibinfo {pages} {3410}}\BibitemShut {NoStop}%
\bibitem [{\citenamefont {Luo}(2005)}]{luo05}%
  \BibitemOpen
  \bibfield  {author} {\bibinfo {author} {\bibnamefont {Luo}, \bibfnamefont
  {S.~L.}}} (\bibinfo {year} {2005}),\ \href {\doibase
  10.1007/s11232-005-0098-6} {\bibfield  {journal} {\bibinfo  {journal}
  {Theoretical and Mathematical Physics}\ }\textbf {\bibinfo {volume}
  {143}}~(\bibinfo {number} {2}),\ \bibinfo {pages} {681}}\BibitemShut
  {NoStop}%
\bibitem [{\citenamefont {Maassen}\ and\ \citenamefont
  {Uffink}(1988)}]{maassen88}%
  \BibitemOpen
  \bibfield  {author} {\bibinfo {author} {\bibnamefont {Maassen}, \bibfnamefont
  {H.}}, \ and\ \bibinfo {author} {\bibfnamefont {J.}~\bibnamefont {Uffink}}}
  (\bibinfo {year} {1988}),\ \href {\doibase 10.1103/PhysRevLett.60.1103}
  {\bibfield  {journal} {\bibinfo  {journal} {Physical Review Letters}\
  }\textbf {\bibinfo {volume} {60}}~(\bibinfo {number} {12}),\ \bibinfo {pages}
  {1103}}\BibitemShut {NoStop}%
\bibitem [{\citenamefont {Maccone}\ and\ \citenamefont
  {Pati}(2014)}]{Maccone14}%
  \BibitemOpen
  \bibfield  {author} {\bibinfo {author} {\bibnamefont {Maccone}, \bibfnamefont
  {L.}}, \ and\ \bibinfo {author} {\bibfnamefont {A.~K.}\ \bibnamefont {Pati}}}
  (\bibinfo {year} {2014}),\ \href {\doibase 10.1103/PhysRevLett.113.260401}
  {\bibfield  {journal} {\bibinfo  {journal} {Physical Review Letters}\
  }\textbf {\bibinfo {volume} {113}}~(\bibinfo {number} {26}),\ \bibinfo
  {pages} {260401}}\BibitemShut {NoStop}%
\bibitem [{\citenamefont {Mandayam}\ and\ \citenamefont
  {Wehner}(2011)}]{mandayam11}%
  \BibitemOpen
  \bibfield  {author} {\bibinfo {author} {\bibnamefont {Mandayam},
  \bibfnamefont {P.}}, \ and\ \bibinfo {author} {\bibfnamefont
  {S.}~\bibnamefont {Wehner}}} (\bibinfo {year} {2011}),\ \href {\doibase
  10.1103/PhysRevA.83.022329} {\bibfield  {journal} {\bibinfo  {journal}
  {Physical Review A}\ }\textbf {\bibinfo {volume} {83}}~(\bibinfo {number}
  {2}),\ \bibinfo {pages} {022329}}\BibitemShut {NoStop}%
\bibitem [{\citenamefont {Mandayam}\ \emph {et~al.}(2010)\citenamefont
  {Mandayam}, \citenamefont {Wehner},\ and\ \citenamefont
  {Balachandran}}]{mandayam10}%
  \BibitemOpen
  \bibfield  {author} {\bibinfo {author} {\bibnamefont {Mandayam},
  \bibfnamefont {P.}}, \bibinfo {author} {\bibfnamefont {S.}~\bibnamefont
  {Wehner}}, \ and\ \bibinfo {author} {\bibfnamefont {N.}~\bibnamefont
  {Balachandran}}} (\bibinfo {year} {2010}),\ \href {\doibase
  10.1063/1.3477319} {\bibfield  {journal} {\bibinfo  {journal} {Journal of
  Mathematical Physics}\ }\textbf {\bibinfo {volume} {51}}~(\bibinfo {number}
  {8}),\ \bibinfo {pages} {082201}}\BibitemShut {NoStop}%
\bibitem [{\citenamefont {Marshall}\ \emph {et~al.}(2011)\citenamefont
  {Marshall}, \citenamefont {Olkin},\ and\ \citenamefont
  {Arnold}}]{marshall11}%
  \BibitemOpen
  \bibfield  {author} {\bibinfo {author} {\bibnamefont {Marshall},
  \bibfnamefont {A.~W.}}, \bibinfo {author} {\bibfnamefont {I.}~\bibnamefont
  {Olkin}}, \ and\ \bibinfo {author} {\bibfnamefont {B.~C.}\ \bibnamefont
  {Arnold}}} (\bibinfo {year} {2011}),\ \href {\doibase
  10.1007/978-0-387-68276-1} {\emph {\bibinfo {title} {{Inequalities: Theory of
  Majorization and Its Applications}}}},\ Springer Series in Statistics\
  (\bibinfo  {publisher} {Springer},\ \bibinfo {address} {New
  York})\BibitemShut {NoStop}%
\bibitem [{\citenamefont {Matthews}\ \emph {et~al.}(2009)\citenamefont
  {Matthews}, \citenamefont {Wehner},\ and\ \citenamefont
  {Winter}}]{matthews09}%
  \BibitemOpen
  \bibfield  {author} {\bibinfo {author} {\bibnamefont {Matthews},
  \bibfnamefont {W.}}, \bibinfo {author} {\bibfnamefont {S.}~\bibnamefont
  {Wehner}}, \ and\ \bibinfo {author} {\bibfnamefont {A.}~\bibnamefont
  {Winter}}} (\bibinfo {year} {2009}),\ \href {\doibase
  10.1007/s00220-009-0890-5} {\bibfield  {journal} {\bibinfo  {journal}
  {Communications in Mathematical Physics}\ }\textbf {\bibinfo {volume}
  {291}}~(\bibinfo {number} {3}),\ \bibinfo {pages} {813}}\BibitemShut
  {NoStop}%
\bibitem [{\citenamefont {Maurer}(1992)}]{maurer92}%
  \BibitemOpen
  \bibfield  {author} {\bibinfo {author} {\bibnamefont {Maurer}, \bibfnamefont
  {U.~M.}}} (\bibinfo {year} {1992}),\ \href {\doibase 10.1007/BF00191321}
  {\bibfield  {journal} {\bibinfo  {journal} {Journal of Cryptology}\ }\textbf
  {\bibinfo {volume} {5}}~(\bibinfo {number} {1}),\ \bibinfo {pages}
  {53}}\BibitemShut {NoStop}%
\bibitem [{\citenamefont {Mayers}(1996)}]{mayers96}%
  \BibitemOpen
  \bibfield  {author} {\bibinfo {author} {\bibnamefont {Mayers}, \bibfnamefont
  {D.}}} (\bibinfo {year} {1996}),\ in\ \href@noop {} {\emph {\bibinfo
  {booktitle} {Proc. CRYPTO 1996}}},\ \bibinfo {series} {LNCS}, Vol.\ \bibinfo
  {volume} {1109}\ (\bibinfo  {publisher} {Springer})\ pp.\ \bibinfo {pages}
  {343--357}\BibitemShut {NoStop}%
\bibitem [{\citenamefont {Mayers}(1997)}]{mayers97}%
  \BibitemOpen
  \bibfield  {author} {\bibinfo {author} {\bibnamefont {Mayers}, \bibfnamefont
  {D.}}} (\bibinfo {year} {1997}),\ \href {\doibase
  10.1103/PhysRevLett.78.3414} {\bibfield  {journal} {\bibinfo  {journal}
  {Physical Review Letters}\ }\textbf {\bibinfo {volume} {78}}~(\bibinfo
  {number} {17}),\ \bibinfo {pages} {3414}}\BibitemShut {NoStop}%
\bibitem [{\citenamefont {Mayers}(2001)}]{mayers01}%
  \BibitemOpen
  \bibfield  {author} {\bibinfo {author} {\bibnamefont {Mayers}, \bibfnamefont
  {D.}}} (\bibinfo {year} {2001}),\ \href {\doibase 10.1145/382780.382781}
  {\bibfield  {journal} {\bibinfo  {journal} {Journal of the ACM}\ }\textbf
  {\bibinfo {volume} {48}}~(\bibinfo {number} {3}),\ \bibinfo {pages}
  {351}}\BibitemShut {NoStop}%
\bibitem [{\citenamefont {Mclnnes}(1987)}]{mcinnes87}%
  \BibitemOpen
  \bibfield  {author} {\bibinfo {author} {\bibnamefont {Mclnnes}, \bibfnamefont
  {J.}}} (\bibinfo {year} {1987}),\ \href@noop {} {\bibinfo  {journal}
  {Technical Report, University of Toronto}\ }\BibitemShut {NoStop}%
\bibitem [{\citenamefont {Miller}\ and\ \citenamefont {Shi}(2014)}]{miller13}%
  \BibitemOpen
\bibfield  {journal} {  }\bibfield  {author} {\bibinfo {author} {\bibnamefont
  {Miller}, \bibfnamefont {C.~A.}}, \ and\ \bibinfo {author} {\bibfnamefont
  {Y.}~\bibnamefont {Shi}}} (\bibinfo {year} {2014}),\ in\ \href {\doibase
  10.1145/2591796.2591843} {\emph {\bibinfo {booktitle} {Proc. ACM STOC
  2014}}},\ pp.\ \bibinfo {pages} {417--426}\BibitemShut {NoStop}%
\bibitem [{\citenamefont {Modi}\ \emph {et~al.}(2012)\citenamefont {Modi},
  \citenamefont {Brodutch}, \citenamefont {Cable}, \citenamefont {Paterek},\
  and\ \citenamefont {Vedral}}]{modi12}%
  \BibitemOpen
  \bibfield  {author} {\bibinfo {author} {\bibnamefont {Modi}, \bibfnamefont
  {K.}}, \bibinfo {author} {\bibfnamefont {A.}~\bibnamefont {Brodutch}},
  \bibinfo {author} {\bibfnamefont {H.}~\bibnamefont {Cable}}, \bibinfo
  {author} {\bibfnamefont {T.}~\bibnamefont {Paterek}}, \ and\ \bibinfo
  {author} {\bibfnamefont {V.}~\bibnamefont {Vedral}}} (\bibinfo {year}
  {2012}),\ \href {\doibase 10.1103/RevModPhys.84.1655} {\bibfield  {journal}
  {\bibinfo  {journal} {Review of Modern Physics}\ }\textbf {\bibinfo {volume}
  {84}}~(\bibinfo {number} {4}),\ \bibinfo {pages} {1655}}\BibitemShut
  {NoStop}%
\bibitem [{\citenamefont {Mosonyi}\ and\ \citenamefont
  {Ogawa}(2015)}]{mosonyiogawa13}%
  \BibitemOpen
  \bibfield  {author} {\bibinfo {author} {\bibnamefont {Mosonyi}, \bibfnamefont
  {M.}}, \ and\ \bibinfo {author} {\bibfnamefont {T.}~\bibnamefont {Ogawa}}}
  (\bibinfo {year} {2015}),\ \href {\doibase 10.1007/s00220-014-2248-x}
  {\bibfield  {journal} {\bibinfo  {journal} {Communications in Mathematical
  Physics}\ }\textbf {\bibinfo {volume} {334}}~(\bibinfo {number} {3}),\
  \bibinfo {pages} {1617}}\BibitemShut {NoStop}%
\bibitem [{\citenamefont {M{\"{u}}ller-Lennert}\ \emph
  {et~al.}(2013)\citenamefont {M{\"{u}}ller-Lennert}, \citenamefont {Dupuis},
  \citenamefont {Szehr}, \citenamefont {Fehr},\ and\ \citenamefont
  {Tomamichel}}]{lennert13}%
  \BibitemOpen
  \bibfield  {author} {\bibinfo {author} {\bibnamefont {M{\"{u}}ller-Lennert},
  \bibfnamefont {M.}}, \bibinfo {author} {\bibfnamefont {F.}~\bibnamefont
  {Dupuis}}, \bibinfo {author} {\bibfnamefont {O.}~\bibnamefont {Szehr}},
  \bibinfo {author} {\bibfnamefont {S.}~\bibnamefont {Fehr}}, \ and\ \bibinfo
  {author} {\bibfnamefont {M.}~\bibnamefont {Tomamichel}}} (\bibinfo {year}
  {2013}),\ \href {\doibase 10.1063/1.4838856} {\bibfield  {journal} {\bibinfo
  {journal} {Journal of Mathematical Physics}\ }\textbf {\bibinfo {volume}
  {54}}~(\bibinfo {number} {12}),\ \bibinfo {pages} {122203}}\BibitemShut
  {NoStop}%
\bibitem [{\citenamefont {Namiki}\ and\ \citenamefont
  {Tokunaga}(2012)}]{namiki12}%
  \BibitemOpen
  \bibfield  {author} {\bibinfo {author} {\bibnamefont {Namiki}, \bibfnamefont
  {R.}}, \ and\ \bibinfo {author} {\bibfnamefont {Y.}~\bibnamefont {Tokunaga}}}
  (\bibinfo {year} {2012}),\ \href {\doibase 10.1103/PhysRevLett.108.230503}
  {\bibfield  {journal} {\bibinfo  {journal} {Physical Review Letters}\
  }\textbf {\bibinfo {volume} {108}}~(\bibinfo {number} {23}),\ \bibinfo
  {pages} {230503}}\BibitemShut {NoStop}%
\bibitem [{\citenamefont {Narasimhachar}\ \emph {et~al.}(2016)\citenamefont
  {Narasimhachar}, \citenamefont {Poostindouz},\ and\ \citenamefont
  {Gour}}]{narasimhachar15}%
  \BibitemOpen
  \bibfield  {author} {\bibinfo {author} {\bibnamefont {Narasimhachar},
  \bibfnamefont {V.}}, \bibinfo {author} {\bibfnamefont {A.}~\bibnamefont
  {Poostindouz}}, \ and\ \bibinfo {author} {\bibfnamefont {G.}~\bibnamefont
  {Gour}}} (\bibinfo {year} {2016}),\ \href {\doibase
  10.1088/1367-2630/18/3/033019} {\bibfield  {journal} {\bibinfo  {journal}
  {New Journal of Physics}\ }\textbf {\bibinfo {volume} {18}}~(\bibinfo
  {number} {3}),\ \bibinfo {pages} {033019}}\BibitemShut {NoStop}%
\bibitem [{\citenamefont {von Neumann}(1932)}]{vonneumann32}%
  \BibitemOpen
  \bibfield  {author} {\bibinfo {author} {\bibnamefont {von Neumann},
  \bibfnamefont {J.}}} (\bibinfo {year} {1932}),\ \href@noop {} {\emph
  {\bibinfo {title} {{Mathematische Grundlagen der Quantenmechanik}}}}\
  (\bibinfo  {publisher} {Springer})\BibitemShut {NoStop}%
\bibitem [{\citenamefont {Ng}\ \emph {et~al.}(2012)\citenamefont {Ng},
  \citenamefont {Berta},\ and\ \citenamefont {Wehner}}]{nelly12}%
  \BibitemOpen
  \bibfield  {author} {\bibinfo {author} {\bibnamefont {Ng}, \bibfnamefont
  {N.~H.~Y.}}, \bibinfo {author} {\bibfnamefont {M.}~\bibnamefont {Berta}}, \
  and\ \bibinfo {author} {\bibfnamefont {S.}~\bibnamefont {Wehner}}} (\bibinfo
  {year} {2012}),\ \href {\doibase 10.1103/PhysRevA.86.042315} {\bibfield
  {journal} {\bibinfo  {journal} {Physical Review A}\ }\textbf {\bibinfo
  {volume} {86}}~(\bibinfo {number} {4}),\ \bibinfo {pages}
  {042315}}\BibitemShut {NoStop}%
\bibitem [{\citenamefont {Ohya}\ and\ \citenamefont {Petz}(1993)}]{ohya93}%
  \BibitemOpen
  \bibfield  {author} {\bibinfo {author} {\bibnamefont {Ohya}, \bibfnamefont
  {M.}}, \ and\ \bibinfo {author} {\bibfnamefont {D.}~\bibnamefont {Petz}}}
  (\bibinfo {year} {1993}),\ \href@noop {} {\emph {\bibinfo {title} {{Quantum
  Entropy and Its Use}}}}\ (\bibinfo  {publisher} {Springer})\BibitemShut
  {NoStop}%
\bibitem [{\citenamefont {Ollivier}\ and\ \citenamefont
  {Zurek}(2001)}]{ollivier01}%
  \BibitemOpen
  \bibfield  {author} {\bibinfo {author} {\bibnamefont {Ollivier},
  \bibfnamefont {H.}}, \ and\ \bibinfo {author} {\bibfnamefont {W.~H.}\
  \bibnamefont {Zurek}}} (\bibinfo {year} {2001}),\ \href {\doibase
  10.1103/PhysRevLett.88.017901} {\bibfield  {journal} {\bibinfo  {journal}
  {Physical Review Letters}\ }\textbf {\bibinfo {volume} {88}}~(\bibinfo
  {number} {1}),\ \bibinfo {pages} {017901}}\BibitemShut {NoStop}%
\bibitem [{\citenamefont {Oppenheim}\ and\ \citenamefont
  {Wehner}(2010)}]{oppenheim10}%
  \BibitemOpen
  \bibfield  {author} {\bibinfo {author} {\bibnamefont {Oppenheim},
  \bibfnamefont {J.}}, \ and\ \bibinfo {author} {\bibfnamefont
  {S.}~\bibnamefont {Wehner}}} (\bibinfo {year} {2010}),\ \href {\doibase
  10.1126/science.1192065} {\bibfield  {journal} {\bibinfo  {journal}
  {Science}\ }\textbf {\bibinfo {volume} {330}}~(\bibinfo {number} {6007}),\
  \bibinfo {pages} {1072}}\BibitemShut {NoStop}%
\bibitem [{\citenamefont {Ozawa}(2003)}]{ozawa03}%
  \BibitemOpen
  \bibfield  {author} {\bibinfo {author} {\bibnamefont {Ozawa}, \bibfnamefont
  {M.}}} (\bibinfo {year} {2003}),\ \href {\doibase 10.1103/PhysRevA.67.042105}
  {\bibfield  {journal} {\bibinfo  {journal} {Physical Review A}\ }\textbf
  {\bibinfo {volume} {67}}~(\bibinfo {number} {4}),\ \bibinfo {pages}
  {042105}}\BibitemShut {NoStop}%
\bibitem [{\citenamefont {Partovi}(1983)}]{partovi83}%
  \BibitemOpen
  \bibfield  {author} {\bibinfo {author} {\bibnamefont {Partovi}, \bibfnamefont
  {M.}}} (\bibinfo {year} {1983}),\ \href {\doibase
  10.1103/PhysRevLett.50.1883} {\bibfield  {journal} {\bibinfo  {journal}
  {Physical Review Letters}\ }\textbf {\bibinfo {volume} {50}}~(\bibinfo
  {number} {24}),\ \bibinfo {pages} {1883}}\BibitemShut {NoStop}%
\bibitem [{\citenamefont {Partovi}(2011)}]{partovi11}%
  \BibitemOpen
  \bibfield  {author} {\bibinfo {author} {\bibnamefont {Partovi}, \bibfnamefont
  {M.~H.}}} (\bibinfo {year} {2011}),\ \href {\doibase
  10.1103/PhysRevA.84.052117} {\bibfield  {journal} {\bibinfo  {journal}
  {Physical Review A}\ }\textbf {\bibinfo {volume} {84}}~(\bibinfo {number}
  {5}),\ \bibinfo {pages} {052117}}\BibitemShut {NoStop}%
\bibitem [{\citenamefont {Pati}\ \emph {et~al.}(2012)\citenamefont {Pati},
  \citenamefont {Wilde}, \citenamefont {Devi}, \citenamefont {Rajagopal},\ and\
  \citenamefont {Sudha}}]{pati12}%
  \BibitemOpen
  \bibfield  {author} {\bibinfo {author} {\bibnamefont {Pati}, \bibfnamefont
  {A.~K.}}, \bibinfo {author} {\bibfnamefont {M.~M.}\ \bibnamefont {Wilde}},
  \bibinfo {author} {\bibfnamefont {A.~R.~U.}\ \bibnamefont {Devi}}, \bibinfo
  {author} {\bibfnamefont {A.~K.}\ \bibnamefont {Rajagopal}}, \ and\ \bibinfo
  {author} {\bibnamefont {Sudha}}} (\bibinfo {year} {2012}),\ \href {\doibase
  10.1103/PhysRevA.86.042105} {\bibfield  {journal} {\bibinfo  {journal}
  {Physical Review A}\ }\textbf {\bibinfo {volume} {86}}~(\bibinfo {number}
  {4}),\ \bibinfo {pages} {042105}}\BibitemShut {NoStop}%
\bibitem [{\citenamefont {Prevedel}\ \emph {et~al.}(2011)\citenamefont
  {Prevedel}, \citenamefont {Hamel}, \citenamefont {Colbeck}, \citenamefont
  {Fisher},\ and\ \citenamefont {Resch}}]{prevedel11}%
  \BibitemOpen
  \bibfield  {author} {\bibinfo {author} {\bibnamefont {Prevedel},
  \bibfnamefont {R.}}, \bibinfo {author} {\bibfnamefont {D.~R.}\ \bibnamefont
  {Hamel}}, \bibinfo {author} {\bibfnamefont {R.}~\bibnamefont {Colbeck}},
  \bibinfo {author} {\bibfnamefont {K.}~\bibnamefont {Fisher}}, \ and\ \bibinfo
  {author} {\bibfnamefont {K.~J.}\ \bibnamefont {Resch}}} (\bibinfo {year}
  {2011}),\ \href {\doibase 10.1038/nphys2048} {\bibfield  {journal} {\bibinfo
  {journal} {Nature Physics}\ }\textbf {\bibinfo {volume} {7}}~(\bibinfo
  {number} {10}),\ \bibinfo {pages} {757}}\BibitemShut {NoStop}%
\bibitem [{\citenamefont {Pucha{\l}a}\ \emph {et~al.}(2015)\citenamefont
  {Pucha{\l}a}, \citenamefont {Rudnicki}, \citenamefont {Chabuda},
  \citenamefont {Paraniak},\ and\ \citenamefont
  {{\.{Z}}yczkowski}}]{puchala15}%
  \BibitemOpen
  \bibfield  {author} {\bibinfo {author} {\bibnamefont {Pucha{\l}a},
  \bibfnamefont {Z.}}, \bibinfo {author} {\bibfnamefont {{\L}.}~\bibnamefont
  {Rudnicki}}, \bibinfo {author} {\bibfnamefont {K.}~\bibnamefont {Chabuda}},
  \bibinfo {author} {\bibfnamefont {M.}~\bibnamefont {Paraniak}}, \ and\
  \bibinfo {author} {\bibfnamefont {K.}~\bibnamefont {{\.{Z}}yczkowski}}}
  (\bibinfo {year} {2015}),\ \href {\doibase 10.1103/PhysRevA.92.032109}
  {\bibfield  {journal} {\bibinfo  {journal} {Physical Review A}\ }\textbf
  {\bibinfo {volume} {92}}~(\bibinfo {number} {3}),\ \bibinfo {pages}
  {032109}}\BibitemShut {NoStop}%
\bibitem [{\citenamefont {Pucha{\l}a}\ \emph {et~al.}(2013)\citenamefont
  {Pucha{\l}a}, \citenamefont {Rudnicki},\ and\ \citenamefont
  {{\.{Z}}yczkowski}}]{puchala13}%
  \BibitemOpen
  \bibfield  {author} {\bibinfo {author} {\bibnamefont {Pucha{\l}a},
  \bibfnamefont {Z.}}, \bibinfo {author} {\bibfnamefont {{\L}.}~\bibnamefont
  {Rudnicki}}, \ and\ \bibinfo {author} {\bibfnamefont {K.}~\bibnamefont
  {{\.{Z}}yczkowski}}} (\bibinfo {year} {2013}),\ \href {\doibase
  10.1088/1751-8113/46/27/272002} {\bibfield  {journal} {\bibinfo  {journal}
  {Journal of Physics A: Mathematical and Theoretical}\ }\textbf {\bibinfo
  {volume} {46}}~(\bibinfo {number} {27}),\ \bibinfo {pages}
  {272002}}\BibitemShut {NoStop}%
\bibitem [{\citenamefont {Rastegin}(2013{\natexlab{a}})}]{rastegin12b}%
  \BibitemOpen
  \bibfield  {author} {\bibinfo {author} {\bibnamefont {Rastegin},
  \bibfnamefont {A.~E.}}} (\bibinfo {year} {2013}{\natexlab{a}}),\ \href
  {\doibase 10.1007/s11128-013-0568-y} {\bibfield  {journal} {\bibinfo
  {journal} {Quantum Information Processing}\ }\textbf {\bibinfo {volume}
  {12}}~(\bibinfo {number} {9}),\ \bibinfo {pages} {2947}}\BibitemShut
  {NoStop}%
\bibitem [{\citenamefont {Rastegin}(2013{\natexlab{b}})}]{rastegin13}%
  \BibitemOpen
  \bibfield  {author} {\bibinfo {author} {\bibnamefont {Rastegin},
  \bibfnamefont {A.~E.}}} (\bibinfo {year} {2013}{\natexlab{b}}),\ \href
  {\doibase 10.1140/epjd/e2013-40453-2} {\bibfield  {journal} {\bibinfo
  {journal} {European Physics Journal D}\ }\textbf {\bibinfo {volume}
  {67}}~(\bibinfo {number} {12}),\ \bibinfo {pages} {269}}\BibitemShut
  {NoStop}%
\bibitem [{\citenamefont {Rastegin}(2014)}]{rastegin14}%
  \BibitemOpen
  \bibfield  {author} {\bibinfo {author} {\bibnamefont {Rastegin},
  \bibfnamefont {A.~E.}}} (\bibinfo {year} {2014}),\ \href {\doibase
  10.1088/0253-6102/61/3/04} {\bibfield  {journal} {\bibinfo  {journal}
  {Communications in Theoretical Physics}\ }\textbf {\bibinfo {volume}
  {61}}~(\bibinfo {number} {3}),\ \bibinfo {pages} {293}}\BibitemShut {NoStop}%
\bibitem [{\citenamefont {Rastegin}(2015{\natexlab{a}})}]{rastegin14b}%
  \BibitemOpen
  \bibfield  {author} {\bibinfo {author} {\bibnamefont {Rastegin},
  \bibfnamefont {A.~E.}}} (\bibinfo {year} {2015}{\natexlab{a}}),\ \href
  {\doibase 10.1007/s11128-014-0869-9} {\bibfield  {journal} {\bibinfo
  {journal} {Quantum Information Processing}\ }\textbf {\bibinfo {volume}
  {14}}~(\bibinfo {number} {2}),\ \bibinfo {pages} {783}}\BibitemShut {NoStop}%
\bibitem [{\citenamefont {Rastegin}(2015{\natexlab{b}})}]{rastegin15}%
  \BibitemOpen
  \bibfield  {author} {\bibinfo {author} {\bibnamefont {Rastegin},
  \bibfnamefont {A.~E.}}} (\bibinfo {year} {2015}{\natexlab{b}}),\ \href
  {\doibase 10.1142/S1230161215500055} {\bibfield  {journal} {\bibinfo
  {journal} {Open Systems {\&} Information Dynamics}\ }\textbf {\bibinfo
  {volume} {22}}~(\bibinfo {number} {01}),\ \bibinfo {pages}
  {1550005}}\BibitemShut {NoStop}%
\bibitem [{\citenamefont {Rastegin}(2015{\natexlab{c}})}]{rastegin15b}%
  \BibitemOpen
  \bibfield  {author} {\bibinfo {author} {\bibnamefont {Rastegin},
  \bibfnamefont {A.~E.}}} (\bibinfo {year} {2015}{\natexlab{c}}),\ \href
  {\doibase 10.1007/s10701-015-9909-2} {\bibfield  {journal} {\bibinfo
  {journal} {Foundations of Physics}\ }\textbf {\bibinfo {volume}
  {45}}~(\bibinfo {number} {8}),\ \bibinfo {pages} {923}}\BibitemShut {NoStop}%
\bibitem [{\citenamefont {Rastegin}\ and\ \citenamefont
  {{\.{Z}}yczkowski}(2016)}]{rastegin16}%
  \BibitemOpen
  \bibfield  {author} {\bibinfo {author} {\bibnamefont {Rastegin},
  \bibfnamefont {A.~E.}}, \ and\ \bibinfo {author} {\bibfnamefont
  {K.}~\bibnamefont {{\.{Z}}yczkowski}}} (\bibinfo {year} {2016}),\ \href
  {\doibase 10.1088/1751-8113/49/35/355301} {\bibfield  {journal} {\bibinfo
  {journal} {Journal of Physics A: Mathematical and Theoretical}\ }\textbf
  {\bibinfo {volume} {49}}~(\bibinfo {number} {35}),\ \bibinfo {pages}
  {355301}}\BibitemShut {NoStop}%
\bibitem [{\citenamefont {Ren}\ and\ \citenamefont {Fan}(2014)}]{ren14}%
  \BibitemOpen
  \bibfield  {author} {\bibinfo {author} {\bibnamefont {Ren}, \bibfnamefont
  {L.-H.}}, \ and\ \bibinfo {author} {\bibfnamefont {H.}~\bibnamefont {Fan}}}
  (\bibinfo {year} {2014}),\ \href {\doibase 10.1103/PhysRevA.90.052110}
  {\bibfield  {journal} {\bibinfo  {journal} {Physical Review A}\ }\textbf
  {\bibinfo {volume} {90}}~(\bibinfo {number} {5}),\ \bibinfo {pages}
  {052110}}\BibitemShut {NoStop}%
\bibitem [{\citenamefont {Renes}\ and\ \citenamefont
  {Boileau}(2009)}]{renes09}%
  \BibitemOpen
  \bibfield  {author} {\bibinfo {author} {\bibnamefont {Renes}, \bibfnamefont
  {J.}}, \ and\ \bibinfo {author} {\bibfnamefont {J.-C.}\ \bibnamefont
  {Boileau}}} (\bibinfo {year} {2009}),\ \href {\doibase
  10.1103/PhysRevLett.103.020402} {\bibfield  {journal} {\bibinfo  {journal}
  {Physical Review Letters}\ }\textbf {\bibinfo {volume} {103}}~(\bibinfo
  {number} {2}),\ \bibinfo {pages} {020402}}\BibitemShut {NoStop}%
\bibitem [{\citenamefont {Renes}\ \emph {et~al.}(2004)\citenamefont {Renes},
  \citenamefont {Blume-Kohout}, \citenamefont {Scott},\ and\ \citenamefont
  {Caves}}]{renes04}%
  \BibitemOpen
  \bibfield  {author} {\bibinfo {author} {\bibnamefont {Renes}, \bibfnamefont
  {J.~M.}}, \bibinfo {author} {\bibfnamefont {R.}~\bibnamefont {Blume-Kohout}},
  \bibinfo {author} {\bibfnamefont {A.~J.}\ \bibnamefont {Scott}}, \ and\
  \bibinfo {author} {\bibfnamefont {C.~M.}\ \bibnamefont {Caves}}} (\bibinfo
  {year} {2004}),\ \href {\doibase 10.1063/1.1737053} {\bibfield  {journal}
  {\bibinfo  {journal} {Journal of Mathematical Physics}\ }\textbf {\bibinfo
  {volume} {45}}~(\bibinfo {number} {6}),\ \bibinfo {pages} {2171}}\BibitemShut
  {NoStop}%
\bibitem [{\citenamefont {Renes}\ and\ \citenamefont
  {Boileau}(2008)}]{renes08}%
  \BibitemOpen
  \bibfield  {author} {\bibinfo {author} {\bibnamefont {Renes}, \bibfnamefont
  {J.~M.}}, \ and\ \bibinfo {author} {\bibfnamefont {J.-C.}\ \bibnamefont
  {Boileau}}} (\bibinfo {year} {2008}),\ \href {\doibase
  10.1103/PhysRevA.78.032335} {\bibfield  {journal} {\bibinfo  {journal}
  {Physical Review A}\ }\textbf {\bibinfo {volume} {78}}~(\bibinfo {number}
  {3}),\ \bibinfo {pages} {032335}}\BibitemShut {NoStop}%
\bibitem [{\citenamefont {Renes}\ \emph {et~al.}(2015)\citenamefont {Renes},
  \citenamefont {Sutter}, \citenamefont {Dupuis},\ and\ \citenamefont
  {Renner}}]{renes15}%
  \BibitemOpen
  \bibfield  {author} {\bibinfo {author} {\bibnamefont {Renes}, \bibfnamefont
  {J.~M.}}, \bibinfo {author} {\bibfnamefont {D.}~\bibnamefont {Sutter}},
  \bibinfo {author} {\bibfnamefont {F.}~\bibnamefont {Dupuis}}, \ and\ \bibinfo
  {author} {\bibfnamefont {R.}~\bibnamefont {Renner}}} (\bibinfo {year}
  {2015}),\ \href {\doibase 10.1109/TIT.2015.2468084} {\bibfield  {journal}
  {\bibinfo  {journal} {IEEE Transactions on Information Theory}\ }\textbf
  {\bibinfo {volume} {61}}~(\bibinfo {number} {11}),\ \bibinfo {pages}
  {6395}}\BibitemShut {NoStop}%
\bibitem [{\citenamefont {Renes}\ and\ \citenamefont {Wilde}(2014)}]{renes14}%
  \BibitemOpen
  \bibfield  {author} {\bibinfo {author} {\bibnamefont {Renes}, \bibfnamefont
  {J.~M.}}, \ and\ \bibinfo {author} {\bibfnamefont {M.~M.}\ \bibnamefont
  {Wilde}}} (\bibinfo {year} {2014}),\ \href {\doibase
  10.1109/TIT.2014.2314463} {\bibfield  {journal} {\bibinfo  {journal} {IEEE
  Transactions on Information Theory}\ }\textbf {\bibinfo {volume}
  {60}}~(\bibinfo {number} {6}),\ \bibinfo {pages} {3090}}\BibitemShut
  {NoStop}%
\bibitem [{\citenamefont {Renner}(2005)}]{renner05}%
  \BibitemOpen
  \bibfield  {author} {\bibinfo {author} {\bibnamefont {Renner}, \bibfnamefont
  {R.}}} (\bibinfo {year} {2005}),\ \emph {\bibinfo {title} {{Security of
  Quantum Key Distribution}}},\ \href {http://arxiv.org/abs/quant-ph/0512258}
  {Ph.D. thesis}\ (\bibinfo  {school} {ETH Zurich})\BibitemShut {NoStop}%
\bibitem [{\citenamefont {Renner}\ and\ \citenamefont
  {K{\"{o}}nig}(2005)}]{rennerkoenig05}%
  \BibitemOpen
  \bibfield  {author} {\bibinfo {author} {\bibnamefont {Renner}, \bibfnamefont
  {R.}}, \ and\ \bibinfo {author} {\bibfnamefont {R.}~\bibnamefont
  {K{\"{o}}nig}}} (\bibinfo {year} {2005}),\ in\ \href {\doibase
  10.1007/978-3-540-30576-7_22} {\emph {\bibinfo {booktitle} {Proc. TCC
  2005}}},\ \bibinfo {series} {LNCS}, Vol.\ \bibinfo {volume} {3378}\ (\bibinfo
  {address} {Cambridge, MA})\ pp.\ \bibinfo {pages} {407--425}\BibitemShut
  {NoStop}%
\bibitem [{\citenamefont {R{\'{e}}nyi}(1961)}]{renyi61}%
  \BibitemOpen
  \bibfield  {author} {\bibinfo {author} {\bibnamefont {R{\'{e}}nyi},
  \bibfnamefont {A.}}} (\bibinfo {year} {1961}),\ in\ \href@noop {} {\emph
  {\bibinfo {booktitle} {Proc. 4th Berkeley Symposium on Mathematical
  Statistics and Probability}}},\ Vol.~\bibinfo {volume} {1}\ (\bibinfo
  {publisher} {University of California Press},\ \bibinfo {address} {Berkeley,
  California, USA})\ pp.\ \bibinfo {pages} {547--561}\BibitemShut {NoStop}%
\bibitem [{\citenamefont {del Rio}\ \emph {et~al.}(2011)\citenamefont {del
  Rio}, \citenamefont {Aberg}, \citenamefont {Renner}, \citenamefont
  {Dahlsten},\ and\ \citenamefont {Vedral}}]{delrio11}%
  \BibitemOpen
  \bibfield  {author} {\bibinfo {author} {\bibnamefont {del Rio}, \bibfnamefont
  {L.}}, \bibinfo {author} {\bibfnamefont {J.}~\bibnamefont {Aberg}}, \bibinfo
  {author} {\bibfnamefont {R.}~\bibnamefont {Renner}}, \bibinfo {author}
  {\bibfnamefont {O.}~\bibnamefont {Dahlsten}}, \ and\ \bibinfo {author}
  {\bibfnamefont {V.}~\bibnamefont {Vedral}}} (\bibinfo {year} {2011}),\ \href
  {\doibase 10.1038/nature10123} {\bibfield  {journal} {\bibinfo  {journal}
  {Nature}\ }\textbf {\bibinfo {volume} {474}}~(\bibinfo {number} {7349}),\
  \bibinfo {pages} {61}}\BibitemShut {NoStop}%
\bibitem [{\citenamefont {Robertson}(1929)}]{robertson29}%
  \BibitemOpen
  \bibfield  {author} {\bibinfo {author} {\bibnamefont {Robertson},
  \bibfnamefont {H.~P.}}} (\bibinfo {year} {1929}),\ \href {\doibase
  10.1103/PhysRev.34.163} {\bibfield  {journal} {\bibinfo  {journal} {Physical
  Review}\ }\textbf {\bibinfo {volume} {34}}~(\bibinfo {number} {1}),\ \bibinfo
  {pages} {163}}\BibitemShut {NoStop}%
\bibitem [{\citenamefont {{Rojas Gonz{\'{a}}lez}}\ \emph
  {et~al.}(1995)\citenamefont {{Rojas Gonz{\'{a}}lez}}, \citenamefont
  {Vaccaro},\ and\ \citenamefont {Barnett}}]{gonzales95}%
  \BibitemOpen
  \bibfield  {author} {\bibinfo {author} {\bibnamefont {{Rojas
  Gonz{\'{a}}lez}}, \bibfnamefont {A.}}, \bibinfo {author} {\bibfnamefont
  {J.~A.}\ \bibnamefont {Vaccaro}}, \ and\ \bibinfo {author} {\bibfnamefont
  {S.~M.}\ \bibnamefont {Barnett}}} (\bibinfo {year} {1995}),\ \href {\doibase
  10.1016/0375-9601(95)00582-N} {\bibfield  {journal} {\bibinfo  {journal}
  {Physics Letters A}\ }\textbf {\bibinfo {volume} {205}}~(\bibinfo {number}
  {4}),\ \bibinfo {pages} {247}}\BibitemShut {NoStop}%
\bibitem [{\citenamefont {Romera}\ and\ \citenamefont
  {Calixto}(2015)}]{romera15}%
  \BibitemOpen
  \bibfield  {author} {\bibinfo {author} {\bibnamefont {Romera}, \bibfnamefont
  {E.}}, \ and\ \bibinfo {author} {\bibfnamefont {M.}~\bibnamefont {Calixto}}}
  (\bibinfo {year} {2015}),\ \href {\doibase 10.1088/0953-8984/27/17/175003}
  {\bibfield  {journal} {\bibinfo  {journal} {Journal of Physics: Condensed
  Matter}\ }\textbf {\bibinfo {volume} {27}}~(\bibinfo {number} {17}),\
  \bibinfo {pages} {175003}}\BibitemShut {NoStop}%
\bibitem [{\citenamefont {Rudnicki}(2011)}]{rudnicki11}%
  \BibitemOpen
  \bibfield  {author} {\bibinfo {author} {\bibnamefont {Rudnicki},
  \bibfnamefont {{\L}.}}} (\bibinfo {year} {2011}),\ \href {\doibase
  10.1007/s10946-011-9227-x} {\bibfield  {journal} {\bibinfo  {journal}
  {Journal of Russian Laser Research}\ }\textbf {\bibinfo {volume}
  {32}}~(\bibinfo {number} {4}),\ \bibinfo {pages} {393}}\BibitemShut {NoStop}%
\bibitem [{\citenamefont {Rudnicki}(2015)}]{rudnicki15}%
  \BibitemOpen
  \bibfield  {author} {\bibinfo {author} {\bibnamefont {Rudnicki},
  \bibfnamefont {{\L}.}}} (\bibinfo {year} {2015}),\ \href {\doibase
  10.1103/PhysRevA.91.032123} {\bibfield  {journal} {\bibinfo  {journal}
  {Physical Review A}\ }\textbf {\bibinfo {volume} {91}}~(\bibinfo {number}
  {3}),\ \bibinfo {pages} {032123}}\BibitemShut {NoStop}%
\bibitem [{\citenamefont {Rudnicki}\ \emph {et~al.}(2014)\citenamefont
  {Rudnicki}, \citenamefont {Pucha{\l}a},\ and\ \citenamefont
  {{\.{Z}}yczkowski}}]{rudnicki14}%
  \BibitemOpen
  \bibfield  {author} {\bibinfo {author} {\bibnamefont {Rudnicki},
  \bibfnamefont {{\L}.}}, \bibinfo {author} {\bibfnamefont {Z.}~\bibnamefont
  {Pucha{\l}a}}, \ and\ \bibinfo {author} {\bibfnamefont {K.}~\bibnamefont
  {{\.{Z}}yczkowski}}} (\bibinfo {year} {2014}),\ \href {\doibase
  10.1103/PhysRevA.89.052115} {\bibfield  {journal} {\bibinfo  {journal}
  {Physical Review A}\ }\textbf {\bibinfo {volume} {89}}~(\bibinfo {number}
  {5}),\ \bibinfo {pages} {052115}}\BibitemShut {NoStop}%
\bibitem [{\citenamefont {Rudnicki}\ \emph {et~al.}(2012)\citenamefont
  {Rudnicki}, \citenamefont {Walborn},\ and\ \citenamefont
  {Toscano}}]{rudnicki12}%
  \BibitemOpen
  \bibfield  {author} {\bibinfo {author} {\bibnamefont {Rudnicki},
  \bibfnamefont {{\L}.}}, \bibinfo {author} {\bibfnamefont {S.~P.}\
  \bibnamefont {Walborn}}, \ and\ \bibinfo {author} {\bibfnamefont
  {F.}~\bibnamefont {Toscano}}} (\bibinfo {year} {2012}),\ \href {\doibase
  10.1103/PhysRevA.85.042115} {\bibfield  {journal} {\bibinfo  {journal}
  {Physical Review A}\ }\textbf {\bibinfo {volume} {85}}~(\bibinfo {number}
  {4}),\ \bibinfo {pages} {042115}}\BibitemShut {NoStop}%
\bibitem [{\citenamefont {Rumin}(2011)}]{rumin11}%
  \BibitemOpen
  \bibfield  {author} {\bibinfo {author} {\bibnamefont {Rumin}, \bibfnamefont
  {M.}}} (\bibinfo {year} {2011}),\ \href {\doibase 10.1215/00127094-1444305}
  {\bibfield  {journal} {\bibinfo  {journal} {Duke Mathematical Journal}\
  }\textbf {\bibinfo {volume} {160}}~(\bibinfo {number} {3}),\ \bibinfo {pages}
  {567}}\BibitemShut {NoStop}%
\bibitem [{\citenamefont {Rumin}(2012)}]{rumin12}%
  \BibitemOpen
  \bibfield  {author} {\bibinfo {author} {\bibnamefont {Rumin}, \bibfnamefont
  {M.}}} (\bibinfo {year} {2012}),\ \href {\doibase 10.1007/s11005-011-0543-4}
  {\bibfield  {journal} {\bibinfo  {journal} {Letters in Mathematical Physics}\
  }\textbf {\bibinfo {volume} {100}}~(\bibinfo {number} {3}),\ \bibinfo {pages}
  {291}}\BibitemShut {NoStop}%
\bibitem [{\citenamefont {Saboia}\ \emph {et~al.}(2011)\citenamefont {Saboia},
  \citenamefont {Toscano},\ and\ \citenamefont {Walborn}}]{saboia11}%
  \BibitemOpen
  \bibfield  {author} {\bibinfo {author} {\bibnamefont {Saboia}, \bibfnamefont
  {A.}}, \bibinfo {author} {\bibfnamefont {F.}~\bibnamefont {Toscano}}, \ and\
  \bibinfo {author} {\bibfnamefont {S.~P.}\ \bibnamefont {Walborn}}} (\bibinfo
  {year} {2011}),\ \href {\doibase 10.1103/PhysRevA.83.032307} {\bibfield
  {journal} {\bibinfo  {journal} {Physical Review A}\ }\textbf {\bibinfo
  {volume} {83}}~(\bibinfo {number} {3}),\ \bibinfo {pages}
  {032307}}\BibitemShut {NoStop}%
\bibitem [{\citenamefont {S{\'{a}}nchez-Ruiz}(1993)}]{sanchez93}%
  \BibitemOpen
  \bibfield  {author} {\bibinfo {author} {\bibnamefont {S{\'{a}}nchez-Ruiz},
  \bibfnamefont {J.}}} (\bibinfo {year} {1993}),\ \href {\doibase
  10.1016/0375-9601(93)90269-6} {\bibfield  {journal} {\bibinfo  {journal}
  {Physics Letters A}\ }\textbf {\bibinfo {volume} {173}}~(\bibinfo {number}
  {3}),\ \bibinfo {pages} {233}}\BibitemShut {NoStop}%
\bibitem [{\citenamefont {S{\'{a}}nchez-Ruiz}(1995)}]{sanchez95}%
  \BibitemOpen
  \bibfield  {author} {\bibinfo {author} {\bibnamefont {S{\'{a}}nchez-Ruiz},
  \bibfnamefont {J.}}} (\bibinfo {year} {1995}),\ \href {\doibase
  10.1016/0375-9601(95)00219-S} {\bibfield  {journal} {\bibinfo  {journal}
  {Physics Letters A}\ }\textbf {\bibinfo {volume} {201}}~(\bibinfo {number}
  {2-3}),\ \bibinfo {pages} {125}}\BibitemShut {NoStop}%
\bibitem [{\citenamefont {S{\'{a}}nchez-Ruiz}(1998)}]{sanchez98}%
  \BibitemOpen
  \bibfield  {author} {\bibinfo {author} {\bibnamefont {S{\'{a}}nchez-Ruiz},
  \bibfnamefont {J.}}} (\bibinfo {year} {1998}),\ \href {\doibase
  10.1016/S0375-9601(98)00292-8} {\bibfield  {journal} {\bibinfo  {journal}
  {Physics Letters A}\ }\textbf {\bibinfo {volume} {244}}~(\bibinfo {number}
  {4}),\ \bibinfo {pages} {189}}\BibitemShut {NoStop}%
\bibitem [{\citenamefont {Scarani}\ \emph {et~al.}(2009)\citenamefont
  {Scarani}, \citenamefont {Bechmann-Pasquinucci}, \citenamefont {Cerf},
  \citenamefont {Dusek}, \citenamefont {L{\"{u}}tkenhaus},\ and\ \citenamefont
  {Peev}}]{scarani09a}%
  \BibitemOpen
  \bibfield  {author} {\bibinfo {author} {\bibnamefont {Scarani}, \bibfnamefont
  {V.}}, \bibinfo {author} {\bibfnamefont {H.}~\bibnamefont
  {Bechmann-Pasquinucci}}, \bibinfo {author} {\bibfnamefont {N.}~\bibnamefont
  {Cerf}}, \bibinfo {author} {\bibfnamefont {M.}~\bibnamefont {Dusek}},
  \bibinfo {author} {\bibfnamefont {N.}~\bibnamefont {L{\"{u}}tkenhaus}}, \
  and\ \bibinfo {author} {\bibfnamefont {M.}~\bibnamefont {Peev}}} (\bibinfo
  {year} {2009}),\ \href {\doibase 10.1103/RevModPhys.81.1301} {\bibfield
  {journal} {\bibinfo  {journal} {Review of Modern Physics}\ }\textbf {\bibinfo
  {volume} {81}}~(\bibinfo {number} {3}),\ \bibinfo {pages} {1301}}\BibitemShut
  {NoStop}%
\bibitem [{\citenamefont {Schaffner}(2007)}]{schaffner07}%
  \BibitemOpen
  \bibfield  {author} {\bibinfo {author} {\bibnamefont {Schaffner},
  \bibfnamefont {C.}}} (\bibinfo {year} {2007}),\ \emph {\bibinfo {title}
  {{Cryptography in the Bounded-Quantum-Storage Model}}},\ \href
  {http://arxiv.org/abs/0709.0289} {\bibinfo {type} {Phd thesis}}\ (\bibinfo
  {school} {University of Aarhus})\BibitemShut {NoStop}%
\bibitem [{\citenamefont {Schneeloch}\ \emph {et~al.}(2014)\citenamefont
  {Schneeloch}, \citenamefont {Broadbent},\ and\ \citenamefont
  {Howell}}]{schneeloch14}%
  \BibitemOpen
  \bibfield  {author} {\bibinfo {author} {\bibnamefont {Schneeloch},
  \bibfnamefont {J.}}, \bibinfo {author} {\bibfnamefont {C.~J.}\ \bibnamefont
  {Broadbent}}, \ and\ \bibinfo {author} {\bibfnamefont {J.~C.}\ \bibnamefont
  {Howell}}} (\bibinfo {year} {2014}),\ \href {\doibase
  10.1103/PhysRevA.90.062119} {\bibfield  {journal} {\bibinfo  {journal}
  {Physical Review A}\ }\textbf {\bibinfo {volume} {90}}~(\bibinfo {number}
  {6}),\ \bibinfo {pages} {062119}}\BibitemShut {NoStop}%
\bibitem [{\citenamefont {Schneeloch}\ \emph {et~al.}(2013)\citenamefont
  {Schneeloch}, \citenamefont {Broadbent}, \citenamefont {Walborn},
  \citenamefont {Cavalcanti},\ and\ \citenamefont {Howell}}]{schneeloch13}%
  \BibitemOpen
  \bibfield  {author} {\bibinfo {author} {\bibnamefont {Schneeloch},
  \bibfnamefont {J.}}, \bibinfo {author} {\bibfnamefont {C.~J.}\ \bibnamefont
  {Broadbent}}, \bibinfo {author} {\bibfnamefont {S.~P.}\ \bibnamefont
  {Walborn}}, \bibinfo {author} {\bibfnamefont {E.~G.}\ \bibnamefont
  {Cavalcanti}}, \ and\ \bibinfo {author} {\bibfnamefont {J.~C.}\ \bibnamefont
  {Howell}}} (\bibinfo {year} {2013}),\ \href {\doibase
  10.1103/PhysRevA.87.062103} {\bibfield  {journal} {\bibinfo  {journal}
  {Physical Review A}\ }\textbf {\bibinfo {volume} {87}}~(\bibinfo {number}
  {6}),\ \bibinfo {pages} {062103}}\BibitemShut {NoStop}%
\bibitem [{\citenamefont {Schr{\"{o}}dinger}(1930)}]{schroedinger30}%
  \BibitemOpen
  \bibfield  {author} {\bibinfo {author} {\bibnamefont {Schr{\"{o}}dinger},
  \bibfnamefont {E.}}} (\bibinfo {year} {1930}),\ \href@noop {} {\bibfield
  {journal} {\bibinfo  {journal} {Proceedings of the Prussian Academy of
  Sciences}\ }\textbf {\bibinfo {volume} {XIX}},\ \bibinfo {pages}
  {296}}\BibitemShut {NoStop}%
\bibitem [{\citenamefont {Schr{\"{o}}dinger}(1935)}]{schroedinger35}%
  \BibitemOpen
  \bibfield  {author} {\bibinfo {author} {\bibnamefont {Schr{\"{o}}dinger},
  \bibfnamefont {E.}}} (\bibinfo {year} {1935}),\ \href {\doibase
  10.1017/S0305004100013554} {\bibfield  {journal} {\bibinfo  {journal}
  {Mathematical Proceedings of the Cambridge Philosophical Society}\ }\textbf
  {\bibinfo {volume} {31}}~(\bibinfo {number} {04}),\ \bibinfo {pages}
  {555}}\BibitemShut {NoStop}%
\bibitem [{\citenamefont {Schumacher}\ and\ \citenamefont
  {Nielsen}(1996)}]{schumacher96}%
  \BibitemOpen
  \bibfield  {author} {\bibinfo {author} {\bibnamefont {Schumacher},
  \bibfnamefont {B.}}, \ and\ \bibinfo {author} {\bibfnamefont {M.~A.}\
  \bibnamefont {Nielsen}}} (\bibinfo {year} {1996}),\ \href {\doibase
  10.1103/PhysRevA.54.2629} {\bibfield  {journal} {\bibinfo  {journal}
  {Physical Review A}\ }\textbf {\bibinfo {volume} {54}}~(\bibinfo {number}
  {4}),\ \bibinfo {pages} {2629}}\BibitemShut {NoStop}%
\bibitem [{\citenamefont {Shannon}(1948)}]{shannon48}%
  \BibitemOpen
  \bibfield  {author} {\bibinfo {author} {\bibnamefont {Shannon}, \bibfnamefont
  {C.}}} (\bibinfo {year} {1948}),\ \href {\doibase
  10.1002/j.1538-7305.1948.tb00917.x} {\bibfield  {journal} {\bibinfo
  {journal} {Bell System Technical Journal}\ }\textbf {\bibinfo {volume}
  {27}},\ \bibinfo {pages} {379}}\BibitemShut {NoStop}%
\bibitem [{\citenamefont {Shor}\ and\ \citenamefont {Preskill}(2000)}]{SP00}%
  \BibitemOpen
  \bibfield  {author} {\bibinfo {author} {\bibnamefont {Shor}, \bibfnamefont
  {P.~W.}}, \ and\ \bibinfo {author} {\bibfnamefont {J.}~\bibnamefont
  {Preskill}}} (\bibinfo {year} {2000}),\ \href {\doibase
  10.1103/PhysRevLett.85.441} {\bibfield  {journal} {\bibinfo  {journal}
  {Physical Review Letters}\ }\textbf {\bibinfo {volume} {85}}~(\bibinfo
  {number} {2}),\ \bibinfo {pages} {441}}\BibitemShut {NoStop}%
\bibitem [{\citenamefont {Slepian}\ and\ \citenamefont
  {Pollak}(1961)}]{slepian64}%
  \BibitemOpen
  \bibfield  {author} {\bibinfo {author} {\bibnamefont {Slepian}, \bibfnamefont
  {D.}}, \ and\ \bibinfo {author} {\bibfnamefont {H.~O.}\ \bibnamefont
  {Pollak}}} (\bibinfo {year} {1961}),\ \href {\doibase
  10.1002/j.1538-7305.1961.tb03976.x} {\bibfield  {journal} {\bibinfo
  {journal} {Bell System Technical Journal}\ }\textbf {\bibinfo {volume}
  {40}}~(\bibinfo {number} {1}),\ \bibinfo {pages} {43}}\BibitemShut {NoStop}%
\bibitem [{\citenamefont {Stinespring}(1955)}]{stinespring54}%
  \BibitemOpen
  \bibfield  {author} {\bibinfo {author} {\bibnamefont {Stinespring},
  \bibfnamefont {W.~F.}}} (\bibinfo {year} {1955}),\ \href {\doibase
  10.1090/S0002-9939-1955-0069403-4} {\bibfield  {journal} {\bibinfo  {journal}
  {Proceedings of the Americal Mathematical Society}\ }\textbf {\bibinfo
  {volume} {6}},\ \bibinfo {pages} {211}}\BibitemShut {NoStop}%
\bibitem [{\citenamefont {Tomamichel}(2012)}]{mythesis}%
  \BibitemOpen
  \bibfield  {author} {\bibinfo {author} {\bibnamefont {Tomamichel},
  \bibfnamefont {M.}}} (\bibinfo {year} {2012}),\ \emph {\bibinfo {title} {{A
  Framework for Non-Asymptotic Quantum Information Theory}}},\ \href
  {http://arxiv.org/abs/1203.2142} {Ph.D. thesis}\ (\bibinfo  {school} {ETH
  Zurich})\BibitemShut {NoStop}%
\bibitem [{\citenamefont {Tomamichel}(2016)}]{mybook}%
  \BibitemOpen
  \bibfield  {author} {\bibinfo {author} {\bibnamefont {Tomamichel},
  \bibfnamefont {M.}}} (\bibinfo {year} {2016}),\ \href {\doibase
  10.1007/978-3-319-21891-5} {\emph {\bibinfo {title} {{Quantum Information
  Processing with Finite Resources --- Mathematical Foundations}}}},\ \bibinfo
  {series} {SpringerBriefs in Mathematical Physics}, Vol.~\bibinfo {volume}
  {5}\ (\bibinfo  {publisher} {Springer International Publishing})\BibitemShut
  {NoStop}%
\bibitem [{\citenamefont {Tomamichel}\ \emph {et~al.}(2014)\citenamefont
  {Tomamichel}, \citenamefont {Berta},\ and\ \citenamefont
  {Hayashi}}]{tomamichel13}%
  \BibitemOpen
  \bibfield  {author} {\bibinfo {author} {\bibnamefont {Tomamichel},
  \bibfnamefont {M.}}, \bibinfo {author} {\bibfnamefont {M.}~\bibnamefont
  {Berta}}, \ and\ \bibinfo {author} {\bibfnamefont {M.}~\bibnamefont
  {Hayashi}}} (\bibinfo {year} {2014}),\ \href {\doibase 10.1063/1.4892761}
  {\bibfield  {journal} {\bibinfo  {journal} {Journal of Mathematical Physics}\
  }\textbf {\bibinfo {volume} {55}}~(\bibinfo {number} {8}),\ \bibinfo {pages}
  {082206}}\BibitemShut {NoStop}%
\bibitem [{\citenamefont {Tomamichel}\ \emph {et~al.}(2009)\citenamefont
  {Tomamichel}, \citenamefont {Colbeck},\ and\ \citenamefont
  {Renner}}]{tomamichel08}%
  \BibitemOpen
  \bibfield  {author} {\bibinfo {author} {\bibnamefont {Tomamichel},
  \bibfnamefont {M.}}, \bibinfo {author} {\bibfnamefont {R.}~\bibnamefont
  {Colbeck}}, \ and\ \bibinfo {author} {\bibfnamefont {R.}~\bibnamefont
  {Renner}}} (\bibinfo {year} {2009}),\ \href {\doibase
  10.1109/TIT.2009.2032797} {\bibfield  {journal} {\bibinfo  {journal} {IEEE
  Transactions on Information Theory}\ }\textbf {\bibinfo {volume}
  {55}}~(\bibinfo {number} {12}),\ \bibinfo {pages} {5840}}\BibitemShut
  {NoStop}%
\bibitem [{\citenamefont {Tomamichel}\ \emph {et~al.}(2010)\citenamefont
  {Tomamichel}, \citenamefont {Colbeck},\ and\ \citenamefont
  {Renner}}]{tomamichel09}%
  \BibitemOpen
  \bibfield  {author} {\bibinfo {author} {\bibnamefont {Tomamichel},
  \bibfnamefont {M.}}, \bibinfo {author} {\bibfnamefont {R.}~\bibnamefont
  {Colbeck}}, \ and\ \bibinfo {author} {\bibfnamefont {R.}~\bibnamefont
  {Renner}}} (\bibinfo {year} {2010}),\ \href {\doibase
  10.1109/TIT.2010.2054130} {\bibfield  {journal} {\bibinfo  {journal} {IEEE
  Transactions on Information Theory}\ }\textbf {\bibinfo {volume}
  {56}}~(\bibinfo {number} {9}),\ \bibinfo {pages} {4674}}\BibitemShut
  {NoStop}%
\bibitem [{\citenamefont {Tomamichel}\ \emph {et~al.}(2013)\citenamefont
  {Tomamichel}, \citenamefont {Fehr}, \citenamefont {Kaniewski},\ and\
  \citenamefont {Wehner}}]{tomamichelfehr12}%
  \BibitemOpen
  \bibfield  {author} {\bibinfo {author} {\bibnamefont {Tomamichel},
  \bibfnamefont {M.}}, \bibinfo {author} {\bibfnamefont {S.}~\bibnamefont
  {Fehr}}, \bibinfo {author} {\bibfnamefont {J.}~\bibnamefont {Kaniewski}}, \
  and\ \bibinfo {author} {\bibfnamefont {S.}~\bibnamefont {Wehner}}} (\bibinfo
  {year} {2013}),\ \href {\doibase 10.1088/1367-2630/15/10/103002} {\bibfield
  {journal} {\bibinfo  {journal} {New Journal of Physics}\ }\textbf {\bibinfo
  {volume} {15}}~(\bibinfo {number} {10}),\ \bibinfo {pages}
  {103002}}\BibitemShut {NoStop}%
\bibitem [{\citenamefont {Tomamichel}\ and\ \citenamefont
  {H{\"{a}}nggi}(2013)}]{haenggi11}%
  \BibitemOpen
  \bibfield  {author} {\bibinfo {author} {\bibnamefont {Tomamichel},
  \bibfnamefont {M.}}, \ and\ \bibinfo {author} {\bibfnamefont
  {E.}~\bibnamefont {H{\"{a}}nggi}}} (\bibinfo {year} {2013}),\ \href {\doibase
  10.1088/1751-8113/46/5/055301} {\bibfield  {journal} {\bibinfo  {journal}
  {Journal of Physics A: Mathematical and Theoretical}\ }\textbf {\bibinfo
  {volume} {46}}~(\bibinfo {number} {5}),\ \bibinfo {pages}
  {055301}}\BibitemShut {NoStop}%
\bibitem [{\citenamefont {Tomamichel}\ \emph {et~al.}(2012)\citenamefont
  {Tomamichel}, \citenamefont {Lim}, \citenamefont {Gisin},\ and\ \citenamefont
  {Renner}}]{tomamichellim11}%
  \BibitemOpen
  \bibfield  {author} {\bibinfo {author} {\bibnamefont {Tomamichel},
  \bibfnamefont {M.}}, \bibinfo {author} {\bibfnamefont {C.~C.~W.}\
  \bibnamefont {Lim}}, \bibinfo {author} {\bibfnamefont {N.}~\bibnamefont
  {Gisin}}, \ and\ \bibinfo {author} {\bibfnamefont {R.}~\bibnamefont
  {Renner}}} (\bibinfo {year} {2012}),\ \href {\doibase 10.1038/ncomms1631}
  {\bibfield  {journal} {\bibinfo  {journal} {Nature Communications}\ }\textbf
  {\bibinfo {volume} {3}},\ \bibinfo {pages} {634}}\BibitemShut {NoStop}%
\bibitem [{\citenamefont {Tomamichel}\ and\ \citenamefont
  {Renner}(2011)}]{tomamichel11}%
  \BibitemOpen
  \bibfield  {author} {\bibinfo {author} {\bibnamefont {Tomamichel},
  \bibfnamefont {M.}}, \ and\ \bibinfo {author} {\bibfnamefont
  {R.}~\bibnamefont {Renner}}} (\bibinfo {year} {2011}),\ \href {\doibase
  10.1103/PhysRevLett.106.110506} {\bibfield  {journal} {\bibinfo  {journal}
  {Physical Review Letters}\ }\textbf {\bibinfo {volume} {106}}~(\bibinfo
  {number} {11}),\ \bibinfo {pages} {110506}}\BibitemShut {NoStop}%
\bibitem [{\citenamefont {Tomamichel}\ \emph {et~al.}(2011)\citenamefont
  {Tomamichel}, \citenamefont {Schaffner}, \citenamefont {Smith},\ and\
  \citenamefont {Renner}}]{tomamichel10}%
  \BibitemOpen
  \bibfield  {author} {\bibinfo {author} {\bibnamefont {Tomamichel},
  \bibfnamefont {M.}}, \bibinfo {author} {\bibfnamefont {C.}~\bibnamefont
  {Schaffner}}, \bibinfo {author} {\bibfnamefont {A.}~\bibnamefont {Smith}}, \
  and\ \bibinfo {author} {\bibfnamefont {R.}~\bibnamefont {Renner}}} (\bibinfo
  {year} {2011}),\ \href {\doibase 10.1109/TIT.2011.2158473} {\bibfield
  {journal} {\bibinfo  {journal} {IEEE Transactions on Information Theory}\
  }\textbf {\bibinfo {volume} {57}}~(\bibinfo {number} {8}),\ \bibinfo {pages}
  {5524}}\BibitemShut {NoStop}%
\bibitem [{\citenamefont {Tsallis}(1988)}]{tsallis88}%
  \BibitemOpen
  \bibfield  {author} {\bibinfo {author} {\bibnamefont {Tsallis}, \bibfnamefont
  {C.}}} (\bibinfo {year} {1988}),\ \href {\doibase 10.1007/BF01016429}
  {\bibfield  {journal} {\bibinfo  {journal} {Journal of Statistical Physics}\
  }\textbf {\bibinfo {volume} {52}}~(\bibinfo {number} {1-2}),\ \bibinfo
  {pages} {479}}\BibitemShut {NoStop}%
\bibitem [{\citenamefont {Uffink}(1990)}]{uffinkThesis}%
  \BibitemOpen
  \bibfield  {author} {\bibinfo {author} {\bibnamefont {Uffink}, \bibfnamefont
  {J.}}} (\bibinfo {year} {1990}),\ \emph {\bibinfo {title} {{Measures of
  Uncertainty and the Uncertainty Principle}}},\ \href
  {http://www.projects.science.uu.nl/igg/jos/publications/proefschrift.pdf}
  {Ph.D. thesis}\ (\bibinfo  {school} {R.U. Utrecht})\BibitemShut {NoStop}%
\bibitem [{\citenamefont {Uhlmann}(1977)}]{uhlmann77}%
  \BibitemOpen
  \bibfield  {author} {\bibinfo {author} {\bibnamefont {Uhlmann}, \bibfnamefont
  {A.}}} (\bibinfo {year} {1977}),\ \href {\doibase 10.1007/BF01609834}
  {\bibfield  {journal} {\bibinfo  {journal} {Communications in Mathematical
  Physics}\ }\textbf {\bibinfo {volume} {54}}~(\bibinfo {number} {1}),\
  \bibinfo {pages} {21}}\BibitemShut {NoStop}%
\bibitem [{\citenamefont {Uhlmann}(1985)}]{uhlmann85}%
  \BibitemOpen
  \bibfield  {author} {\bibinfo {author} {\bibnamefont {Uhlmann}, \bibfnamefont
  {A.}}} (\bibinfo {year} {1985}),\ \href {\doibase 10.1002/andp.19854970419}
  {\bibfield  {journal} {\bibinfo  {journal} {Annals of Physics}\ }\textbf
  {\bibinfo {volume} {497}}~(\bibinfo {number} {4}),\ \bibinfo {pages}
  {524}}\BibitemShut {NoStop}%
\bibitem [{\citenamefont {Umegaki}(1962)}]{umegaki62}%
  \BibitemOpen
  \bibfield  {author} {\bibinfo {author} {\bibnamefont {Umegaki}, \bibfnamefont
  {H.}}} (\bibinfo {year} {1962}),\ \href@noop {} {\bibfield  {journal}
  {\bibinfo  {journal} {Kodai Math. Sem. Rep.}\ }\textbf {\bibinfo {volume}
  {14}},\ \bibinfo {pages} {59}}\BibitemShut {NoStop}%
\bibitem [{\citenamefont {Unruh}(2010)}]{unruh10}%
  \BibitemOpen
  \bibfield  {author} {\bibinfo {author} {\bibnamefont {Unruh}, \bibfnamefont
  {D.}}} (\bibinfo {year} {2010}),\ in\ \href {\doibase
  10.1007/978-3-642-13190-5} {\emph {\bibinfo {booktitle} {Proc. EUROCRYPT
  2010}}},\ \bibinfo {series} {LNCS}, Vol.\ \bibinfo {volume} {6110}\ (\bibinfo
   {publisher} {Springer})\ pp.\ \bibinfo {pages} {486--505}\BibitemShut
  {NoStop}%
\bibitem [{\citenamefont {Vaccaro}(2012)}]{vaccaro11}%
  \BibitemOpen
  \bibfield  {author} {\bibinfo {author} {\bibnamefont {Vaccaro}, \bibfnamefont
  {J.~A.}}} (\bibinfo {year} {2012}),\ \href {\doibase 10.1098/rspa.2011.0271}
  {\bibfield  {journal} {\bibinfo  {journal} {Proceedings of the Royal Society
  A: Mathematical, Physical and Engineering Sciences}\ }\textbf {\bibinfo
  {volume} {468}}~(\bibinfo {number} {2140}),\ \bibinfo {pages}
  {1065}}\BibitemShut {NoStop}%
\bibitem [{\citenamefont {Vadhan}(2012)}]{vadhan12}%
  \BibitemOpen
  \bibfield  {author} {\bibinfo {author} {\bibnamefont {Vadhan}, \bibfnamefont
  {S.~P.}}} (\bibinfo {year} {2012}),\ \href {\doibase 10.1561/0400000010}
  {\bibfield  {journal} {\bibinfo  {journal} {Foundations and Trends in
  Theoretical Computer Science}\ }\textbf {\bibinfo {volume} {7}}~(\bibinfo
  {number} {1-3}),\ \bibinfo {pages} {1}}\BibitemShut {NoStop}%
\bibitem [{\citenamefont {Vallone}\ \emph {et~al.}(2014)\citenamefont
  {Vallone}, \citenamefont {Marangon}, \citenamefont {Tomasin},\ and\
  \citenamefont {Villoresi}}]{vallone14}%
  \BibitemOpen
  \bibfield  {author} {\bibinfo {author} {\bibnamefont {Vallone}, \bibfnamefont
  {G.}}, \bibinfo {author} {\bibfnamefont {D.~G.}\ \bibnamefont {Marangon}},
  \bibinfo {author} {\bibfnamefont {M.}~\bibnamefont {Tomasin}}, \ and\
  \bibinfo {author} {\bibfnamefont {P.}~\bibnamefont {Villoresi}}} (\bibinfo
  {year} {2014}),\ \href {\doibase 10.1103/PhysRevA.90.052327} {\bibfield
  {journal} {\bibinfo  {journal} {Physical Review A}\ }\textbf {\bibinfo
  {volume} {90}}~(\bibinfo {number} {5}),\ \bibinfo {pages}
  {052327}}\BibitemShut {NoStop}%
\bibitem [{\citenamefont {{Ver Steeg}}\ and\ \citenamefont
  {Wehner}(2009)}]{versteeg09}%
  \BibitemOpen
  \bibfield  {author} {\bibinfo {author} {\bibnamefont {{Ver Steeg}},
  \bibfnamefont {G.}}, \ and\ \bibinfo {author} {\bibfnamefont
  {S.}~\bibnamefont {Wehner}}} (\bibinfo {year} {2009}),\ \href@noop {}
  {\bibfield  {journal} {\bibinfo  {journal} {Quantum Information and
  Computation}\ }\textbf {\bibinfo {volume} {9}}~(\bibinfo {number}
  {9{\&}10}),\ \bibinfo {pages} {0801}}\BibitemShut {NoStop}%
\bibitem [{\citenamefont {de~Vicente}\ and\ \citenamefont
  {S{\'{a}}nchez-Ruiz}(2008)}]{devicente08}%
  \BibitemOpen
  \bibfield  {author} {\bibinfo {author} {\bibnamefont {de~Vicente},
  \bibfnamefont {J.}}, \ and\ \bibinfo {author} {\bibfnamefont
  {J.}~\bibnamefont {S{\'{a}}nchez-Ruiz}}} (\bibinfo {year} {2008}),\ \href
  {\doibase 10.1103/PhysRevA.77.042110} {\bibfield  {journal} {\bibinfo
  {journal} {Physical Review A}\ }\textbf {\bibinfo {volume} {77}}~(\bibinfo
  {number} {4}),\ \bibinfo {pages} {042110}}\BibitemShut {NoStop}%
\bibitem [{\citenamefont {Walborn}\ \emph {et~al.}(2011)\citenamefont
  {Walborn}, \citenamefont {Salles}, \citenamefont {Gomes}, \citenamefont
  {Toscano},\ and\ \citenamefont {{Souto Ribeiro}}}]{walborn11}%
  \BibitemOpen
  \bibfield  {author} {\bibinfo {author} {\bibnamefont {Walborn}, \bibfnamefont
  {S.~P.}}, \bibinfo {author} {\bibfnamefont {A.}~\bibnamefont {Salles}},
  \bibinfo {author} {\bibfnamefont {R.~M.}\ \bibnamefont {Gomes}}, \bibinfo
  {author} {\bibfnamefont {F.}~\bibnamefont {Toscano}}, \ and\ \bibinfo
  {author} {\bibfnamefont {P.~H.}\ \bibnamefont {{Souto Ribeiro}}}} (\bibinfo
  {year} {2011}),\ \href {\doibase 10.1103/PhysRevLett.106.130402} {\bibfield
  {journal} {\bibinfo  {journal} {Physical Review Letters}\ }\textbf {\bibinfo
  {volume} {106}}~(\bibinfo {number} {13}),\ \bibinfo {pages}
  {130402}}\BibitemShut {NoStop}%
\bibitem [{\citenamefont {Walborn}\ \emph {et~al.}(2009)\citenamefont
  {Walborn}, \citenamefont {Taketani}, \citenamefont {Salles}, \citenamefont
  {Toscano},\ and\ \citenamefont {{de Matos Filho}}}]{walborn09}%
  \BibitemOpen
  \bibfield  {author} {\bibinfo {author} {\bibnamefont {Walborn}, \bibfnamefont
  {S.~P.}}, \bibinfo {author} {\bibfnamefont {B.~G.}\ \bibnamefont {Taketani}},
  \bibinfo {author} {\bibfnamefont {A.}~\bibnamefont {Salles}}, \bibinfo
  {author} {\bibfnamefont {F.}~\bibnamefont {Toscano}}, \ and\ \bibinfo
  {author} {\bibfnamefont {R.~L.}\ \bibnamefont {{de Matos Filho}}}} (\bibinfo
  {year} {2009}),\ \href {\doibase 10.1103/PhysRevLett.103.160505} {\bibfield
  {journal} {\bibinfo  {journal} {Physical Review Letters}\ }\textbf {\bibinfo
  {volume} {103}}~(\bibinfo {number} {16}),\ \bibinfo {pages}
  {160505}}\BibitemShut {NoStop}%
\bibitem [{\citenamefont {Weedbrook}\ \emph {et~al.}(2012)\citenamefont
  {Weedbrook}, \citenamefont {Pirandola}, \citenamefont
  {Garc{\'{i}}a-Patr{\'{o}}n}, \citenamefont {Cerf}, \citenamefont {Ralph},
  \citenamefont {Shapiro},\ and\ \citenamefont {Lloyd}}]{weedbrook11}%
  \BibitemOpen
  \bibfield  {author} {\bibinfo {author} {\bibnamefont {Weedbrook},
  \bibfnamefont {C.}}, \bibinfo {author} {\bibfnamefont {S.}~\bibnamefont
  {Pirandola}}, \bibinfo {author} {\bibfnamefont {R.}~\bibnamefont
  {Garc{\'{i}}a-Patr{\'{o}}n}}, \bibinfo {author} {\bibfnamefont
  {N.}~\bibnamefont {Cerf}}, \bibinfo {author} {\bibfnamefont {T.}~\bibnamefont
  {Ralph}}, \bibinfo {author} {\bibfnamefont {J.}~\bibnamefont {Shapiro}}, \
  and\ \bibinfo {author} {\bibfnamefont {S.}~\bibnamefont {Lloyd}}} (\bibinfo
  {year} {2012}),\ \href {\doibase 10.1103/RevModPhys.84.621} {\bibfield
  {journal} {\bibinfo  {journal} {Review of Modern Physics}\ }\textbf {\bibinfo
  {volume} {84}}~(\bibinfo {number} {2}),\ \bibinfo {pages} {621}}\BibitemShut
  {NoStop}%
\bibitem [{\citenamefont {Wehner}\ \emph {et~al.}(2008)\citenamefont {Wehner},
  \citenamefont {Schaffner},\ and\ \citenamefont {Terhal}}]{wehner08}%
  \BibitemOpen
  \bibfield  {author} {\bibinfo {author} {\bibnamefont {Wehner}, \bibfnamefont
  {S.}}, \bibinfo {author} {\bibfnamefont {C.}~\bibnamefont {Schaffner}}, \
  and\ \bibinfo {author} {\bibfnamefont {B.~M.}\ \bibnamefont {Terhal}}}
  (\bibinfo {year} {2008}),\ \href {\doibase 10.1103/PhysRevLett.100.220502}
  {\bibfield  {journal} {\bibinfo  {journal} {Physical Review Letters}\
  }\textbf {\bibinfo {volume} {100}}~(\bibinfo {number} {22}),\ \bibinfo
  {pages} {220502}}\BibitemShut {NoStop}%
\bibitem [{\citenamefont {Wehner}\ and\ \citenamefont
  {Winter}(2008)}]{wehnerwinter08}%
  \BibitemOpen
  \bibfield  {author} {\bibinfo {author} {\bibnamefont {Wehner}, \bibfnamefont
  {S.}}, \ and\ \bibinfo {author} {\bibfnamefont {A.}~\bibnamefont {Winter}}}
  (\bibinfo {year} {2008}),\ \href {\doibase 10.1063/1.2943685} {\bibfield
  {journal} {\bibinfo  {journal} {Journal of Mathematical Physics}\ }\textbf
  {\bibinfo {volume} {49}}~(\bibinfo {number} {6}),\ \bibinfo {pages}
  {062105}}\BibitemShut {NoStop}%
\bibitem [{\citenamefont {Wehner}\ and\ \citenamefont
  {Winter}(2010)}]{wehner09}%
  \BibitemOpen
  \bibfield  {author} {\bibinfo {author} {\bibnamefont {Wehner}, \bibfnamefont
  {S.}}, \ and\ \bibinfo {author} {\bibfnamefont {A.}~\bibnamefont {Winter}}}
  (\bibinfo {year} {2010}),\ \href {\doibase 10.1088/1367-2630/12/2/025009}
  {\bibfield  {journal} {\bibinfo  {journal} {New Journal of Physics}\ }\textbf
  {\bibinfo {volume} {12}}~(\bibinfo {number} {2}),\ \bibinfo {pages}
  {025009}}\BibitemShut {NoStop}%
\bibitem [{\citenamefont {Wehrl}(1978)}]{wehrl78}%
  \BibitemOpen
  \bibfield  {author} {\bibinfo {author} {\bibnamefont {Wehrl}, \bibfnamefont
  {A.}}} (\bibinfo {year} {1978}),\ \href {\doibase 10.1103/RevModPhys.50.221}
  {\bibfield  {journal} {\bibinfo  {journal} {Review of Modern Physics}\
  }\textbf {\bibinfo {volume} {50}}~(\bibinfo {number} {2}),\ \bibinfo {pages}
  {221}}\BibitemShut {NoStop}%
\bibitem [{\citenamefont {Weyl}(1928)}]{weyl28}%
  \BibitemOpen
  \bibfield  {author} {\bibinfo {author} {\bibnamefont {Weyl}, \bibfnamefont
  {H.}}} (\bibinfo {year} {1928}),\ \href@noop {} {\emph {\bibinfo {title}
  {{Gruppentheorie und Quantenmechanik}}}}\ (\bibinfo  {publisher} {Hirzel},\
  \bibinfo {address} {Leipzig})\BibitemShut {NoStop}%
\bibitem [{\citenamefont {Wilde}\ \emph {et~al.}(2014)\citenamefont {Wilde},
  \citenamefont {Winter},\ and\ \citenamefont {Yang}}]{wilde13}%
  \BibitemOpen
  \bibfield  {author} {\bibinfo {author} {\bibnamefont {Wilde}, \bibfnamefont
  {M.~M.}}, \bibinfo {author} {\bibfnamefont {A.}~\bibnamefont {Winter}}, \
  and\ \bibinfo {author} {\bibfnamefont {D.}~\bibnamefont {Yang}}} (\bibinfo
  {year} {2014}),\ \href {\doibase 10.1007/s00220-014-2122-x} {\bibfield
  {journal} {\bibinfo  {journal} {Communications in Mathematical Physics}\
  }\textbf {\bibinfo {volume} {331}}~(\bibinfo {number} {2}),\ \bibinfo {pages}
  {593}}\BibitemShut {NoStop}%
\bibitem [{\citenamefont {Winter}\ and\ \citenamefont {Yang}(2016)}]{winter15}%
  \BibitemOpen
  \bibfield  {author} {\bibinfo {author} {\bibnamefont {Winter}, \bibfnamefont
  {A.}}, \ and\ \bibinfo {author} {\bibfnamefont {D.}~\bibnamefont {Yang}}}
  (\bibinfo {year} {2016}),\ \href {\doibase 10.1103/PhysRevLett.116.120404}
  {\bibfield  {journal} {\bibinfo  {journal} {Physical Review Letters}\
  }\textbf {\bibinfo {volume} {116}}~(\bibinfo {number} {12}),\ \bibinfo
  {pages} {120404}}\BibitemShut {NoStop}%
\bibitem [{\citenamefont {Wiseman}\ \emph {et~al.}(2007)\citenamefont
  {Wiseman}, \citenamefont {Jones},\ and\ \citenamefont {Doherty}}]{wiseman07}%
  \BibitemOpen
  \bibfield  {author} {\bibinfo {author} {\bibnamefont {Wiseman}, \bibfnamefont
  {H.~M.}}, \bibinfo {author} {\bibfnamefont {S.~J.}\ \bibnamefont {Jones}}, \
  and\ \bibinfo {author} {\bibfnamefont {A.~C.}\ \bibnamefont {Doherty}}}
  (\bibinfo {year} {2007}),\ \href {\doibase 10.1103/PhysRevLett.98.140402}
  {\bibfield  {journal} {\bibinfo  {journal} {Physical Review Letters}\
  }\textbf {\bibinfo {volume} {98}}~(\bibinfo {number} {14}),\ \bibinfo {pages}
  {140402}}\BibitemShut {NoStop}%
\bibitem [{\citenamefont {Wootters}\ and\ \citenamefont
  {Zurek}(1979)}]{wootters79}%
  \BibitemOpen
  \bibfield  {author} {\bibinfo {author} {\bibnamefont {Wootters},
  \bibfnamefont {W.}}, \ and\ \bibinfo {author} {\bibfnamefont {W.~H.}\
  \bibnamefont {Zurek}}} (\bibinfo {year} {1979}),\ \href {\doibase
  10.1103/PhysRevD.19.473} {\bibfield  {journal} {\bibinfo  {journal} {Physical
  Review D}\ }\textbf {\bibinfo {volume} {19}}~(\bibinfo {number} {2}),\
  \bibinfo {pages} {473}}\BibitemShut {NoStop}%
\bibitem [{\citenamefont {Wootters}\ and\ \citenamefont
  {Fields}(1989)}]{wootters89}%
  \BibitemOpen
  \bibfield  {author} {\bibinfo {author} {\bibnamefont {Wootters},
  \bibfnamefont {W.~K.}}, \ and\ \bibinfo {author} {\bibfnamefont {B.~D.}\
  \bibnamefont {Fields}}} (\bibinfo {year} {1989}),\ \href {\doibase
  10.1016/0003-4916(89)90322-9} {\bibfield  {journal} {\bibinfo  {journal}
  {Annals of Physics}\ }\textbf {\bibinfo {volume} {191}}~(\bibinfo {number}
  {2}),\ \bibinfo {pages} {363}}\BibitemShut {NoStop}%
\bibitem [{\citenamefont {Wootters}\ and\ \citenamefont
  {Sussman}(2007)}]{wooters07}%
  \BibitemOpen
  \bibfield  {author} {\bibinfo {author} {\bibnamefont {Wootters},
  \bibfnamefont {W.~K.}}, \ and\ \bibinfo {author} {\bibfnamefont {D.~M.}\
  \bibnamefont {Sussman}}} (\bibinfo {year} {2007}),\ in\ \href@noop {} {\emph
  {\bibinfo {booktitle} {Proc. QCMC 2007}}},\ p.\ \bibinfo {pages} {269},\
  \Eprint {http://arxiv.org/abs/0704.1277} {arXiv:0704.1277} \BibitemShut
  {NoStop}%
\bibitem [{\citenamefont {Wootters}\ and\ \citenamefont
  {Zurek}(1982)}]{wootters82}%
  \BibitemOpen
  \bibfield  {author} {\bibinfo {author} {\bibnamefont {Wootters},
  \bibfnamefont {W.~K.}}, \ and\ \bibinfo {author} {\bibfnamefont {W.~H.}\
  \bibnamefont {Zurek}}} (\bibinfo {year} {1982}),\ \href {\doibase
  10.1038/299802a0} {\bibfield  {journal} {\bibinfo  {journal} {Nature}\
  }\textbf {\bibinfo {volume} {299}}~(\bibinfo {number} {5886}),\ \bibinfo
  {pages} {802}}\BibitemShut {NoStop}%
\bibitem [{\citenamefont {Wu}\ \emph {et~al.}(2009)\citenamefont {Wu},
  \citenamefont {Yu},\ and\ \citenamefont {Molmer}}]{wu09}%
  \BibitemOpen
  \bibfield  {author} {\bibinfo {author} {\bibnamefont {Wu}, \bibfnamefont
  {S.}}, \bibinfo {author} {\bibfnamefont {S.}~\bibnamefont {Yu}}, \ and\
  \bibinfo {author} {\bibfnamefont {K.}~\bibnamefont {Molmer}}} (\bibinfo
  {year} {2009}),\ \href {\doibase 10.1103/PhysRevA.79.022104} {\bibfield
  {journal} {\bibinfo  {journal} {Physical Review A}\ }\textbf {\bibinfo
  {volume} {79}}~(\bibinfo {number} {2}),\ \bibinfo {pages}
  {022104}}\BibitemShut {NoStop}%
\bibitem [{\citenamefont {Zozor}\ \emph {et~al.}(2013)\citenamefont {Zozor},
  \citenamefont {Bosyk},\ and\ \citenamefont {Portesi}}]{zozor13}%
  \BibitemOpen
  \bibfield  {author} {\bibinfo {author} {\bibnamefont {Zozor}, \bibfnamefont
  {S.}}, \bibinfo {author} {\bibfnamefont {G.~M.}\ \bibnamefont {Bosyk}}, \
  and\ \bibinfo {author} {\bibfnamefont {M.}~\bibnamefont {Portesi}}} (\bibinfo
  {year} {2013}),\ \href {\doibase 10.1088/1751-8113/46/46/465301} {\bibfield
  {journal} {\bibinfo  {journal} {Journal of Physics A: Mathematical and
  Theoretical}\ }\textbf {\bibinfo {volume} {46}}~(\bibinfo {number} {46}),\
  \bibinfo {pages} {465301}}\BibitemShut {NoStop}%
\bibitem [{\citenamefont {Zozor}\ \emph {et~al.}(2014)\citenamefont {Zozor},
  \citenamefont {Bosyk},\ and\ \citenamefont {Portesi}}]{zozor14}%
  \BibitemOpen
  \bibfield  {author} {\bibinfo {author} {\bibnamefont {Zozor}, \bibfnamefont
  {S.}}, \bibinfo {author} {\bibfnamefont {G.~M.}\ \bibnamefont {Bosyk}}, \
  and\ \bibinfo {author} {\bibfnamefont {M.}~\bibnamefont {Portesi}}} (\bibinfo
  {year} {2014}),\ \href {\doibase 10.1088/1751-8113/47/49/495302} {\bibfield
  {journal} {\bibinfo  {journal} {Journal of Physics A: Mathematical and
  Theoretical}\ }\textbf {\bibinfo {volume} {47}}~(\bibinfo {number} {49}),\
  \bibinfo {pages} {495302}}\BibitemShut {NoStop}%
\bibitem [{\citenamefont {Zurek}(2003)}]{zurek03}%
  \BibitemOpen
  \bibfield  {author} {\bibinfo {author} {\bibnamefont {Zurek}, \bibfnamefont
  {W.~H.}}} (\bibinfo {year} {2003}),\ \href {\doibase
  10.1103/RevModPhys.75.715} {\bibfield  {journal} {\bibinfo  {journal} {Review
  of Modern Physics}\ }\textbf {\bibinfo {volume} {75}}~(\bibinfo {number}
  {3}),\ \bibinfo {pages} {715}}\BibitemShut {NoStop}%
\bibitem [{\citenamefont {{\.{Z}}yczkowski}\ and\ \citenamefont
  {Bengtsson}(2004)}]{zyczkowski04}%
  \BibitemOpen
  \bibfield  {author} {\bibinfo {author} {\bibnamefont {{\.{Z}}yczkowski},
  \bibfnamefont {K.}}, \ and\ \bibinfo {author} {\bibfnamefont
  {I.}~\bibnamefont {Bengtsson}}} (\bibinfo {year} {2004}),\ \href {\doibase
  10.1023/B:OPSY.0000024753.05661.c2} {\bibfield  {journal} {\bibinfo
  {journal} {Open Systems {\&} Information Dynamics}\ }\textbf {\bibinfo
  {volume} {11}}~(\bibinfo {number} {01}),\ \bibinfo {pages} {3}}\BibitemShut
  {NoStop}%
\end{thebibliography}
\end{document}